\documentclass[fleqn,usenatbib]{mnras}

\usepackage{newtxtext,newtxmath}
\usepackage[T1]{fontenc}

\DeclareRobustCommand{\VAN}[3]{#2}
\let\VANthebibliography\thebibliography
\def\thebibliography{\DeclareRobustCommand{\VAN}[3]{##3}\VANthebibliography}

\usepackage{graphicx}	% Including figure files
\usepackage{amsmath}	% Advanced maths commands
\usepackage{subcaption}
\usepackage{threeparttable}
\usepackage{longtable}
\usepackage{multirow}
\usepackage{wasysym}

\newcommand{\unsim}{\mathord{\sim}}

\newcommand{\msun}{M$_\odot$}
\newcommand{\rsun}{R$_\odot$}
\newcommand{\mjup}{M$_{\jupiter}$}
\newcommand{\rjup}{R$_{\jupiter}$}
\newcommand{\gcm}{$\rm g\,cm^{-3}$}
\newcommand{\kms}{$\mathrm{km}\,\mathrm{s}^{-1}$}
\newcommand{\ms}{$\mathrm{m}\,\mathrm{s}^{-1}$}

\newcommand{\alm}{\alpha_{\ell,m}}
\newcommand{\blm}{\beta_{\ell,m}}
\newcommand{\glm}{\gamma_{\ell,m}}

\newcommand{\benjamin}[1]{{\color{black}{{#1}}}}
\newcommand{\jf}[1]{{\color{black}{{#1}}}}
\newcommand{\ben}[1]{{\color{black}{{#1}}}}
\newcommand{\coaut}[1]{{\color{black}{{#1}}}}
\newcommand{\last}[1]{{\color{black}{{#1}}}}
\newcommand{\lastt}[1]{{\color{black}{{#1}}}}
\newcommand{\finalreview}[1]{{\color{black}{{#1}}}}

\title[Monitoring V1298 Tau with SPIRou]{Monitoring the young planet host V1298 Tau with SPIRou: planetary system and \coaut{evolving} large-scale magnetic field}

\author[B. Finociety et al.]{B. Finociety$^{1}$ \thanks{E-mail: benjamin.finociety@irap.omp.eu},
J.-F. Donati$^{1}$, P.I. Cristofari$^{2}$, C. Moutou$^{1}$, C. Cadieux$^{3}$, N.J. Cook$^{3}$, E. Artigau$^{3}$, \newauthor C. Baruteau$^{1}$, F. Debras$^{1}$, P. Fouqué$^{1}$, J. Bouvier$^{4}$, S.H.P Alencar$^{5}$, X. Delfosse$^{4}$, K. Grankin$^{6}$, \newauthor A. Carmona$^{4}$, P. Petit$^{1}$, Á. Kóspál$^{7}$ and the SLS/SPICE consortium
\\
$^{1}$ Université de Toulouse, CNRS, IRAP, 14 av. Belin, 31400 Toulouse, France \\
$^{2}$ Center for Astrophysics | Harvard \& Smithsonian, 60 Garden Street, MS-15, Cambridge, MA 02138, USA \\
$^{3}$ Université de Montréal, Département de Physique, IREX, Montréal, QC H3C 3J7, Canada \\
$^{4}$ Université Grenoble Alpes, CNRS, IPAG, F-38000 Grenoble, France \\
$^{5}$ Departamento de Fisica - ICEx - UFMG, Av. Antonio Carlos 6627, 30270-901 Belo Horizonte, MG, Brazil  \\
$^{6}$ Crimean Astrophysical Observatory, 298409 Nauchny, Republic of Crimea \\
$^{7}$ Konkoly Observatory, HUN-REN Research Centre for Astronomy and Earth Sciences, CSFK, MTA Centre of Excellence, Konkoly-Thege Mikl\'os \'ut 15-17,\\   1121 Budapest, Hungary
}

\date{Accepted 2023 October 2. Received 2023 September 26; in original form 2023 August 23}

\pubyear{2023}

\begin{document}
\label{firstpage}
\pagerange{\pageref{firstpage}--\pageref{lastpage}}
\maketitle

\begin{abstract}
We report results of a spectropolarimetric monitoring of the young Sun-like star V1298~Tau based on data collected with the near-infrared spectropolarimeter SPIRou at the Canada-France-Hawaii Telescope between late 2019 and early 2023. Using Zeeman-Doppler Imaging and the Time-dependent Imaging of Magnetic Stars methods on circularly polarized spectra, we reconstructed the large-scale magnetic topology of the star (and its temporal evolution), found to be mainly poloidal and axisymmetric with an average strength varying from 90 to 170~G over the $\unsim3.5$~years of monitoring. The magnetic field features a dipole whose strength evolves from 85 to 245~G, and whose inclination with respect to the stellar rotation axis remains stable until 2023 where we observe a sudden change, suggesting that the field may undergo a polarity reversal, potentially similar to those periodically experienced by the Sun. Our data suggest that the differential rotation shearing the surface of V1298~Tau is about 1.5 times stronger than that of the Sun. \coaut{When coupling our data with previous photometric results from \finalreview{K2 and TESS} and assuming circular orbits for all four planets, we report a $3.9\sigma$ detection of the radial velocity signature of the outermost planet (e), associated with a \jf{most probable} mass, density and orbital period of $M_e=0.95^{+0.33}_{-0.24}$~\mjup, $\rho_e=1.66^{+0.61}_{-0.48}$~\gcm\ and $P_e=53.0039\pm0.0001$~d, respectively.} %Regarding the 4-planet system and assuming circular orbits for all planets, we report a $3.9\sigma$ detection of the radial velocity signature of the outermost planet (e), associated with a \jf{most probable} mass, density and orbital period of $M_e=0.96^{+0.33}_{-0.25}$~\mjup, $\rho_e=1.70^{+0.63}_{-0.49}$~\gcm\ and $P_e=53.5\pm0.4$~d, respectively. When coupled with previous photometric results from TESS \citep{feinstein22}, we find that the most likely parameters for planet e are $M_e=0.90^{+0.31}_{-0.23}$~\mjup, $\rho_e=1.58^{+0.58}_{-0.46}$~\gcm\ and $P_e=53.0039\pm0.0001$~d.
For the 3 inner planets, we only derive 99\% confidence upper limits on their mass of $0.44$~\mjup, $0.22$~\mjup\ and $0.25$~\mjup, for b, c and d, respectively.  
\end{abstract}


%%%%%%%%%%%%%%%%%%%%%%%%%%%%%%%%%%%%%%%%%%%%%%%%%%

%%%%%%%%%%%%%%%%% BODY OF PAPER %%%%%%%%%%%%%%%%%%

\section{Introduction}

The detection and characterization of exoplanets around young Sun-like stars is key for deriving observational constraints on theoretical models of the early phases of star / planet formation. In particular, it is essential to estimate the radii and masses, and thereby the bulk densities, of young planets to refine the mass-radius relations for planets younger than 30 Myr \citep{mann16}. In addition, characterizing the orbital configuration of young close-in giants will bring new clues on the inward migration that giant planets are subject to as they form within the protoplanetary discs of their host stars, and thereby on the impact migrating giant planets have on the dynamical architecture of early planetary systems \citep{baruteau14}. However, only few such planets have yet been detected, partly due to the high activity levels that Pre-Main Sequence (PMS) stars exhibit, especially due to their short rotation period. Such active stars indeed generate photometric and radial velocity (RV) signatures that are much larger than those expected from transiting planets, even in the case of close-in giant ones. To date, only 9 planets younger than 25 Myr have well-measured radii thanks to the detection of photometric transits whose depths are typically 10 times smaller than the photometric variations caused by activity: the Super-Neptune K2-33~b \citep{david16}; the two warm Neptunes AU Mic~b and c \citep{plavchan20,martioli21,szabo22}; the four Jupiter-size planets V1298~Tau~b, c, d and e \citep{david19b,feinstein22}; the hot Jupiter HIP~67522~b \citep{rizzuto20}; and the warm Jupiter HD~114082~b \citep{zakhozhay22}. Constraining the mass of these planets is important to refine the mass-radius relations at such early stages of evolution but also to better characterize these young planets in terms of density and composition. However, detecting these planets through velocimetric monitoring is still challenging given that the amplitude of the RV fluctuations induced by the star itself can reach more than $\unsim10\times$ the amplitude of the planet signatures (e.g. \citealt{donati23}).

This stellar activity is strongly related to the magnetic field hosted by PMS stars. It is therefore essential to characterize the large-scale magnetic field of such stars to better understand the underlying dynamo processes at work in the stellar interior, but also to analyse star-planet interactions (e.g. \citealt{strugarek15}). In particular, Zeeman-Doppler Imaging (ZDI; \citealt{semel89, donati06}), an efficient tomographic method, already allowed one to reconstruct the large-scale magnetic topology of active PMS stars from time series of circularly polarized spectra (e.g. \citealt{donati16,donati23,yu17,yu19,finociety21,finociety23}). New tomographic methods, such as Time-dependent Imaging of Magnetic Stars (TIMeS; \citealt{finociety22}), are under development to take into account the temporal evolution of stellar magnetic fields, which would help to unveil the long-term evolution of the field and to identify whether these stars feature a magnetic cycle or not.

In this paper we focus on V1298~Tau, an early-K post T~Tauri star with no reported accretion \citep{david19a}, known to be the youngest solar-mass star with a multiplanetary system for which the radii of Jupiter-like planets have been precisely measured to date. V1298~Tau is located in the Taurus constellation at a distance of $108.0\pm0.2$~pc \citep{dr3}. \last{From Gaia measurements, the effective temperature and the logarithmic surface gravity are found to be equal to $T_{\rm eff} = 4941_{-16}^{+31}$~K and $\log{g} = 4.227_{-0.008}^{+0.010}$ \citep{dr3}, respectively, in agreement with previous values derived by \cite{suarez-mascareno21} and \cite{david19a}, of $5050\pm100$~K / $4.25\pm0.20$ and $4970\pm120$~K / $4.246\pm0.034$, respectively. \cite{suarez-mascareno21} also outlined that the effective temperature is} consistent with that inferred from the $JHK_S$ photometry obtained with the two Micron All-Sky Survey (2MASS) and from the Johnson $V$ photometry collected with the AAVSO Photometric All-Sky Survey (APASS) when assuming an extinction $E(B - V ) = 0.061$, \last{corresponding to an extinction $A_v=0.19\pm0.10$ following the standard extinction law. A previous independent estimate of the color excess from
the multicolor Maidanak observations yields consistent value [$E(V-R_j) = 0.06$ and $A_v =0.24$~mag; \citealt{grankin13}].} 

\begin{figure}
    \centering
    \includegraphics[scale=0.28,trim={0.5cm 2cm 0cm 2cm},clip]{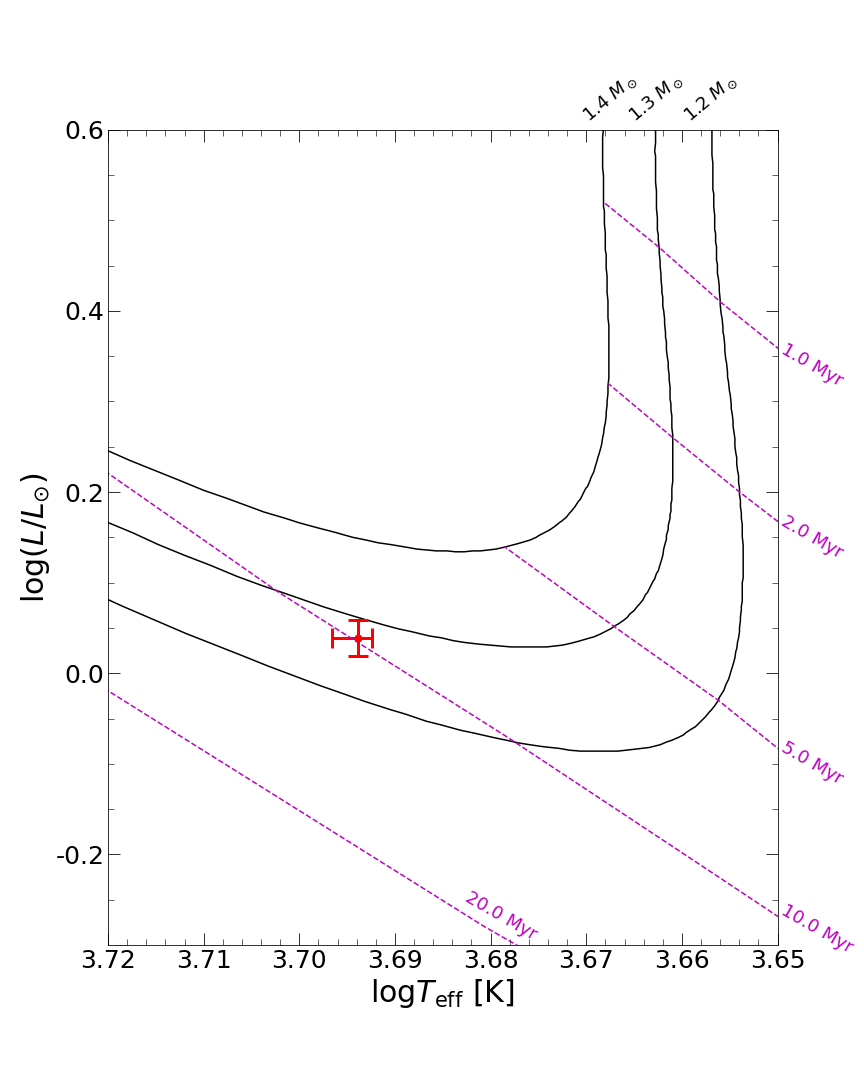}
    \caption{Location of V1298~Tau in the HR diagram, using the updated $T_{\rm eff}$ and $R_*$ from Table~\ref{tab:stellar_parameters}. The black lines correspond to evolutionary tracks from models of \citet{baraffe15} for the stellar mass indicated above the curves while isochrones are depicted as magenta dashed lines.}
    \label{fig:hr_diagram}
\end{figure}

%Using observed magnitudes and evolutionary models such as those of \cite{baraffe15}, \cite{suarez-mascareno21} found that the mass and radius of V1298~Tau are equal to $1.170\pm0.060$~\msun\ and $1.278\pm0.070$~\rsun, respectively. 
Given the stellar rotation period and the line-of-sight projected equatorial velocity we obtained (Sec.~\ref{sec:zdi}), one can infer a relatively accurate estimate of the stellar radius, equal to $R_*=1.43\pm0.03$~\rsun, assuming the stellar rotation axis is perpendicular to the line of sight. Using the Gaia estimate for the $\log{g}$, one therefore infers a stellar mass $M_*=1.26\pm0.06$~\msun. \lastt{From $T_{\rm eff}$ and $R_*$, we also estimate the luminosity relative to the Sun to be equal to $L_*/L_\odot=1.094\pm0.050$. Our updated values of the stellar parameters are summarized in Table~\ref{tab:stellar_parameters}. These estimates are consistent with those previously determined by \cite{suarez-mascareno21}, equal to $1.170\pm0.060$~\msun, $1.278\pm0.070$~\rsun\ and $0.954\pm0.040$, for the mass, radius and relative luminosity, respectively, from absolute magnitudes and evolutionary models. From its position in the temperature-luminosity diagram, using our updated stellar parameters, we find that the age of V1298~Tau is $\unsim10$--15~Myr (using the models of \citealt{siess2000} or \citealt{baraffe15}; Fig.~\ref{fig:hr_diagram}), making V1298~Tau younger than previously estimated (e.g. \citealt{david19a,suarez-mascareno21}). More recent studies focussing on the kinematics of V1298~Tau, based on Gaia DR3, suggest that the star may belong to the D2 \citep{krolikowksi21} or D3 \citep{gaidos22} subgroups of the Taurus star forming region, both being younger than 10 Myr. This comes as further argument in favour of V1298~Tau being younger than initially estimated. The evolutionary models of \cite{siess2000} and \cite{baraffe15} also predict that the depth of the convective envelope of V1298~Tau is about 40\% of the stellar radius, similar to that of AB~Dor ($M_*=1$~\msun) or LQ~Hya ($M_*=0.95$~\msun), two young solar-mass stars slightly older than V1298~Tau (40 - 50~Myr; \citealt{donati03c}).}

\begin{table}
    \centering
    \caption{Stellar properties used in our study.}
    \label{tab:stellar_parameters}
    \begin{tabular}{lll}
    \hline
         distance (pc) & $108.0\pm0.2$ & \citet{dr3} \\
         $T_{\rm eff}$ (K) & $4941^{+31}_{-16}$ & \citet{dr3}  \\
         %$T_{\rm eff}$ (K) & $5236\pm49$ & from SPIRou data  \\
         $\log g$ (dex) & $4.227^{+0.010}_{-0.008}$ &  \citet{dr3}\\
         %$\log g$ (dex) & $4.48 \pm 0.06$ &  from SPIRou data\\
         %[M/H] (dex) & $0.10\pm0.15$ &  \citet{suarez-mascareno21}\\
         %$\rm [M/H]$ (dex) & $0.38\pm0.03$& from SPIRou data \\
         %$M_*$ (\msun) & $1.10\pm0.02$ & from SPIRou data \\
         $R_*$ (\rsun) & \last{$1.43 \pm 0.03$} & \last{from $v\sin{i}$ and $P_{\rm rot}$} \\
         $M_*$ (\msun) & \last{$1.26\pm0.06$} & \last{from $R_*$ and $\log{g}$} \\
         $L/L_\odot$ & $1.094\pm0.05$ & from $T_{\rm eff}$ and $R_*$ \\
         %$R_*$ (\rsun) & $1.19\pm0.03$ & from SPIRou data \\
         $P_{\rm rot}$ (d) & 2.91 & period used to phase data \\
         $P_{\rm rot}$ (d) & $2.910 \pm 0.005$ & from $B_\ell$ \\
         $P_{\rm rot}$ (d) & $2.909 \pm 0.009$ &  from RVs \\
         $v\sin{i}$ ($\mathrm{km\,s^{-1}}$) & $24.9\pm0.5$ & from ZDI \\
         $\Omega_{\rm eq}$ ($\rm mrad\,d^{-1}$) & $2185.1 \pm 0.5$ & from ZDI \\
         d$\Omega$ ($\rm mrad\,d^{-1}$)& $82.0\pm2.0$ & derived from ZDI \\
         age (Myr) & $10\pm1$ & from of \citet{baraffe15} \\
        age (Myr) & $14\pm1$ & from of \citet{siess2000} \\
         
    \hline
    \end{tabular}

\end{table}

V1298~Tau is a target of great interest to study planetary systems at very early stages of their evolution. In the last few years, many studies therefore focused on this system to further characterize the planet properties. In particular, from transit events observed in the K2 and TESS light-curves, the orbital periods of the three innermost planets (b, c and d) were precisely measured \citep{david19b,feinstein22}. However, and with only 2 transits reported so far, the orbital period of planet e is less well determined, with \jf{17 possible values ranging from 42.7 to 62.8~d \citep{feinstein22}, 7 of them being more plausible, between 43.3 and 55.4~d \citep{sikora23}.} From velocimetric measurements, \cite{suarez-mascareno21} estimated the mass of planets b and e to be equal to $0.64\pm0.19$ and $1.16\pm0.30$~\mjup, respectively, while \cite{sikora23} found (i) a lower mass for planet e ($0.66\pm0.26$~\mjup), (ii) only an upper limit for planet b's mass of 0.50~\mjup\ and (iii) a low-significance $2.2\sigma$ \jf{candidate} detection of planet c with a mass of $0.062^{+0.029}_{-0.028}$~\mjup. A more recent study of \cite{turrini23}, based on the same RV dataset as in \cite{suarez-mascareno21}, reports a slightly larger mass for planet e, equal to $1.23^{+0.35}_{-0.48}$~\mjup, more consistent with the estimate of \cite{suarez-mascareno21}. In addition, the obliquity of the orbital axis of planets b and c with respect to the stellar rotation axis is found to be low, suggesting that these planets underwent a smooth migration inwards within the protoplanetary disc \citep{feinstein21,gaidos22}. V1298~Tau also offers the opportunity to study atmospheres of young planets, by measuring or constraining the atmospheric evaporation of the innermost planets resulting from the high activity of the host star \citep{poppenhaeger21,vissapragada21, maggio22}.

V1298~Tau has been monitored between late 2019 and early 2023 with the near-infrared (NIR) spectropolarimeter SPIRou in the framework of the SPIRou Legacy Survey (SLS), a large programme allocated 310 nights at \coaut{the} Canada-France-Hawaii Telescope (CFHT), within a PI programme, and most recently, within the SPICE large programme (174 nights \coaut{allocated} between late 2022 and mid 2024) at CFHT, aiming at consolidating and enhancing the results provided by the SLS. We start this paper with a detailed description of our spectropolarimetric data set (Sec.~\ref{sec:observations}). We then present the results concerning the large-scale magnetic field of \coaut{V1298~Tau}, reconstructed thanks to ZDI and TIMeS (Sec.~\ref{sec:magnetic_field}). In Sec.~\ref{sec:characterization_system}, we analyse our velocimetric measurements to constrain the mass of the planets hosted by V1298~Tau and the orbital period of planet e. In Sec.~\ref{sec:chromospheric_activity}, we focus on the stellar activity using three activity proxies in the NIR domain (the \ion{He}{i} triplet at 1083~nm and the Paschen $\beta$ and Brackett $\gamma$ lines). We finally summarize and discuss our results in Sec.~\ref{sec:summary}.

\section{Observations}
\label{sec:observations}
We extensively observed V1298 Tau with SPIRou between late 2019 and early 2023, collecting 181 high-resolution spectra ranging from 950 to 2500~nm at a spectral resolving power of $\unsim$70,000 \citep{donati20}. Altogether, 97 of these observations were collected in the framework of the SLS (2019 Oct 02 - 2021 Jan 08, 2022 Feb 01), more specifically within the WP3 package focusing on the magnetic topology of PMS stars, 52 within the PI Programme of Benjamin Finociety (run ID 21BF21, 2021 Sep 17 - 2022 Jan 30) and the 32 remaining \jf{ones} within the SPICE large programme (2022 Nov 02 - 2023 Feb 09). %More specifically, the observations are spread over four observing seasons: (i) 2019 Oct 02 - 2020 Feb 19 ($\unsim140$~d); (ii) 2020 Aug 26 - 2021 Jan 08 ($\unsim135$~d); (iii) 2021 Sep 17 - 2022 Feb 01 ($\unsim137$~d); and (iv) 2022 Nov 02 - 2023 Feb 09 ($\unsim99$~d).
\coaut{More specifically, the observations are spread over four observing seasons: (i) $\unsim140$~d from 2019 Oct to 2020 Feb, (ii) $\unsim135$~d from
2020 Aug to 2021 Jan, (iii) $\unsim137$~d from 2021 Sep to 2022 Feb; and (iv) $\unsim99$~d from 2022 Nov to 2023 Feb.}
Each observation is composed of a sequence of 4 subexposures of $\unsim500$~s taken at different azimuths of the polarimeter retarder in order to get rid of potential spurious signals in the polarization and systematic errors at first order \citep{donati97}. We show the journal of observations in Appendix~\ref{sec:log_journal}.

Data reduction was first carried out with a version of the \texttt{Libre-ESpRIT} pipeline, orginally developped for ESPADOnS data \citep{donati97}, adapted for SPIRou observations \citep{donati20}. In addition, we corrected the spectra from telluric lines using a PCA approach similar to the one developped in \cite{artigau14}, providing us 176 usable telluric-corrected spectra in both unpolarized (Stokes~$I$) and circularly polarized (Stokes~$V$) \coaut{observations}, leaving out 5 \coaut{spectra} whose quality is too low for our spectropolarimetric analysis \jf{(Table~\ref{tab:log_ZDI})}. The signal-to-noise ratio (SNR) per pixel in the $H$ band of \coaut{the usable} spectra range from 80 to 253 (median of 212). We then applied Least-Squares Deconvolution (LSD; \citealt{donati97}) on the 176 spectra, using a mask generated with the VALD-3 database \citep{vald}, featuring an effective temperature and a $\log g$ of 5000~K and 4.5, respectively, and containing moderate to strong atomic lines only (relative depth $>10$\% of the continuum). The resulting Stokes~$V$ LSD profiles show clear Zeeman signatures whose typical peak-to-peak amplitude is of 0.2\%, with SNRs ranging from \coaut{1400 to 5700 (median of 4770)}. \jf{As for the other similar young planet-hosting star AU~Mic \citep{donati23}, we} however do not observe clear distorsions of the Stokes~$I$ LSD profiles, suggesting that our spectroscopic data are only weakly affected by brightness features at the surface of the star, in agreement with the small amplitude of the photometric fluctuations in the TESS light curve (2021 Sep 16 - Nov 06) of the order of a few percent\footnote{The TESS light curve is publicly available from the Mikulski Archive for Space Telescopes (MAST) website.}. We will therefore only focus on the reconstruction of the large-scale magnetic field from our LSD profiles in the following (Sec.~\ref{sec:magnetic_field}).

We also processed the 181 observations with the nominal SPIRou pipeline, called \texttt{APERO} \citep{cook22}, \benjamin{much} better optimised for RVs than \texttt{Libre-ESpRIT}. Applying the line-by-line (LBL) method \citep{artigau22} on each reduced spectrum yielded 174 accurate RV values, with a \jf{typical error bar of 14~\ms (Table~\ref{tab:journal_apero}). Seven spectra were rejected (on 2019 Dec 08, 2019 Dec 13, 2020 Jan 26, 2020 Jan 27, 2020 Jan 28, 2020 Nov 06 and 2022 Feb 01)}, affected by instrumental effects and/or bad weather conditions, \benjamin{which prevented} us to obtain accurate \benjamin{RVs} at these dates.

\section{Magnetic field of V1298 Tau}
\label{sec:magnetic_field}

\subsection{Longitudinal field}
\label{sec:longitudinal_field}
\begin{figure*}
    \centering
    \begin{subfigure}{\textwidth}
         \centering
         
         \includegraphics[scale=0.32,trim={0cm 1cm 0cm 3cm},clip]{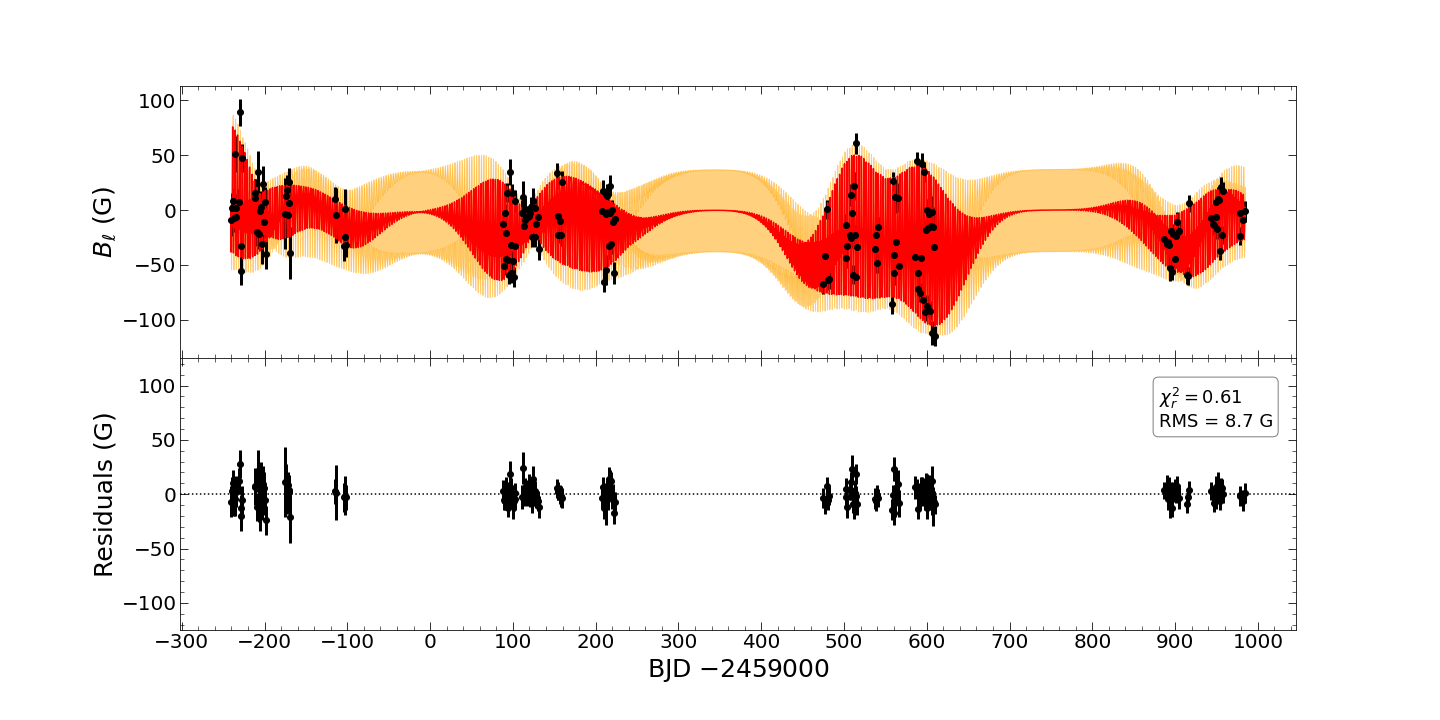}
         %\caption{}
    \end{subfigure}  

    \begin{subfigure}{\textwidth}
         \centering
         
         \includegraphics[scale=0.32,trim={0cm 1cm 0cm 2.cm},clip]{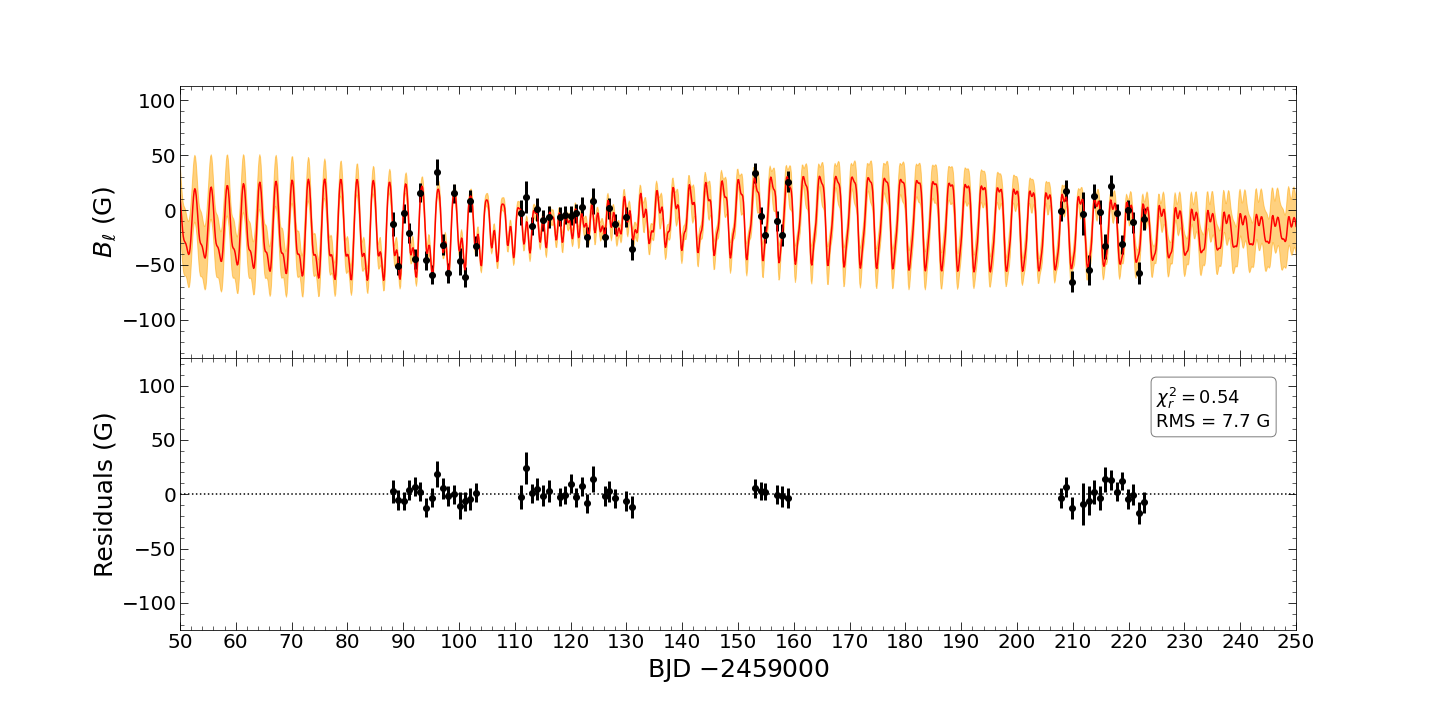}
         
    \end{subfigure}  

    \begin{subfigure}{\textwidth}
         \centering
         \includegraphics[scale=0.32,trim={0cm 1cm 0cm 2.cm},clip]{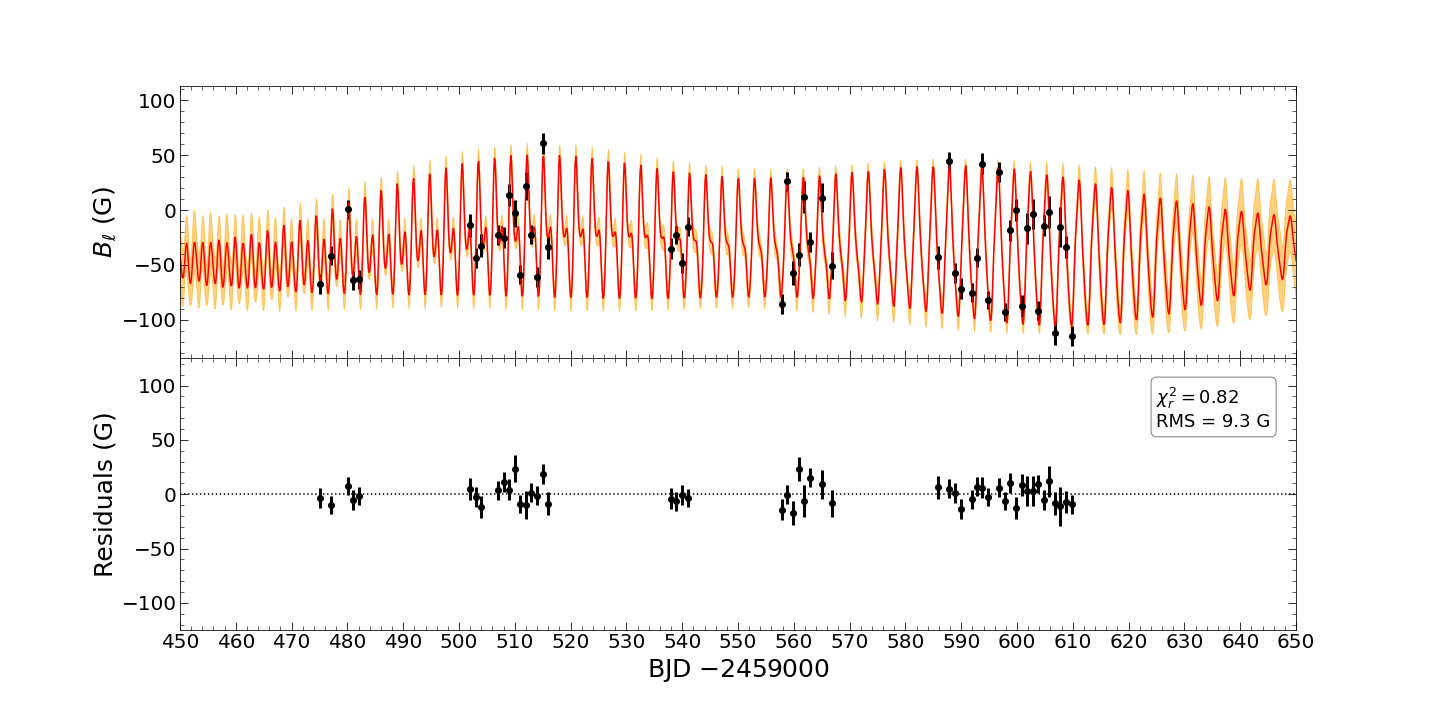}
    \end{subfigure}  
    \caption{Longitudinal field measurements of V1298~Tau \coaut{obtained} from SPIRou data \benjamin{over our observing period (top panel) with a zoom on the 2020 Aug - 2021 Jan (middle panel) and 2021 Sep - 2022 Feb (bottom panel) seasons.} \benjamin{For each panel, the top plot shows he raw measurements as black dots with their associated error bars while the red curve represents the quasi-periodic GP fit with the 1$\sigma$ confidence area in light orange. The bottom plots show the residuals between the raw measurements and the fit, \coaut{showing a RMS dispersion of 8.7~G, i.e. smallest than the typical error bar of 9.5~G, and a $\chi^2_r=0.61$.}  }}
    \label{fig:bl}
\end{figure*}

We computed the longitudinal field $B_\ell$, i.e. the line-of-sight-projected component of the vector magnetic field averaged over the visible hemisphere, as the first moment of our 176 Stokes~$V$ LSD profiles \citep{donati97}. \coaut{We find that $B_\ell$ ranges from $-115$ to 89~G between 2019 and 2023, with a typical uncertainty of 9.5~G.} \coaut{As we expect $B_\ell$ to be rotationally modulated, we modelled these values using a quasi-periodic (QP) Gaussian Process (GP; \citealt{rasmussen06}), following the definition given by \cite{rajpaul15}:

\begin{equation}
    k(t,t') = \theta_1^2 \, \exp\left[-\frac{(t-t')^2}{2\,\theta_2^2} - \frac{\sin^2\left( \frac{(t-t')\pi}{\theta_3}\right)}{2\,\theta_4^2}\right]
    \label{eq:QP_kernel}
\end{equation}

where $t$ and $t'$ correspond to the dates of two observations. $\theta_1$ is the amplitude of the GP, $\theta_2$ is the exponential decay time-scale (estimating the typical lifetime of the features generating the signal, e.g. spots), $\theta_3$ is the recurrence period (close to the stellar rotation period $P_{\rm rot}$) and $\theta_4$ is the smoothing parameter setting the amount of short-term variations included in the fit. We also added another parameter, called $\theta_5$, representing a potential excess of uncorrelated noise, not taken into account in our measured error bars (e.g. due to instrinsic variability). We then want to maximize the log likelihood function $\log \mathcal{L}$ defined in Eq.~\eqref{eq:log_likelihood} :

\begin{equation}
    \log \mathcal{L} = -\frac{1}{2}\left(N_0 \log{2\pi} + \log{|\boldsymbol{\mathsf{K}}+\boldsymbol{{\Sigma}}+\boldsymbol{\mathsf{S}}| + \boldsymbol{y}^T(\boldsymbol{\mathsf{K}}+\boldsymbol{{\Sigma}}+\boldsymbol{\mathsf{S}})^{-1}\boldsymbol{y}}  \right) 
    \label{eq:log_likelihood}
\end{equation}

where $\boldsymbol{y}$ is the vector of the measured $B_\ell$, whose length is equal to $N_0$ (i.e. to the number of observations). $\boldsymbol{\mathsf{K}}$ corresponds to the covariance matrix associated with the quasi-periodic kernel, $\boldsymbol{{\Sigma}}$ represents a diagonal matrix containing the variance of our measurements and $\boldsymbol{\mathsf{S}}=\theta_5^2\,\boldsymbol{\mathsf{I}}$ with $\boldsymbol{\mathsf{I}}$ being the identity matrix.

We used a MCMC approach, thanks to the \textsc{emcee python} module \citep{emcee}, to explore the parameter domains and sample the posterior distribution of our 5 parameters, based on 20000 iterations of 150 walkers. We then estimated the optimal set of parameters (maximizing $\log \mathcal{L}$) from these posterior distributions (see Table~\ref{tab:bl_parameters}), after removing a burn-in period of 5000 iterations. 

We find that the recurrence period is well constrained ($\theta_3=2.910\pm0.005$~d) and consistent with the estimate of the stellar rotation period provided by \cite{suarez-mascareno21}. $B_\ell$ data are fitted down to $\chi^2_r=0.61$, the RMS of the residuals being smaller than the typical error bar (see Fig.~\ref{fig:bl}), suggesting that these data are not significantly affected by intrinsic variability as demonstrated by a value of $\theta_5$ smaller than the typical uncertainty and compatible with 0 within $\unsim2\sigma$. }

As expected by the low value of the decay time-scale, close to one month ($\theta_2=36\pm5$~d, i.e. about 4 times smaller than a typical observing season), we observe a clear evolution of the longitudinal field between \benjamin{late} 2019 and early 2023, and within each season. In particular, we note that the full amplitude of the curve has a minimum in 2020 ($\unsim30$~G) and a maximum in 2021 ($\unsim130$~G). This evolution already indicates that we need to split the dataset of each season into several subsets to apply ZDI that assumes a static configuration of the magnetic topology (Sec.~\ref{sec:zdi}).

Running GP regression on individual observing season, we find that the decay time-scale is much less constrained and equal to \coaut{$68^{+70}_{-35}$~d, $56^{+22}_{-16}$~d, $102^{+59}_{-38}$~d and $41^{+15}_{-11}$~d during 2019 Oct - 2020 Feb, 2020 Aug - 2021 Jan, 2021 Sep - 2022 Feb and 2022 Nov - 2023 Feb}, respectively, reflecting a slower evolution of the magnetic field during the 2021-2022 observing season.

\begin{table}
    \centering
    \caption{\jf{Optimal value obtained for the GP modeling of the longitudinal field $B_\ell$ from a MCMC approach. The knee of the modified Jeffreys priors is set to the typical uncertainty on the the longitudinal field, noted $\sigma_{B_\ell}$.}}
    \label{tab:bl_parameters}
    %\resizebox{0.49\textwidth}{!}{
    \begin{tabular}{lll}
    \hline
         Hyperparameter   & Prior & Optimal value \\ \hline
         GP amplitude (G), $\theta_1$  & mod Jeffreys ($\sigma_{B_\ell}$) & $37\pm5$ \\
         Decay time-scale (d), $\theta_2$ & Uniform (0, 200) & $36\pm5$ \\
         Recurrence period (d), $\theta_3$ & Gaussian (2.91, 0.1) & $2.910\pm0.005$ \\
         Smoothing, $\theta_4$ & Uniform (0,3) & $0.47\pm0.06$ \\
         Uncorrelated noise (G), $\theta_5$ & mod Jeffreys ($\sigma_{B_\ell}$) & $3.2^{+3.4}_{-1.7}$ \\ \hline
    
    \end{tabular}
    %}
\end{table}

We note that we were not able to measure the surface field of V1298~Tau, unlike what has been done for the other young star AU~Mic, given the large line-of-sight projected equatorial velocity ($v\sin{i}$) value ($\unsim25$~\kms, see Sec.~\ref{sec:zdi}) that prevented us from obtaining reliable measurements from rotationally broadened unpolarized line profiles.

\subsection{Zeeman-Doppler Imaging}
\label{sec:zdi}
In order to reconstruct the large-scale magnetic field of V1298~Tau, we used ZDI \citep{semel89,brown91,donatibrown97,donati06}, a tomographic method allowing one to invert Stokes~$I$ and $V$ LSD profiles into a large-scale magnetic topology at the surface of an active star. As this problem is ill-posed (infinite number of topologies compatible with the data \benjamin{at a given $\chi^2_r$ level}), ZDI uses the principle of maximum entropy to \benjamin{select} the simplest solution, i.e. \benjamin{the one} with the minimal amount of information needed to fit the data down to \benjamin{the requested $\chi^2_r$ level}, by iteratively adding magnetic features at the surface of the star and comparing the associated LSD profiles with the observed ones.

In practice, we divide the stellar surface into a grid of 3000 cells, initially with no magnetic field. At each iteration, the local Stokes~$I$ and/or $V$ LSD profiles are computed from each cell using the Unno-Rachkovsky’s solution of the polarized radiative transfer equations in a plane-parallel Milne Eddington atmosphere (e.g. \citealt{landi04}), and then integrated over the visible stellar hemisphere, providing the synthetic LSD profiles corresponding to the reconstructed image. As mentioned in \cite{finociety23}, we replaced the built-in prescription for the limb-darkening law by a linear relation for the continuum only, choosing a limb-darkening coefficient $\epsilon=0.3$ compatible with $T_{\rm eff}=5000$~K and $\log g = 4.5$ in the $H$ band \citep{claret11}. 

The large-scale magnetic field is described by the sum of a poloidal and a toroidal component, both described as spherical harmonic expansions using the formalism of \cite{donati06}\footnote{As mentioned in \cite{finociety22}, the original formalism was slightly modified to provide a more consistent description of the magnetic field.}. In particular, the poloidal component is characterized by sets of complex coefficients noted $\alpha_{\ell,m}$ and $\beta_{\ell,m}$ while the toroidal component is fully described by a set of complex coefficients $\gamma_{\ell,m}$, where $\ell$ and $m$ denote the degree and order of the spherical harmonic mode in the expansion. We limited our study to the spherical harmonic modes up to a degree $\ell_{\max}=10$ as adding higher degree modes does not significantly improve the reconstruction.

ZDI also allows one to reconstruct the distribution of dark and bright features at the surface of active stars from distorsions of the Stokes~$I$ LSD profiles. \coaut{However, we find that, for V1298~Tau, brightness inhomogeneities have only a minor impact on our data (and our reconstructed magnetic maps) as only very weak variations, due to spots and plages, are observed in our Stokes~$I$ LSD profiles}.

\last{We usually assume that the star rotates a solid body in a first step, then include surface differential rotation (DR) in the reconstruction as described in Sec.~\ref{sec:dr}. All results shown below were obtained taking into account the mean DR we estimated for V1298~Tau, following the process detailed in Sec.~\ref{sec:dr}.}

Before applying this method to our dataset, we first computed the rotation cycles $c$ of our observations using the following ephemeris:

\begin{equation}
    \mathrm{BJD\;(d)} = 2459500 + P_{\rm rot}\,c 
\end{equation}

where the initial date ($\rm BJD_0 =2459500$) was arbitrarily chosen and $P_{\rm rot}=2.91$~d, consistent with our estimates from $B_\ell$ and RVs (see Sec.~\ref{sec:characterization_system}) and that of \cite{suarez-mascareno21}. Given the short decay time-scale of magnetic features, we were forced to split our dataset into 10 subsets, gathering between 12 and 22 LSD profiles spread over a maximum of 40~d (see Table~\ref{tab:log_ZDI} for the definition of the subsets\footnote{The temporal distribution of the observations was too sparse at some epochs, preventing us to use 16 out of the 149 LSD profiles (e.g. isolated observations in 2020 Nov).}). We then applied ZDI on each of these subsets, assuming a line model featuring a mean wavelength, Doppler width and Landé factor of 1750~nm, 3.4~\kms and 1.2, respectively. \jf{Assuming that the orbital axis of the transiting planets coincides with the stellar rotation axis $i$}, we should have $i$ close to 90\degr.  However, we set $i=80^\circ$ in ZDI to reduce the degeneracy between the northern and southern hemispheres in the reconstruction process. \finalreview{Using the magnetic topology of 2020 Aug-Sep (see Fig.~\ref{fig:zdi_maps}), we simulated dataset with the same phase coverage, SNRs and $v\sin{i}$ for an inclination of $i=80^\circ$ and $90^\circ$. Applying ZDI with $i=80^\circ$ for both datasets shows that the impact on the reconstruction of the 10$^\circ$ difference for $i$ is minimal with respect to the original distribution, and helps to remove part of the north/south degeneracy, therefore justifying our choice. We derive $v\sin{i}=24.9\pm0.5$~\kms\ as part of the imaging process.} %All physical parameters of V1298 Tau used in this study are shown in Table~\ref{tab:stellar_parameters}.

For each of the 10 subsets, both Stokes~$I$ and $V$ LSD profiles were fitted down to a unit $\chi^2_r$; we show the reconstructed magnetic topology in Fig.~\ref{fig:zdi_maps_2019} (for 2019) and Fig.~\ref{fig:zdi_maps} (for 2020-2023) while the brightness maps are shown in Fig~\ref{fig:zdi_Q}. As expected from the weak distorsions of the Stokes~$I$ LSD profiles, the brightness maps show only a few low contrasted features, associated with a low spot coverage of about 3\%, with an \last{episodic spot showing up at some epochs in the the polar region of the northern hemisphere.} We note that the SNRs of the Stokes~$V$ LSD profiles associated with \benjamin{the} 2019 SPIRou data are about twice smaller than \benjamin{those of} the other epochs (partly due to a lower exposure time), \benjamin{thereby} reducing our ability to clearly detect Zeeman signatures, especially in 2019 Nov-Dec (see Fig.~\ref{fig:ZDI_LSD}). \benjamin{It suggests} that the associated reconstructed images are less \benjamin{reliable than those at} later epochs. In the following, we will therefore only \benjamin{discuss} magnetic reconstructions \benjamin{derived} from data collected between 2020 Aug and 2023 Feb.

\begin{figure*}
    \centering
    \centering
    %\hspace*{-0.4cm}
    \begin{subfigure}{\textwidth}
         \centering
         \includegraphics[scale=0.15,trim={0cm 0cm 0cm 0cm},clip]{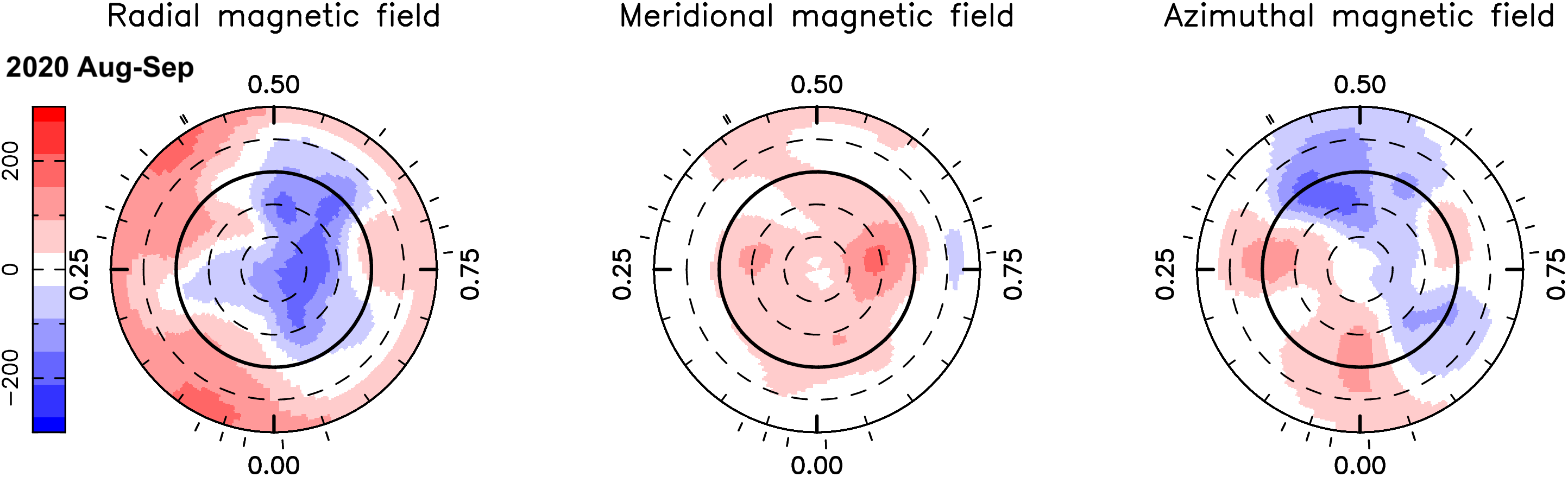}
         
    \end{subfigure}
    \hfill
    %\hspace*{-0.4cm}
    \begin{subfigure}{\textwidth}
         \centering
         \includegraphics[scale=0.15,trim={0cm 0cm 0cm 2.3cm},clip]{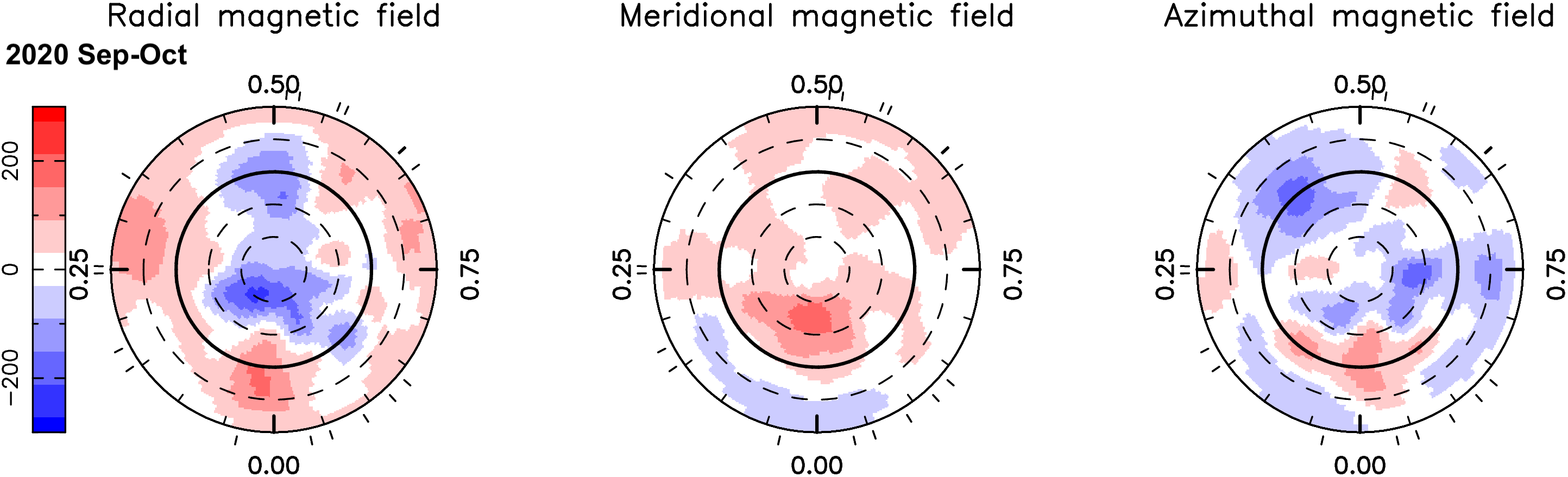}
         
    \end{subfigure}
    \hfill
    %\hspace*{-0.4cm}
    \begin{subfigure}{\textwidth}
         \centering
         \includegraphics[scale=0.15,trim={0cm 0cm 0cm 2.3cm},clip]{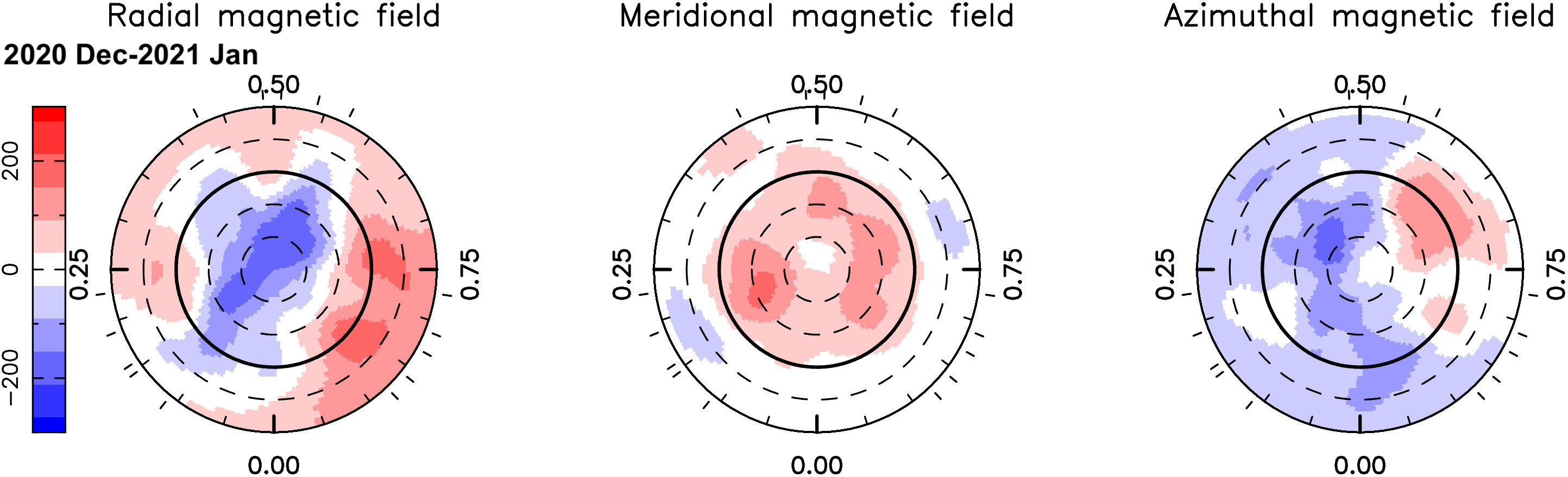}
         
    \end{subfigure}
    \begin{subfigure}{\textwidth}
         \centering
         \includegraphics[scale=0.15,trim={0cm 0cm 0cm 2.3cm},clip]{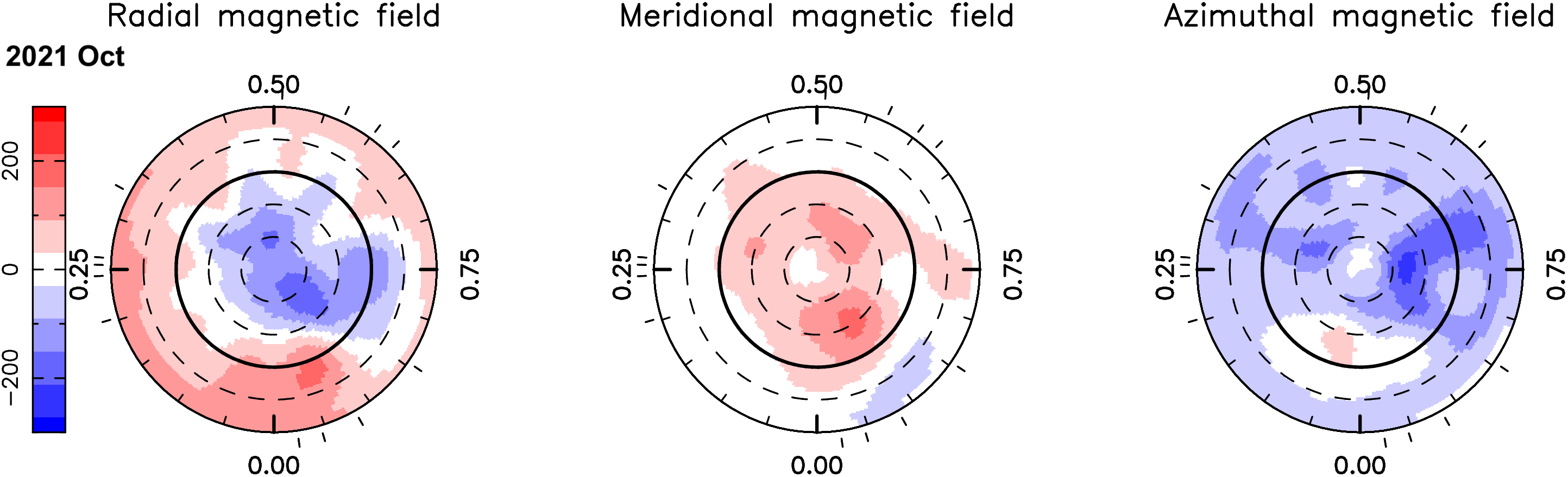}
         
    \end{subfigure}
    \begin{subfigure}{\textwidth}
         \centering
         \includegraphics[scale=0.15,trim={0cm 0cm 0cm 2.3cm},clip]{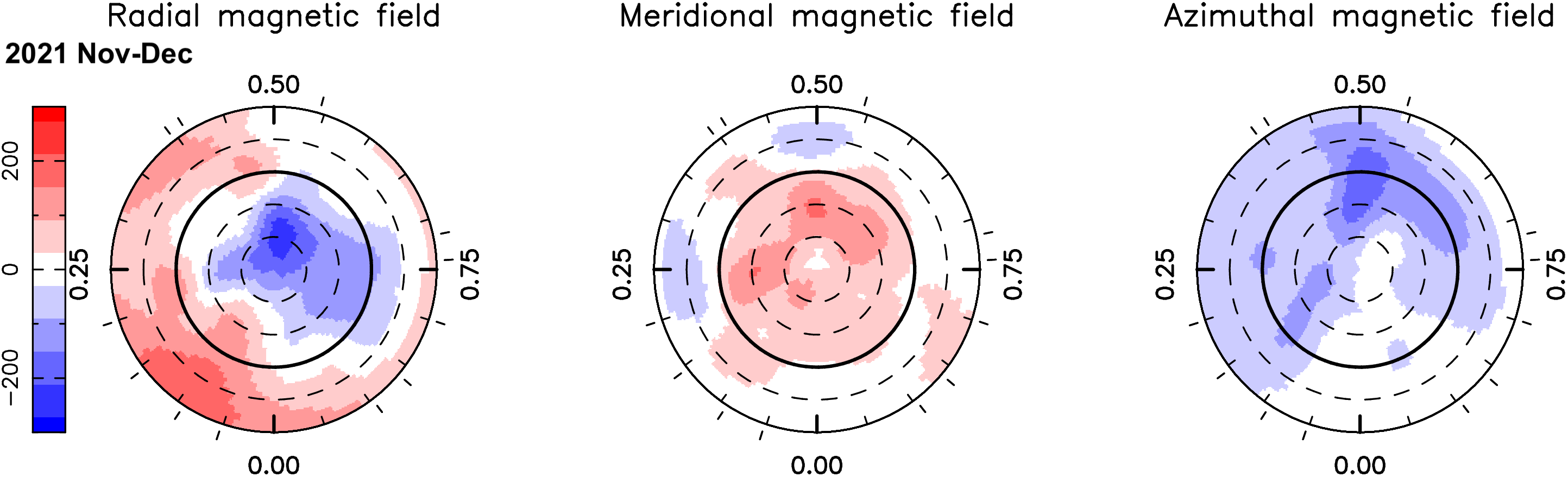}
         
    \end{subfigure}
    
    \caption{ZDI maps of the radial (left), meridional (middle) and azimuthal (right) magnetic field components for the 8 subsets associated with the following epochs: 2020 Aug-Sep, 2020 Sep-Oct 2020, 2020 Dec - 2021 Jan, 2021 Oct, 2021 Nov-Dec, 2022 Jan, 2022 Nov and 2023 Jan. The star is shown in a flattened polar view down to $-60^\circ$, with the pole at the center, the equator depicted as a bold circle and the \benjamin{$-30^\circ$, 30$^\circ$ and $60^\circ$ latitude parallels} represented by dashed circles. Positive radial, meridional and azimuthal fields are shown in red and point outwards, polewards and counter-clockwise, respectively. The ticks around the star refer to the phases of the SPIRou spectropolarimetric observations used in ZDI. These maps were obtained \last{assuming} the mean differential rotation derived in Sec.~\ref{sec:dr} and are shown at mid-time throughout the corresponding subset.}
    \label{fig:zdi_maps}
\end{figure*}

\begin{figure*} \ContinuedFloat
    \centering
    \centering
    %\hspace*{-0.4cm}
    \begin{subfigure}{\textwidth}
         \centering
         \includegraphics[scale=0.15,trim={0cm 0cm 0cm 0cm},clip]{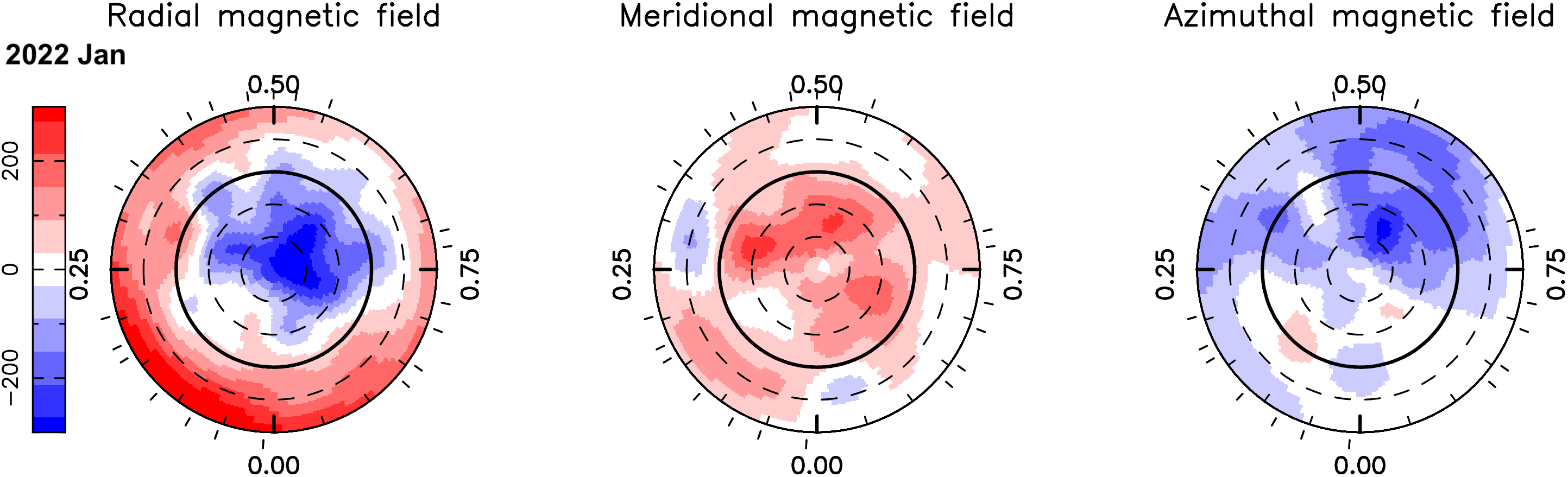}
         
    \end{subfigure}
    \hfill
    %\hspace*{-0.4cm}
    \begin{subfigure}{\textwidth}
         \centering
         \includegraphics[scale=0.15,trim={0cm 0cm 0cm 2.3cm},clip]{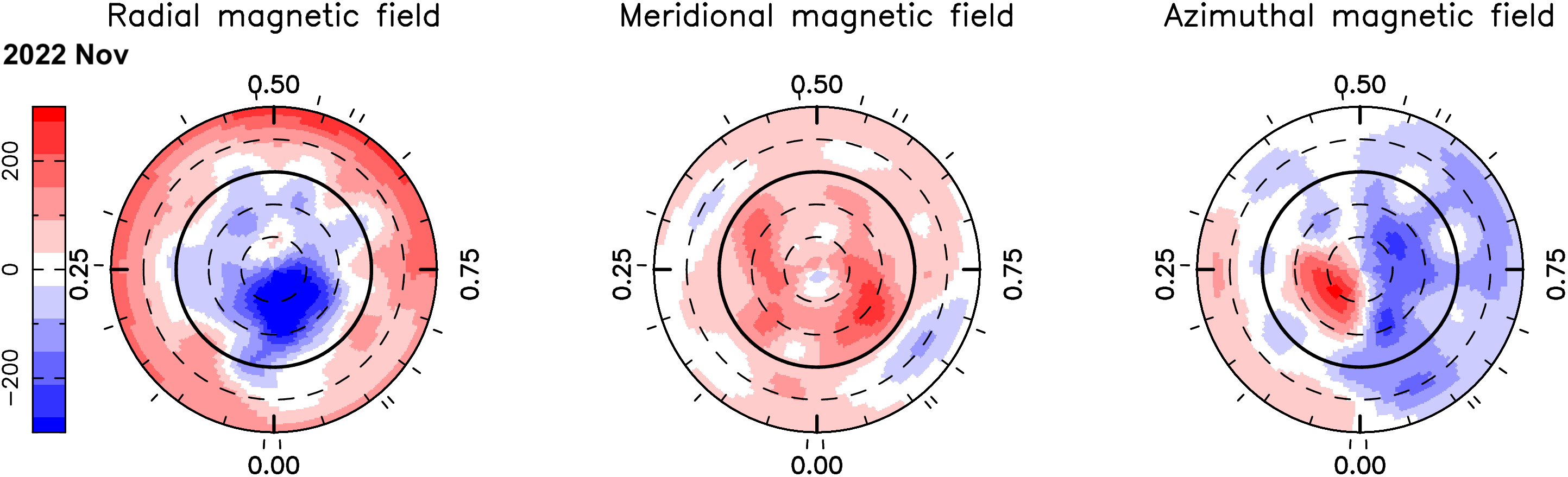}
         
    \end{subfigure}
    \hfill
    %\hspace*{-0.4cm}
    \begin{subfigure}{\textwidth}
         \centering
         \includegraphics[scale=0.15,trim={0cm 0cm 0cm 2.3cm},clip]{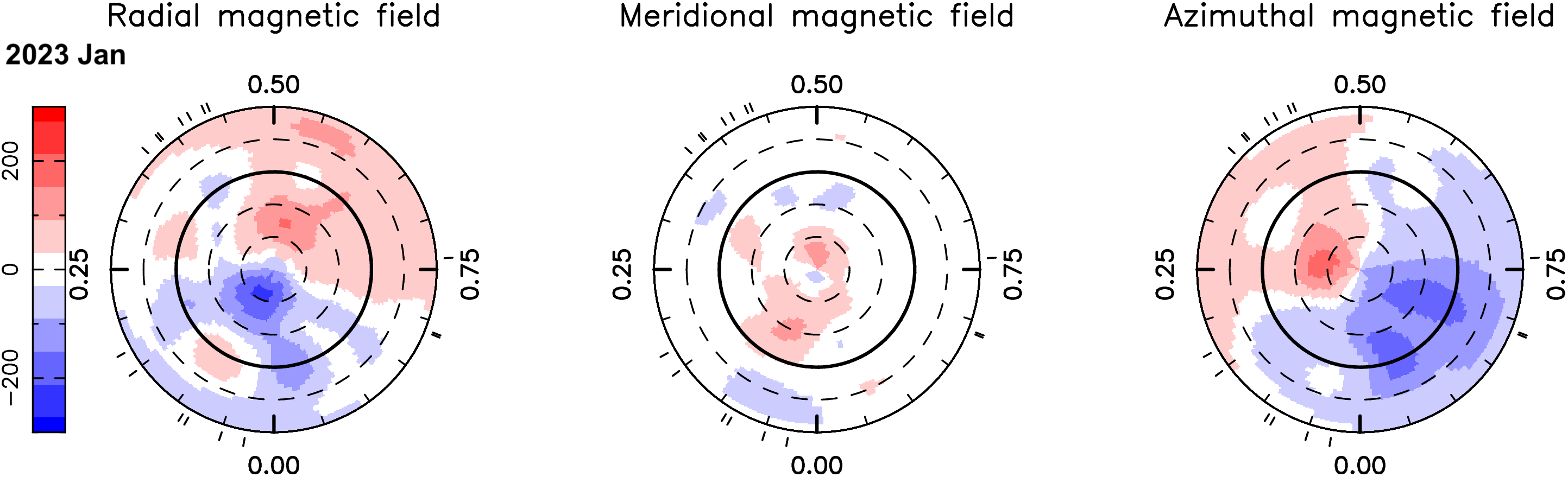}
         
    \end{subfigure}

    \caption{Continued.}
    %\label{fig:my_label}
\end{figure*}

Despite a rapid evolution of the longitudinal field, we find that the magnetic topology of this star remains mainly poloidal, this component enclosing between \coaut{65}\% (2021 Oct) and \coaut{90}\% (2020 Aug-Sep) of the reconstructed magnetic energy, more than \coaut{45\% of which being concentrated in axisymmetric modes}, until 2023 Jan (see Table~\ref{tab:magnetic_properties}). This field component mainly consists of a dipole, whose inclination with respect to the rotation axis is low ($<35^\circ$) until 2023 Jan and whose strength ranges from \coaut{85 to 245~G}. In 2023 Jan, we see a change in the topology, especially for the radial component of the field, with the dipole component of the poloidal field being now tilted at \coaut{75$^\circ$}, potentially suggesting the beginning of a polarity reversal. To investigate this point further and find out whether this evolution is real or rather caused by, e.g., limited data sampling, we carried out the following experiment. Using the ZDI map of 2022 Nov, we simulated a dataset with the same phase coverage and SNR as that of our 2023 Jan data, and applied ZDI to these simulated data. As the reconstructed large-scale magnetic topology is close to the input field, we can conclude that the strong tilt of the magnetic topology that we report from out 2023 Jan data is likely real.

The large-scale magnetic topology of V1298~Tau also features a significant toroidal field, enclosing between \coaut{10}\% (2020 Aug-Sep) and \coaut{35}\% (2021 Oct) of the reconstructed magnetic energy, and found to be mainly axisymmetric and dipolar. We also find that the quadratically-averaged large-scale magnetic flux over the stellar surface, noted <$B_V$>, \coaut{varies by $\unsim80$\% between 2020 Aug and 2022 Nov (from 100 to 180~G).}

\begin{table}
	\centering
	\caption{Properties of the large-scale magnetic field of V1298 Tau between 2020 and 2023, from ZDI reconstruction \coaut{(including differential rotation)} of 8 datasets. Columns 1--2 show the number of the dataset and the associated epoch. Columns 3 to 5 list the quadratically-averaged large-scale magnetic flux over the stellar surface, the strength and tilt, with respect to the stellar rotation axis, of the dipole (of the poloidal component). The two last columns gather the fraction of reconstructed magnetic energy enclosed by the poloidal/toroidal components of the field and in the axisymmetric modes of these components. The typical error bars on these values are of 5-10\%.}
	\label{tab:magnetic_properties}
    \resizebox{0.49\textwidth}{!}{
	\begin{tabular}{llcccccc}
    \hline
         Dataset & Epoch & <$B_V$> & $B_d$ & Tilt & Poloidal & Toroidal  \\
         & & (G) & (G) & ($^\circ$) & (\%) \\ \hline
         \#1 & 2019 Oct & 70 & 30 & 80 & 85 / 5 & 15 / 20 \\
         \#2 & 2019 Nov-Dec & 40 & 15 & 25 & 90 / 10 & 10 / 15 \\
         \#3 & 2020 Aug-Sep & 120 & 145 & 30 & 90 / 50 & 10 / 15 \\
         \#4 & 2020 Sep-Oct & 100 & 90 & 15 & 80 / 45 & 20 / 55 \\
         \#5 & 2020 Dec - 2021 Jan& 125 & 130 & 30 & 85 / 55 & 15 / 70 \\
         \#6 & 2021 Oct & 125 & 125 & 25 & 65 / 65 & 35 / 85 \\
         \#7 & 2021 Nov-Dec & 120 & 145 & 35 & 75 / 60 & 25 / 90 \\
         \#8 & 2022 Jan & 180 & 245 & 30 & 85 / 70 & 20 / 75 \\
         \#9 & 2022 Nov & 165 & 210 & 15 & 90 / 65 & 10 / 70 \\
         \#10 & 2023 Jan & 100 & 85 & 75 & 80 / 10 & 20 / 55 \\
    \hline
    \end{tabular}
    }
\end{table}

\begin{figure*}
    \centering
    \includegraphics[scale=0.5,trim={0cm 0cm 0cm 0cm},clip]{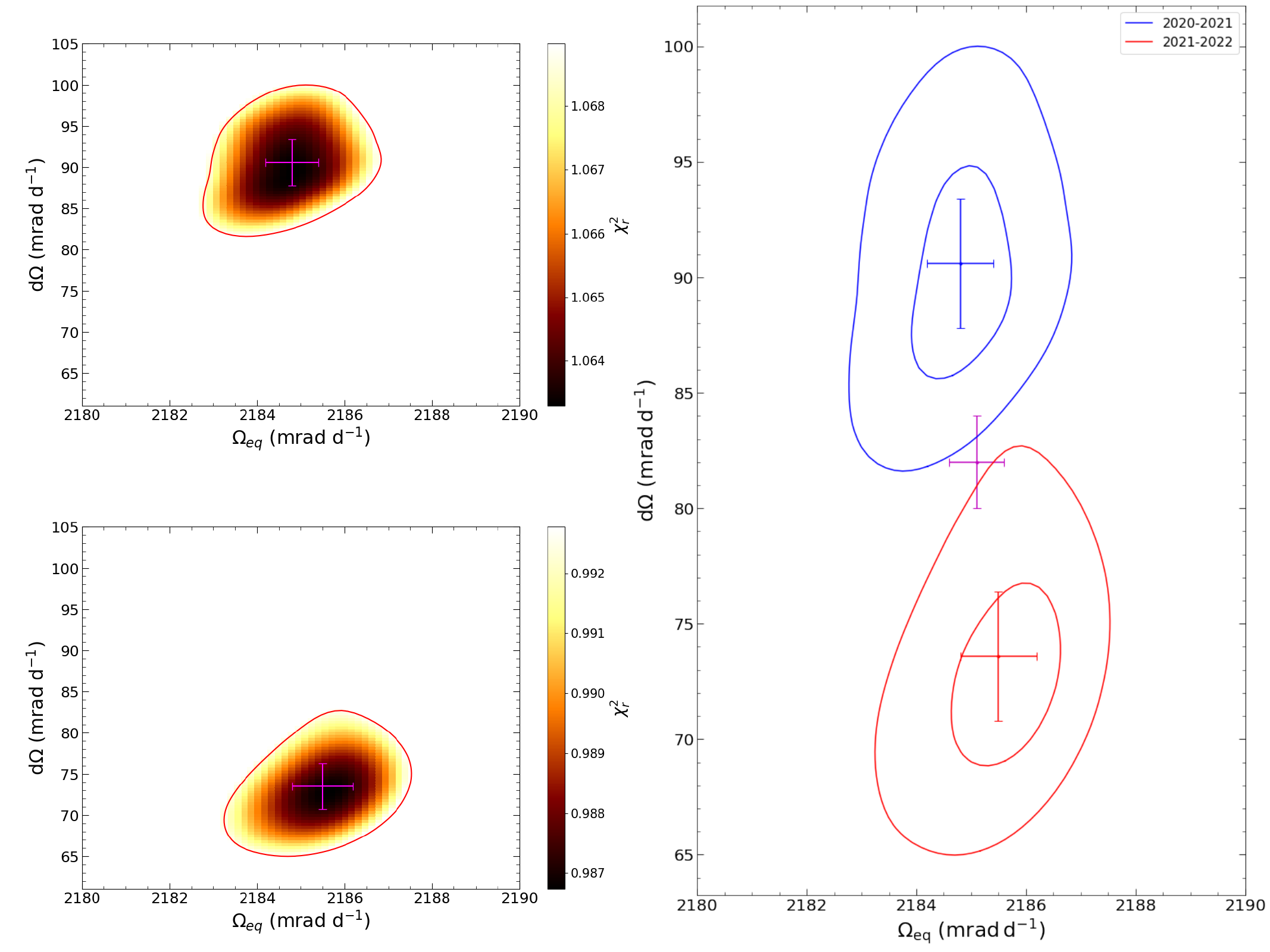}
    
    \caption{Surface differential rotation of V1298~Tau as measured from SPIRou data \benjamin{during the 2020 Aug - 2021 Jan (upper left) and 2021 Sep - 2022 Feb (bottom left) seasons}. $\chi^2_r$ maps computed from Stokes~$V$ LSD profiles over a grid of parameters $\Omega_{\rm eq}$, the rotation rate at the equator, and d$\Omega$ the difference in rotation rate between the pole and the equator. Red \benjamin{contours} show the $3\sigma$ confidence \benjamin{interval}, while the magenta crosses depict the optimal value with its associated error bars, derived from the fit of a 2D paraboloid close to the minimum of the grid. The right panel shows the contours associated with $1\sigma$ and $3\sigma$ confidence \benjamin{intervals} obtained from both observing seasons (\benjamin{2020 Aug - 2021 Jan: blue; 2021 Sep - 2022 Feb: red}). The magenta cross depicts the mean parameters.  }
    \label{fig:dr}
\end{figure*}

\subsection{Differential rotation}
\label{sec:dr}
From ZDI reconstructions, we see that our data show some variability within each observing season. Part of this variability can potentially be due to surface differential rotation. One can use ZDI to estimate DR by assuming that the surface shear follows a solar-like law given by:

\begin{equation}
    \Omega(\theta) = \Omega_{\rm eq} - \left(\cos{\theta}\right)^2\,\mathrm{d}\Omega
\end{equation}

where $\Omega_{\rm eq}$ and d$\Omega$ correspond to the parameters of the DR law, i.e. the rotation rate at the equator and the pole-to-equator rotation rate difference, respectively, and $\theta$ represents the colatitude.

As our subsets cover up to $\unsim40$~d, we need to merge some of them to diagnose subtle temporal variations of the Stokes~$V$ LSD profiles under the effect of DR. We thus estimated DR during the 2020 Aug - 2021 Jan season from subsets \#3 to \#5 (\coaut{49 Stokes~$V$ profiles spanning 135~d}, i.e. $\unsim46$ rotation cycles), and during the 2021 Sep - 2022 Feb season from subsets \#6 to \#8 (\coaut{47 observations spanning 109~d, i.e. $\unsim37$ rotation cycles}). The Zeeman signatures collected between 2019 Oct and Dec are \jf{both too noisy and too sparse} to provide us with reliable estimates of DR, while mediocre phase coverage in 2023 and the sudden evolution of the overall topology between the first half and the second half of the 2022 Nov - 2023 Feb season prevents us to \jf{estimate} DR at \jf{this time}.   

In practice, we fitted these \jf{two} subsets of Stokes~$V$ LSD profiles at a given amount of information (i.e. magnetic energy), over a grid of DR parameters. This process yielded $\chi^2_r$ maps from which we estimated the optimal values for both $\Omega_{\rm eq}$ and d$\Omega$ and their associated error bars by adjusting a 2D paraboloid close to the minimum \citep{donati2000,petit02,donati03,finociety21,finociety23}.

\coaut{From Stokes~$V$ LSD profiles collected during the 2020 Aug - 2021 Jan season (subsets \#3 to \#5), we find $\Omega_{\rm eq}=2184.8 \pm 0.6$~$\rm mrad\,d^{-1}$ and d$\Omega=90.6\pm2.8$~$\rm mrad\,d^{-1}$. This implies that the rotation period ranges from $2.876 \pm 0.001$~d at the equator to $3.000\pm0.004$~d at the pole. For the 2021 Sep - 2022 Feb season (subsets \#6 to \#8), we find a weaker level of DR with $\Omega_{\rm eq}=2185.5 \pm 0.7$~$\rm mrad\,d^{-1}$ and d$\Omega=73.5\pm2.8$~$\rm mrad\,d^{-1}$, corresponding to a period at the equator and at the pole of $2.875\pm0.001$ and $2.975\pm0.004$~d, respectively. 

While both estimates of $\Omega_{\rm eq}$ are consistent, estimates of d$\Omega$ differ \coaut{by} more than $3\sigma$, suggesting that the DR may vary at the surface of V1298~Tau between the two consecutive observing seasons (see Fig.~\ref{fig:dr}). In addition, for both observing seasons, the confidence \benjamin{contours} are slightly distorted, reflecting that the $\chi^2_r$ maps are not perfect 2D paraboloids, especially when \benjamin{moving} away from the minimum of the maps, which likely increases the size of the error bars.}

The higher level of DR measured from \benjamin{2020 Aug - 2021 Jan} data is consistent with the evolution of the longitudinal field. \benjamin{We indeed} estimate a lower decay time-scale of $B_\ell$ during this observing season, indicating a faster evolution of the \benjamin{large-scale field that may partly reflect a stronger DR}.

We chose a unique set of parameters to describe the DR at the surface of V1298~Tau, computed as the weighted means of the estimates provided by both considered observing seasons, \coaut{i.e. $\Omega_{\rm eq}=2185.1 \pm 0.5$~$\rm mrad\,d^{-1}$ and d$\Omega=82.0\pm2.0$~$\rm mrad\,d^{-1}$. These values imply that the equator of the star rotates with a period of $2.875 \pm 0.001$~d while the pole rotates in $2.988\pm0.003$~d, the equator lapping the pole by one cycle in $\unsim76$~d, i.e. about half an observing season. Using these values, we find that both our $B_\ell$ (Sec.~\ref{sec:longitudinal_field}) and RV data (Sec.~\ref{sec:planet_masses}), with a period of $2.910\pm0.005$ and $2.909\pm0.009$~d, respectively, probe the same latitudes of about 30$^\circ$ (between 31--37$^\circ$ and 28--39$^\circ$, respectively).}

Splitting each data sets in two subsets, each covering about half the observing season, and carrying the same analysis yield discrepant DR parameters, with error bars up to 3 times larger than \benjamin{those} mentioned above, demonstrating that a large number of Stokes~$V$ profiles collected over the whole observing season are needed to reliably estimate the DR parameters.

\subsection{Time-dependent Imaging of Magnetic Stars}

As our data are spread over several years, we also applied the new \coaut{tomographic} method named Time-dependent Imaging of Magnetic Stars (TIMeS) \benjamin{outlined} in \cite{finociety22}. As for ZDI, this method aims at finding the simplest large-scale magnetic topology consistent with the data (i.e. Stokes~$V$ LSD profiles) but this time allowing it to evolve with time. TIMeS uses sparse approximations to identify as few spherical harmonic modes (associated with $\alm$, $\blm$ and $\glm$ coefficients) as possible to reconstruct the magnetic topology and GPs to model the time-dependence of each identified coefficients. This method can only be applied in the cases where the magnetic field is not too strong, typically with a magnetic flux lower than $\unsim3$~kG for a $v\sin{i}=25$~\kms. Given the $v\sin{i}$ of V1298~Tau and the values of <$B_V$> \coaut{and of the local field} inferred from the ZDI reconstructions in the previous sections ($\ll3$~kG), we therefore assume that TIMeS can be applied to our data.

In practice, thanks to the principle of sparsity, we first look for the simplest linear combination of known magnetic topologies that \benjamin{provides a satisfactory fit to subsets} of $n$ consecutive profiles, assuming that the \benjamin{large-scale} magnetic field does not significantly evolve over the time span covered by the $n$ profiles. We did not \benjamin{consider} the subsets for which 2 consecutive profiles are separated by more than twice the decay time-scale of the longitudinal field. As mentioned in \cite{finociety22}, the $n$ profiles should sample the rotation cycle while \benjamin{typically} covering 10 to 20\% of the decay time-scale of the longitudinal field, explaining why we set $n=6$ in this paper (corresponding to $\unsim20$\% of the $B_\ell$ decay time-scale). This first step yields the spherical harmonic modes that significantly contribute to the overall topology over the whole dataset, with a value of the corresponding $\alm$, $\blm$ and $\glm$ coefficients for each subset. TIMeS then models these time series of coefficients using GPs with a squared exponential kernel, allowing one to compute the magnetic maps and associated Stokes~$V$ LSD profiles for each observed rotation cycle.

\begin{figure*} %\ContinuedFloat
    \centering
    
    \begin{subfigure}{\textwidth}
         \centering
         \includegraphics[scale=0.15,trim={0cm 0cm 0cm 0cm},clip]{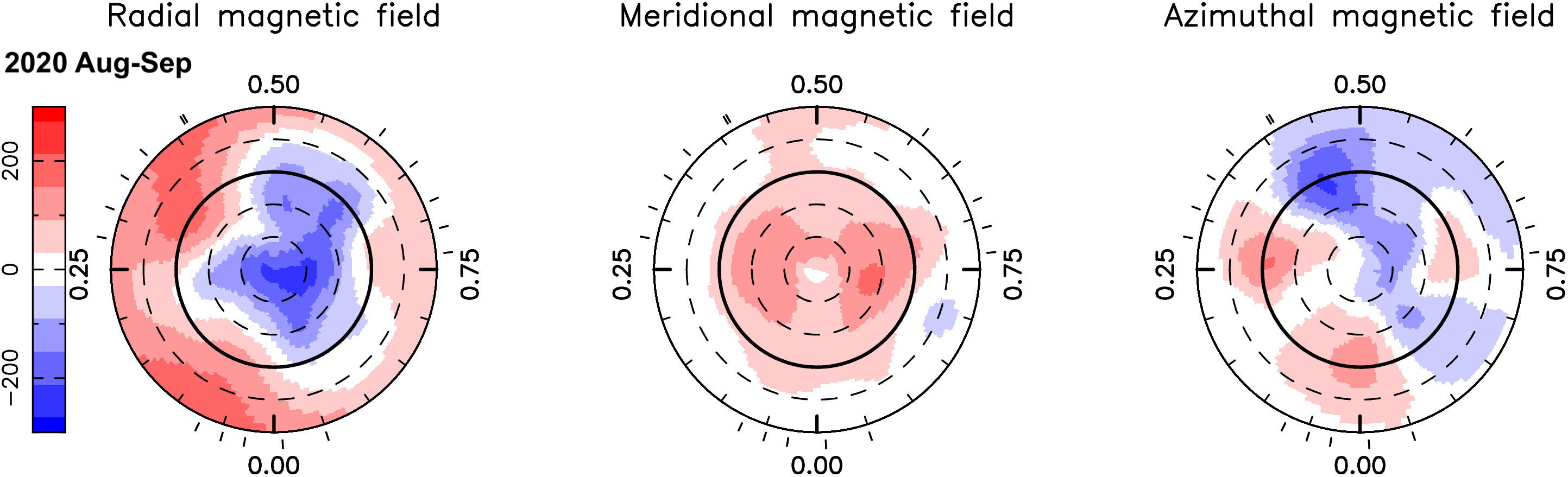}
         
    \end{subfigure}
    \hfill
    %\hspace*{-0.4cm}
    \begin{subfigure}{\textwidth}
         \centering
         \includegraphics[scale=0.15,trim={0cm 0cm 0cm 2.3cm},clip]{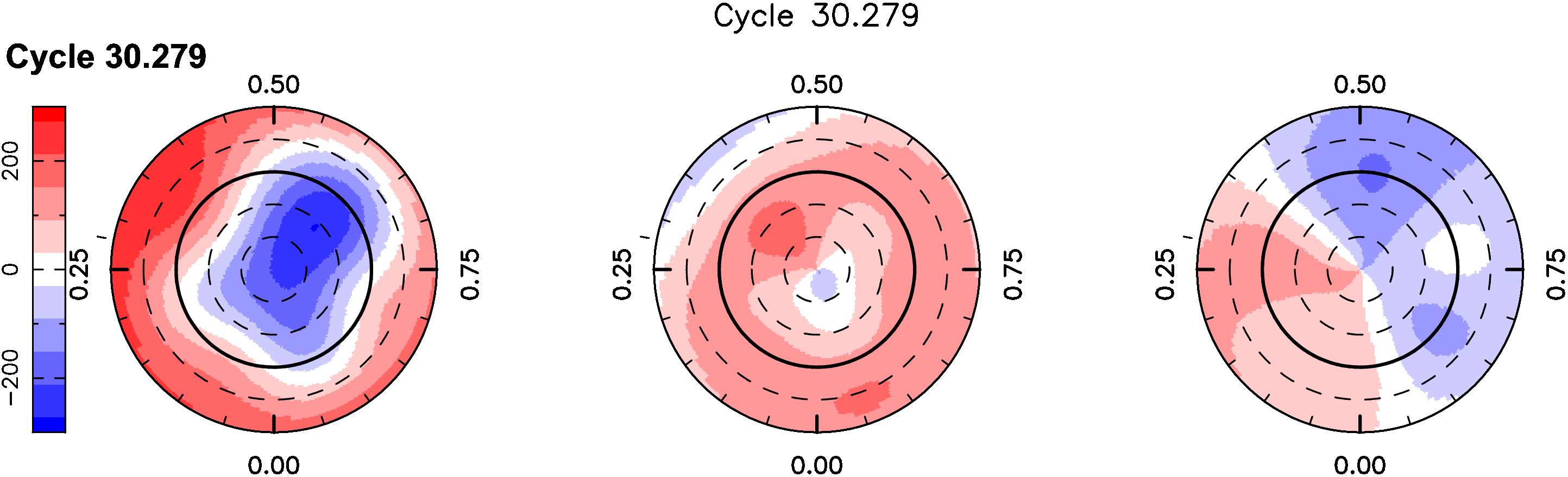}
         
    \end{subfigure}
    \hfill
    %\hspace*{-0.4cm}
    \begin{subfigure}{\textwidth}
         \centering
         \includegraphics[scale=0.15,trim={0cm 0cm 0cm 2.5cm},clip]{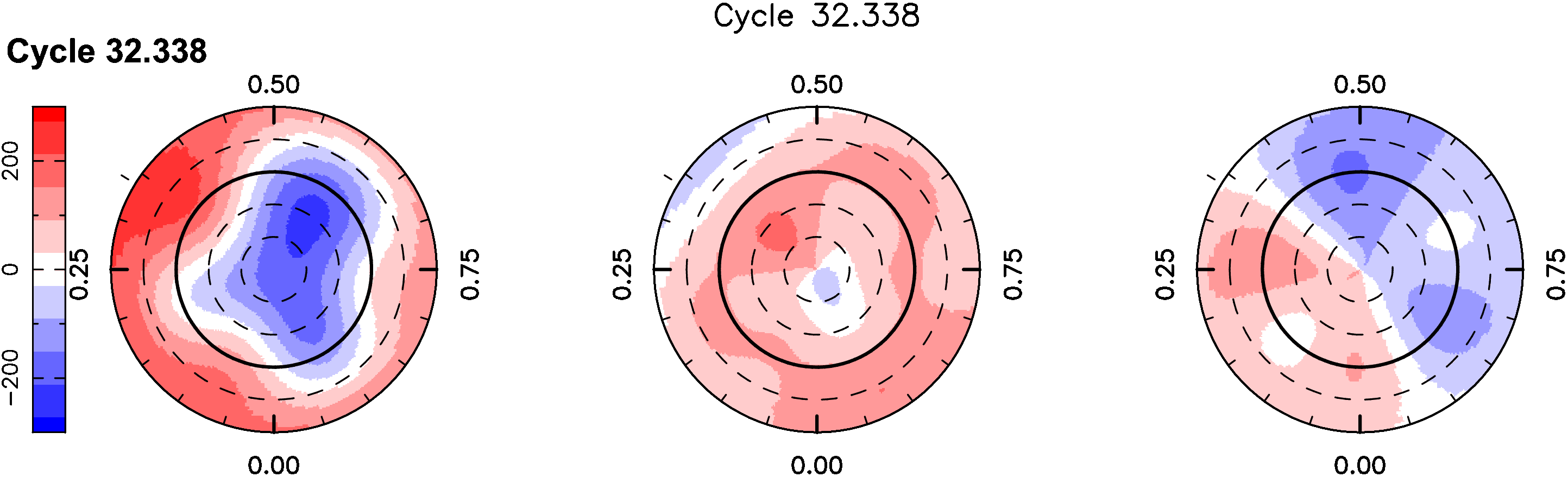}
         
    \end{subfigure} \hfill

    \begin{subfigure}{\textwidth}
         \centering
         \includegraphics[scale=0.15,trim={0cm 0cm 0cm 2.3cm},clip]{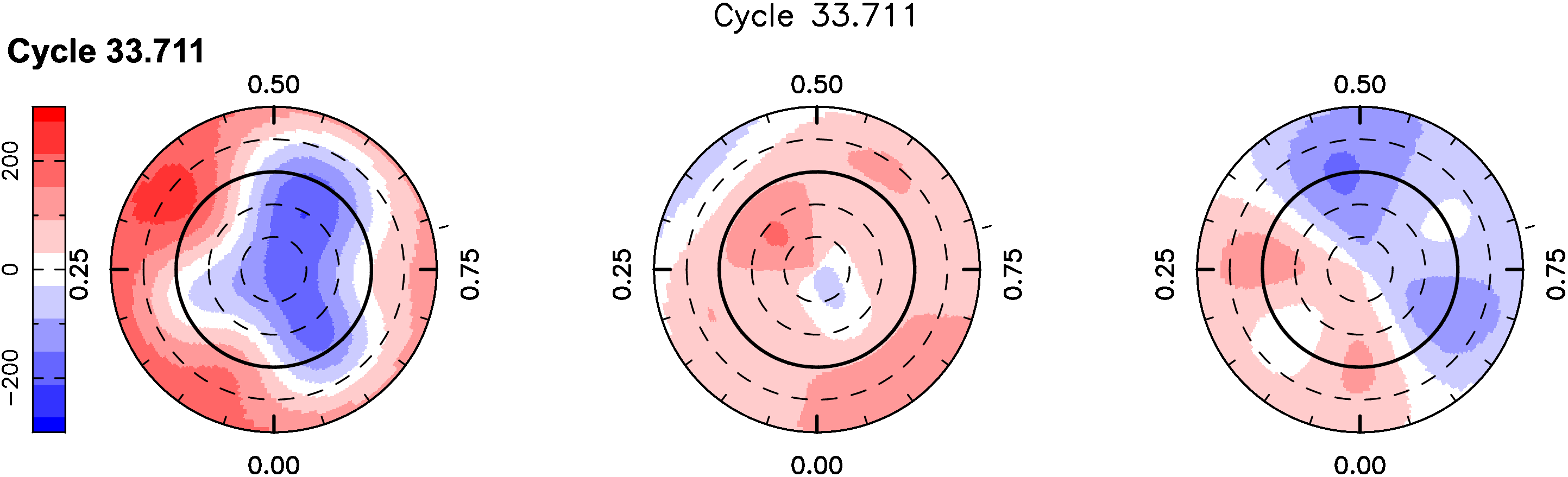}
         
    \end{subfigure}
    \hfill
    \begin{subfigure}{\textwidth}
         \centering
         \includegraphics[scale=0.15,trim={0cm 0cm 0cm 2.3cm},clip]{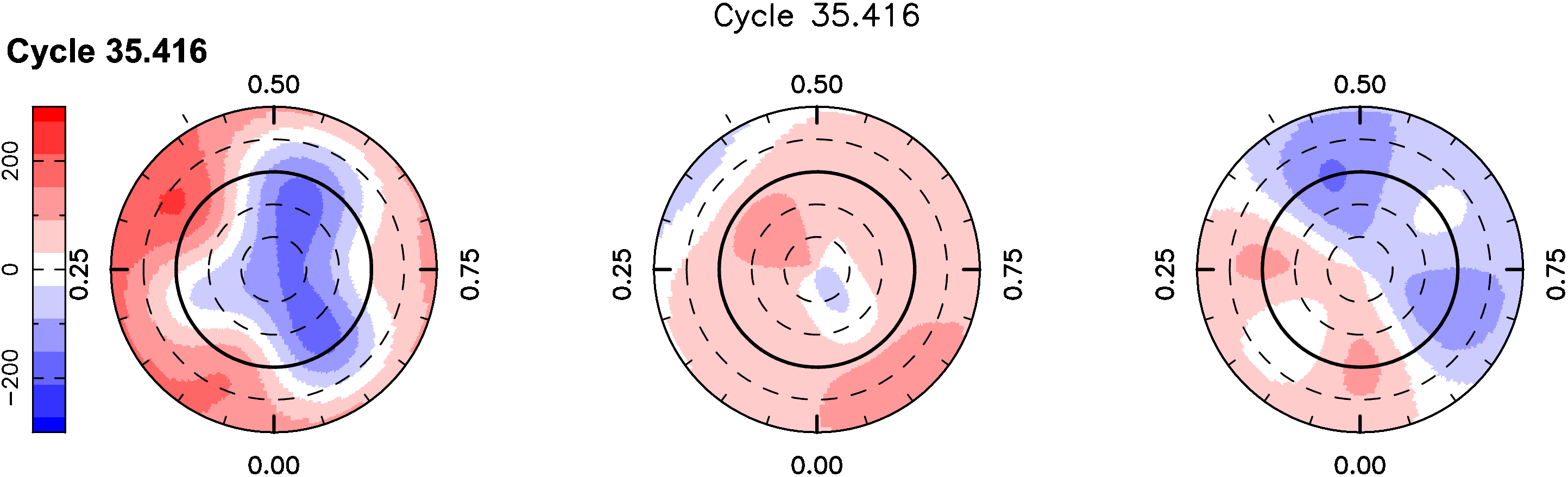}
         
    \end{subfigure}
    
    \caption{Magnetic maps in 2020 Aug-Sep reconstructed with ZDI without differential rotation (first row) and TIMeS (second \coaut{to fifth} rows). As TIMeS allows the reconstruction of a map at each observed date, we only show \coaut{four} specific epochs indicated on the left. See Fig.~\ref{fig:zdi_maps} for a detailed description of the Figure.}
    \label{fig:map_TIMeS_aug20}
\end{figure*}

\begin{figure*} %\ContinuedFloat
    \centering
    
    \begin{subfigure}{\textwidth}
         \centering
         \includegraphics[scale=0.15,trim={0cm 0cm 0cm 0cm},clip]{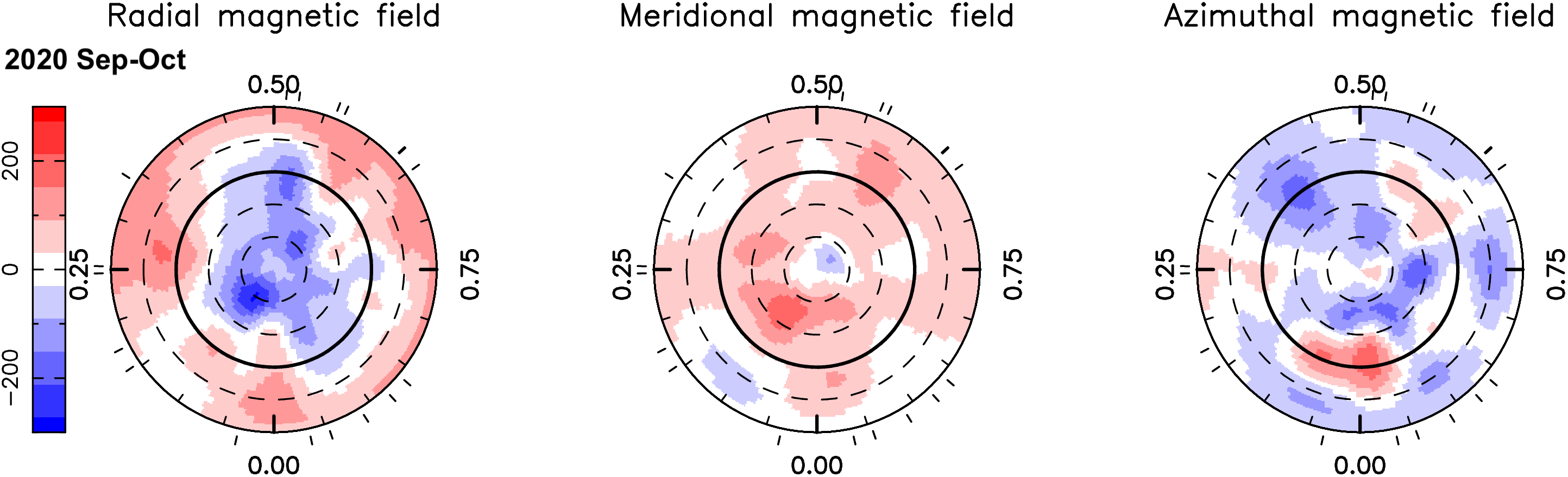}
         
    \end{subfigure}
    \hfill
    %\hspace*{-0.4cm}
    \begin{subfigure}{\textwidth}
         \centering
         \includegraphics[scale=0.15,trim={0cm 0cm 0cm 2.3cm},clip]{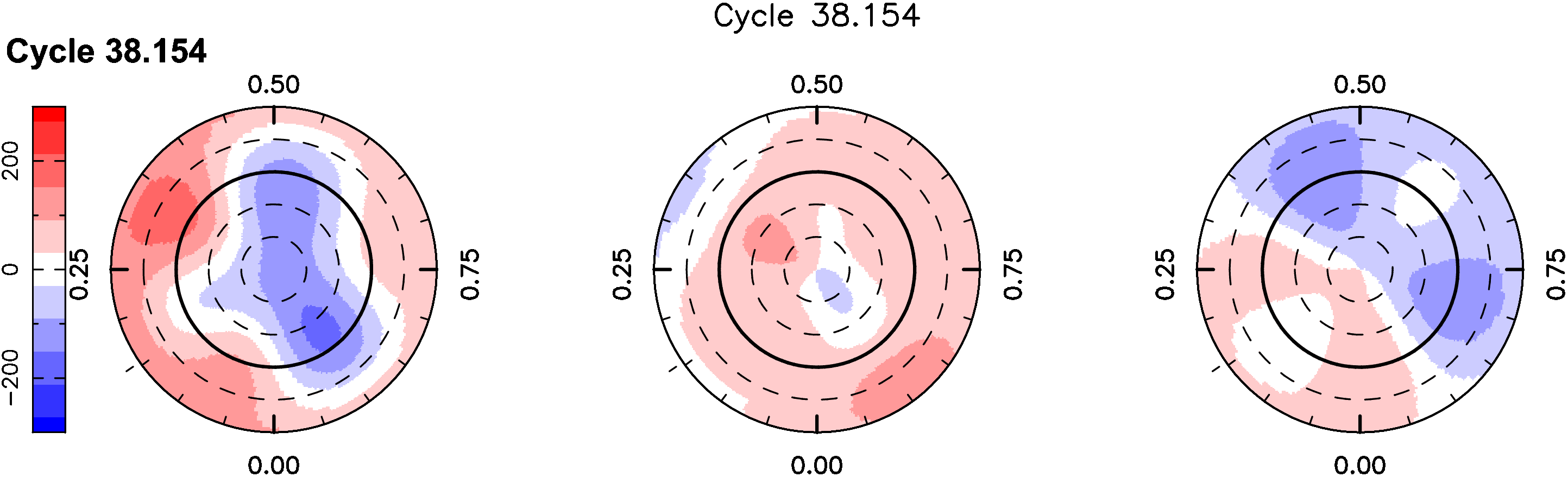}
         
    \end{subfigure}
    \hfill
    %\hspace*{-0.4cm}
    \begin{subfigure}{\textwidth}
         \centering
         \includegraphics[scale=0.15,trim={0cm 0cm 0cm 2.3cm},clip]{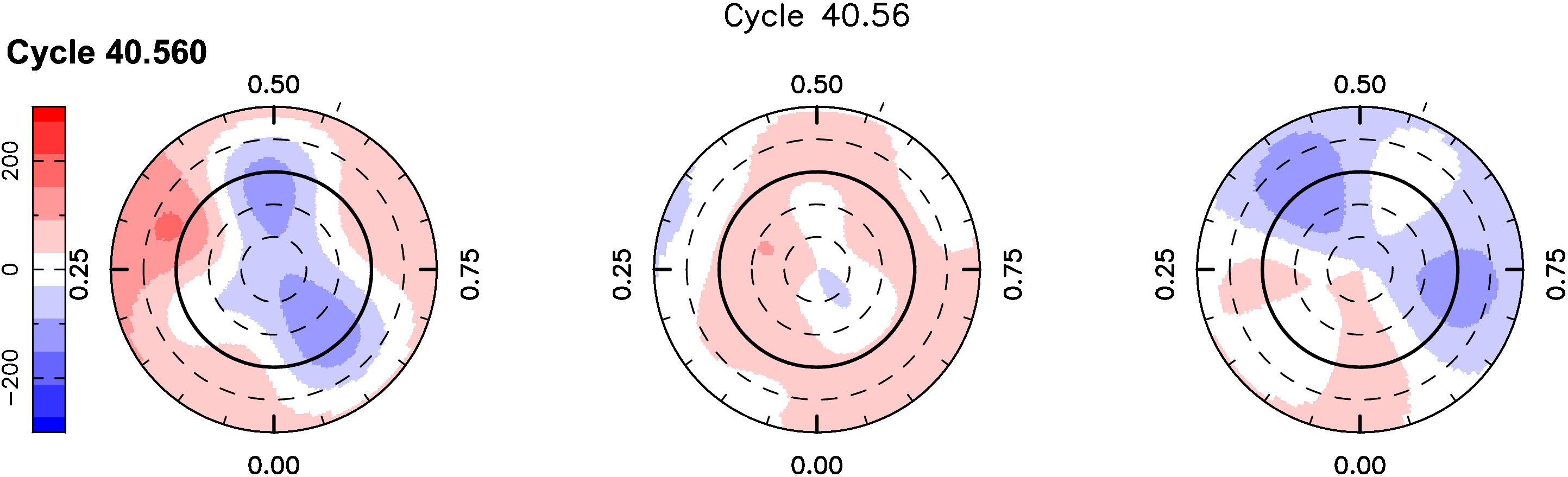}

    \end{subfigure}

    \hfill
    %\hspace*{-0.4cm}
    \begin{subfigure}{\textwidth}
         \centering
         \includegraphics[scale=0.15,trim={0cm 0cm 0cm 2.3cm},clip]{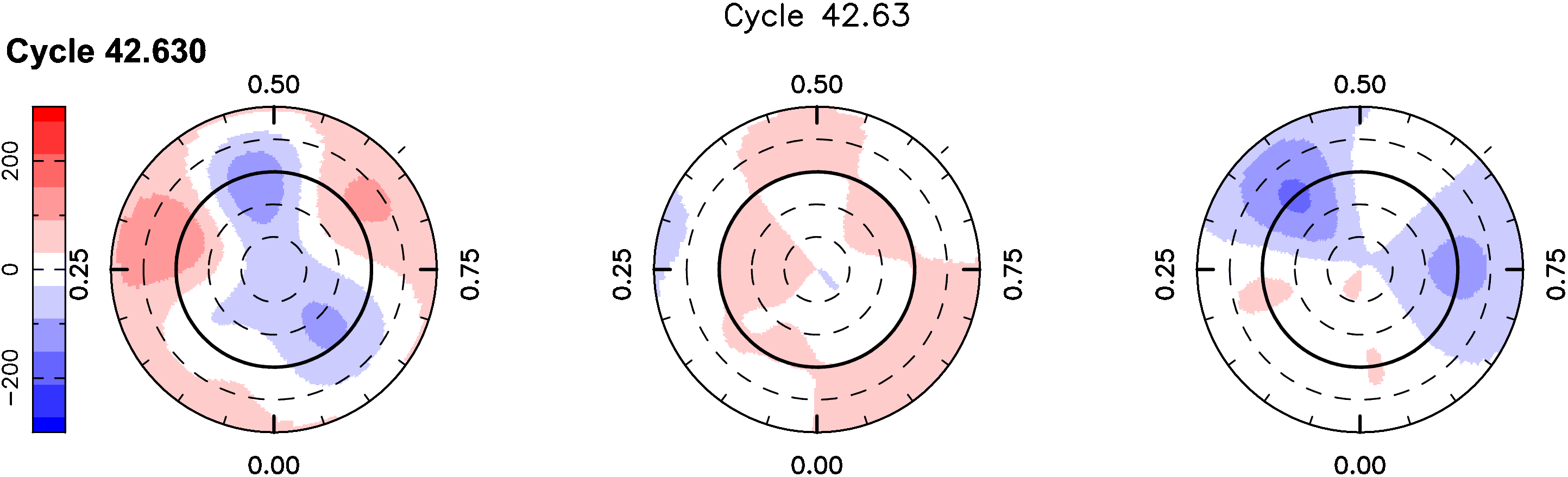}

    \end{subfigure}

    \hfill
    %\hspace*{-0.4cm}
    \begin{subfigure}{\textwidth}
         \centering
         \includegraphics[scale=0.15,trim={0cm 0cm 0cm 2.3cm},clip]{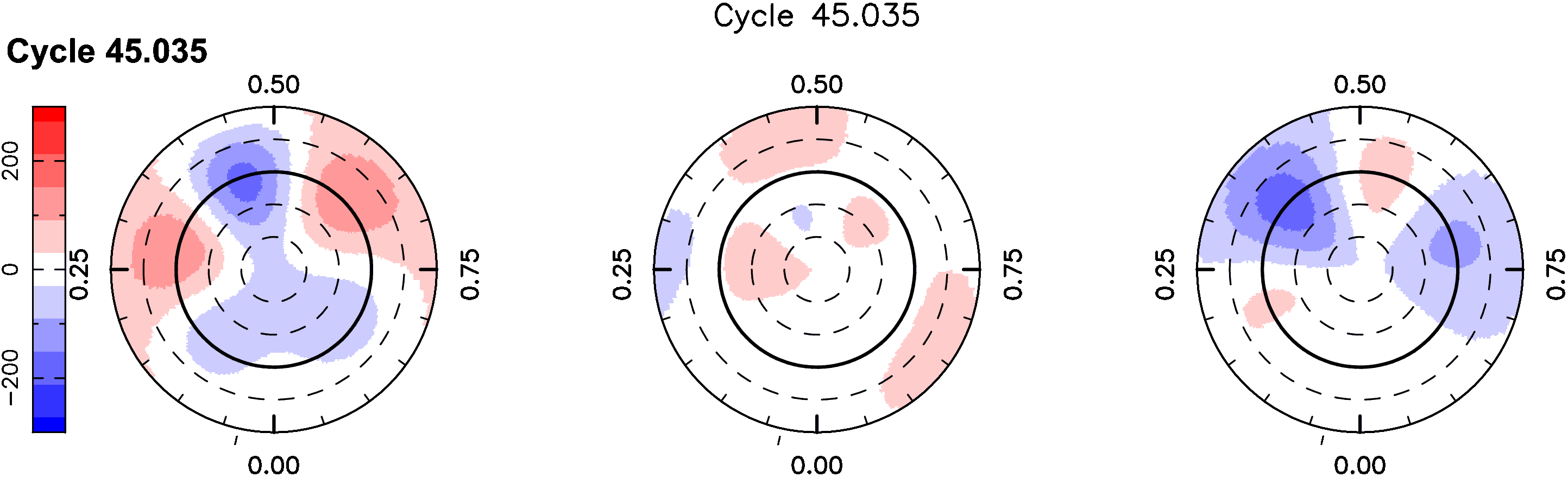}

    \end{subfigure}
    
    \caption{Same as Fig.~\ref{fig:map_TIMeS_aug20} for 2020 Sep-Oct.}
    \label{fig:map_TIMeS_oct20}
\end{figure*}

\begin{figure*} %\ContinuedFloat
    \centering
    
    \begin{subfigure}{\textwidth}
         \centering
         \includegraphics[scale=0.15,trim={0cm 0cm 0cm 0cm},clip]{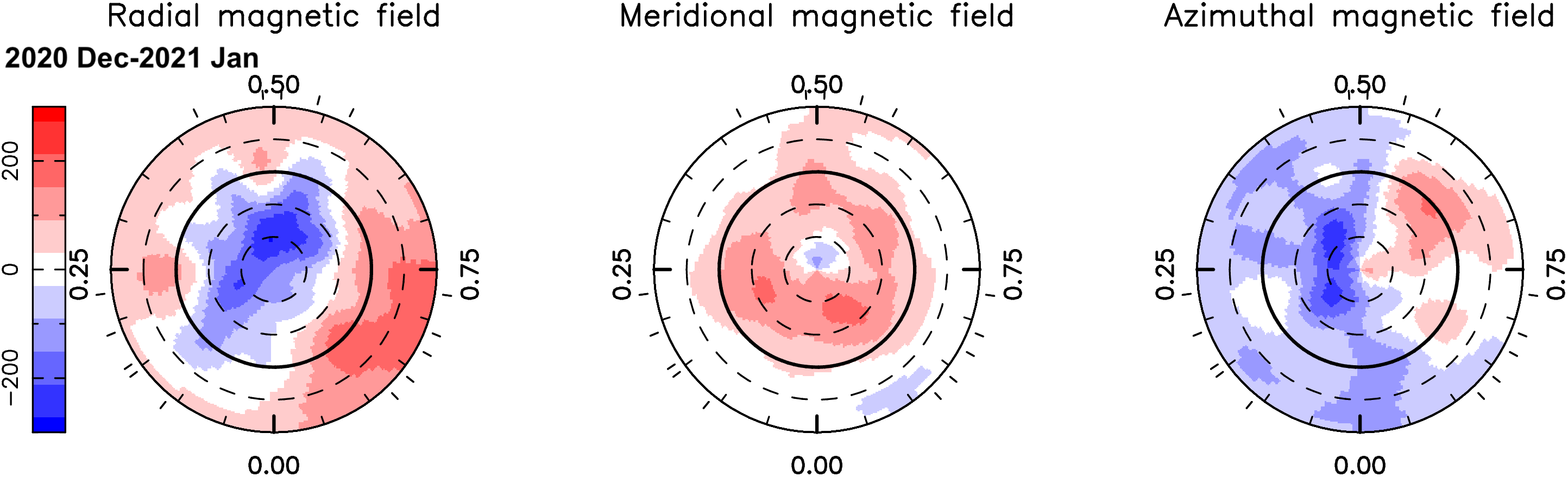}
         
    \end{subfigure}
    \hfill
    %\hspace*{-0.4cm}
    \begin{subfigure}{\textwidth}
         \centering
         \includegraphics[scale=0.15,trim={0cm 0cm 0cm 2.3cm},clip]{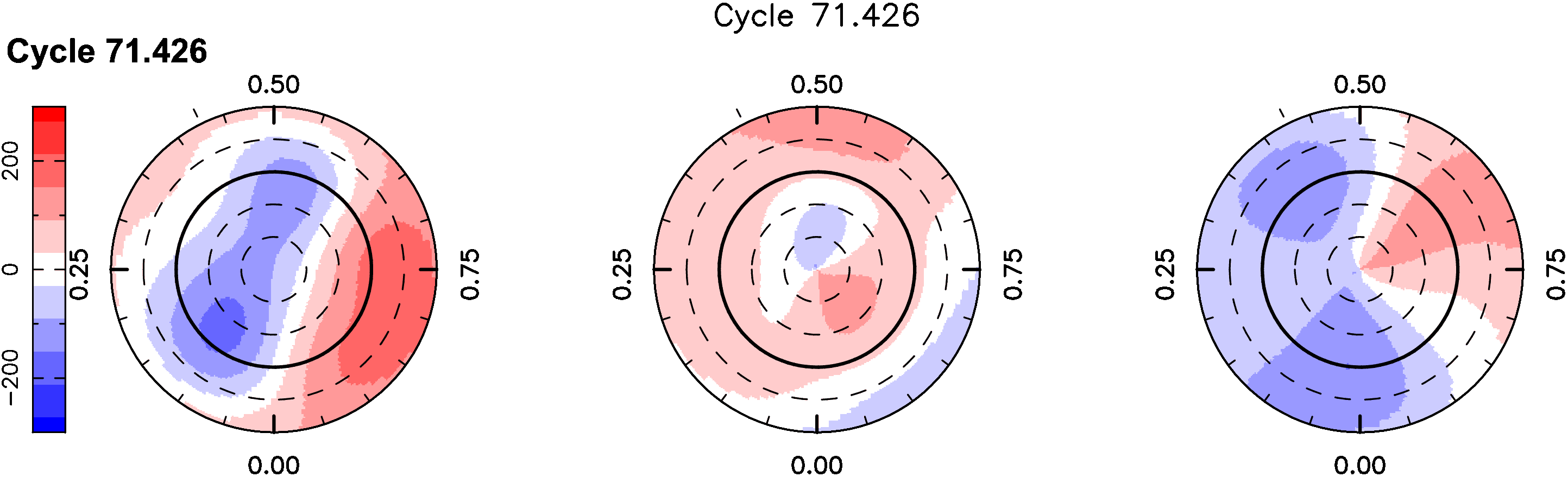}
         
    \end{subfigure}
    \hfill
    %\hspace*{-0.4cm}
    \begin{subfigure}{\textwidth}
         \centering
         \includegraphics[scale=0.15,trim={0cm 0cm 0cm 2.3cm},clip]{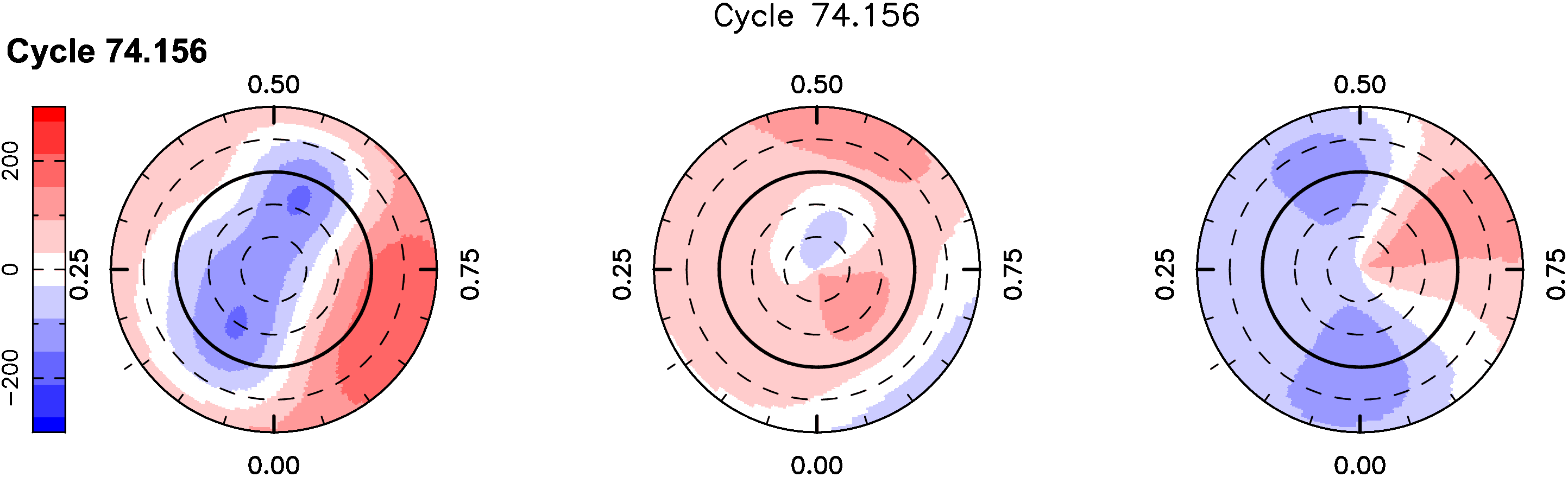}
         
    \end{subfigure}

    \hfill
    %\hspace*{-0.4cm}
    \begin{subfigure}{\textwidth}
         \centering
         \includegraphics[scale=0.15,trim={0cm 0cm 0cm 2.3cm},clip]{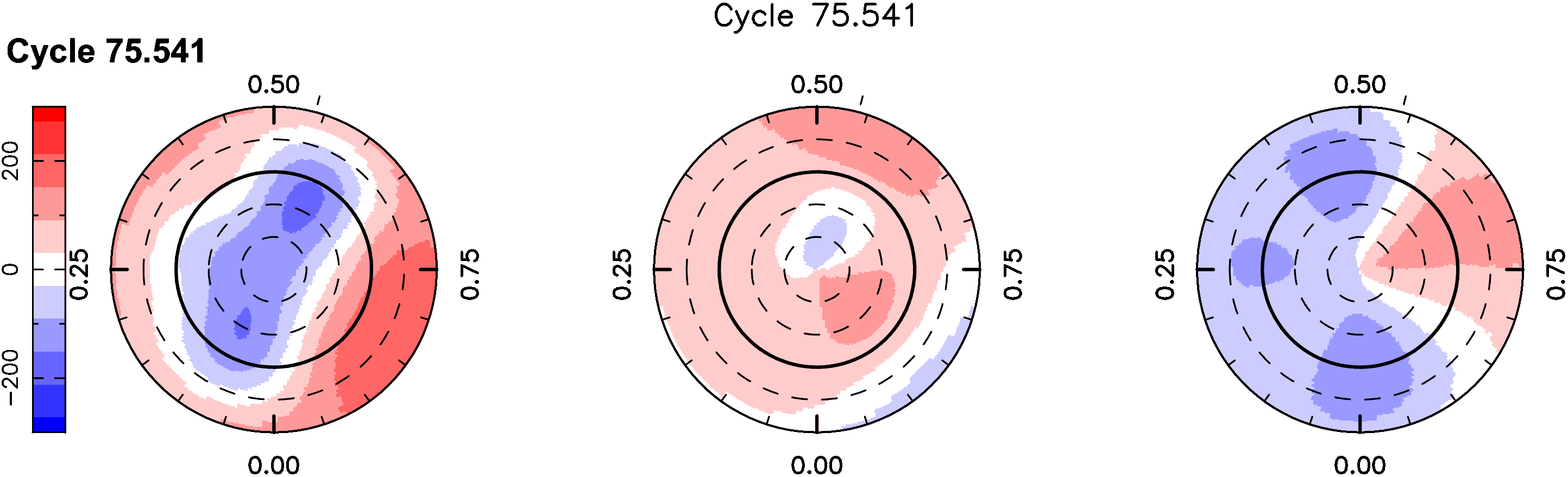}
         
    \end{subfigure}

    \hfill
    %\hspace*{-0.4cm}
    \begin{subfigure}{\textwidth}
         \centering
         \includegraphics[scale=0.15,trim={0cm 0cm 0cm 2.3cm},clip]{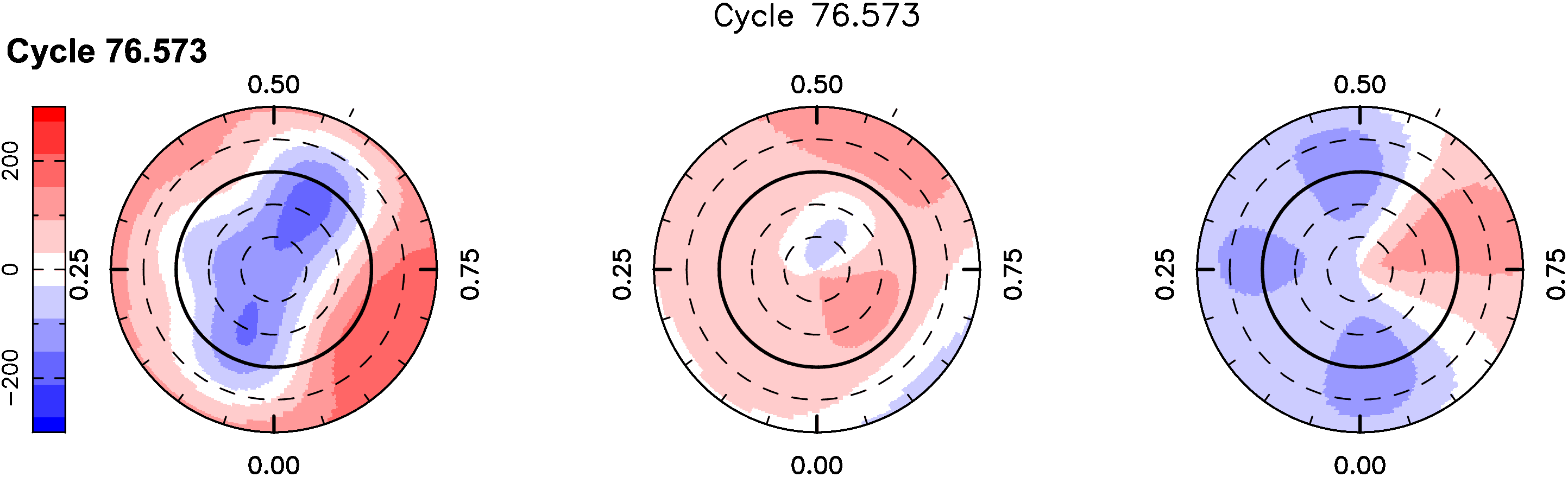}
         
    \end{subfigure}
    
    \caption{Same as Fig.~\ref{fig:map_TIMeS_aug20} for 2020 Dec - 2021 Jan.}
    \label{fig:map_TIMeS_dec20}
\end{figure*}

\benjamin{We applied TIMeS to all the Stokes~$V$ LSD profiles covering \benjamin{2020 Aug - 2023 Feb} (i.e. subsets \#3 to \#10) at once, \benjamin{allowing all spherical harmonic modes up to $\ell_{\rm max}=10$ (as in ZDI, see Sec.~\ref{sec:zdi}) and} assuming $i=70^\circ$ to better identify axisymmetric structures\footnote{\coaut{As outlined in \cite{finociety22}, TIMeS is more sensitive to the inclination than ZDI and setting $i<80^\circ$ is preferable, explaining why we chose $i=70^\circ$ for this analysis. Note that the ZDI reconstructions assuming this inclination are not significantly modified.}} (e.g. dipole) that are missed when using $i=80^\circ$. We were able to fit the data down to $\chi^2_r=1.49$, revealing that the model does not succeed in fitting the data down to the noise level, likely due to the limited available resolution at the surface of the star, the model only featuring 11 modes up to $\ell=4$, even though $\ell_{\max}=10$.} We show the reconstructed magnetic maps during the \benjamin{2020 Aug - 2021 Jan season} at specific rotation cycles in Figs.~\ref{fig:map_TIMeS_aug20}, \ref{fig:map_TIMeS_oct20} and \ref{fig:map_TIMeS_dec20} while those associated with \benjamin{2021 Sep - 2022 Feb and 2022 Nov - 2023 Feb} data can be seen in Appendix~\ref{sec:TIMeS_reconstruction_appendix} (Figs.~\ref{fig:map_TIMeS_oct21} to \ref{fig:map_TIMeS_jan23}). The synthetic Stokes~$V$ LSD profiles obtained with TIMeS are also shown in Fig.~\ref{fig:TIMeS_LSD}. Despite some loss of \benjamin{spatial resolution}, we find that the reconstructed maps are \benjamin{quite} similar to those obtained with ZDI at the same epochs, with only a subtle evolution within each subset considered in the previous sections (as expected from the unit $\chi^2_r$ obtained with ZDI). In particular, we identify the same topology for the radial field and the \benjamin{similar} large \benjamin{features} in the azimuthal field (e.g. around phases 0.35-0.45 in \benjamin{late} 2020). 

\begin{figure}
    \centering \hspace*{-0.35cm}
    \includegraphics[scale=0.32]{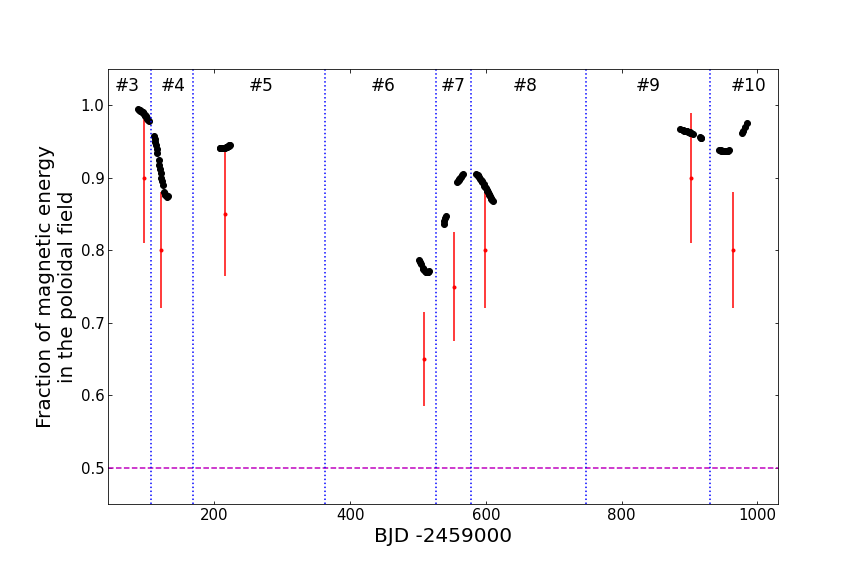}
    \caption{Evolution with time of the poloidal component of the field between 2020 and 2023. In black, we show the fraction of the reconstructed magnetic energy enclosed in the poloidal field, estimated using TIMeS. The red points correspond to the value reconstructed from ZDI applied on each subsets, \coaut{assuming typical error bars of 10\% (see Table~\ref{tab:magnetic_properties})}. The vertical blue dotted lines depict the limits between each subsets used in ZDI (with their number at the top of the plot) while the magenta dashed line denotes the limit above which the magnetic field is mainly poloidal (i.e. more than 50\% of the reconstructed magnetic energy is enclosed in the poloidal component). }
    \label{fig:poloidal_evolution}
\end{figure}

\coaut{This first application of TIMeS \benjamin{to real data} therefore demonstrates that this method is efficient at reconstructing reliable magnetic topologies even if the data \benjamin{span} several years, with only a small amount of information that is not reconstructed (especially the smallest features). %Once again, we find that the field is mainly poloidal, except in 2023 February, \benjamin{as a potential result of the field polarity reversal that may have begun in 2023 January}. 
In addition, we find that the fraction of the magnetic energy enclosed in the poloidal field is globally consistent with that derived from ZDI applied on each subset, except for subsets \#6 and \#10 which likely results from the loss of \benjamin{spatial resolution} with TIMeS (Fig.~\ref{fig:poloidal_evolution}).} 

Using Lomb-Scargle periodograms \citep{zechmeister09} with the \textsc{astropy python} module\footnote{The description of the LombScargle class can be found at \url{https://docs.astropy.org/en/stable/timeseries/lombscargle.html}.} on each of the $\alm$, $\blm$ and $\glm$ coefficients found with TIMeS does not reveal any common periodicity (see Fig.~\ref{fig:periodogram_TIMeS}), suggesting that we \benjamin{did not yet detect a magnetic cycle in our data, which is not surprising given that only a hint of a first polarity switch has been observed up to now.} Future observations of V1298~Tau will \benjamin{allow us to conclude whether the putative change in the poloidal field polarity is confirmed, and is part of a cycle much longer than the time span of our current observations.}

\section{Characterizing the multi-planet system}
\label{sec:characterization_system}
\subsection{Orbital period of planet e}

We analysed the RVs obtained from the data reduction performed with \texttt{APERO} using the LBL method \citep{artigau22} in order to look for the RV signatures induced by the 4 transiting planets hosted by V1298~Tau. The orbital period and transit times of these planets were accurately determined by \cite{feinstein22}, except for planet e for which the orbital period is still unknown (but longer than 40~d; \citealt{suarez-mascareno21,feinstein22,sikora23}). Using these values (see Table~\ref{tab:RV_results}) and assuming circular orbits for all planets (consistent with the results of \citealt{suarez-mascareno21} and \citealt{sikora23}), we only need to retrieve the semi-amplitude of the planet RV signatures ($K_b$, $K_c$, $K_d$ and $K_e$ for planet b, c, d and e, respectively) as well as the orbital period of planet e, noted $P_e$, from our data.

In practice, we used a model featuring a QP GP to model the RV signal induced by stellar activity, 4 Keplerian curves associated with the planet signatures (assuming circular orbits) and an additional white noise (e.g. due to intrinsic variability). As for the longitudinal field, we sampled the posterior distribution of each parameter (planet and GP) using a MCMC method to estimate their optimal value.%, chosen as the value maximizing the posterior distribution. 

We note that the posterior distribution of $P_e$ is multimodal if we impose a uniform prior on the orbital period of planet e (between 42 and 65~d, following the list of probable periods of \citealt{feinstein22}), showing 1 global maximum at about 53.5~d and three local maxima at 42.7~d, 46.4~d and 60.6~d (see Fig.~\ref{fig:corner_plot_rv_uniform}). We therefore run our process again, each time with a narrow Gaussian prior centred on one of these periods with a standard deviation of 1~d to investigate which one is the most likely. We then computed the marginal logarithmic likelihood $\log \mathcal{L_M}$ following the method described by \cite{chib01}, in order to estimate the significance of one solution with respect to the others from the difference in $\log \mathcal{L_M}$ (i.e. logarithmic Bayes Factor / log BF). Considering the model featuring an orbital period for planet e at $53.5$~d as the reference, we find that the log BF for the 3 other models (associated with an orbital period of $42.7$, $46.4$ and $60.6$~d, respectively) is equal to $-1.4$, $-2.4$ and $-4.1$, suggesting that the model featuring a prior centred on an orbital period of $53.5$~d is more likely than the three others. 
From the MCMC approach, we therefore estimate the most likely orbital period to be $P_e=53.5\pm0.4$~d and we thus use this model to further characterize the 4 planets in the following.

\subsection{Planet masses}
\label{sec:planet_masses}
With our reference model, we are able to fit our data down to $\chi^2_r=5.97$ with the residuals exhibiting a RMS dispersion of 34.8~\ms. This value for the $\chi^2_r$ suggest that discrepancies exist between the model and our measurements and that our error bars (estimated from photon noise) are likely underestimated, e.g. due to intrinsic variability that has not be taken into account. More generally, our detections of the planet-induced RV variations are limited by the high level of stellar activity compared to the recovered planet RV signatures (GP amplitude of $\unsim120$~\ms, i.e. 2.6 to 28 times larger than the semi-amplitude of the planet signatures) and of intrinsic variability (modeled by the additional white noise, equal to $46\pm5$~\ms, i.e. $3\times$ larger than the typical error bar of 14~\ms) of V1298~Tau. In particular, we find that the decay time-scale is equal to $32^{+6}_{-5}$~d, compatible with the one derived from $B_\ell$ measurements, indicating that the surface of the star evolves rapidly. The GP is also modulated by the stellar rotation with a period equal to $2.909 \pm 0.009$~d, \coaut{similar} to the one found from $B_\ell$ measurements. All these results clearly illustrate the need for robust filtering methods given how difficult the detection of planets around young active stars is, even when their ephemerides (orbital periods, transit times) are well determined.

Despite this intense activity, we find that planet e is best detected, at a $3.9\sigma$ level, with a RV semi-amplitude $K_e=47^{+16}_{-12}$~\ms, associated with a mass $M_e=1.01^{+0.35}_{-0.26}$~\mjup\ (for an orbital period of $53.5\pm0.4$~d). Assuming a radius of $0.89\pm0.04$~\rjup, following \cite{feinstein22}, the planet density is found to be $\rho_e=1.79^{+0.66}_{-0.51}$~\gcm.

\begin{figure*}
    \centering
    \includegraphics[scale=0.4,trim={1cm 5cm 1cm 5cm},clip]{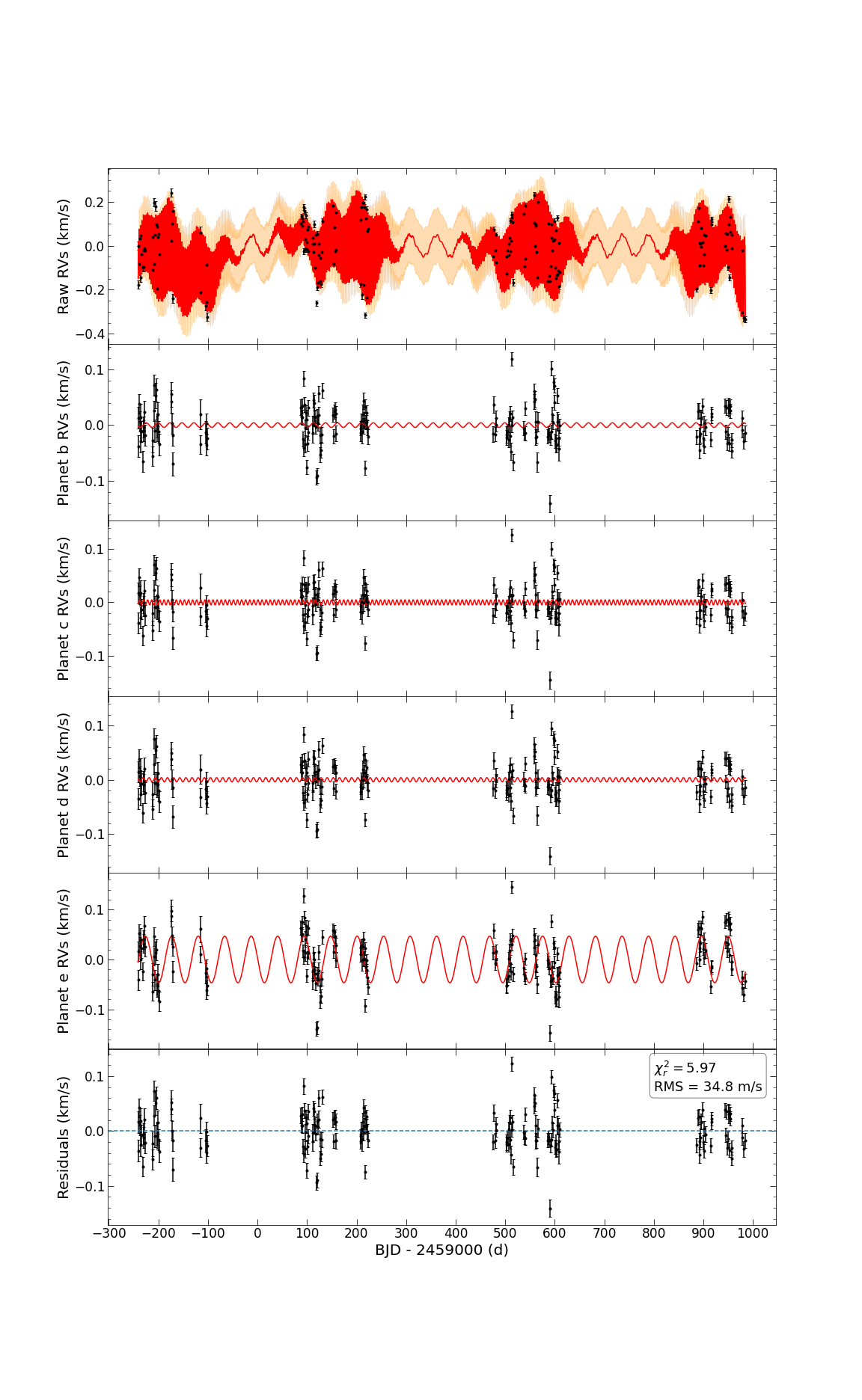}
    \caption{RVs of V1298~Tau between end 2019 and early 2023. The first panel shows the raw RVs (black dots) along with the best model (including the activity jitter and the planet sginatures) in red and the $1\sigma$ confidence level in light orange. 2$^{\rm nd}$ to 5$^{\rm th}$ panels show the retrieved RV signatures of planets b, c, d and e, respectively. The last panel shows the residuals RVs, having a RMS dispersion of 24.3~\ms, i.e. close to the typical error bar of our measurements.}
    \label{fig:RV_fit}
\end{figure*}

\begin{figure}
    \centering
    
    \begin{subfigure}{0.49\textwidth}
         \centering
         \includegraphics[scale=0.3,trim={0cm 1cm 0cm 2.3cm},clip]{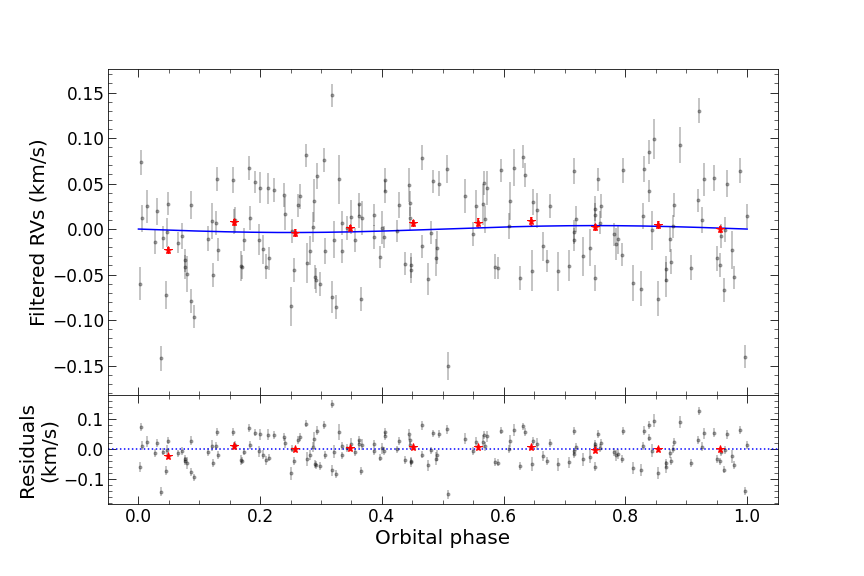}
    \end{subfigure}
    \hfill
    
    \begin{subfigure}{0.49\textwidth}
         \centering
         \includegraphics[scale=0.3,trim={0cm 1cm 0cm 2.3cm},clip]{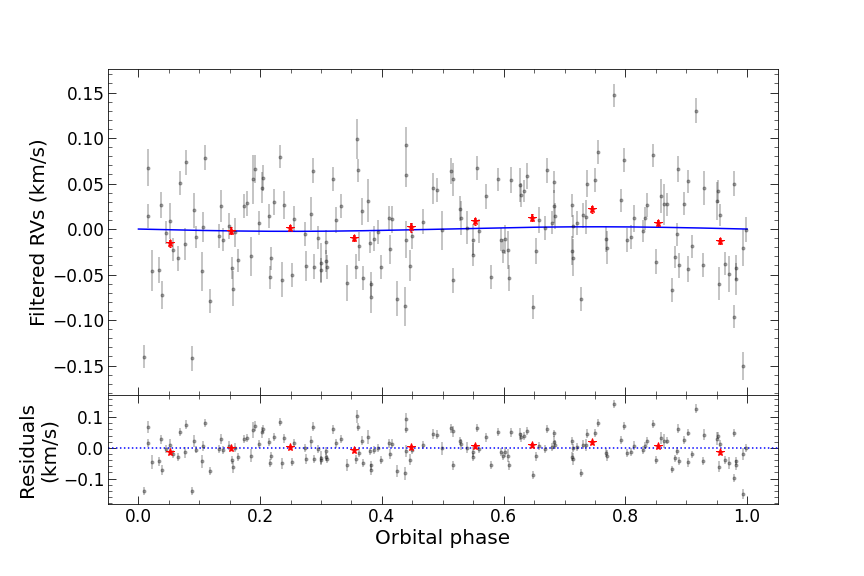}   
    \end{subfigure}
    \hfill
    
    \begin{subfigure}{0.49\textwidth}
         \centering
         \includegraphics[scale=0.3,trim={0cm 0cm 0cm 2.3cm},clip]{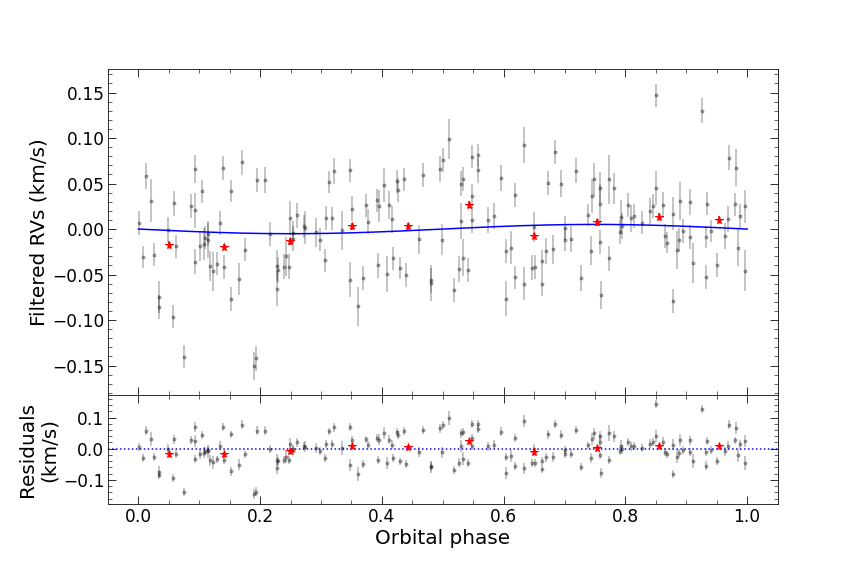}
    \end{subfigure}
    
    \begin{subfigure}{0.49\textwidth}
         \centering
         \includegraphics[scale=0.3,trim={0cm 1cm 0cm 2.3cm},clip]{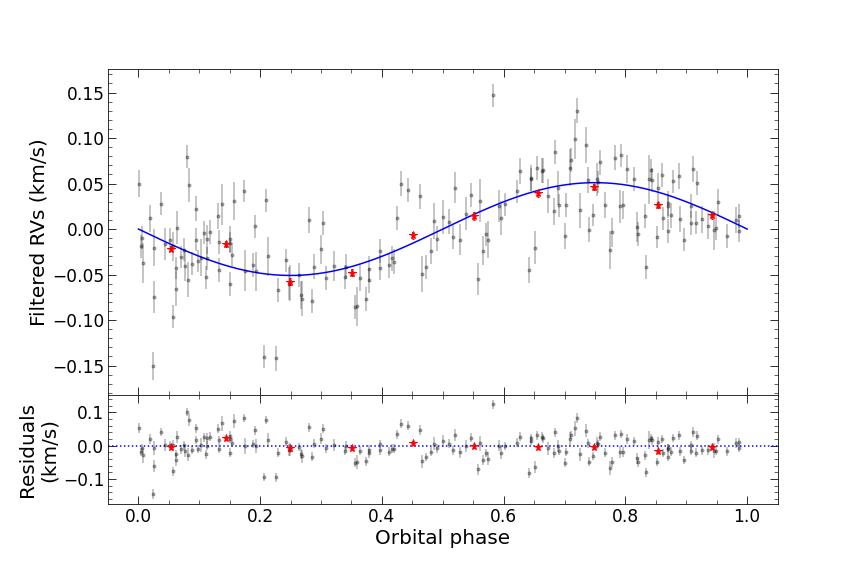}
    \end{subfigure}

    \caption{Phase-folded filtered (top panels) and residuals (bottom panels) RVs for planets b (1$^{\rm st}$ panel), c (2$^{\rm nd}$ panel), d (3$^{\rm rd}$ panel) and e (4$^{\rm th}$ panel). The 174 measured RVs are shown as black dots (with their error bar) while the red stars correspond to RVs averaged over 0.1 orbital cycle. The RV signature (derived from our MCMC process) associated with each planet is shown as a blue solid line. The RMS dispersion of the residuals is equal to 34.8~\ms\ (see Fig.~\ref{fig:RV_fit}).}
    \label{fig:rv_phase_folded}
\end{figure}

For the three innermost planets, we do not obtain clear detections as we find $K_b=4.2\pm10.7$~\ms, $K_c=4.8\pm6.2$~\ms\ and $K_d=4.9\pm6.7$~\ms, corresponding to masses of $M_b=0.07\pm0.18$~\mjup, $M_c=0.05\pm0.07$~\mjup\ and $M_d=0.05\pm0.08$~\mjup, and densities of $\rho_b=0.14\pm0.36$~\gcm, $\rho_c=0.76\pm0.99$~\gcm\ and $\rho_d=0.40\pm0.66$, respectively. Based on the 99\% confidence level interval associated with the posterior distribution of each planet mass \jf{derived} with the MCMC approach (after converting the sampled semi-amplitudes into masses), we derive upper limits on these masses of $M_b<0.44$~\mjup, $M_c<0.22$~\mjup\ and $M_d<0.25$~\mjup. 
We summarize all the planet parameters in Table~\ref{tab:RV_results} for the 4 best periods and we show the associated corner plots in Appendix~\ref{sec:mcmc_results}. We also show the best fit (i.e. for $P_e=53.5$~d) and the associated phase-folded filtered RVs in Figs.~\ref{fig:RV_fit} and~\ref{fig:rv_phase_folded}, respectively.

Assuming an eccentric orbit for planet e\footnote{As in \cite{donati23}, we introduced the variables $\sqrt{e}\cos{\omega}$ and $\sqrt{e}\sin{\omega}$ in our model ($e$ and $\omega$ being the eccentricity and the angle of periastron, respectively) with Gaussian priors whose mean and standard deviation are equal to 0 and 0.3 to account for the distribution of eccentricities in multi-planet systems \citep{vaneylen19}.} (in the case of the most likely orbital period of 53.5 d), yields no improvement in confidence level ($\log{\rm BF}=-0.7$ with respect to our best model) and an eccentricity consistent with 0 ($0.03\pm0.05$), indicating a posteriori that our previous assumption of circular orbits was justified to model our data.

\begin{figure*}
    \includegraphics[scale=0.32,trim={2.5cm 2.5cm 1cm 1cm},clip]{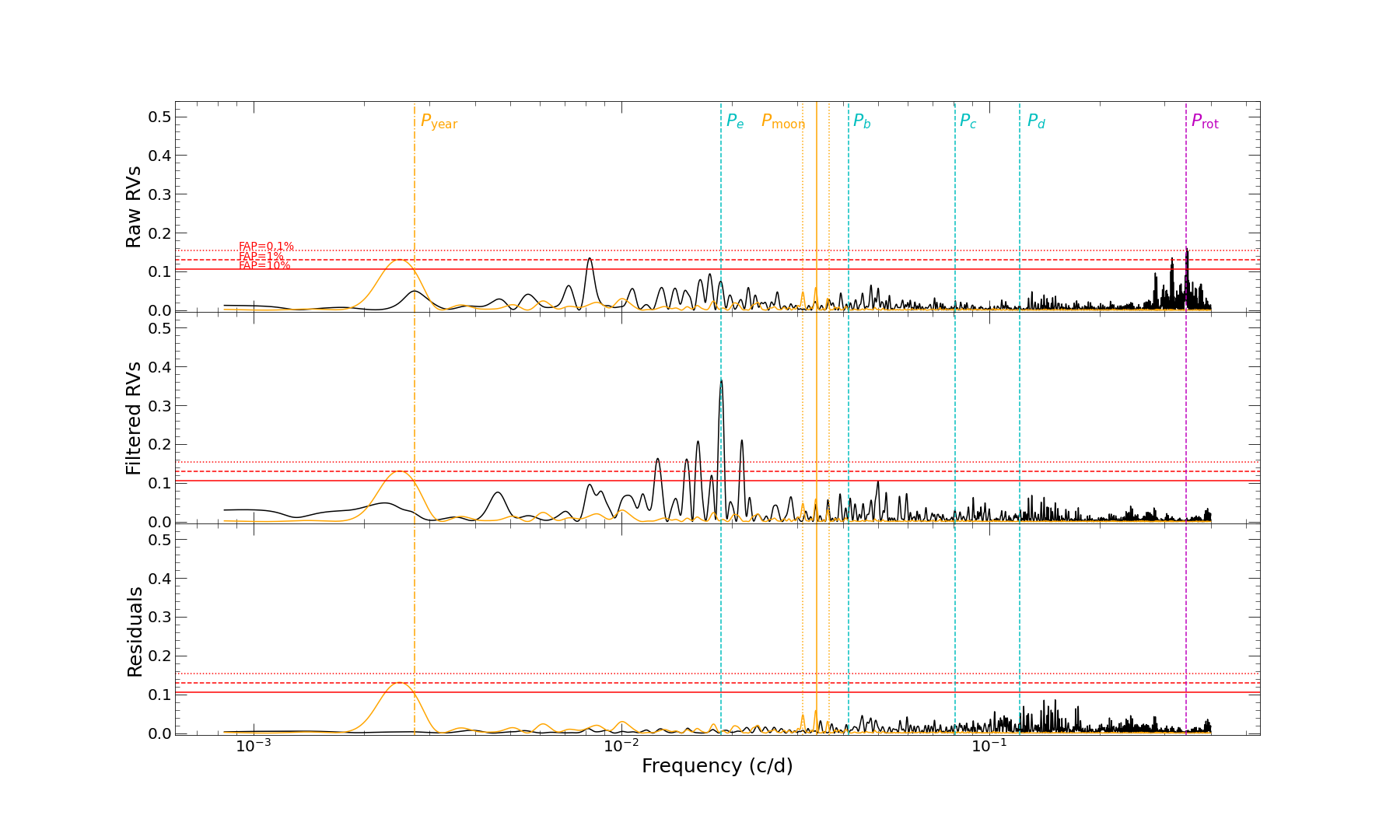}
    \caption{Lomb-Scargle periodograms of the raw RVs (top), filtered RVs (middle) and residuals (bottom) computed using the \textsc{astropy python} module. The horizontal dotted, dashed and solid red lines depict the FAPs of 0.1, 1 and 10\%, respectively. The vertical dashed magenta line corresponds to the stellar rotation period while the dashed cyan lines show the orbital periods of planets e, b, c and d (from left to right). \benjamin{The orange curve depicts the window function, the orange solid and dotted vertical lines correspond to the synodic period of the Moon and its 1-yr aliases, respectively, and the orange dashed vertical line outlines the 1-yr period. }}
    \label{fig:periodogram_RV}
\end{figure*}

In addition, we computed the Lomb-Scargle periodograms of the raw RVs, filtered RVs and residuals RVs to look for potential additional planets that may have not been detected before (e.g. non-transiting planets). \last{Looking at the raw RVs, we do not see any significant peak at 1~yr (with a FAP lower than 10\%), suggesting that the contamination of our data from telluric lines is insignificant for our purpose. When focussing on the filtered RVs (in which only activity was removed), we see a significant power at the orbital period of planet e, but no clear peak in the periodogram of the residual RVs (see Fig.~\ref{fig:periodogram_RV}).} \last{We simulated RV datasets with the same temporal distribution as our actual SPIRou data using our best model on which we added the signature of a potential fifth planet beyond the orbit of V1298~Tau~e. Using these datasets, we estimate that only planets more massive than $2.7$ and 5.9~\mjup\ at 0.7 and 1.4~au (corresponding to an orbital period of 6 months and $1.5$~yr), respectively, could be reliably detected with our dataset.}

\begin{table*}
	\centering
	\caption{MCMC results for the analysis of the RV data. 1$^{\rm st}$ and 2$^{\rm nd}$ columns list the parameters and their associated priors corresponding to a model featuring an activity signal and four planet-related RV signatures. The 4 last columns show the best value for each parameter, in the case of a narrow prior centred on 3 different values of $P_{e_0}$ for the orbital period of planet e. \jf{The knee of the modified Jeffreys priors is set to the typical RV uncertainty, noted $\sigma_{RV}$.} Note that the $\log{\rm BF}$ is computed with respect to the marginal likelihood associated with $P_{e_0}=53.5$~d.  }
	\label{tab:RV_results}
    %\resizebox{0.49\textwidth}{!}{
	\begin{tabular}{llcccc}
    \hline
    Parameter & Prior & $P_{e_0}=53.5$~d  & \jf{$P_{e_0}=42.7$~d} & \jf{$P_{e_0}=46.4$~d} & $P_{e_0}=60.6$~d \\
    \hline

    $\theta_1$ (\kms) & mod Jeffreys ($\sigma_{RV}$) & $0.12\pm0.01$ & $0.13\pm0.01$ & $0.13\pm0.01$ & $0.12\pm0.01$  \\
    $\theta_2$ (d) & log Gaussian ($\log 30$, $\log 2$) & \jf{$32^{+6}_{-5}$}  & \jf{$28^{+5}_{-4}$} & \jf{$31^{+5}_{-4}$}  & \jf{$30^{+6}_{-5}$} \\
    $\theta_3$ (d) & Gaussian (2.91, 0.1) & \jf{$2.909\pm0.009$}  & \jf{$2.921\pm0.010$}  & \jf{$2.904\pm0.010$} & \jf{$2.923\pm0.008$} \\
    $\theta_4$ & Uniform (0, 3) & \jf{$0.28\pm0.05$}  & \jf{$0.28\pm0.06$} &  \jf{$0.32\pm0.06$} & \jf{$0.24\pm0.05$} \\
    $\theta_5$ (\ms) & mod Jeffreys ($\sigma_{RV}$) & \jf{$46\pm5$}  & \jf{$45^{+5}_{-4}$} & \jf{$46\pm5$} & \jf{$46\pm5$} \\ \hline

    $K_b$ (\ms) & mod Jeffreys ($\sigma_{RV}$) & \jf{$4.2 \pm 10.7$}  & \jf{$6.3\pm10.6$} & \jf{$5.1\pm11.6$} & \jf{$3.3 \pm 10.9$}\\[1mm]
    $P_b$ (d) & Fixed from \citet{feinstein22} & 24.1315 & 24.1315 & 24.1315 & 24.1315 \\
    $T_b$ (2459000+) & Fixed from \citet{feinstein22}  & 481.0902 & 481.0902 & 481.0902 & 481.0902 \\
    $R_b$ (\rjup) & Fixed from \citet{feinstein22} & 0.85 & 0.85 & 0.85 & 0.85 \\
    $M_b$ (\mjup) & Derived from $K_b$, $P_b$ and $M_*$ & \jf{$0.07\pm0.18$} & \jf{$0.11\pm0.18$}  & \jf{$0.08\pm0.19$} & \jf{$0.05\pm0.18$}  \\[1mm]
    $\rho_b$ (\gcm)& Derived from $M_b$ and $R_b$  & \jf{$0.14\pm0.36$}  & \jf{$0.21\pm0.36$} & \jf{$0.17\pm0.39$} & \jf{$0.11\pm0.37$} \\ \hline

    $K_c$ (\ms) & mod Jeffreys ($\sigma_{RV}$) & \jf{$4.8\pm6.2$}  & \jf{$3.8\pm6.3$} &  \jf{$5.5\pm6.3$} & \jf{$3.8 \pm 6.3$} \\[1mm]
    $P_c$ (d) & Fixed from \citet{feinstein22} & 8.2438 & 8.2438 & 8.2438 & 8.2438 \\
    $T_c$ (2459000+) & Fixed from \citet{feinstein22}  & 481.1664 & 481.1664 & 481.1664 & 481.1664 \\
    $R_c$ (\rjup)& Fixed from \citet{feinstein22} & 0.45 & 0.45 & 0.45 & 0.45 \\
    $M_c$ (\mjup) & Derived from $K_c$, $P_c$ and $M_*$ & \jf{$0.05\pm0.07$} & \jf{$0.04\pm0.07$}  & \jf{$0.06\pm0.07$} & \jf{$0.04\pm0.07$}  \\[1mm]
    $\rho_c$ (\gcm)& Derived from $M_c$ and $R_c$  & \jf{$0.76\pm0.99$}  & \jf{$0.60\pm1.00$} & \jf{$0.86\pm1.00$} & \jf{$0.61\pm1.00$} \\ \hline

    $K_d$ (\ms) & mod Jeffreys ($\sigma_{RV}$) & \jf{$4.9\pm6.7$}  & \jf{$4.4\pm6.5$} & \jf{$6.0\pm6.6$} & \jf{$5.8 \pm 7.0$} \\[1mm]
    $P_d$ (d) & Fixed from \citet{feinstein22} &  12.3960 &  12.3960 & 12.3960 & 12.3960\\
    $T_d$ (2459000+) & Fixed from \citet{feinstein22}  & 478.4149 & 478.4149 & 478.4149 & 478.4149 \\
    $R_d$ (\rjup)& Fixed from \citet{feinstein22} & 0.55 & 0.55 & 0.55 & 0.55 \\
    $M_d$ (\mjup) & Derived from $K_d$, $P_d$ and $M_*$ & \jf{$0.05\pm0.08$} & \jf{$0.06\pm0.08$}  & \jf{$0.08\pm0.08$} & \jf{$0.07\pm0.09$}  \\[1mm]
    $\rho_d$ (\gcm)& Derived from $M_d$ and $R_d$  & \jf{$0.40\pm0.66$}  & \jf{$0.43\pm0.65$} & \jf{$0.60\pm0.66$} & \jf{$0.58\pm0.70$} \\ \hline

    $K_e$ (\ms) & mod Jeffreys ($\sigma_{RV}$) & \jf{$47^{+16}_{-12}$}  & \jf{$44^{+21}_{-14}$} & \jf{$36^{+18}_{-12}$} & \jf{$25^{+24}_{-12}$} \\[1mm]
    $P_e$ (d) & Gaussian ($P_{e_0}$, 1) & \jf{$53.5\pm0.4$}  & \jf{$42.7   \pm0.3$} & \jf{$46.4\pm0.4$} & \jf{$60.6\pm0.7$} \\
    $T_e$ (2459000+) & Fixed from \citet{feinstein22}  & 481.7967 & 481.7967 & 481.7967 & 481.7967 \\
    $R_e$ (\rjup)& Fixed from \citet{feinstein22} & 0.89 & 0.89 & 0.89 & 0.89 \\
    $M_e$ (\mjup) & Derived from $K_e$, $P_e$ and $M_*$ & \jf{$1.01^{+0.35}_{-0.26}$}  & \jf{$0.87^{+0.41}_{-0.28}$} & \jf{$0.75^{+0.37}_{-0.25}$} & \jf{$0.56^{+0.54}_{-0.27}$} \\[1mm]
    $\rho_e$ (\gcm)& Derived from $M_e$ and $R_e$ & \jf{$1.79^{+0.66}_{-0.51}$}  & \jf{$1.53^{+0.76}_{-0.54}$} & \jf{$1.31^{+0.67}_{-0.47}$} & \jf{$0.99^{+0.96}_{-0.50}$} \\ \hline

    $\chi^2_r$ & & \jf{5.97}  & \jf{5.53} & \jf{5.92} & \jf{5.74} \\
    RMS (\ms) & & \jf{34.8}  & \jf{33.8} & \jf{34.8} & \jf{34.4} \\ 
    $\log \mathcal{L_M}$ & & \jf{$-215.5$} & \jf{$-216.9$} & $\jf{-217.9}$  & \jf{$-219.6$} \\
    log BF & & 0 & \jf{$-1.4$}  & \jf{$-2.4$} & \jf{$-4.1$} \\ \hline
    %log BF & & -1.3 & -1.7 & 0 \\ \hline
    \end{tabular}
    %}
\end{table*}

%\section{Chromospheric activity}
\subsection{Additional constraints from previous studies}
\label{sec:contraints_feinstein}
From the transit times of planet e in the TESS and K2 light curves, \cite{feinstein22} derived 17 probable orbital periods for this planet. We therefore modeled our data, \jf{assuming circular orbits,} this time imposing a Gaussian prior on the orbital period of this planet centered on each of these values with a standard deviation of 0.0001, corresponding to the typical error bar reported by \cite{feinstein22}. 

From the $\log \mathcal{L_M}$, we find that the most likely period is $P_e=53.0039\pm0.0001$~d, \jf{yielding a fit to our RV data associated with a $\chi^2_r=6.30$ and a RMS dispersion of the residuals of 35.8~\ms. For this period, we report a 4$\sigma$ detection of the RV signature of planet e ($K_e=44^{+15}_{-11}$~\ms) associated with a mass of $M_e=0.95^{+0.33}_{-0.24}$, while the 3 other planets remain undetected.}

Using this value as a reference, we computed the log BF between this model and the 16 others. From this criterion, we cannot \jf{definitely} rule out any of the 17 periods provided by \cite{feinstein22} as all periods yield a $| \log{\rm BF} | < 6.4$. We however distinguish 3 groups: the most likely periods ($|\log{\rm BF}|< 2.5$), the plausible periods ($2.5<|\log{\rm BF}|<5$) and the \jf{least} likely periods ($|\log{\rm BF}|\geq 5$). 
In addition to $P_e=53.0039$~d, 3 other values belong to the most likely group ($P_e=54.2085$, 46.7681 and 45.8687~d, associated with log BF of $-1.1$, $-2.0$ and $-2.4$ respectively). The intermediate plausible group contains 6 orbital periods ($P_e=45.0033$, 47.7035, 48.6770, 59.6294, 61.1583 and 62.7678~d), that are still considered as potential values, even though the derived mass for planet e can be more than twice smaller than that obtained in Sec.~\ref{sec:planet_masses} or in the most likely group. The least likely group therefore gathers the 7 remaining values ($P_e=44,1699$, 49.6911, 50.7484, 51.8516, 55.4692, 56.7899 and 58.1750~d). We summarized the results for the 4 most likely orbital periods in Table~\ref{tab:RV_results_feinstein}, while those associated with the plausible and least likely groups are given in Tables~\ref{tab:RV_results_feinstein_less_likely} and \ref{tab:RV_results_feinstein_not_likely}, respectively.

As for the previous case in Sec.~\ref{sec:planet_masses}, estimating the eccentricity of planet e as part of the process does not signifcantly improve the results (e.g. $\log{\rm BF} = 0.2$).% between the models featuring $P_e=53.0039$~d with and without eccentric orbit for planet e). We still find an eccentricity consistent with 0 ($0.04\pm0.06$), justifying, once again, our assumption of circular orbits. 

\begin{table*}
	\centering
	\caption{MCMC results for the analysis of the RV data for the 4 most likely planet e orbital periods derived by \citet{feinstein22}. 1$^{\rm st}$ and 2$^{\rm nd}$ columns list the parameters and their associated priors corresponding to a model featuring an activity signal and four planet-related RV signatures. The 4 last columns show the best value for each parameter, in the case of a narrow prior centred on different values of $P_{e_0}$ for the orbital period of planet e. \jf{The knee of the modified Jeffreys priors is set to the typical RV uncertainty, noted $\sigma_{RV}$.} Note that the $\log{\rm BF}$ is computed with respect to the marginal logarithmic likelihood associated with $P_{e_0}=53.0039$~d.  }
	\label{tab:RV_results_feinstein}
    %\resizebox{0.49\textwidth}{!}{
	\begin{tabular}{llccccc}
    \hline
    Parameter & Prior & $P_{e_0}=53.0039$~d & $P_{e_0}=54.2085$~d & $P_{e_0}=46.7681$~d & $P_{e_0}=45.8687$~d   \\
    \hline

    $\theta_1$ (\kms) & mod Jeffreys ($\sigma_{RV}$) & $0.13\pm0.01$ & $0.12\pm0.01$ & $0.13\pm0.01$ & $0.13\pm0.01$  \\
    $\theta_2$ (d) & log Gaussian ($\log 30$, $\log 2$) & \jf{$33^{+6}_{-5}$} & \jf{$30^{+6}_{-5}$} & \jf{$32^{+5}_{-4}$} & \jf{$29^{+5}_{-4}$}   \\
    $\theta_3$ (d) & Gaussian (2.91, 0.1) & \jf{$2.907\pm0.009$} & \jf{$2.917\pm0.009$}  & \jf{$2.906\pm0.008$} & \jf{$2.906\pm0.012$} \\
    $\theta_4$ & Uniform (0, 3) & \jf{$0.30\pm0.06$} & \jf{$0.25\pm0.05$} &  \jf{$0.31\pm0.06$} & \jf{$0.32\pm0.06$} \\
    $\theta_5$ (\ms) & mod Jeffreys ($\sigma_{RV}$) & \jf{$47\pm5$} & \jf{$45\pm5$} & \jf{$45\pm5$} & \jf{$47\pm5$}  \\ \hline

    $K_b$ (\ms) & mod Jeffreys ($\sigma_{RV}$) & \jf{$4.1 \pm 10.8$} & \jf{$3.5\pm11.0$} & \jf{$6.1\pm11.6$} & \jf{$3.7 \pm 11.5$} \\[1mm]
    $P_b$ (d) & Fixed from \citet{feinstein22} & 24.1315 & 24.1315 & 24.1315 & 24.1315  \\
    $T_b$ (2459000+) & Fixed from \citet{feinstein22}  & 481.0902 & 481.0902 & 481.0902 & 481.0902  \\
    $R_b$ (\rjup) & Fixed from \citet{feinstein22} & 0.85 & 0.85 & 0.85 & 0.85  \\
    $M_b$ (\mjup) & Derived from $K_b$, $P_b$ and $M_*$ & \jf{$0.07\pm0.18$} & \jf{$0.06\pm0.18$}  & \jf{$0.11\pm0.19$} & \jf{$0.06\pm0.19$}  \\[1mm]
    $\rho_b$ (\gcm)& Derived from $M_b$ and $R_b$  & \jf{$0.14\pm0.36$}  & \jf{$0.12\pm0.37$} & \jf{$0.20\pm0.39$} & \jf{$0.13\pm0.39$} \\ \hline

    $K_c$ (\ms) & mod Jeffreys ($\sigma_{RV}$) & \jf{$5.7\pm6.5$} & \jf{$4.1\pm6.2$} &  \jf{$5.7\pm6.3$} &  \jf{$4.9\pm6.1$} \\[1mm]
    $P_c$ (d) & Fixed from \citet{feinstein22} & 8.2438 & 8.2438 & 8.2438 & 8.2438  \\
    $T_c$ (2459000+) & Fixed from \citet{feinstein22}  & 481.1664 & 481.1664 & 481.1664 & 481.1664  \\
    $R_c$ (\rjup)& Fixed from \citet{feinstein22} & 0.45 & 0.45  & 0.45 & 0.45 \\
    $M_c$ (\mjup) & Derived from $K_c$, $P_c$ and $M_*$ & \jf{$0.06\pm0.07$} & \jf{$0.04\pm0.07$}  & \jf{$0.06\pm0.07$} & \jf{$0.05\pm0.07$}  \\[1mm]
    $\rho_c$ (\gcm)& Derived from $M_c$ and $R_c$  & \jf{$0.90\pm1.03$}  & \jf{$0.64\pm0.98$} & \jf{$0.89\pm1.00$} & \jf{$0.78\pm0.97$} \\ \hline

    $K_d$ (\ms) & mod Jeffreys ($\sigma_{RV}$) & \jf{$3.7\pm7.0$} & \jf{$4.9\pm6.7$} & \jf{$5.8\pm6.7$} & \jf{$5.7\pm6.7$}   \\[1mm]
    $P_d$ (d) & Fixed from \citet{feinstein22} &  12.3960 & 12.3960 & 12.3960 & 12.3960 \\
    $T_d$ (2459000+) & Fixed from \citet{feinstein22}  & 478.4149 & 478.4149 & 478.4149 & 478.4149 \\
    $R_d$ (\rjup)& Fixed from \citet{feinstein22} & 0.55 & 0.55 & 0.55 & 0.55 \\
    $M_d$ (\mjup) & Derived from $K_d$, $P_d$ and $M_*$ & \jf{$0.05\pm0.09$} & \jf{$0.06\pm0.08$}  & \jf{$0.07\pm0.08$} & \jf{$0.07\pm0.08$}  \\[1mm]
    $\rho_d$ (\gcm)& Derived from $M_d$ and $R_d$  & \jf{$0.37\pm0.69$}  & \jf{$0.48\pm0.66$} & \jf{$0.58\pm0.67$} & \jf{$0.57\pm0.67$} \\ \hline

    $K_e$ (\ms) & mod Jeffreys ($\sigma_{RV}$) & \jf{$44^{+15}_{-11}$} & \jf{$43^{+20}_{-14}$} & \jf{$36^{+17}_{-11}$} & \jf{$33^{+15}_{-11}$} \\[1mm]
    $P_e$ (d) & Gaussian ($P_{e_0}$, 0.0001) & $53.0039\pm0.0001$ & $54.2085\pm0.0001$ & $46.7681\pm0.0001$ & $45.8687\pm0.0001$  \\
    $T_e$ (2459000+) & Fixed from \citet{feinstein22}  & 481.7967 & 481.7967 & 481.7967 & 481.7967 \\
    $R_e$ (\rjup)& Fixed from \citet{feinstein22} & 0.89 & 0.89 & 0.89 & 0.89  \\
    $M_e$ (\mjup) & Derived from $K_e$, $P_e$ and $M_*$ & \jf{$0.95^{+0.33}_{-0.24}$}  & \jf{$0.92^{+0.43}_{-0.29}$} & \jf{$0.74^{+0.35}_{-0.23}$} & \jf{$0.66^{+0.32}_{-0.21}$} \\[1mm]
    $\rho_e$ (\gcm)& Derived from $M_e$ and $R_e$ & \jf{$1.66^{+0.61}_{-0.48}$}  & \jf{$1.70^{+0.80}_{-0.57}$} & \jf{$1.30^{+0.63}_{-0.45}$} & \jf{$1.18^{+0.58}_{-0.41}$} \\\hline

    $\chi^2_r$ & & \jf{6.30} & \jf{5.66} & \jf{5.76} & \jf{6.16}   \\
    RMS (\ms) & & \jf{35.8} & \jf{33.9} & \jf{34.4} &  \jf{35.5} \\ 
    $\log \mathcal{L_M}$ & & \jf{$-216.2$} & $\jf{-217.3}$  & \jf{$-218.2$} & \jf{$-218.6$}\\
    log BF & & \jf{$0$} & $\jf{-1.1}$  & \jf{$-2.0$} & \jf{$-2.4$}\\ \hline
    %log BF & & -1.3 & -1.7 & 0 \\ \hline
    \end{tabular}
    %}
\end{table*}

\section{Chromospheric activity}
\label{sec:chromospheric_activity}

\begin{figure}
    \centering
    \begin{subfigure}{0.47\textwidth}
         \centering \hspace*{-0.6cm}
         \includegraphics[scale=0.25,trim={1cm 0cm 0cm 0cm},clip]{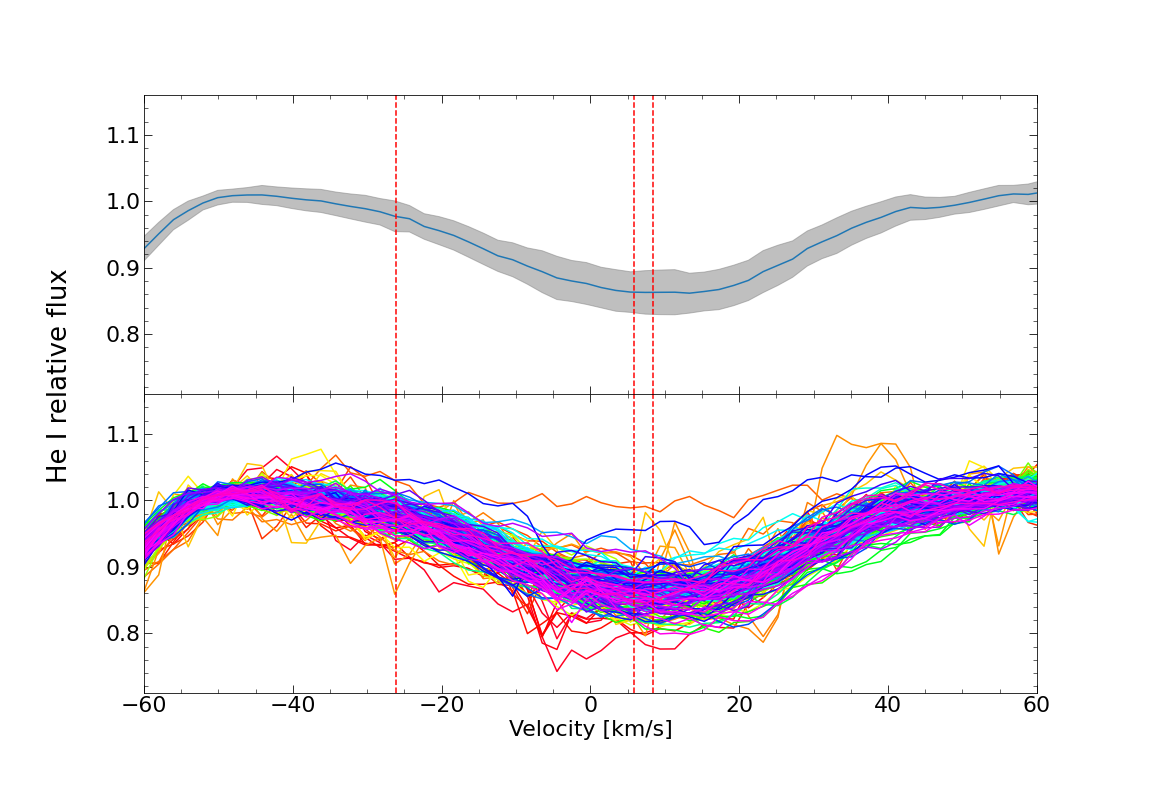}
         
    \end{subfigure}
    \hfill
    \begin{subfigure}{0.47\textwidth}
         \centering \hspace*{-0.6cm}
         \includegraphics[scale=0.25,trim={1cm 0cm 0cm 0cm},clip]{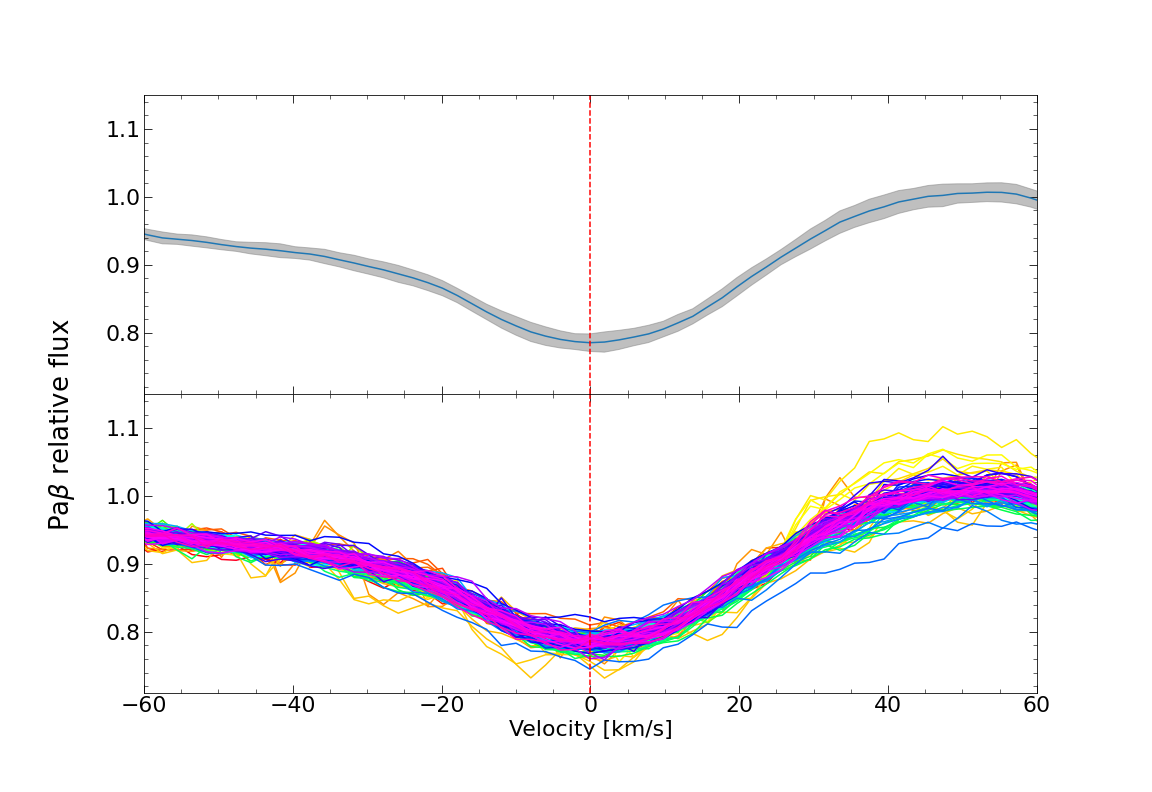}
         
    \end{subfigure}
    \begin{subfigure}{0.47\textwidth}
         \centering \hspace*{-0.6cm}
         \includegraphics[scale=0.25,trim={1cm 0cm 0cm 0cm},clip]{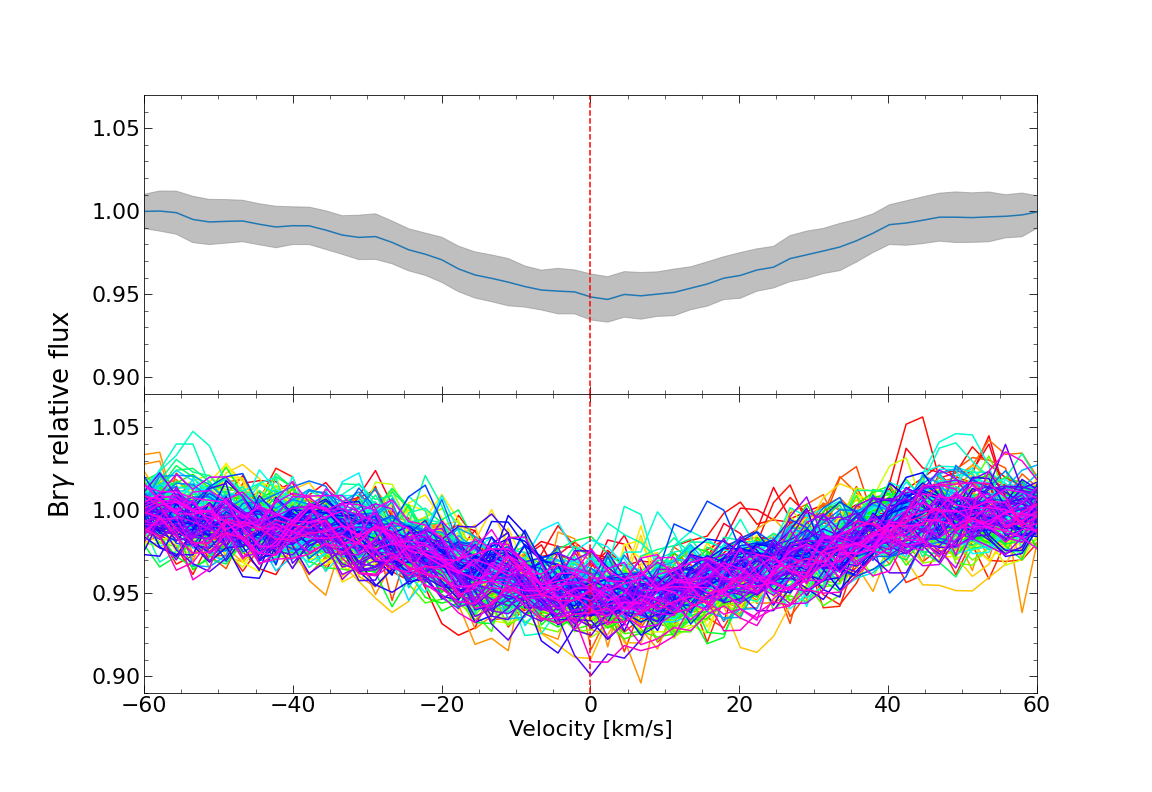}
         
    \end{subfigure}
    \caption{\coaut{Spectra of V1298~Tau for the \ion{He}{I} triplet (top), Pa$\beta$ (middle) and Br$\gamma$ (bottom). For each panel, the top curve shows the median profile with the dispersion in each velocity bin (in light grey) while the bottom curves depict the superposition of all individual telluric-corrected spectra. The red vertical dashed lines show the location of the lines. \last{The blue wing of Pa$\beta$ is likely blended with a \ion{Ca}{i} line.}} }
    \label{fig:median_spectra}
\end{figure}

\coaut{We focussed on three lines known to probe stellar activity in the NIR domain, namely the \ion{He}{I} triplet at 1083~nm, and the hydrogen lines Paschen beta (Pa$\beta$) at 1282~nm and Brackett gamma (Br$\gamma$) at 2165~nm \citep{zirin82,short98}. We show the individual spectra and the median one in Fig.~\ref{fig:median_spectra}. \last{We note that the extreme blue wing of Pa$\beta$ does not reach the continuum because of the nearby \ion{Ca}{i} line that blends with it. However, this feature does not vary more than the continuum and should thereferore have no significant impact on our analysis. }

We quantified the variations of the lines with respect to the median computing the equivalent width variations (EWVs) as done for several other young stars (e.g. \citealt{finociety21,finociety23,donati23}). In practice, we divided each telluric-corrected spectra by the median one and we computed the equivalent width of the residual spectra by integrating \last{between $-40$ and $+40$~\kms\ in the stellar rest frame}. \last{The EWVs show a dispersion of 2.70, 1.26 and 1.15~\kms, for the \ion{He}{i}, Pa$\beta$ and Br$\gamma$ lines, respectively.} We find that these indices are only weakly correlated with the stellar rotation. Fitting the EWVs of each line using a QP GP, we note that only the \ion{He}{I} EWVs are rotationnaly modulated with a period $P=2.907 \pm 0.018$~d, close to the values found with our $B_\ell$ and RV measurements. For this indicator, the excess of uncorrelated noise, tracing the intrinsic variability, is about $22\times$ larger than the formal photon noise error bar for each indicator (median of 0.10~\kms), reflecting the intense activity triggered by V1298~Tau, as already suggested by our RV fit. In particular, we note that the amplitude of this weak modulation (i.e. of the GP) is equal to 2.23~\kms\, i.e. about 65\% that of the excess of white noise. }

\section{Summary and \ben{Discussion}}
\label{sec:summary}

We presented the analysis of NIR spectropolarimetric \benjamin{observations of V1298~Tau} collected with SPIRou between 2019 Oct 02 and 2023 Feb 06, \benjamin{aimed at characterizing the large-scale magnetic topology of the host star and at further constraining the masses and densities of the 4 transiting planets.}

\subsection{Evolution of the magnetic field of V1298~Tau}

Our Stokes~$V$ LSD profiles show clear Zeeman signatures demonstrating that V1298~Tau hosts a strong large-scale magnetic field. The longitudinal field $B_\ell$, tracing this large-scale component, is modulated by the stellar rotation with a period of $2.910\pm0.005$~d, \jf{consistent with, though much more accurate than the} estimate of $2.91\pm0.05$~d inferred from RV measurements by \cite{suarez-mascareno21}. $B_\ell$ evolves rapidly, over a time-scale of $36\pm5$~d (the \jf{full} amplitude of the pattern varying from $\unsim30$~G in 2020 to $\unsim130$~G in 2021), suggesting that the large-scale magnetic topology changes both within an observing season and between two consecutive seasons. 

Given the fast evolution of $B_\ell$, we split our data into several subsets per season on which ZDI was applied independently to reconstruct the large-scale magnetic topology of V1298~Tau.
The large-scale magnetic topology of V1298~Tau is found to be mainly poloidal and axisymmetric with a toroidal component becoming less axisymmetric from 2020 to 2023. Comparing the magnetic reconstructions of this star with those of \coaut{two young solar-mass stars similar to V1298~Tau, namely} AB~Dor (e.g. \citealt{donati03}) and LQ~Hya (e.g. \citealt{lehtinen22}), we find that V1298~Tau hosts a simpler large-scale magnetic field, with a mostly axisymmetric radial field mainly characterized by the dipole component of the poloidal field unlike both other stars. All three stars also continuously show azimuthal features at their surface, that \jf{most likely} indicate that the underlying dynamo processes operate throughout the entire convective zone of these stars, as suggested by \cite{donati03}. In addition, ZDI reconstructions from our latest \coaut{observing epoch} show a rapid change in the radial field, with the dipole component (of the poloidal field) becoming much more tilted in early 2023 (from 20--30$^\circ$ to 75$^\circ$), \jf{and possibly tracing} a polarity reversal, as \jf{that recently reported} for LQ~Hya \citep{lehtinen22}. These changes \jf{suggest} that the magnetic field of V1298~Tau follows a long-term evolution \jf{potentially similar} to the 11-year solar cycle. 

Part of the observed magnetic variability is likely \jf{attributable} to DR at the surface of the star, found to be equal to \coaut{$82.0\pm2.0$~$\rm mrad\,d^{-1}$, i.e. $\unsim1.5$ times that of the Sun. This value implies a typical lap time of $\unsim76$~d for the equator to lap the pole by one rotation}, implying that DR clearly distorts the regular modulation of spectropolarimetric data over a typical observing season spanning about 4.5 months (i.e. $\unsim135$~d, see Table~\ref{tab:log_ZDI}). This level of DR is similar to that of young solar-mass stars whose stellar structure resembles that of V1298 Tau, such as AB Dor \jf{(d$\Omega_{\rm AB Dor}$ of the order of $70\,\mathrm{mrad\,d^{-1}}$; \citealt{jeffers07})} and LQ~Hya (d$\Omega_{\rm LQ Hya}$ of the order of $100\,\mathrm{mrad\,d^{-1}}$; \citealt{donati03}).
In addition, our results suggest that the DR at the surface of V1298~Tau may be varying with time, \coaut{passing from $90.6\pm2.8$~$\rm mrad\,d^{-1}$ in 2020 Aug - 2021 Jan to $73.5\pm2.8$~$\rm mrad\,d^{-1}$ in 2021 Sep - 2022 Feb.} This variation of $\unsim 15\,\mathrm{mrad\,d^{-1}}$ is larger than the one observed for the Sun ($<1~\mathrm{mrad\,d^{-1}}$) but is in good agreement with the observations of AB~Dor and LQ~Hya for which large variations of DR have also been reported from year to year, with, in particular, \jf{d$\Omega_{\rm AB Dor}$ ranging from 60 to 75~$\rm mrad\,d^{-1}$ \citep{jeffers07}} and d$\Omega_{\rm LQ Hya}$ ranging from 14 to 200~$\rm mrad\,d^{-1}$ \citep{donati03,kovari04,lehtinen22}. The global increase in the strength of the magnetic field of V1298~Tau between 2020 Aug - 2021 Jan and 2021 Sep - 2022 Feb (see the evolution of <$B_V$> in Sec.~\ref{sec:zdi}) may help to slow down the DR and may therefore contribute to the observed decrease in d$\Omega$ between both seasons. \coaut{The DR parameters of the Sun vary periodically with a period corresponding to that of the solar cycle and this peridocity is correlated with solar activity \citep{poljanvic22}. However, with only two measurements of DR at the surface of V1298~Tau so far and no reported magnetic cycle, we cannot conclude whether DR parameters for V1298~Tau follow the same trend as those of the Sun.}

We used, for the first time on actual data, the new tomographic method TIMeS \citep{finociety22} allowing one to reconstruct the long-term temporal evolution of magnetic fields without splitting the whole data set. We fitted all the Stokes~$V$ LSD profiles down to $\chi^2_r=1.49$, suggesting that discrepancies still exist between the synthetic model and the observed data. This is not surprising as \cite{finociety22} mentioned in the original paper that the method is likely to miss the smallest features that only contribute little to the observed data. One major difference between TIMeS and ZDI is that the model derived with TIMeS only uses some spherical harmonics modes that strongly contribute to the magnetic topology over the whole time span of the observations to describe the magnetic field while the model derived with ZDI uses all spherical harmonic modes up to a maximum degree $\ell_{\rm max}$. We therefore suppose that the TIMeS method failed at fitting the data down to a unit $\chi^2_r$ because some high degree modes do not have a strong contribution to the magnetic field over all epochs (e.g. small features that may have a small lifetime) and they were therefore not selected for the model (only modes up to $\ell=4$ were indeed identified by the method). Future improvements of the method to increase the flexibility in the choice of the modes should help to reduce these discrepancies \coaut{(e.g. through new criteria of selection, implementation of DR in the model)}, but the results we obtained with this first application on real data are already quite promising. We indeed find an evolution of the magnetic topology with TIMeS similar to that reconstructed from the application of ZDI on data subsets. In addition, TIMeS allows one to access to the temporal evolution of the large-scale field and its decomposition into poloidal and toroidal components. Comparing the fraction of reconstructed magnetic energy enclosed in the poloidal field derived from ZDI and TIMeS, we see that we have a global consistency, except at some epochs \jf{at the beginning or end of a season} \coaut{(typically 2021 Oct, 2023 Jan)}. %We also note that the magnetic field reconstructed with TIMeS becomes mainly toroidal in 2023 February, which may announce \jf{a global polarity switch as it does on the Sun}. We nevertheless caution that this result may be an edge effect \jf{and that further} spectropolarimetric observations of this star are needed to \jf{confirm this conclusion}. 
To investigate whether we can distinguish a magnetic cycle within our data set, we computed periodograms of the coefficients reconstructed with TIMeS. We however do not see any significant period shared by all coefficients, and therefore conclude that, if this star has a magnetic cycle, it must be longer than the time span of our 2020-2023 observations (897~d, i.e. $\unsim2.5$~years), as already suggested by the fact that only the beginning of a potential reversal has been identified in our ZDI reconstruction. Thorough spectropolarimetric monitorings of V1298~Tau are thus clearly needed in the \jf{coming} years, \jf{to} help us identify whether the dynamo processes at work in this star are similar to \benjamin{those operating in} the Sun (e.g. $\alpha \Omega$ processes) or more complex \jf{ones} (e.g. $\alpha^2$, $\alpha^2 \Omega$ processes, see \citealt{rincon19} for a review). %\coaut{In addition, although none of the 4 planets are rocky, there might be unknown Earth-like planets in the system, and the potential magnetic cycle of V1298~Tau may have consequences for the habitability of the planets (e.g., \citealt{gallet17}).}

\subsection{Constraining the planet parameters}

Using the LBL method \citep{artigau22}, we computed 174 accurate RV values, over the 3.5 years of SPIRou monitoring, from which we estimated the mass and orbital period of the outermost planet (e) and upper limits on the mass of the three innermost planets. 
Assuming circular orbits, our SPIRou RV measurements yield 4 probable \jf{orbital periods for planet e} at $42.7\pm0.3$~d, $46.4\pm0.4$~d, $53.5\pm0.4$~d and $60.6\pm0.7$~d. Computing the $\log{\mathcal{L_M}}$ associated with each model, we find that, among these solutions, the most likely model is the one featuring $P_e=53.5\pm0.4$~d. We note that our best value disagrees with those found found by \cite{sikora23} and \citet{turrini23}, equal to \jf{$46.768131 \pm 000076$~d} and $45.46^{+1.10}_{-0.23}$~d, respectively, but these values remain compatible with one of our less likely solutions ($46.4\pm0.3$~d) within $1\sigma$ and $2.4\sigma$, respectively. Although the log BF suggests that the period $P_e=42.7\pm0.3$~d is the second most likely, we consider it to be less probable as this value corresponds to the lower limit provided by the TESS monitoring \citep{feinstein22}. We also consider that our largest period ($60.6\pm0.7$~d) is also less probable given the log BF with respect to our best model ($\log{\rm BF} =-4.1$) and the most plausible interval ($43.3 < P_e/\mathrm{d}<55.4$) derived by \cite{sikora23}.

Using our most likely model, i.e. the one featuring a period of 53.5~d, we obtain the most significant detection of the RV signature of planet e since the new constraints on its orbital period provided by the TESS monitoring. We indeed report a 3.9$\sigma$ detection of the RV signature of planet e, yielding a mass of $1.01^{+0.35}_{-0.26}$~\mjup. This value is intermediate between those obtained by \cite{sikora23}, \cite{suarez-mascareno21} and \cite{turrini23}, equal to $0.66\pm0.26$, $1.16\pm0.30$~\mjup\ and $1.23^{+0.35}_{-0.48}$~\mjup, respectively, but still compatible within $1\sigma$ with all values. From our mass value and the updated radius of planet e provided by \cite{feinstein22}, we estimate the density of this planet to be equal to $1.79^{+0.66}_{-0.51}$~\gcm, slightly lower than those given in \cite{feinstein22} and \cite{turrini23}, using similar planet radius, but still compatible within $1\sigma$. As already noted by previous studies, this density is suprisingly high for a young planet, the implications of this value being discussed in the next Section. 

The RV signatures of the three innermost planets are undetected from our SPIRou measurements. We therefore estimate the following upper limits on these masses, based on the 99\% confidence interval of their posterior distributions sampled through the MCMC approach: $M_b<0.44$~\mjup, $M_c<0.22$~\mjup\ and $M_d<0.25$~\mjup. These results illustrate the need for (i) powerful filtering methods to get rid of activity jitter in RV measurements to clearly unveil the planet properties and (ii) more observations to better sample the RV signature induced by each planet of this system. 

\last{Our results are generally consistent with those provided by \cite{sikora23} but less so with those of \cite{suarez-mascareno21}. We outline that the masses derived by \cite{suarez-mascareno21} may be unreliable as a result of a potential overfit of their RV data as recently suggested by \cite{blunt23}. Although our results do not allow us to strongly constrain the mass of the three innermost planets, we note that our estimate of $K_b=4.2\pm10.7$~\ms\ disagrees at a $\unsim3\sigma$ level with the value of $41\pm12$~\ms\ previously reported in \cite{suarez-mascareno21}, suggesting that planet b may be several times less massive. This result is further strengthened by the upper limit of 0.08~\mjup\ provided by the HST transmission spectra of V1298~Tau~b (Barat et al., submitted, 2023), fully consistent with our estimate of $M_b=0.07\pm0.17$~\mjup.} %In addition, the upper limit we derived on the mass of this planet is slightly lower than the upper limit of 0.50~\mjup\ estimated by \cite{sikora23} and therefore tends to confirm that planet b is lighter than first reported by \cite{suarez-mascareno21}. 

\begin{figure*}
    \centering
    \includegraphics[scale=0.5,trim={2.cm 4cm 2cm 4.cm},clip]{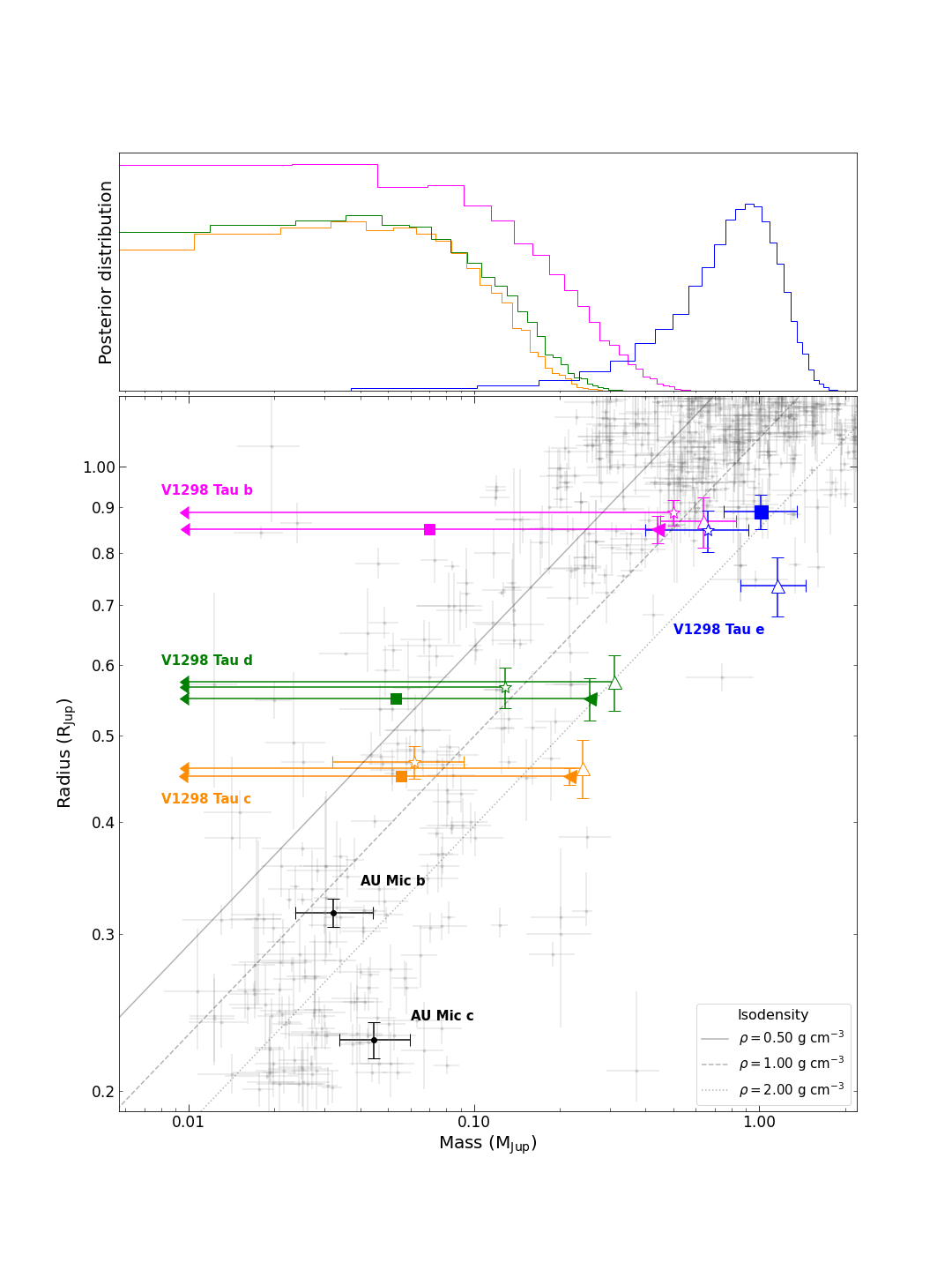}
    \caption{Mass-radius diagram. \benjamin{\textit{Top panel}: Posterior distributions of the planet masses derived from the MCMC approach. V1298 Tau b, c, d and e are shown in magenta, orange, green and blue, respectively. \textit{Bottom panel}: The light grey dots correspond to confirmed exoplanets with radius and masses known with a precision of 30\% or lower. The full squares show the position of the 4 planets according to the mass and error bars derived in the present paper (assuming the radii of \citealt{feinstein22}), while the left-pointing triangles depict the upper limit associated with the 99\% confidence level. The open up-pointing triangles show the estimates and $3\sigma$ upper limits from \citet{suarez-mascareno21} while the open stars correspond to recent estimates and $2\sigma$ upper limits from \citet{sikora23}. Upper limits are identified by an arrow pointing to the left. The same color code as for the upper panel is used to identify the 4 planets. For comparison, the black dots correspond to the position of the two young planets AU Mic b and c from the estimates of \citet{donati23}. We also plot 3 isodensities as thin grey lines, corresponding to a density of 0.50 (solid), 1.00 (dashed) and 2.00 (dotted) \gcm.}}
    \label{fig:mass_radius_diagram}
\end{figure*}

Our estimate of planet c's mass \last{($M_c=0.05\pm0.07$~\mjup)} is similar to the estimate of $19.8^{+9.3}_{-8.9}$~$M_\oplus$ \coaut{(i.e. $0.062^{+0.029}_{-0.028}$~\mjup)} derived by \cite{sikora23}, but with error bars $2.3\times$ larger, both results being fully consistent within $1\sigma$. Our RV data does not allow us to bring stronger constraints on the mass of planet d with respect to the previous studies, our upper limit being $\unsim1.6\times$ larger than that of \cite{sikora23} but still $\unsim1.6\times$ lower than that of \cite{suarez-mascareno21}.
The derived upper mass limits yield \ben{upper limits for the planet densities} of $\rho_b<0.89$~\gcm, $\rho_c<2.94$~\gcm\ and $\rho_d<1.89$~\gcm. Future observations of V1298~Tau will help to refine these densities and suggest what their internal structures could be.
To compare the results from both previous velocimetric studies and ours, we show the position of the four planets in a mass-radius diagram in Fig.~\ref{fig:mass_radius_diagram}. 

We further confirm that the assumption of circular orbits is best adapted to model our RV data, in particular for planet e for which we derive an eccentricity consistent with 0 (with an error bar of 0.05).   

Using the transit times of planet e derived from the K2 and TESS light curves \cite{feinstein22} listed 17 probable values for $P_e$ with a precision of 0.0001~d. We therefore used these values as a prior on $P_e$ for the modeling of our RV data. As outlined in Sec.~\ref{sec:contraints_feinstein}, the models associated with these periods can be sorted according to the value of their $\log{\mathcal{L_M}}$. As expected from our main analysis, we find that, among these 17 periods, $P_e=53.0039\pm0.0001$~d is the most likely, this value being the closest to our best period and consistent within 1.3$\sigma$. Three periods yield a $\log{\mathcal{L_M}}$ slightly smaller than that of the model featuring $P_e=53.0039$~d (i.e. $|\Delta \log{\mathcal{L_M}}|<2.5$). Among these 3 periods, one remains close to 53.5~d (i.e. $P_e=54.2085$~d) while the 2 others ($P_e=46.7681$ and $45.8687$~d) are in good agreement with one of the other solutions of our main analysis, i.e. $P_e=46.4\pm0.4$. These results thus strengthen our \ben{initial conclusion} of an orbital period close to 53~d for the outermost planet of the V1298~Tau system. Future photometric transits of planet e are needed to further pinpoint whether its actual orbital period is $\unsim46$ or $\unsim53$~d.% However, given the log BF we computed for each of the 17 values, we find that we cannot firmly rule out any of these periods. We nevertheless note that 7 of them are much less likely than the others ($|\log{\rm BF}|>5$), with, in particular, the mass of planet e being greatly underestimated ($\leq 0.25$~\mjup) with respect to that derived by our best models and previous studies \citep{suarez-mascareno21,sikora23,turrini23}.

With an orbital period of $P_e=53.0039$~d, the RV semi-amplitude of planet e is found to be equal to $K_e=44^{+15}_{-11}$~\ms, corresponding to a mass and density of $0.95^{+0.33}_{-0.24}$~\mjup\ and $\rho_e=1.66^{+0.61}_{-0.48}$~\gcm, respectively. These values are therefore fully consistent with and only sligthly lower than those derived \ben{in} our main analysis, as expected from the \ben{small} difference in the orbital period of planet e between both analyses. 
  
\subsection{Formation and evolution of the V1298 Tau system}

According to the orbital periods of the four planets, the V1298~Tau system is likely not in a resonant chain \ben{with only planet b, c and d having respective period ratios close to 3:2 (for c vs d) and 2:1 (for d vs b)}. A resonant chain would indeed favour an orbital period of planet e close to 48 or 49~d rather than 53~d (or even 46~d), which seems unlikely given our data (see Tables~\ref{tab:RV_results_feinstein_less_likely} and \ref{tab:RV_results_feinstein_not_likely}). Our finding is consistent with previous studies focussing on the formation of this planetary system \citep{tejada22,turrini23}. This implies that either the V1298~Tau system did not form in a resonant chain, or if it did, it must have undergone a rapid phase of dynamical instabilities that broke the resonant chain after the disc dispersal (i.e. in 10-15~Myr given the young age of V1298~Tau). 

%We caution that there is apparently no reason for the system to be formed in a resonant chain with a strict commensurability of the planet orbital periods, as several processes occuring in the circumstellar disc may affect the evolution of the planets (e.g. planet-disc interactions ; \citealt{baruteau13}). 
\ben{Although theoretical models indicate that disc migration in a multiplanetary system tends to favour the planets to be in a resonant chain, several processes, such as planet-disc interactions, may prevent the formation of such chains (e.g. interaction between a planet and the wake of another one; \citealt{baruteau13}). We also outline that the turbulence of the circumstellar disc can play a role on the disc-planet interactions that eventually disrupt the strict mean motion resonance between the planets \citep{pierens11,paardekooper13}.}
\cite{turrini23} discussed several mechanisms that may break a resonant chain after its formation. The first one is the eccentricity damping through tidal dissipation \citep{batygin13} but this scenario seems unlikely as the timescale on which it would occur for planet c is 2 orders of magnitude larger than the age of the system \citep{turrini23}. Photoevaporation of planets c and d may also contribute to the break of resonance \citep{goldberg22} but, once again, such a phenomenon would occur over more than 100~Myr, inconsistent with the age of the star \citep{turrini23}. More plausible explanations involve planetesimal scattering or planet-planet scattering. However, planetesimal scattering seems unlikely for this system as simulations show that a planetesimal belt of 50~$M_\oplus$ would not have a significant impact on the V1298~Tau system  and a much more massive belt would be required to break the resonance \citep{turrini23}. \cite{turrini21} suggest that the most likely scenario to explain the non-resonance between planets b and e relates to planet-planet scattering, involving the presence of one or more planets on an eccentric orbit beyond the orbit of planet e. We however consider that this scenario is quite unlikely given that the eccentricity of planet e we derived from our data is consistent with 0. In addition, such additional planets would be difficult to detect through photometric transits and RV monitoring due to their long orbital periods. In particular, no hint of a fifth planet in this system has been reported so far and, using GAIA and SPHERE observations, \cite{turrini23} indicate that no planet more massive than $2$~\mjup\ should orbit beyond 50~au from the host star (i.e. with an orbital period longer than 300~yr). \last{From simulations, we also estimated that our data excludes the presence of planets more massive than 2.7~\mjup\ and 5.9~\mjup\ orbiting at 0.7 and 1.4~au (i.e. corresponding to orbital periods of 6 months or 1.5~yr), respectively, in good agreement with the results of \cite{turrini23} for similar orbital periods. }  %As a result, planet-planet scattering, involving the presence of one or more planets on an eccentric orbit beyond the orbit of planet e, is apparently the most likely scenario. Such planets will be difficult to detect through photometric transits and RV monitorings due to their long orbital periods. In particular,%, we do not detect the RV signature of additional planets in our data (see Sec.~\ref{sec:planet_masses}), suggesting that a fifth planet, if any, should have period longer than $\unsim3.3$~yr (i.e. at a distance larger than 2.3~ua).
% \cite{turrini23} only exclude the presence of planets more massive than $2$~\mjup\ beyond 50~ua (i.e. orbital period longer than 300~yr).

According to our best model, planet e has a suprisingly high density given its young age\footnote{\ben{Assuming solar metallicity and no irradiation effects, the models of \cite{baraffe08} predict that a 1~\mjup\ planet of 10-15~Myr orbiting around a solar-mass star should have a radius of $\unsim1.3$~\rjup, implying a mean density of about $0.56$~\gcm\ (i.e. $3.2\times$ smaller than the one we derived for V1298~Tau~e). Accounting for the irradation from a Sun at 0.045~au, the predicted density is even smaller, close to 0.45~\gcm.}}. In particular, the mass and density we derived for this planet are consistent with the plausible composition inferred by \cite{sikora23} from models of \cite{fortney07} accounting for the irradiation from a Sun-like host star at a distance of 0.1~au\footnote{The semi-major axis of the planets hosted by V1298~Tau are \ben{equal to $0.172\pm0.003$, $0.084\pm0.001$, $0.110\pm0.002$ and $0.293\pm0.005$~au, for planet b, c, d and e, respectively, assuming the orbital periods of \cite{feinstein22} for the 3 inner planets and an orbital period of $53.5\pm0.4$~d for planet e. Assuming an orbital period of $53.0039\pm0.0001$~d for planet e yields a consistent semi-major axis of $0.291\pm0.005$~au.}}, i.e. a 100~$M_\oplus$ core composed of a 50/50 by mass ice-rock mixture surrounded by a H/He envelope. In addition, even though our derived mass is slightly lower than the one inferred from their recent study, \cite{turrini23} suggest that \ben{the ice-rock core may be more massive than 100~$M_\oplus$ with about 2/3 of planet e composed of heavy elements and 1/3 of hydrogen and helium. %planet e may contain up to 110~$M_\oplus$ of heavy elements, i.e. about 35\% of the total mass (using the value derived in our paper). 
This would mean a relative core mass $\unsim6\times$ larger than that of Jupiter according to formation models \citep{venturini20}, i.e. a planet composition closer to that of Neptune than that of Jupiter.} This result is in good agreement with the models of \cite{baraffe08}, indicating that planet e should be composed of more than 50\% of heavy elements (up to 90\%) to match the measured radius (0.89~\rjup) at this young an age (10-15~Myr). \benjamin{However}, these models predict that, assuming a metallicity of 0.10 (as in the host star; \citealt{suarez-mascareno21}), the planet should be \ben{older} than 10~Gyr given the measured radius, which is clearly unrealistic.

To date, the \ben{cause of such a high density of planet e} is still speculative. One explanation \ben{relates to the critical mass for the onset of runaway gas accretion onto a protoplanetary core.} In the case of planet e, this mass would be larger than that generally admitted (10 to 20~$M_\oplus$) possibly due to uncertainties on the hypotheses on which models rely such as the size, mass and composition of the dust, having a direct impact on the opacity and on the cooling of the gas that can be accreted by the planet. Alternative explanations for the high density of V1298~Tau~e may be related to a high accretion \ben{rate} of pebbles, enriching the content of heavy elements of the planet, through different mechanisms. \ben{One generally assumes that a planet on a circular orbit reaches its pebble isolation mass (PIM, i.e. the mass above which the planet cannot accrete more pebbles) for $M$~$\unsim10 - 20~$~$M_\oplus$ depending on the disc's aspect ratio around the planet \citep[e.g.,][]{ataiee18}}. However, in some conditions, a planet on an eccentric orbit can reach a much higher PIM ($>100$~$M_\oplus$ for moderate eccentricity, e.g. $e>0.04$, and a turbulent visosity $\alpha=0.005$; \citealt{chametla22}), which could be the case of V1298~Tau~e, according to the estimate we derived ($0.03\pm0.05$). The accretion of dust and gas through a vertical (or meridional) circulation \citep{binkert21,szulgayi22} is another process that can partly contribute to the increase in the heavy elements content in V1298~Tau~e even if the planet has reached its PIM, that deserves to be further investigated.

\cite{turrini23}, focussing on the evolution of the V1298~Tau system, also discussed other possible explanations for the high density observed for planet e. \ben{One is that this planet formed beyond the CO$_2$ snowline and then underwent a large-scale migration within a massive circumstellar disc in such a way to accrete heavy elements along its pathway (as probably for HAT-P-20 b; \citealt{thorngren16}), e.g. through the accretion of an intense planetesimal flux preventing runaway accretion of the disc gas to occur \citep{turrini23}.} As also noted by \cite{turrini23}, another option is that this density results from an accretion of vapor coming from an enhancement of the volatile content of the circumstellar disc due to drifting and evaporating pebbles when they reach their ice lines (e.g. \citealt{booth19,schneider21}). Characterizing the atmosphere of planet e through transit observations would help to discriminate between both hypotheses as the accretion of planetesimal is expected to yield a lower refractory-to-volatile ratio as well as a different C/O ratio than the second process \citep{turrini21,schneider21,pacetti22}. We however caution that these two processes \ben{may not be consistent with what we know of V1298~Tau}. In particular, the metallicity of the host star is close to solar ($0.10\pm0.15$; \citealt{suarez-mascareno21}), and does not hint at a high content of heavy elements in the circumstellar disc that would lead towards the second hypothesis.
Though much older ($\unsim1.5$~Gyr; \citealt{ment18}), HAT-P-2 b ($M=8.62^{+0.39}_{-0.55}$~\mjup, $R=0.951^{+0.039}_{-0.053}$~\rjup; \citealt{loeillet08}) is another example of a hot Jupiter with an even higher density ($\rho=12.5^{+2.6}_{-3.6}$~\gcm; \citealt{loeillet08}) that may be related to a large content of heavy elements (up to 500~$M_\oplus$; \citealt{baraffe08}). The most likely scenario to explain the formation of HAT-P-2~b is that the planet is actually the result of several collisions with planetary embryos \citep{baraffe08}. In this scenario, the absolute mass of heavy elements in the planet increases and the gas is ejected from the planet, therefore increasing the relative mass of heavy elements with respect to the initial mass of the planet. This mechanism may be an alternative explanation for the formation of V1298~Tau~e, which deserves to be also investigated in future studies.

\cite{maggio22} mentioned that the two outermost planets (e and b) should not be affected by a strong photoevaporation given their large mass with respect to their radius. However, this conclusion was drawn from the mass of planet b inferred by \cite{suarez-mascareno21}, \ben{which} is larger than the upper limit that we report in the current paper and that of \cite{sikora23}, \ben{suggesting that the conclusion may be revised taking into account our updated mass upper limit.} %, indicating that planet b may actually suffer from atmospheric mass loss, if its actual mass is confirmed to be much smaller than $0.64$~\mjup. 
In addition, \cite{turrini23} indicate that, during its formation, planet b may have undergone a migration with heavy element enrichment (as for planet e). However, this conclusion was also drawn from a density derived assuming the mass obtained by \cite{suarez-mascareno21} and thus likely overestimated and inconsistent with our data. In particular, \cite{turrini23} suggest that planet b may contain up to 90~$M_\oplus$ (i.e. $\unsim0.28$~\mjup) of heavy elements, \ben{representing 64\% of our mass upper limit (0.44~\mjup), which thus seems unlikely given our upper limit on the density of planet b ($<0.89$~\gcm).} %which seems very unlikely given the mass we obtained from our RV data ($M_b=0.07\pm0.17$~\mjup). 
The scenario for the formation of planet b can therefore be revised when the actual mass of this planet is further constrained.

According to \cite{maggio22}, planets c and d may undergo strong atmospheric evaporation if their masses is lower than $\unsim40$~$M_\oplus$ ($\unsim0.13$~\mjup). \ben{Our mass} estimates are thus in favour of strong atmospheric photoevaporation, strengthening the conclusion of \cite{sikora23} about planet c. \ben{The V1298~Tau system, hosting 4 planets younger than 30~Myr, is an ideal laboratory to study star-planet interactions at early stages of the evolution. Future monitoring of this target will therefore help us to better understand the impact of such interactions on the formation and evolution of young planets.}

\last{In a forthcoming work, we plan to couple all existing RV measurements (i.e. collected with HARPS-N, MAROON-X, SPIRou and HIRES; \citealt{suarez-mascareno21,sikora23,blunt23}) to perform a consistent analysis and hopefully to refine the planet masses. Such a study should include two potentially coupled GPs to simultaneously and consistently model the activity signals in the optical and NIR domains. In parallel, we will pursue our spectropolarimetric monitoring of V1298~Tau thanks to the SPICE Large Programme in the next semesters.} In light of the results reported in the present paper, these future observations will help us to fulfill two major goals. The first one is to reconstruct the large-scale magnetic topology of the star using ZDI and TIMeS from new data to better characterize the dynamo process at the origin of the observed field and its evolution (e.g. potential magnetic cycle subject to $\alpha \Omega$ processes). The second goal is obviously to refine estimates of the mass of the four planets, but also to firmly ascertain the orbital period of planet e, a quite \coaut{challenging} task given the amplitude of the stellar activity jitter more than $2.6\times$ larger than the semi-amplitude of the presumably \benjamin{most} massive planet (i.e. V1298~Tau~e). Obtaining accurate measurements of the mass of the four planets of this system is an \ben{essential} step to further constrain the mass-radius relation at early stages of evolution, as only \ben{few other planets aged of less than 30~Myr have precise measured radius and mass (AU Mic~b and c, HD~114082~b and HIP~67522~b; \citealt{donati23,zakhozhay22,rizzuto20})}, but also to better understand the evolution of young planets and the associated star-planet interaction.
Coupling spectroscopic and high-precision photometric observations (e.g. TESS, PLATO), especially during transit events, will clearly help us to refine the orbital parameters, the atmospheric escape, if any, (e.g. study of the \ion{He}{i} line during transit events) and the mass of the 4 planets. More generally, we would need multi-instruments campaigns (e.g. combination of optical/NIR spectropolarimeters, such as ESPaDOnS, SPIRou, CRIRES+) would provide huge benefits to better characterize the magnetic topology of V1298~Tau and the properties of its planetary system.

\section*{Acknowledgements}

This work includes data collected in the framework of the SPIRou Legacy Survey (SLS), an international large programme that was allocated on the Canada-France-Hawaii Telescope (CFHT) at the summit of Maunakea by the Institut National des Sciences de l’Univers of the Centre National de la Recherche Scientifique of France, the National Research Council of Canada, and the University of Hawaii. We acknowledge funding by the European Research Council (ERC) under the H2020 research \& innovation programme (grant agreement \#740651 NewWorlds). \coaut{SHPA acknowledges financial support from CNPq, CAPES and Fapemig. A.C. acknowledges funding from the French ANR under contract number ANR\-18\-CE31\-0019 (SPlaSH). This work is supported by the French National Research Agency in the framework of the Investissements d'Avenir program (ANR-15-IDEX-02), through the funding of the ``Origin of Life" project of the Grenoble-Alpes University.}

%%%%%%%%%%%%%%%%%%%%%%%%%%%%%%%%%%%%%%%%%%%%%%%%%%
\section*{Data Availability}

The data collected in the framework of the SLS will be publicly available from the Canadian Astronomy Data Center (CADC) by 2024, while the PI data are already public.



\bibliographystyle{mnras}
\bibliography{main} % if your bibtex file is called example.bib

%\begin{thebibliography}{99}
%\end{thebibliography}

%%%%%%%%%%%%%%%%%%%%%%%%%%%%%%%%%%%%%%%%%%%%%%%%%%

%%%%%%%%%%%%%%%%% APPENDICES %%%%%%%%%%%%%%%%%%%%%

\appendix

\section{Journals of observations}
\label{sec:log_journal}
We provide a full journal of the observations reduced with the \texttt{Libre-ESpRIT} pipeline adapted for SPIRou in Table~\ref{tab:log_ZDI} and the journals of observations reduced with \texttt{APERO} in Tables~\ref{tab:journal_apero}.% and \ref{tab:journal_apero_deltaT}.

\begin{table*} 
	\centering
	\caption{Spectropolarimetric observations collected between 2019 and 2023, reduced with the \texttt{Libre-ESpRIT} pipeline adapted for SPIRou data. For each observation, we list the date (day, hour), the barycentric Julian date (BJD), the rotation cycle (as defined in Sec.~\ref{sec:zdi}), the SNR in the $H$ band, the noise level in the LSD profile, the longitudinal field $B_\ell$ and the number of the subset used in ZDI (see Sec.~\ref{sec:zdi}).}
	\label{tab:log_ZDI}
	\begin{tabular}{cccccccc}
    \hline
     Date & UTC & BJD & Cycle & SNR & $\sigma_V$ & $B_\ell$ & \# ZDI \\ 
 &  & 2459000+ &  &  & ($10^{-4}\,I_c$) & (G) &  \\ \hline
2019 Oct 02 & 13:36:45 & -240.92942 & -82.794 & 160 & 3.17 & $-9.4 \pm 14.2$& 1 \\
2019 Oct 03 & 13:57:59 & -239.91458 & -82.445 & 161 & 2.85 & $2.0 \pm 13.2$& 1 \\
2019 Oct 05 & 11:03:45 & -238.03544 & -81.799 & 176 & 2.64 & $8.1 \pm 11.8$& 1 \\
2019 Oct 06 & 10:49:33 & -237.04522 & -81.459 & 184 & 2.72 & $-7.6 \pm 12.0$& 1 \\
2019 Oct 07 & 14:52:09 & -235.87667 & -81.057 & 139 & 3.75 & $51.0 \pm 16.3$& 1 \\
2019 Oct 08 & 09:24:54 & -235.10387 & -80.792 & 171 & 3.33 & $-6.6 \pm 14.1$& 1 \\
2019 Oct 09 & 12:11:43 & -233.98794 & -80.408 & 164 & 2.99 & $1.4 \pm 13.1$& 1 \\
2019 Oct 12 & 11:46:45 & -231.00506 & -79.383 & 170 & 2.68 & $7.0 \pm 12.0$& 1 \\
2019 Oct 13 & 14:06:44 & -229.90778 & -79.006 & 170 & 2.77 & $88.8 \pm 12.5$& 1 \\
2019 Oct 14 & 11:09:21 & -229.03090 & -78.705 & 170 & 2.85 & $-55.9 \pm 12.6$& 1 \\
2019 Oct 15 & 12:57:33 & -227.95569 & -78.335 & 161 & 3.30 & $-33.0 \pm 14.3$& 1 \\
2019 Oct 16 & 11:07:36 & -227.03199 & -78.018 & 185 & 2.85 & $47.2 \pm 12.6$& 1 \\
2019 Oct 31 & 10:50:51 & -212.04282 & -72.867 & 139 & 4.01 & $10.5 \pm 17.4$& 2 \\
2019 Nov 01 & 11:04:03 & -211.03361 & -72.520 & 177 & 3.05 & $15.2 \pm 13.3$& 2 \\
2019 Nov 03 & 10:19:43 & -209.06433 & -71.843 & 176 & 2.92 & $-19.9 \pm 12.7$& 2 \\
2019 Nov 04 & 11:29:17 & -208.01597 & -71.483 & 123 & 4.53 & $34.2 \pm 19.4$& 2 \\
2019 Nov 05 & 13:44:58 & -206.92171 & -71.107 & 162 & 3.22 & $-21.6 \pm 14.2$& 2 \\
2019 Nov 07 & 10:27:08 & -205.05904 & -70.467 & 144 & 4.25 & $-1.1 \pm 17.9$& 2 \\
2019 Nov 08 & 12:25:03 & -203.97711 & -70.095 & 159 & 3.34 & $2.8 \pm 14.5$& 2 \\
2019 Nov 09 & 09:37:13 & -203.09364 & -69.792 & 139 & 4.27 & $-30.8 \pm 18.0$& 2 \\
2019 Nov 10 & 10:19:47 & -202.06405 & -69.438 & 156 & 3.93 & $23.7 \pm 16.5$& 2 \\
2019 Nov 11 & 10:06:46 & -201.07307 & -69.097 & 165 & 3.48 & $-10.9 \pm 14.8$& 2 \\
2019 Nov 13 & 09:12:58 & -199.11039 & -68.423 & 170 & 3.26 & $7.2 \pm 14.0$& 2 \\
2019 Nov 14 & 09:36:55 & -198.09374 & -68.073 & 171 & 2.99 & $-40.4 \pm 13.1$& 2 \\
2019 Dec 07 & 09:52:47 & -175.08273 & -60.166 & 80 & 7.52 & $-3.3 \pm 32.2$& 2 \\
2019 Dec 08 & 08:47:45 & -174.12792 & -59.838 & 145 & 3.71 & $12.7 \pm 16.0$& 2 \\
2019 Dec 09 & 08:59:53 & -173.11951 & -59.491 & 157 & 3.13 & $18.4 \pm 13.8$& 2 \\
2019 Dec 10 & 07:53:49 & -172.16542 & -59.163 & 181 & 2.68 & $-4.7 \pm 11.7$& 2 \\
2019 Dec 11 & 10:42:40 & -171.04819 & -58.779 & 145 & 3.61 & $6.4 \pm 15.3$& 2 \\
2019 Dec 12 & 09:00:04 & -170.11947 & -58.460 & 151 & 2.84 & $25.2 \pm 12.7$& 2 \\
2019 Dec 13 & 08:38:10 & -169.13471 & -58.122 & 107 & 5.89 & $-39.3 \pm 23.9$& 2 \\
2020 Feb 05 & 05:27:09 & -115.27106 & -39.612 & 161 & 2.51 & $9.9 \pm 11.1$& - \\
2020 Feb 06 & 05:54:59 & -114.25182 & -39.262 & 102 & 5.92 & $-4.6 \pm 25.1$& - \\
2020 Feb 16 & 05:23:19 & -104.27480 & -35.833 & 152 & 3.06 & $-32.6 \pm 13.5$& - \\
2020 Feb 17 & 05:30:02 & -103.27023 & -35.488 & 132 & 4.26 & $1.4 \pm 17.9$& - \\
2020 Feb 18 & 05:26:10 & -102.27301 & -35.145 & 175 & 2.78 & $-24.4 \pm 12.1$& - \\
2020 Feb 19 & 05:25:58 & -101.27325 & -34.802 & 180 & 2.63 & $-31.5 \pm 11.6$& - \\
2020 Aug 26 & 14:42:57 & 88.11322 & 30.279 & 197 & 2.34 & $-12.5 \pm 10.8$& 3 \\
2020 Aug 27 & 13:33:22 & 89.06498 & 30.607 & 234 & 1.92 & $-50.5 \pm 8.9$& 3 \\
2020 Aug 28 & 13:34:37 & 90.06596 & 30.951 & 221 & 1.80 & $-3.0 \pm 8.5$& 3 \\
2020 Aug 29 & 14:29:19 & 91.10404 & 31.307 & 212 & 1.92 & $-20.9 \pm 8.9$& 3 \\
2020 Aug 30 & 14:13:30 & 92.09315 & 31.647 & 224 & 1.94 & $-44.5 \pm 8.9$& 3 \\
2020 Aug 31 & 14:17:54 & 93.09630 & 31.992 & 194 & 1.84 & $15.5 \pm 8.6$& 3 \\
2020 Sep 01 & 14:29:26 & 94.10442 & 32.338 & 192 & 1.87 & $-45.8 \pm 8.9$& 3 \\
2020 Sep 02 & 14:39:14 & 95.11132 & 32.684 & 214 & 1.91 & $-58.8 \pm 8.9$& 3 \\
2020 Sep 03 & 14:34:19 & 96.10800 & 33.027 & 178 & 2.53 & $34.9 \pm 11.7$& 3 \\
2020 Sep 04 & 14:18:22 & 97.09702 & 33.367 & 211 & 2.21 & $-31.4 \pm 9.9$& 3 \\
2020 Sep 05 & 14:20:39 & 98.09870 & 33.711 & 212 & 2.03 & $-57.5 \pm 9.3$& 3 \\
2020 Sep 06 & 13:54:00 & 99.08029 & 34.048 & 215 & 1.94 & $15.2 \pm 9.0$& 3 \\
2020 Sep 07 & 15:21:37 & 100.14124 & 34.413 & 171 & 2.71 & $-46.7 \pm 12.5$& 3 \\
2020 Sep 08 & 13:49:35 & 101.07742 & 34.735 & 221 & 1.93 & $-60.8 \pm 8.9$& 3 \\
2020 Sep 09 & 13:24:07 & 102.05982 & 35.072 & 210 & 2.16 & $8.4 \pm 9.9$& 3 \\
2020 Sep 10 & 13:23:01 & 103.05916 & 35.416 & 218 & 1.94 & $-32.8 \pm 8.9$& 3 \\ \hline
\end{tabular}
\end{table*}

\begin{table*}  \ContinuedFloat
	\centering
	\caption{Continued}
	%\label{tab:example_table}
	\begin{tabular}{cccccccc}
    \hline
Date & UTC & BJD & Cycle & SNR & $\sigma_V$ & $B_\ell$ & \# ZDI \\ 
 &  & 2459000+ &  &  & ($10^{-4}\,I_c$) & (G) &  \\ \hline
2020 Sep 18 & 12:36:29 & 111.02759 & 38.154 & 203 & 2.63 & $-2.4 \pm 11.4$& 4 \\
2020 Sep 19 & 13:35:35 & 112.06872 & 38.512 & 182 & 3.25 & $11.5 \pm 14.7$& 4 \\
2020 Sep 20 & 14:03:47 & 113.08839 & 38.862 & 226 & 1.82 & $-14.4 \pm 8.5$& 4 \\
2020 Sep 21 & 11:48:57 & 113.99484 & 39.173 & 192 & 2.27 & $0.7 \pm 10.5$& 4 \\
2020 Sep 22 & 12:15:13 & 115.01317 & 39.523 & 220 & 2.12 & $-9.2 \pm 9.5$& 4 \\
2020 Sep 23 & 13:14:26 & 116.05438 & 39.881 & 206 & 2.15 & $-6.2 \pm 9.8$& 4 \\
2020 Sep 25 & 12:40:16 & 118.03083 & 40.560 & 239 & 1.82 & $-6.0 \pm 8.4$& 4 \\
2020 Sep 26 & 13:12:27 & 119.05327 & 40.912 & 232 & 1.73 & $-4.9 \pm 8.2$& 4 \\
2020 Sep 27 & 12:34:52 & 120.02724 & 41.246 & 226 & 1.90 & $-5.3 \pm 8.8$& 4 \\
2020 Sep 28 & 11:02:51 & 120.96342 & 41.568 & 234 & 1.88 & $-2.7 \pm 8.7$& 4 \\
2020 Sep 29 & 13:29:38 & 122.06545 & 41.947 & 232 & 1.90 & $3.2 \pm 8.8$& 4 \\
2020 Sep 30 & 10:52:30 & 122.95640 & 42.253 & 232 & 2.00 & $-24.8 \pm 8.9$& 4 \\
2020 Oct 01 & 13:12:29 & 124.05370 & 42.630 & 184 & 2.69 & $8.6 \pm 11.9$& 4 \\
2020 Oct 03 & 14:46:34 & 126.11920 & 43.340 & 220 & 2.00 & $-24.3 \pm 9.2$& 4 \\
2020 Oct 04 & 11:06:44 & 126.96660 & 43.631 & 226 & 2.08 & $1.6 \pm 9.4$& 4 \\
2020 Oct 05 & 10:28:17 & 127.93997 & 43.966 & 236 & 2.02 & $-12.7 \pm 9.1$& 4 \\
2020 Oct 07 & 10:17:42 & 129.93278 & 44.650 & 217 & 2.06 & $-6.3 \pm 9.3$& 4 \\
2020 Oct 08 & 13:07:34 & 131.05082 & 45.035 & 209 & 2.22 & $-35.2 \pm 10.1$& 4 \\
2020 Oct 30 & 12:18:21 & 153.01791 & 52.583 & 221 & 1.93 & $33.9 \pm 8.9$& - \\
2020 Oct 31 & 12:50:06 & 154.04002 & 52.935 & 243 & 1.67 & $-5.1 \pm 7.9$& - \\
2020 Nov 01 & 09:56:44 & 154.91965 & 53.237 & 252 & 1.75 & $-22.5 \pm 8.0$& - \\
2020 Nov 03 & 10:47:20 & 156.95487 & 53.936 & 232 & 2.02 & $-9.7 \pm 9.1$& - \\
2020 Nov 04 & 10:53:09 & 157.95894 & 54.281 & 216 & 2.17 & $-22.6 \pm 9.8$& - \\
2020 Nov 05 & 09:00:19 & 158.88062 & 54.598 & 233 & 2.07 & $25.9 \pm 9.3$& - \\
2020 Dec 24 & 08:17:18 & 207.85031 & 71.426 & 209 & 2.04 & $-0.6 \pm 9.3$& 5 \\
2020 Dec 25 & 08:32:59 & 208.86115 & 71.774 & 222 & 2.07 & $17.7 \pm 9.3$& 5 \\
2020 Dec 26 & 07:58:47 & 209.83735 & 72.109 & 222 & 2.14 & $-65.3 \pm 9.7$& 5 \\
2020 Dec 28 & 08:59:53 & 211.87967 & 72.811 & 146 & 4.34 & $-3.2 \pm 19.2$& 5 \\
2020 Dec 29 & 08:32:00 & 212.86025 & 73.148 & 173 & 3.07 & $-54.4 \pm 13.5$& 5 \\
2020 Dec 30 & 08:25:44 & 213.85584 & 73.490 & 207 & 2.40 & $12.5 \pm 10.8$& 5 \\
2020 Dec 31 & 08:11:27 & 214.84586 & 73.830 & 203 & 2.47 & $-1.8 \pm 11.1$& 5 \\
2021 Jan 01 & 06:58:05 & 215.79486 & 74.156 & 198 & 2.51 & $-33.0 \pm 11.2$& 5 \\
2021 Jan 02 & 07:26:51 & 216.81477 & 74.507 & 210 & 2.04 & $22.3 \pm 9.2$& 5 \\
2021 Jan 03 & 07:58:28 & 217.83666 & 74.858 & 223 & 1.87 & $-2.8 \pm 8.7$& 5 \\
2021 Jan 04 & 07:08:44 & 218.80206 & 75.190 & 235 & 1.86 & $-31.4 \pm 8.4$& 5 \\
2021 Jan 05 & 07:41:19 & 219.82462 & 75.541 & 218 & 1.98 & $0.1 \pm 9.2$& 5 \\
2021 Jan 06 & 07:15:27 & 220.80659 & 75.879 & 221 & 2.05 & $-11.1 \pm 9.4$& 5 \\
2021 Jan 07 & 07:37:07 & 221.82157 & 76.227 & 211 & 2.22 & $-57.5 \pm 10.1$& 5 \\
2021 Jan 08 & 07:46:46 & 222.82820 & 76.573 & 206 & 2.17 & $-7.7 \pm 10.0$& 5 \\
2021 Sep 17 & 13:43:20 & 475.07388 & 163.256 & 234 & 2.07 & $-67.2 \pm 9.3$& - \\
2021 Sep 19 & 11:59:44 & 477.00211 & 163.918 & 223 & 1.89 & $-41.6 \pm 8.6$& - \\
2021 Sep 22 & 13:05:24 & 480.04798 & 164.965 & 215 & 1.82 & $0.7 \pm 8.5$& - \\
2021 Sep 23 & 13:30:54 & 481.06578 & 165.315 & 219 & 1.93 & $-64.0 \pm 8.8$& - \\
2021 Sep 24 & 13:30:41 & 482.06572 & 165.658 & 208 & 1.88 & $-62.9 \pm 8.7$& - \\
2021 Oct 14 & 11:50:19 & 501.99755 & 172.508 & 210 & 2.32 & $-14.0 \pm 10.3$& 6 \\
2021 Oct 15 & 11:28:00 & 502.98211 & 172.846 & 227 & 1.98 & $-43.3 \pm 9.1$& 6 \\
2021 Oct 16 & 12:21:26 & 504.01928 & 173.203 & 203 & 2.27 & $-32.3 \pm 10.3$& 6 \\
2021 Oct 19 & 13:05:35 & 507.05014 & 174.244 & 240 & 2.05 & $-22.8 \pm 9.1$& 6 \\
2021 Oct 20 & 11:45:53 & 507.99484 & 174.569 & 228 & 2.08 & $-25.1 \pm 9.3$& 6 \\
2021 Oct 21 & 12:06:24 & 509.00915 & 174.917 & 226 & 2.24 & $13.9 \pm 9.9$& 6 \\
2021 Oct 22 & 12:07:18 & 510.00982 & 175.261 & 168 & 2.78 & $-2.9 \pm 12.3$& 6 \\
2021 Oct 23 & 11:51:29 & 510.99890 & 175.601 & 248 & 1.87 & $-58.8 \pm 8.4$& 6 \\
2021 Oct 24 & 12:30:07 & 512.02578 & 175.954 & 177 & 2.88 & $22.3 \pm 12.7$& 6 \\
2021 Oct 25 & 09:28:49 & 512.89993 & 176.254 & 243 & 1.88 & $-22.4 \pm 8.6$& 6 \\
2021 Oct 26 & 11:13:30 & 513.97267 & 176.623 & 240 & 1.98 & $-60.9 \pm 8.9$& 6 \\
2021 Oct 27 & 11:57:19 & 515.00315 & 176.977 & 216 & 2.06 & $60.7 \pm 9.3$& 6 \\
2021 Oct 28 & 12:28:43 & 516.02500 & 177.328 & 196 & 2.30 & $-33.8 \pm 10.5$& 6 \\ \hline

\end{tabular}
\end{table*}

\begin{table*}  \ContinuedFloat
	\centering
	\caption{Continued}
	%\label{tab:example_table}
	\begin{tabular}{cccccccc}
    \hline
    Date & UTC & BJD & Cycle & SNR & $\sigma_V$ & $B_\ell$ & \# ZDI \\ 
 &  & 2459000+ &  &  & ($10^{-4}\,I_c$) & (G) &  \\ \hline
2021 Nov 19 & 10:35:48 & 537.94721 & 184.862 & 203 & 2.16 & $-35.1 \pm 9.7$& 7 \\
2021 Nov 20 & 10:11:48 & 538.93054 & 185.199 & 229 & 1.84 & $-22.5 \pm 8.4$& 7 \\
2021 Nov 21 & 10:18:53 & 539.93548 & 185.545 & 197 & 1.96 & $-48.1 \pm 9.0$& 7 \\
2021 Nov 22 & 10:53:29 & 540.95951 & 185.897 & 231 & 1.87 & $-15.3 \pm 8.6$& 7 \\
2021 Dec 09 & 09:20:44 & 557.89492 & 191.716 & 216 & 2.10 & $-85.6 \pm 9.5$& 7 \\
2021 Dec 10 & 08:45:09 & 558.87018 & 192.052 & 230 & 1.92 & $26.0 \pm 8.7$& 7 \\
2021 Dec 11 & 08:49:02 & 559.87286 & 192.396 & 206 & 2.37 & $-57.4 \pm 10.7$& 7 \\
2021 Dec 12 & 08:54:58 & 560.87694 & 192.741 & 199 & 2.36 & $-40.6 \pm 10.7$& 7 \\
2021 Dec 13 & 09:28:59 & 561.90053 & 193.093 & 158 & 3.26 & $12.2 \pm 14.5$& 7 \\
2021 Dec 14 & 07:38:37 & 562.82387 & 193.410 & 231 & 1.97 & $-28.8 \pm 9.0$& 7 \\
2021 Dec 16 & 11:02:16 & 564.96521 & 194.146 & 170 & 2.96 & $11.2 \pm 13.2$& 7 \\
2021 Dec 18 & 08:40:17 & 566.86653 & 194.799 & 175 & 2.87 & $-50.5 \pm 12.7$& 7 \\
2022 Jan 06 & 08:34:54 & 585.86176 & 201.327 & 235 & 2.30 & $-43.2 \pm 10.4$& 8 \\
2022 Jan 08 & 08:15:43 & 587.84829 & 202.010 & 238 & 1.89 & $44.3 \pm 8.7$& 8 \\
2022 Jan 09 & 08:17:20 & 588.84935 & 202.354 & 234 & 2.00 & $-57.5 \pm 9.1$& 8 \\
2022 Jan 10 & 09:26:45 & 589.89748 & 202.714 & 238 & 2.06 & $-71.5 \pm 9.5$& 8 \\
2022 Jan 12 & 08:52:09 & 591.87330 & 203.393 & 241 & 1.84 & $-75.1 \pm 8.4$& 8 \\
2022 Jan 13 & 08:20:54 & 592.85153 & 203.729 & 236 & 1.97 & $-43.5 \pm 8.6$& 8 \\
2022 Jan 14 & 06:59:34 & 593.79498 & 204.053 & 219 & 2.11 & $42.2 \pm 9.4$& 8 \\
2022 Jan 15 & 08:36:53 & 594.86247 & 204.420 & 247 & 1.87 & $-82.1 \pm 8.4$& 8 \\
2022 Jan 17 & 06:50:18 & 596.78831 & 205.082 & 209 & 2.05 & $34.6 \pm 9.2$& 8 \\
2022 Jan 18 & 07:58:19 & 597.83545 & 205.442 & 239 & 1.81 & $-92.9 \pm 8.3$& 8 \\
2022 Jan 19 & 07:40:35 & 598.82305 & 205.781 & 225 & 2.05 & $-18.6 \pm 9.3$& 8 \\
2022 Jan 20 & 07:22:44 & 599.81058 & 206.120 & 224 & 2.24 & $0.1 \pm 10.2$& 8 \\
2022 Jan 21 & 08:13:25 & 600.84569 & 206.476 & 239 & 2.23 & $-87.3 \pm 10.0$& 8 \\
2022 Jan 22 & 07:18:30 & 601.80746 & 206.807 & 165 & 3.07 & $-16.7 \pm 13.8$& 8 \\
2022 Jan 23 & 08:11:59 & 602.84451 & 207.163 & 193 & 3.12 & $-3.8 \pm 13.8$& 8 \\
2022 Jan 24 & 07:43:09 & 603.82441 & 207.500 & 240 & 1.83 & $-91.5 \pm 8.4$& 8 \\
2022 Jan 25 & 07:51:28 & 604.83009 & 207.845 & 219 & 2.02 & $-14.1 \pm 9.3$& 8 \\
2022 Jan 26 & 06:58:15 & 605.79305 & 208.176 & 160 & 3.20 & $-1.8 \pm 14.3$& 8 \\
2022 Jan 27 & 07:00:00 & 606.79418 & 208.520 & 211 & 2.41 & $-112.2 \pm 11.0$& 8 \\
2022 Jan 28 & 07:00:14 & 607.79425 & 208.864 & 108 & 3.98 & $-15.4 \pm 17.8$& 8 \\
2022 Jan 29 & 06:50:19 & 608.78727 & 209.205 & 199 & 2.19 & $-33.8 \pm 10.0$& 8 \\
2022 Jan 30 & 07:04:52 & 609.79728 & 209.552 & 235 & 1.94 & $-114.6 \pm 9.0$& 8 \\
2022 Nov 02 & 13:01:59 & 886.04829 & 304.484 & 238 & 1.65 & $-26.1 \pm 7.7$& 9 \\
2022 Nov 06 & 12:18:09 & 890.01799 & 305.848 & 219 & 2.04 & $-30.9 \pm 9.4$& 9 \\
2022 Nov 07 & 12:36:23 & 891.03068 & 306.196 & 225 & 1.91 & $-31.2 \pm 8.9$& 9 \\
2022 Nov 08 & 12:43:07 & 892.03538 & 306.541 & 202 & 2.23 & $-29.1 \pm 10.2$& 9 \\
2022 Nov 09 & 13:06:42 & 893.05179 & 306.891 & 208 & 2.12 & $-32.3 \pm 9.8$& 9 \\
2022 Nov 10 & 14:29:05 & 894.10903 & 307.254 & 189 & 2.50 & $-52.9 \pm 11.4$& 9 \\
2022 Nov 11 & 13:31:54 & 895.06934 & 307.584 & 207 & 2.07 & $-19.0 \pm 9.6$& 9 \\
2022 Nov 12 & 10:27:35 & 895.94137 & 307.884 & 253 & 1.64 & $-56.1 \pm 7.6$& 9 \\
2022 Nov 14 & 10:39:27 & 897.94965 & 308.574 & 208 & 1.96 & $-21.7 \pm 9.1$& 9 \\
2022 Nov 16 & 13:32:58 & 900.07018 & 309.302 & 228 & 1.85 & $-44.5 \pm 8.6$& 9 \\
2022 Nov 17 & 12:54:51 & 901.04373 & 309.637 & 240 & 1.70 & $-23.7 \pm 7.8$& 9 \\
2022 Nov 18 & 13:52:13 & 902.08358 & 309.994 & 222 & 1.92 & $-10.7 \pm 8.9$& 9 \\
2022 Nov 21 & 12:43:21 & 905.03577 & 311.009 & 197 & 2.17 & $-18.9 \pm 10.1$& 9 \\
2022 Dec 01 & 10:31:23 & 914.94412 & 314.414 & 245 & 1.83 & $-59.6 \pm 8.1$& 9 \\
2022 Dec 02 & 14:00:46 & 916.08950 & 314.807 & 240 & 1.81 & $-60.0 \pm 8.3$& 9 \\
2022 Dec 03 & 11:47:38 & 916.99703 & 315.119 & 249 & 1.78 & $6.0 \pm 8.0$& 9 \\
2022 Dec 30 & 11:08:30 & 943.96886 & 324.388 & 246 & 1.85 & $-7.7 \pm 8.3$& 10 \\
2023 Jan 01 & 09:20:08 & 945.89349 & 325.049 & 230 & 1.86 & $-13.1 \pm 8.6$& 10 \\
2023 Jan 02 & 10:34:18 & 946.94493 & 325.411 & 234 & 1.89 & $-13.8 \pm 8.8$& 10 \\
2023 Jan 04 & 09:33:26 & 948.90254 & 326.083 & 243 & 1.63 & $-6.3 \pm 7.6$& 10 \\
2023 Jan 05 & 10:02:50 & 949.92289 & 326.434 & 212 & 2.15 & $6.9 \pm 9.7$& 10 \\
2023 Jan 06 & 07:22:54 & 950.81176 & 326.739 & 210 & 2.09 & $-17.1 \pm 9.5$& 10 \\
2023 Jan 07 & 07:49:05 & 951.82987 & 327.089 & 191 & 2.48 & $10.3 \pm 11.1$& 10 \\
2023 Jan 08 & 08:13:35 & 952.84681 & 327.439 & 222 & 1.96 & $8.8 \pm 8.9$& 10 \\
2023 Jan 09 & 10:17:58 & 953.93312 & 327.812 & 237 & 1.84 & $-37.2 \pm 8.3$& 10 \\
2023 Jan 10 & 09:49:31 & 954.91329 & 328.149 & 232 & 1.92 & $21.3 \pm 8.8$& 10 \\
2023 Jan 12 & 08:00:02 & 956.83712 & 328.810 & 240 & 1.77 & $-22.5 \pm 8.3$& 10 \\
2023 Jan 13 & 09:03:30 & 957.88112 & 329.169 & 240 & 1.90 & $17.1 \pm 8.6$& 10 \\
2023 Feb 02 & 08:05:52 & 977.83937 & 336.027 & 243 & 1.88 & $-23.8 \pm 8.4$& 10 \\
2023 Feb 03 & 07:57:59 & 978.83380 & 336.369 & 243 & 1.87 & $-2.7 \pm 8.3$& 10 \\
2023 Feb 06 & 06:57:19 & 981.79138 & 337.385 & 220 & 2.19 & $-9.4 \pm 9.7$& 10 \\
2023 Feb 09 & 07:11:40 & 984.80106 & 338.420 & 217 & 2.08 & $-1.1 \pm 9.5$& 10 \\ \hline

\end{tabular}
\end{table*}

\begin{table*}  %\ContinuedFloat
	\centering
	\caption{Spectropolarimetric observations collected between 2019 and 2023, reduced with the APERO pipeline. For each observation, we list the date, the barycentric Julian date (BJD) and the associated RV computed from the LBL approach of \citet{artigau22}.}
	\label{tab:journal_apero}
	\begin{tabular}{ccc|ccc|ccc}
    \hline
   Date & BJD & RV & Date & BJD & RV & Date & BJD & RV  \\
 & +2459000 & ($\mathrm{m\,s^{-1}}$) & & +2459000 & ($\mathrm{m\,s^{-1}}$) & & +2459000 & ($\mathrm{m\,s^{-1}}$) \\ \hline
2019 Oct 02 & -240.92770 & $-173.0 \pm 19.7$ & 2020 Sep 26 & 119.05388 & $-257.4 \pm 12.8$ & 2021 Dec 13 & 561.90135 & $100.1 \pm 18.8$  \\
2019 Oct 03 & -239.91417 & $3.3 \pm 19.4$ & 2020 Sep 27 & 120.02753 & $-183.5 \pm 13.9$ & 2021 Dec 14 & 562.82473 & $-26.3 \pm 13.1$  \\
2019 Oct 05 & -238.03456 & $-78.0 \pm 17.3$ & 2020 Sep 28 & 120.96423 & $58.1 \pm 13.0$ & 2021 Dec 16 & 564.96708 & $-141.4 \pm 17.2$  \\
2019 Oct 06 & -237.04434 & $35.4 \pm 18.0$ & 2020 Sep 29 & 122.06575 & $-164.9 \pm 13.3$ & 2021 Dec 18 & 566.86757 & $201.9 \pm 17.3$  \\
2019 Oct 07 & -235.87655 & $49.6 \pm 23.3$ & 2020 Sep 30 & 122.95714 & $-39.3 \pm 14.2$ & 2022 Jan 06 & 585.86248 & $-155.9 \pm 13.0$  \\
2019 Oct 08 & -235.10344 & $-140.6 \pm 20.2$ & 2020 Oct 01 & 124.05417 & $14.2 \pm 16.1$ & 2022 Jan 08 & 587.84907 & $105.6 \pm 12.1$  \\
2019 Oct 09 & -233.98644 & $-33.4 \pm 18.9$ & 2020 Oct 03 & 126.11995 & $-156.9 \pm 13.6$ & 2022 Jan 09 & 588.85007 & $-132.8 \pm 12.8$  \\
2019 Oct 12 & -231.00453 & $-93.9 \pm 18.2$ & 2020 Oct 04 & 126.96732 & $-51.8 \pm 13.1$ & 2022 Jan 10 & 589.89933 & $-157.0 \pm 15.7$  \\
2019 Oct 13 & -229.90697 & $-19.6 \pm 18.7$ & 2020 Oct 05 & 127.94074 & $-170.4 \pm 13.6$ & 2022 Jan 12 & 591.87405 & $-89.7 \pm 12.1$  \\
2019 Oct 14 & -229.03016 & $-94.1 \pm 18.6$ & 2020 Oct 07 & 129.93348 & $-24.8 \pm 13.8$ & 2022 Jan 13 & 592.85215 & $100.4 \pm 12.6$  \\
2019 Oct 15 & -227.95457 & $-8.9 \pm 20.2$ & 2020 Oct 08 & 131.05194 & $116.3 \pm 13.5$ & 2022 Jan 14 & 593.79602 & $41.5 \pm 12.4$  \\
2019 Oct 16 & -227.03129 & $-18.2 \pm 19.1$ & 2020 Oct 30 & 153.01912 & $185.1 \pm 13.1$ & 2022 Jan 15 & 594.86326 & $-16.7 \pm 12.8$  \\
2019 Oct 31 & -212.04262 & $-105.8 \pm 22.5$ & 2020 Oct 31 & 154.04080 & $-77.9 \pm 12.3$ & 2022 Jan 17 & 596.78913 & $25.9 \pm 12.7$  \\
2019 Nov 01 & -211.03317 & $43.3 \pm 18.0$ & 2020 Nov 01 & 154.92044 & $88.7 \pm 12.5$ & 2022 Jan 18 & 597.83624 & $29.4 \pm 12.3$  \\
2019 Nov 03 & -209.06372 & $-77.7 \pm 18.2$ & 2020 Nov 03 & 156.95591 & $-22.7 \pm 13.4$ & 2022 Jan 19 & 598.82397 & $115.0 \pm 12.9$  \\
2019 Nov 04 & -207.98978 & $201.1 \pm 18.8$ & 2020 Nov 04 & 157.95988 & $-16.0 \pm 14.2$ & 2022 Jan 20 & 599.81147 & $34.3 \pm 12.9$  \\
2019 Nov 05 & -206.92075 & $52.8 \pm 20.4$ & 2020 Nov 05 & 158.88105 & $154.7 \pm 14.0$ & 2022 Jan 21 & 600.84655 & $-95.5 \pm 12.9$  \\
2019 Nov 07 & -205.05744 & $184.4 \pm 22.7$ & 2020 Dec 24 & 207.85127 & $120.6 \pm 12.9$ & 2022 Jan 22 & 601.80756 & $72.3 \pm 17.7$  \\
2019 Nov 08 & -203.97627 & $102.2 \pm 20.6$ & 2020 Dec 25 & 208.86203 & $-170.1 \pm 13.9$ & 2022 Jan 23 & 602.84476 & $-142.5 \pm 15.5$  \\
2019 Nov 09 & -203.09302 & $-192.6 \pm 22.8$ & 2020 Dec 26 & 209.83772 & $180.4 \pm 13.2$ & 2022 Jan 24 & 603.82528 & $-129.6 \pm 12.8$  \\
2019 Nov 10 & -202.06264 & $90.3 \pm 21.2$ & 2020 Dec 28 & 211.88035 & $-217.3 \pm 21.4$ & 2022 Jan 25 & 604.83094 & $130.5 \pm 13.0$  \\
2019 Nov 11 & -201.07236 & $43.8 \pm 21.2$ & 2020 Dec 29 & 212.86158 & $200.8 \pm 17.2$ & 2022 Jan 26 & 605.79593 & $-128.0 \pm 17.0$  \\
2019 Nov 13 & -199.10934 & $56.3 \pm 19.5$ & 2020 Dec 30 & 213.85724 & $200.4 \pm 14.4$ & 2022 Jan 27 & 606.79591 & $-111.5 \pm 14.1$  \\
2019 Nov 14 & -198.09287 & $-34.0 \pm 19.3$ & 2020 Dec 31 & 214.84685 & $-176.2 \pm 14.7$ & 2022 Jan 28 & 607.79565 & $16.7 \pm 21.1$  \\
2019 Dec 08 & -174.12800 & $24.6 \pm 21.9$ & 2021 Jan 01 & 215.79469 & $178.7 \pm 14.9$ & 2022 Jan 29 & 608.78903 & $-180.4 \pm 14.1$  \\
2019 Dec 09 & -173.11839 & $245.4 \pm 19.8$ & 2021 Jan 02 & 216.81589 & $225.8 \pm 12.8$ & 2022 Jan 30 & 609.79823 & $-117.0 \pm 12.7$  \\
2019 Dec 10 & -172.16480 & $104.7 \pm 17.9$ & 2021 Jan 03 & 217.83753 & $-312.0 \pm 12.7$ & 2022 Nov 02 & 886.04909 & $-193.1 \pm 12.9$  \\
2019 Dec 11 & -171.04722 & $-235.8 \pm 20.8$ & 2021 Jan 04 & 218.80285 & $128.6 \pm 12.6$ & 2022 Nov 06 & 890.01754 & $115.0 \pm 14.2$  \\
2019 Dec 12 & -170.11785 & $160.8 \pm 18.3$ & 2021 Jan 05 & 219.82561 & $169.8 \pm 12.6$ & 2022 Nov 07 & 891.03186 & $175.0 \pm 13.5$  \\
2020 Feb 05 & -115.26991 & $-209.2 \pm 16.8$ & 2021 Jan 06 & 220.80741 & $-231.7 \pm 12.4$ & 2022 Nov 08 & 892.03611 & $-86.5 \pm 14.8$  \\
2020 Feb 06 & -114.25273 & $64.1 \pm 26.6$ & 2021 Jan 07 & 221.82268 & $70.6 \pm 13.3$ & 2022 Nov 09 & 893.05207 & $17.5 \pm 14.3$  \\
2020 Feb 16 & -104.27375 & $-268.7 \pm 20.3$ & 2021 Jan 08 & 222.82881 & $81.0 \pm 13.3$ & 2022 Nov 10 & 894.10944 & $12.7 \pm 16.8$  \\
2020 Feb 17 & -103.25826 & $-84.4 \pm 17.9$ & 2021 Sep 17 & 475.07447 & $-44.8 \pm 13.8$ & 2022 Nov 11 & 895.07057 & $0.3 \pm 14.2$  \\
2020 Feb 18 & -102.27201 & $-253.8 \pm 18.9$ & 2021 Sep 19 & 477.00284 & $76.1 \pm 14.6$ & 2022 Nov 12 & 895.94206 & $85.1 \pm 11.9$  \\
2020 Feb 19 & -101.27236 & $-320.1 \pm 19.4$ & 2021 Sep 22 & 480.04888 & $-16.4 \pm 13.1$ & 2022 Nov 14 & 897.95194 & $18.1 \pm 12.7$  \\
2020 Aug 26 & 88.11057 & $103.5 \pm 14.3$ & 2021 Sep 23 & 481.06655 & $-72.9 \pm 13.1$ & 2022 Nov 16 & 900.07114 & $-86.4 \pm 12.9$  \\
2020 Aug 27 & 89.06586 & $137.5 \pm 12.8$ & 2021 Sep 24 & 482.06653 & $52.0 \pm 12.6$ & 2022 Nov 17 & 901.04443 & $-34.2 \pm 12.2$  \\
2020 Aug 28 & 90.06677 & $129.8 \pm 13.5$ & 2021 Oct 14 & 501.99866 & $-124.4 \pm 14.4$ & 2022 Nov 18 & 902.08416 & $51.8 \pm 13.8$  \\
2020 Aug 29 & 91.10476 & $94.8 \pm 13.0$ & 2021 Oct 15 & 502.98309 & $-46.5 \pm 13.5$ & 2022 Nov 21 & 905.03656 & $48.6 \pm 14.7$  \\
2020 Aug 30 & 92.09409 & $113.5 \pm 12.9$ & 2021 Oct 16 & 504.02054 & $-18.0 \pm 13.5$ & 2022 Dec 01 & 914.94482 & $-198.2 \pm 12.5$  \\
2020 Aug 31 & 93.09734 & $180.6 \pm 13.8$ & 2021 Oct 19 & 507.05093 & $-37.0 \pm 12.9$ & 2022 Dec 02 & 916.09044 & $101.6 \pm 12.9$  \\
2020 Sep 01 & 94.10528 & $-21.6 \pm 13.9$ & 2021 Oct 20 & 507.99552 & $14.0 \pm 13.6$ & 2022 Dec 03 & 916.99788 & $123.5 \pm 12.0$  \\
2020 Sep 02 & 95.11229 & $161.8 \pm 13.4$ & 2021 Oct 21 & 509.01032 & $38.8 \pm 13.6$ & 2022 Dec 30 & 943.96972 & $2.2 \pm 12.8$  \\
2020 Sep 03 & 96.10971 & $32.7 \pm 15.9$ & 2021 Oct 22 & 510.01080 & $-20.0 \pm 16.7$ & 2023 Jan 01 & 945.89484 & $61.0 \pm 12.7$  \\
2020 Sep 04 & 97.09765 & $-9.9 \pm 15.2$ & 2021 Oct 23 & 510.99954 & $116.6 \pm 13.2$ & 2023 Jan 02 & 946.94596 & $-16.3 \pm 13.1$  \\
2020 Sep 05 & 98.09947 & $141.5 \pm 13.3$ & 2021 Oct 24 & 512.02744 & $26.7 \pm 15.9$ & 2023 Jan 04 & 948.90342 & $55.6 \pm 12.3$  \\
2020 Sep 06 & 99.08083 & $-10.6 \pm 13.2$ & 2021 Oct 25 & 512.90057 & $145.1 \pm 12.5$ & 2023 Jan 05 & 949.92427 & $-93.1 \pm 13.8$  \\
2020 Sep 07 & 100.14238 & $20.5 \pm 17.1$ & 2021 Oct 26 & 513.97343 & $130.4 \pm 12.5$ & 2023 Jan 06 & 950.81330 & $216.8 \pm 13.7$  \\
2020 Sep 08 & 101.07823 & $78.9 \pm 13.0$ & 2021 Oct 27 & 515.00383 & $111.1 \pm 12.6$ & 2023 Jan 07 & 951.83052 & $102.0 \pm 15.8$  \\
2020 Sep 09 & 102.06046 & $90.1 \pm 14.1$ & 2021 Oct 28 & 516.02592 & $-162.9 \pm 14.4$ & 2023 Jan 08 & 952.84767 & $-139.8 \pm 12.9$  \\
2020 Sep 10 & 103.06001 & $-20.5 \pm 13.0$ & 2021 Nov 19 & 537.94856 & $147.1 \pm 13.0$ & 2023 Jan 09 & 953.93397 & $135.5 \pm 12.6$  \\
2020 Sep 18 & 111.02850 & $13.9 \pm 16.0$ & 2021 Nov 20 & 538.93158 & $-90.4 \pm 12.3$ & 2023 Jan 10 & 954.91419 & $65.9 \pm 12.8$  \\
2020 Sep 19 & 112.06913 & $38.3 \pm 17.4$ & 2021 Nov 21 & 539.93672 & $-56.6 \pm 12.5$ & 2023 Jan 12 & 956.83812 & $53.2 \pm 12.5$  \\
2020 Sep 20 & 113.08884 & $-24.8 \pm 13.3$ & 2021 Nov 22 & 540.96028 & $186.3 \pm 12.5$ & 2023 Jan 13 & 957.88180 & $-8.1 \pm 12.9$  \\
2020 Sep 21 & 113.99413 & $13.4 \pm 16.3$ & 2021 Dec 09 & 557.89568 & $107.6 \pm 13.3$ & 2023 Feb 02 & 977.84013 & $-16.7 \pm 13.4$  \\
2020 Sep 22 & 115.01378 & $102.7 \pm 14.2$ & 2021 Dec 10 & 558.87091 & $234.9 \pm 13.2$ & 2023 Feb 03 & 978.83460 & $-300.2 \pm 13.1$  \\
2020 Sep 23 & 116.05571 & $-96.2 \pm 13.9$ & 2021 Dec 11 & 559.87428 & $2.3 \pm 14.2$ & 2023 Feb 06 & 981.79224 & $-325.7 \pm 14.1$  \\
2020 Sep 25 & 118.03139 & $53.5 \pm 12.7$ & 2021 Dec 12 & 560.87938 & $94.4 \pm 14.3$ & 2023 Feb 09 & 984.80233 & $-331.5 \pm 13.7$  \\
     \hline
\end{tabular}
\end{table*}

%%%%%%%%%%%%%%%%%%%%%%%%%%%%%%%%%%%%%%%%%%%%%%%%%%

\section{ZDI reconstructions}
\label{ap:zdi}
\begin{figure*}
    \centering
    \centering
    %\hspace*{-0.4cm}
    \begin{subfigure}{\textwidth}
         \centering
         \includegraphics[scale=0.15,trim={0cm 0cm 0cm 0cm},clip]{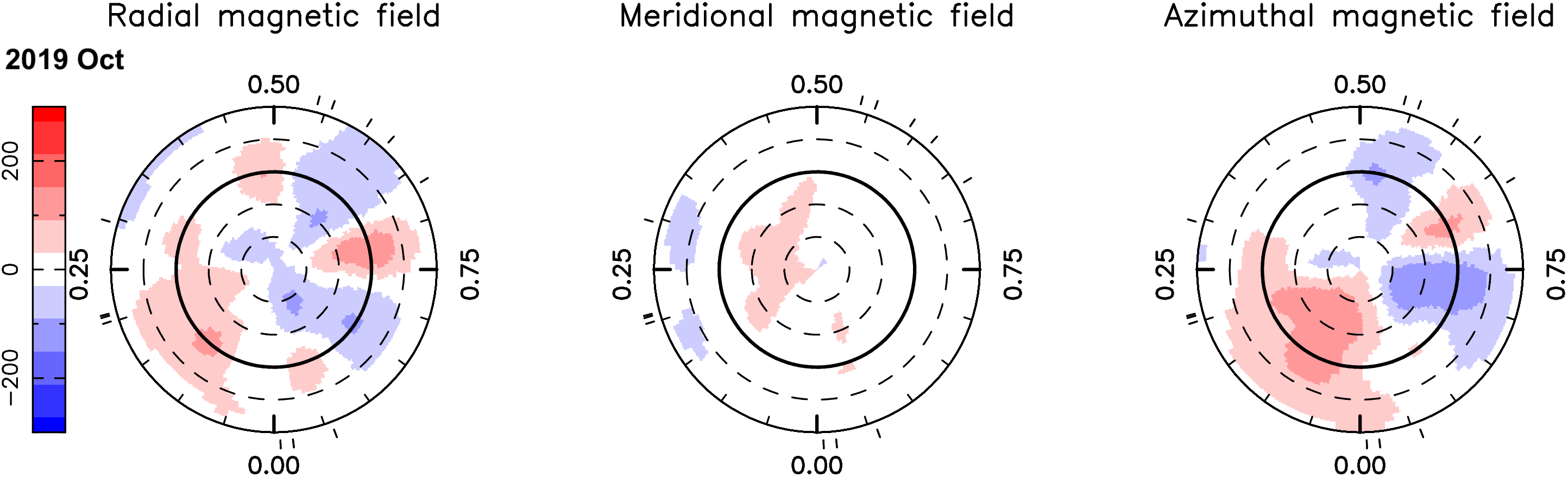}
         
    \end{subfigure}
    \hfill
    %\hspace*{-0.4cm}
    \begin{subfigure}{\textwidth}
         \centering
         \includegraphics[scale=0.15,trim={0cm 0cm 0cm 2.3cm},clip]{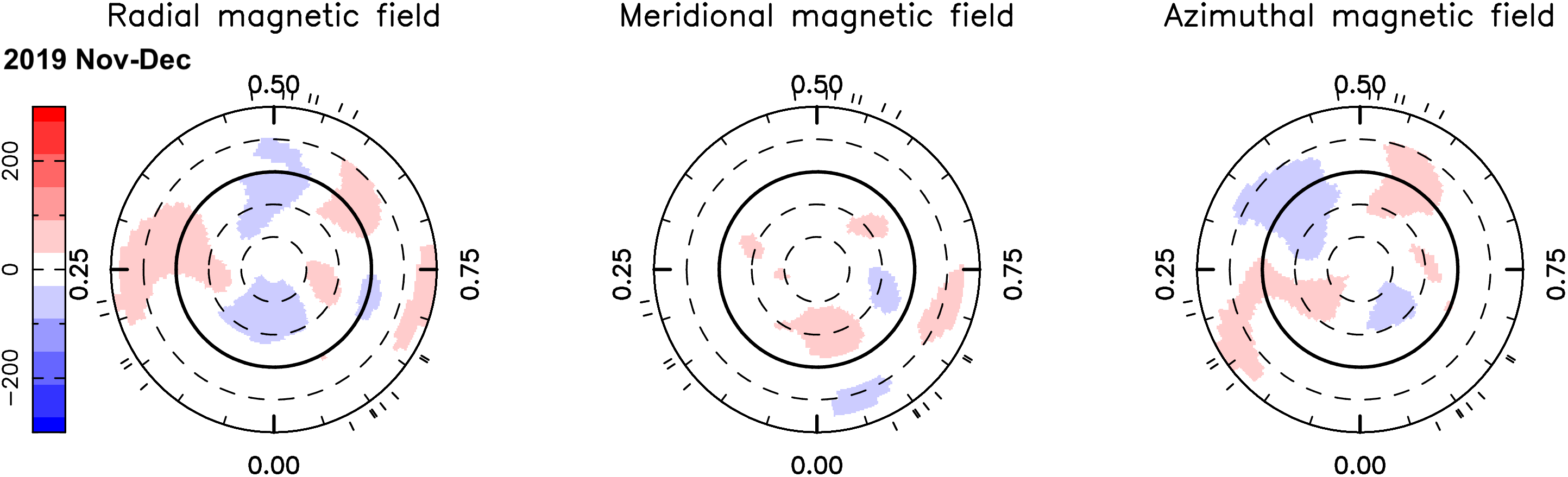}
         
    \end{subfigure}
    \caption{Maps reconstructed with ZDI from SPIRou data collected in 2019 Oct (top) and 2019 Nov-Dec (bottom). See Fig.~\ref{fig:zdi_maps} for a detailed description. These maps are less reliable than those of Fig.~\ref{fig:zdi_maps} due to a SNR of the Stokes~$V$ LSD profiles about twice smaller \coaut{than in the other observing epochs.}}
    \label{fig:zdi_maps_2019}
\end{figure*}

In Fig.~\ref{fig:zdi_maps_2019}, we show the magnetic maps reconstructed with ZDI from 2019 data whose SNR was twice lower than at future epochs, yielding less reliable reconstructions. \coaut{We show the reconstructed brightness maps in Fig.~\ref{fig:zdi_Q}}. We also show the synthetic Stokes~$V$ LSD profiles for each subset in Fig.~\ref{fig:ZDI_LSD}.

\begin{figure*}
    \centering
    \centering
    %\hspace*{-0.4cm}
    \begin{subfigure}{0.33\textwidth}
         \centering
         \includegraphics[scale=0.1,trim={0cm 0cm 0cm 3.5cm},clip]{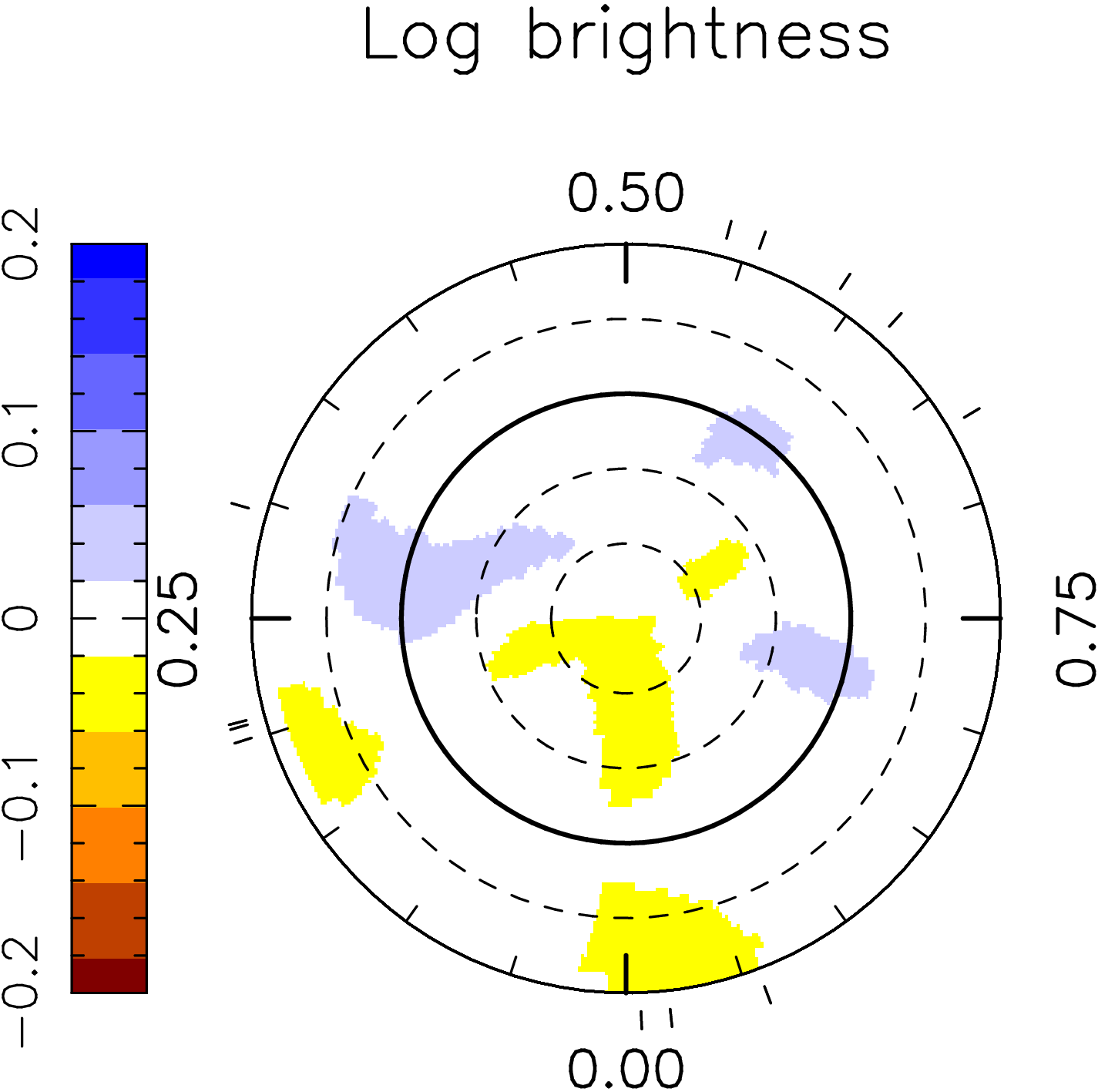}
         \caption{2019 Oct}
    \end{subfigure}
    \hfill
    %\hspace*{-0.4cm}
    \begin{subfigure}{0.33\textwidth}
         \centering
         \includegraphics[scale=0.1,trim={0cm 0cm 0cm 3.5cm},clip]{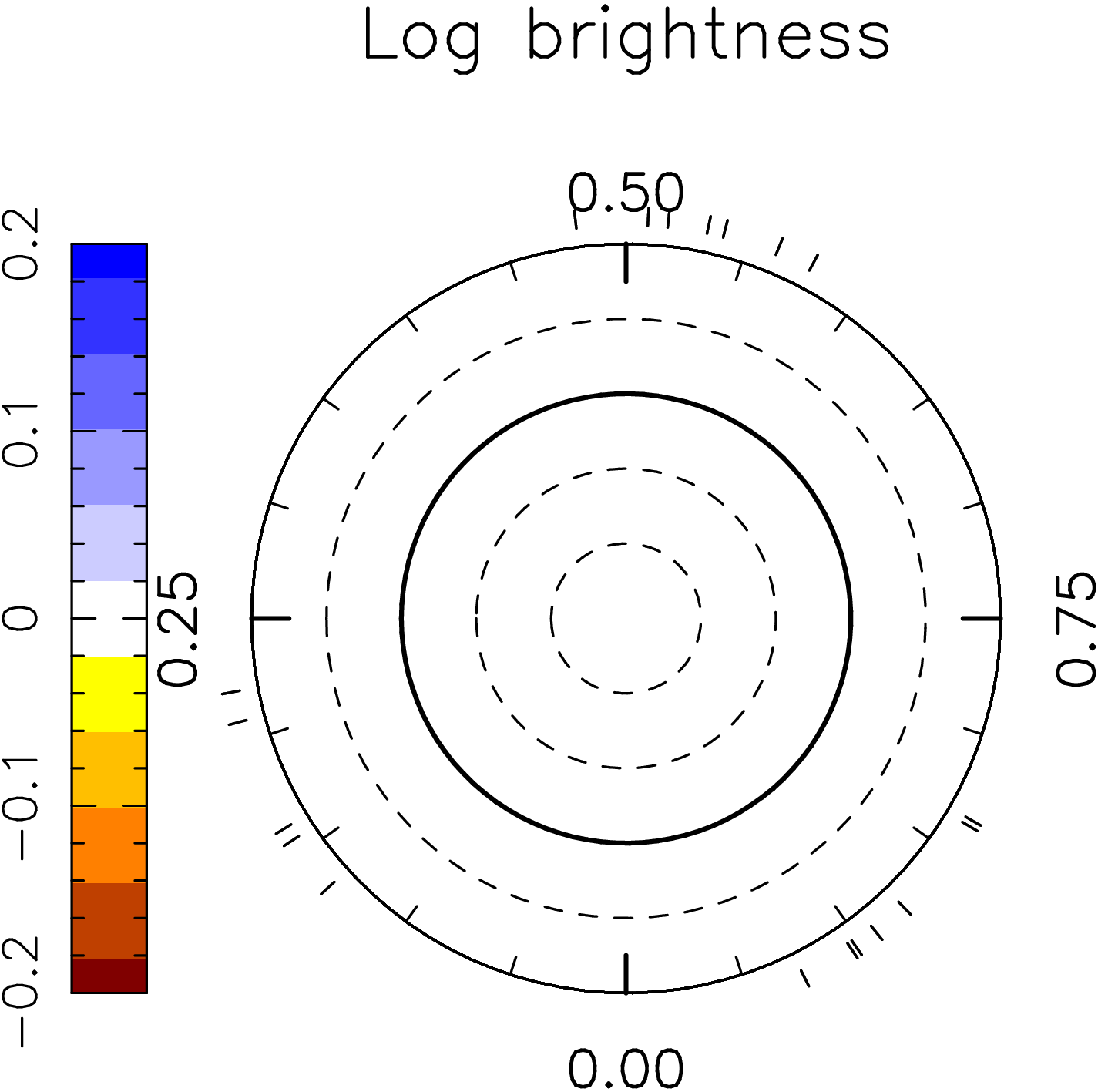}
         \caption{2019 Nov-Dec}
    \end{subfigure}
    \hfill
    %\hspace*{-0.4cm}
    \begin{subfigure}{0.33\textwidth}
         \centering
         \includegraphics[scale=0.1,trim={0cm 0cm 0cm 3.5cm},clip]{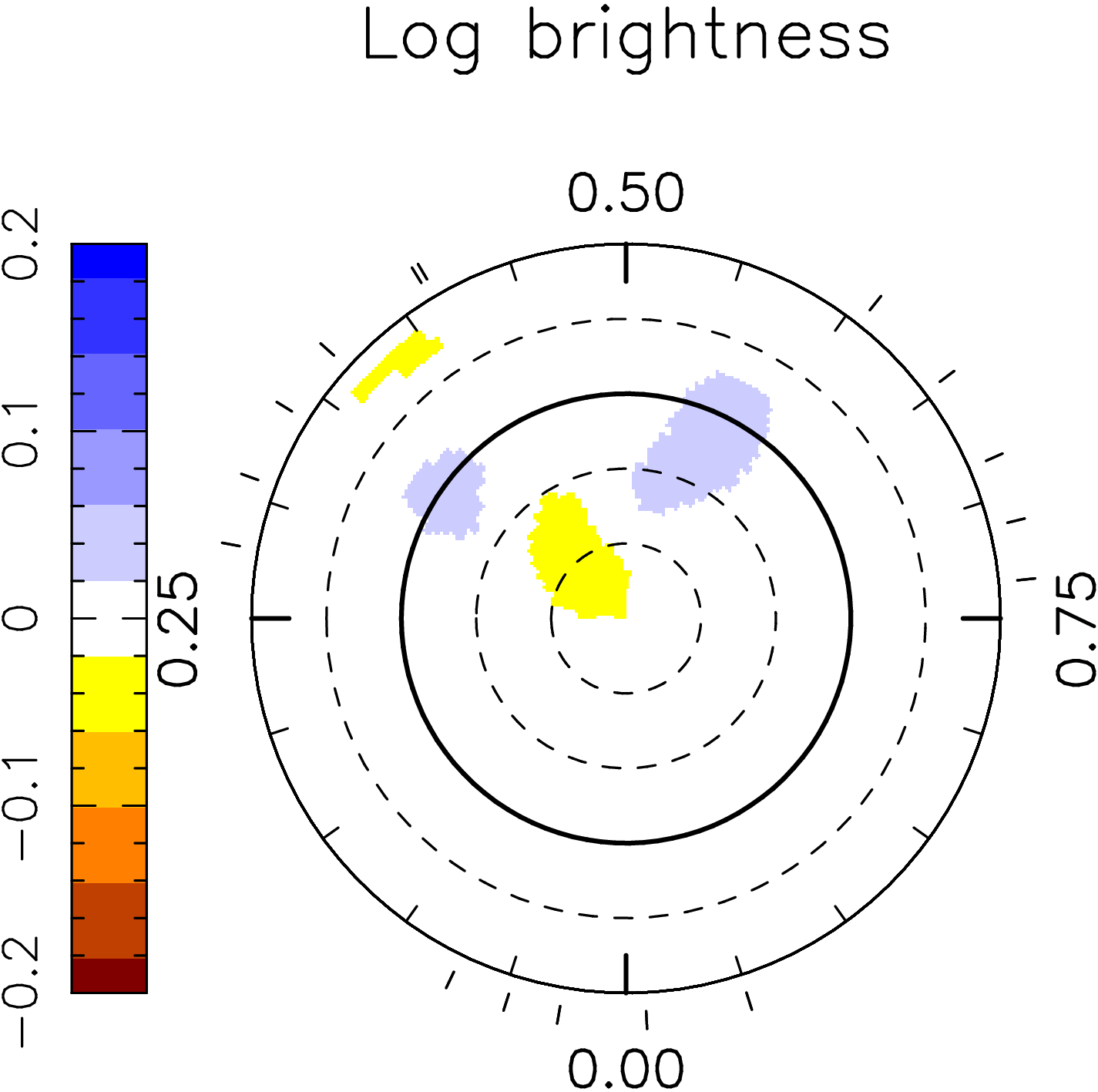}
         \caption{2020 Aug-Sep}
    \end{subfigure}
    \hfill
    \begin{subfigure}{0.33\textwidth}
         \centering
         \includegraphics[scale=0.1,trim={0cm 0cm 0cm 3.5cm},clip]{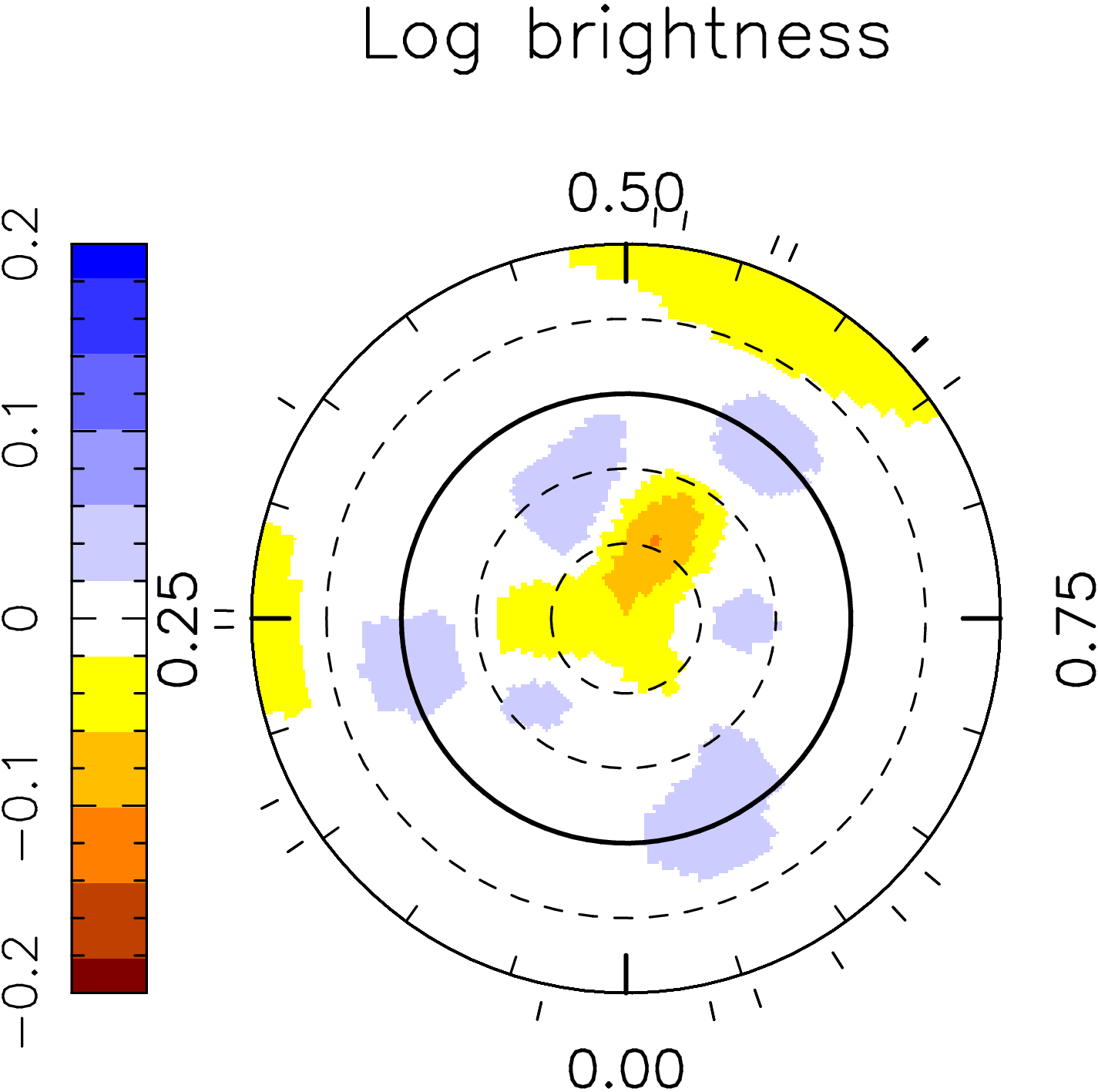}
         \caption{2020 Sep-Oct}
    \end{subfigure}
    \hfill
    %\hspace*{-0.4cm}
    \begin{subfigure}{0.33\textwidth}
         \centering
         \includegraphics[scale=0.1,trim={0cm 0cm 0cm 3.5cm},clip]{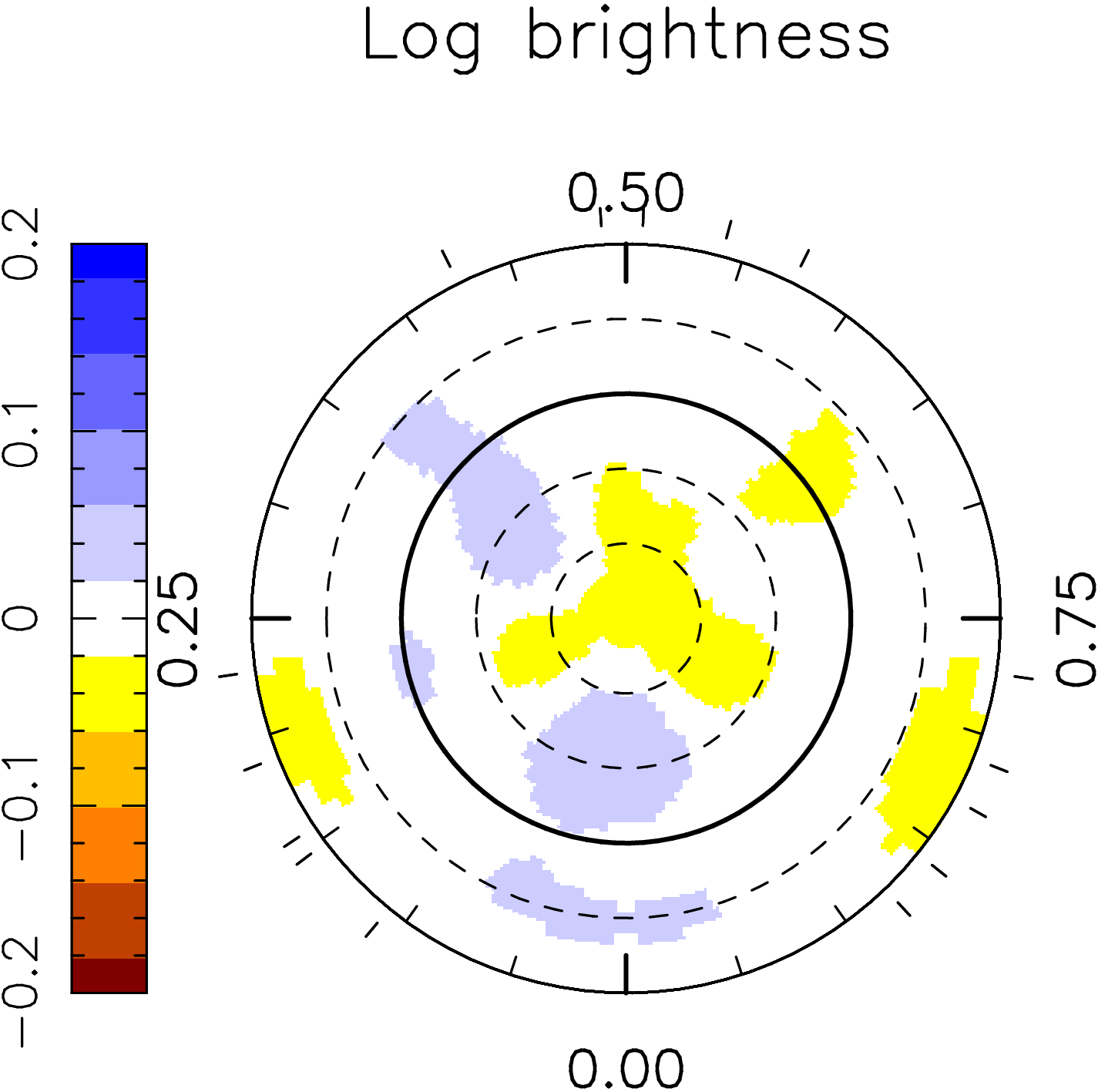}
         \caption{2020 Dec - 2021 Jan}
    \end{subfigure}
    \hfill
    %\hspace*{-0.4cm}
    \begin{subfigure}{0.33\textwidth}
         \centering
         \includegraphics[scale=0.1,trim={0cm 0cm 0cm 3.5cm},clip]{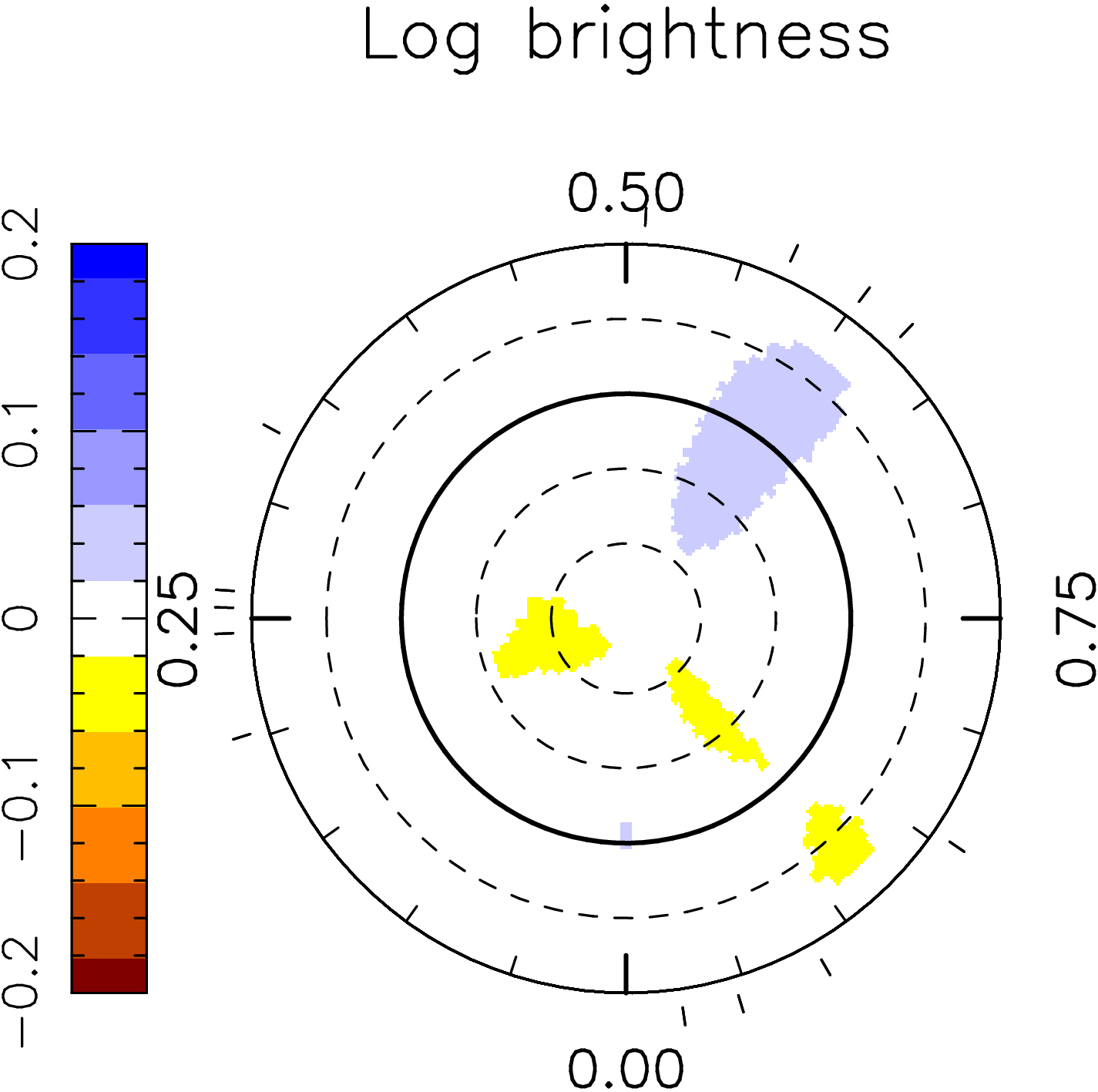}
         \caption{2021 Oct}
    \end{subfigure}
    \caption{\coaut{Logarithmic relative surface brightness maps reconstructed with ZDI for each of the 10 subsets. Yellow regions depict dark cool spots while the bright plages are shown in blue.}}
    \label{fig:zdi_Q}
\end{figure*}

\begin{figure*} \ContinuedFloat
    \centering
    \centering
    %\hspace*{-0.4cm}
    \begin{subfigure}{0.33\textwidth}
         \centering
         \includegraphics[scale=0.1,trim={0cm 0cm 0cm 3.5cm},clip]{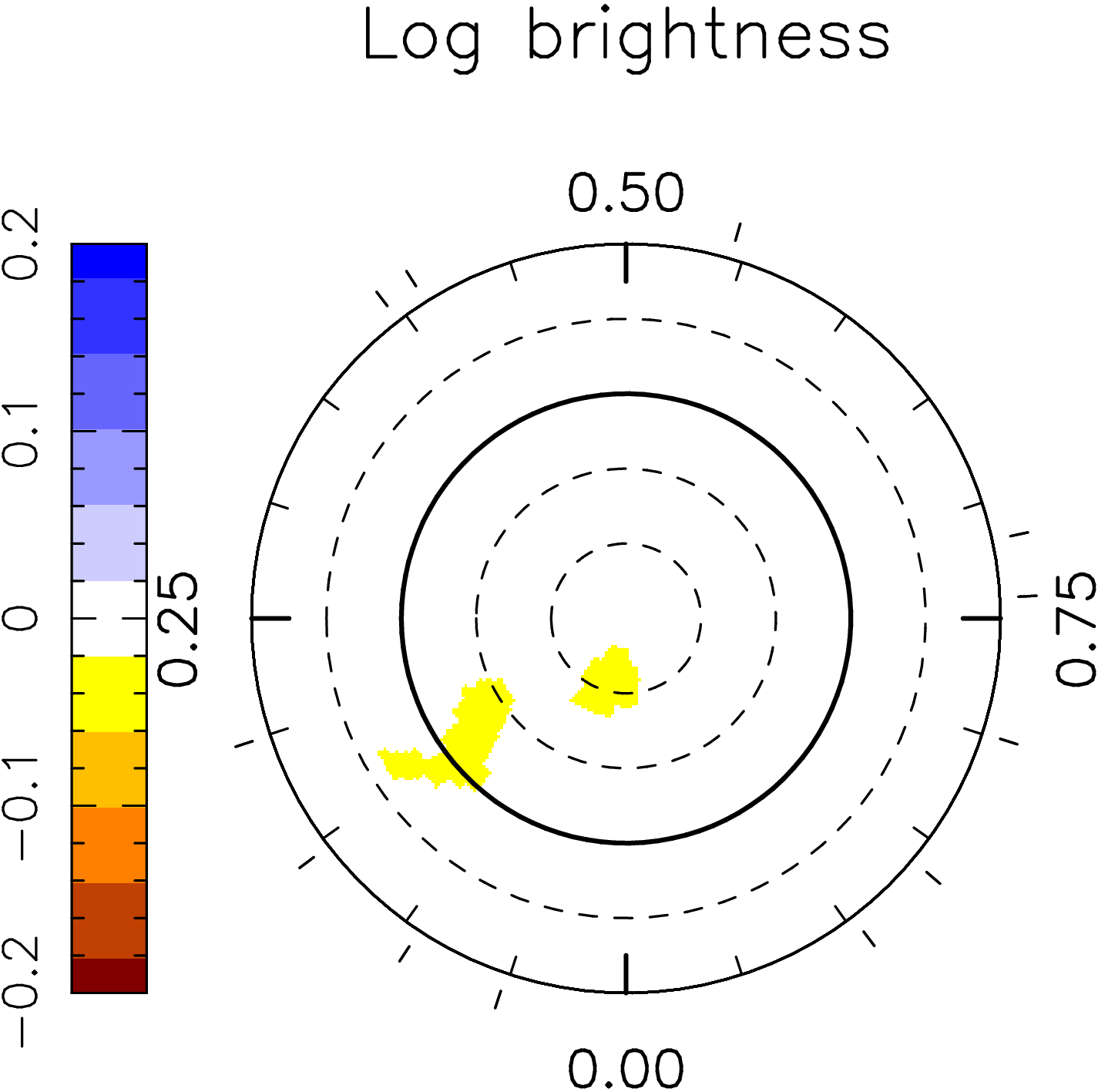}
         \caption{2021 Nov-Dec}
    \end{subfigure}
    \hfill
    %\hspace*{-0.4cm}
    \begin{subfigure}{0.33\textwidth}
         \centering
         \includegraphics[scale=0.1,trim={0cm 0cm 0cm 3.5cm},clip]{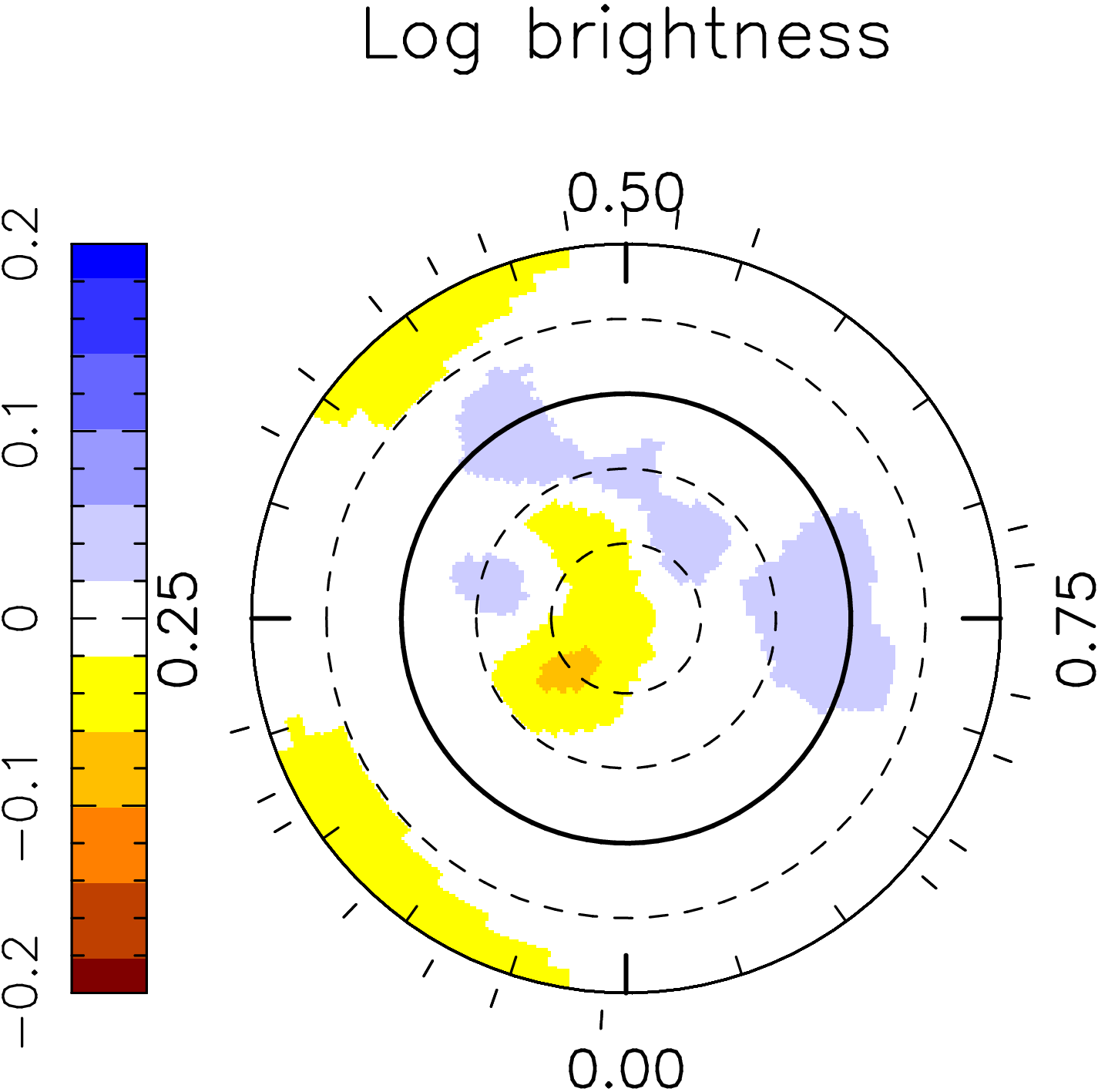}
         \caption{2022 Jan}
    \end{subfigure}
    \hfill
    %\hspace*{-0.4cm}
    \begin{subfigure}{0.33\textwidth}
         \centering
         \includegraphics[scale=0.1,trim={0cm 0cm 0cm 3.5cm},clip]{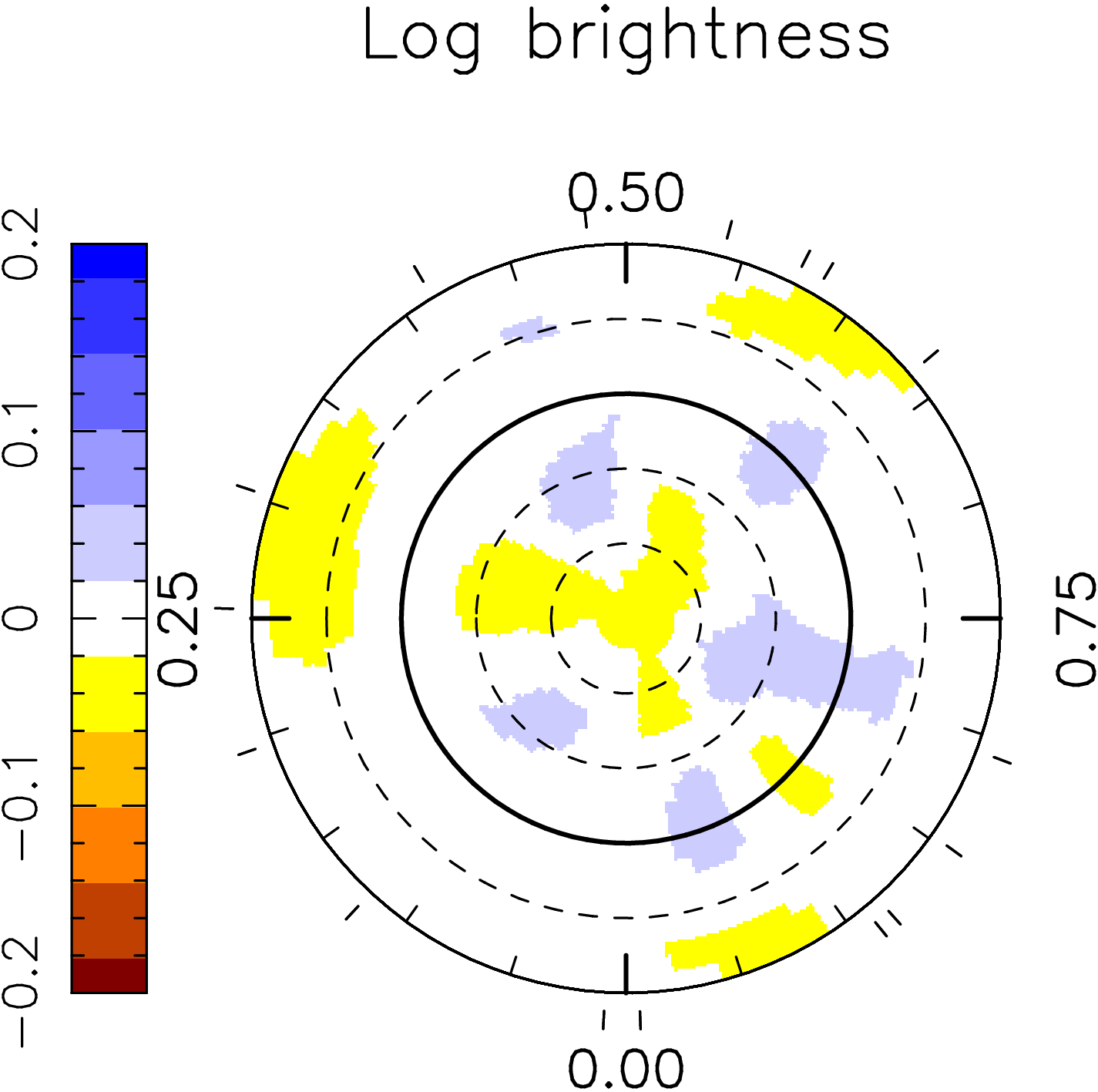}
         \caption{2022 Nov}
    \end{subfigure}
    \hfill
    \begin{subfigure}{0.33\textwidth}
         \centering
         \includegraphics[scale=0.1,trim={0cm 0cm 0cm 3.5cm},clip]{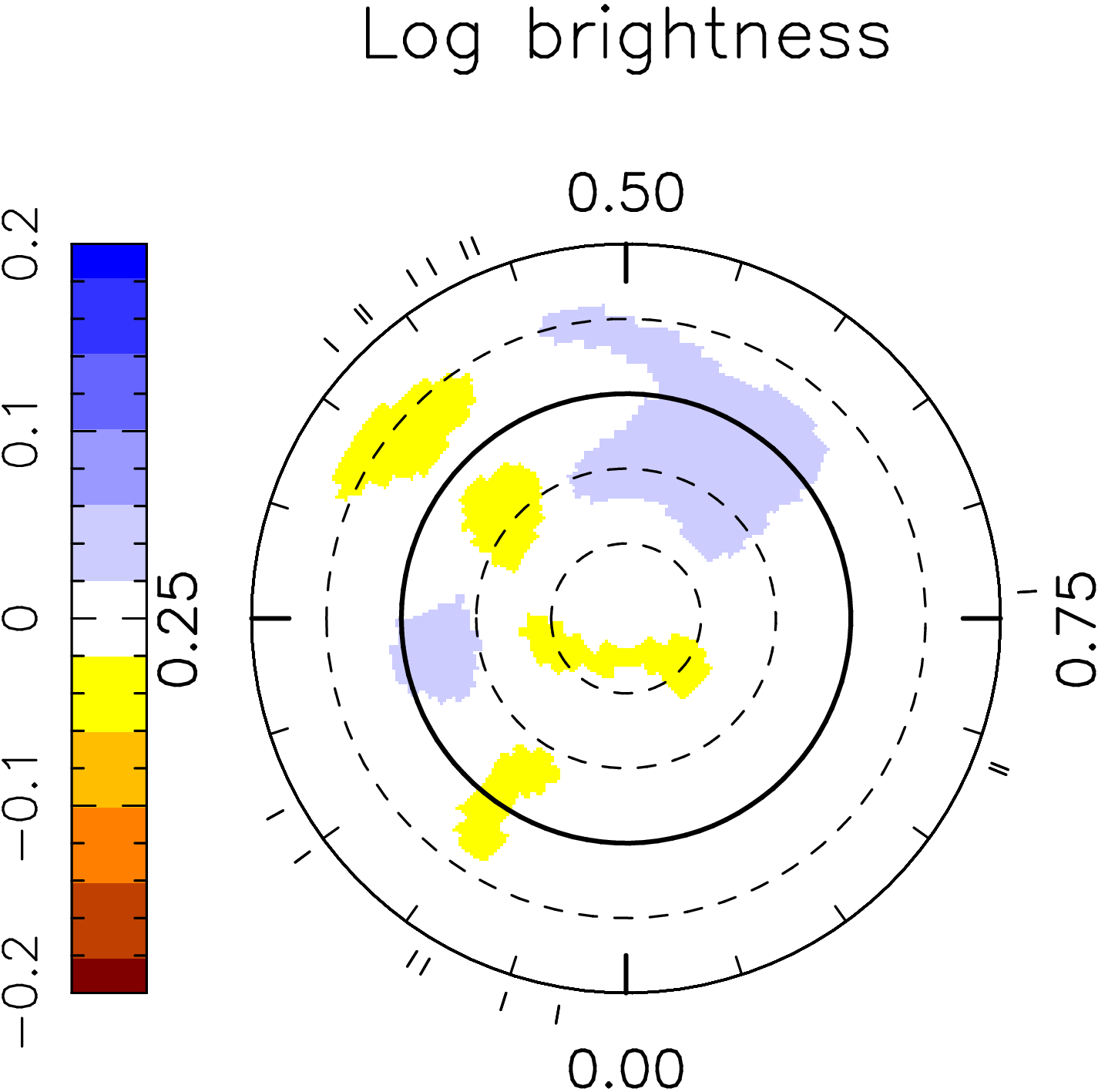}
         \caption{2023 Jan}
    \end{subfigure}
    
    \caption{\coaut{Continued.}}
    %\label{fig:zdi_Q}
\end{figure*}

\begin{figure*}
    \centering
    \centering
    %\hspace*{-0.4cm}
    \begin{subfigure}{0.33\textwidth}
         \centering
         \includegraphics[scale=0.15,trim={0cm 0cm 0cm 0cm},clip]{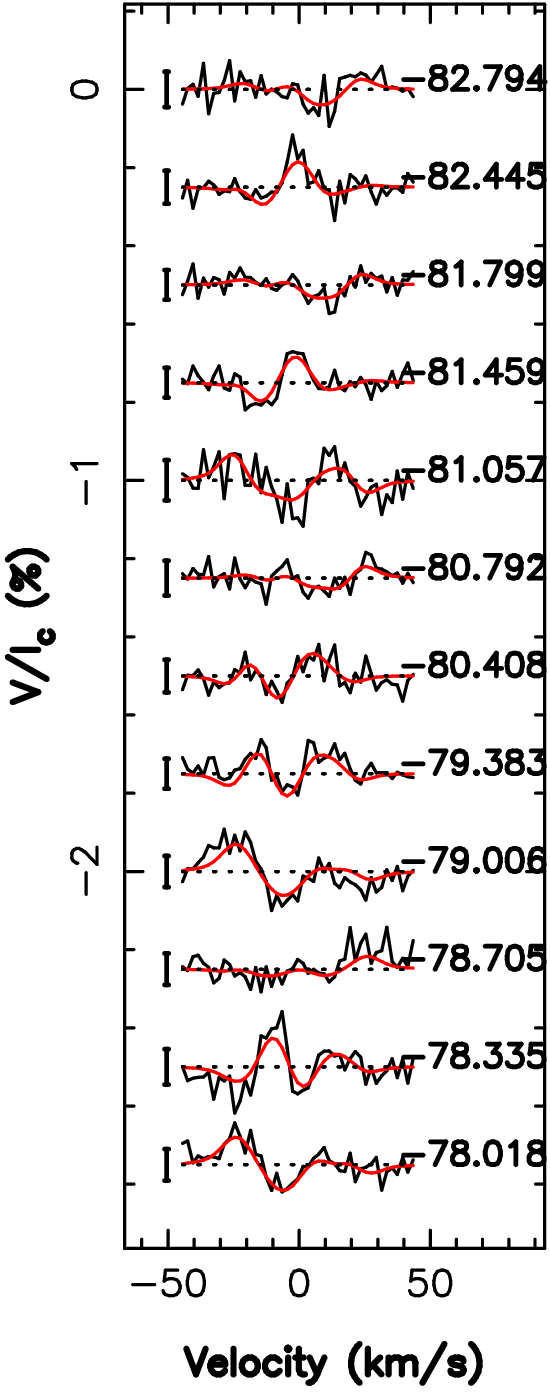}
         \caption{2019 Oct}
    \end{subfigure}
    %\hfill
    %\hspace*{-0.4cm}
    \begin{subfigure}{0.33\textwidth}
         \centering
         \includegraphics[scale=0.15,trim={0cm 0cm 0cm 0cm},clip]{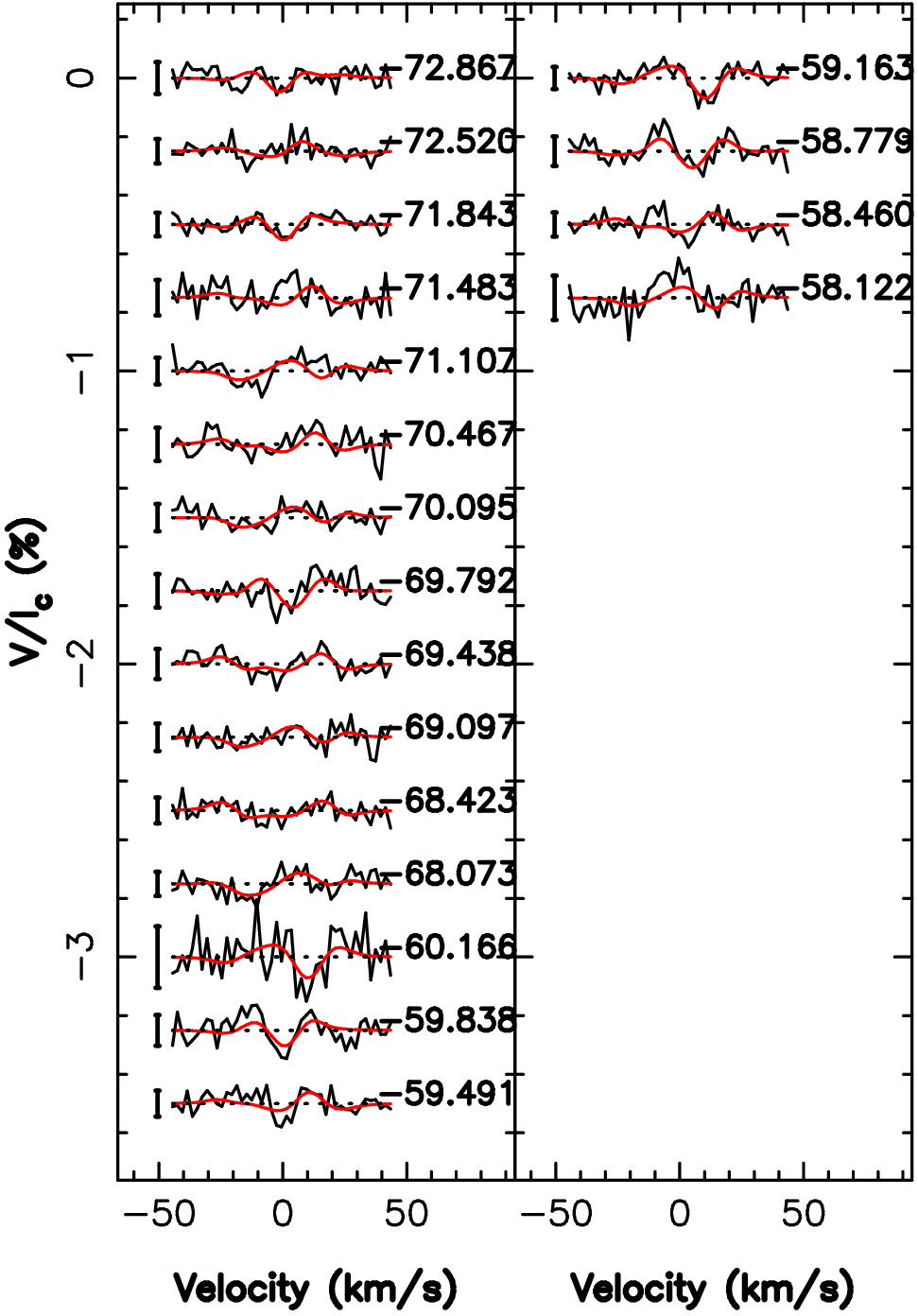}
         \caption{2019 Nov - Dec}
    \end{subfigure}
    %\hfill
    \begin{subfigure}{0.33\textwidth}
         \centering
         \includegraphics[scale=0.15,trim={0cm 0cm 0cm 0cm},clip]{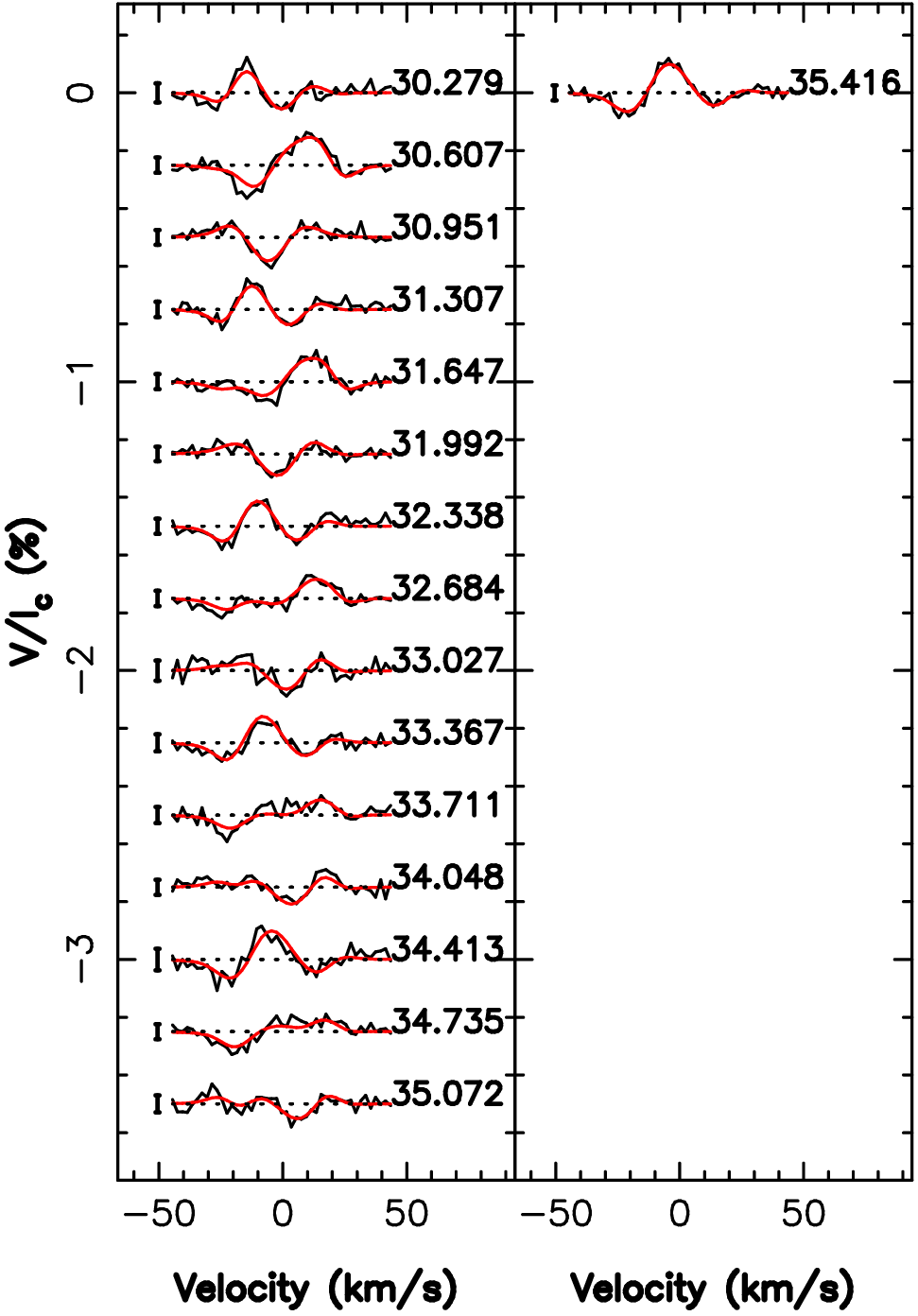}
         \caption{2020 Aug - Sep}
    \end{subfigure}
    %\hfill
    %\hspace*{-0.4cm}

    \caption{Stokes~$V$ LSD profiles associated with the spectra collected between 2019 and 2023. The observed data are in black while the reconstructed profiles derived with ZDI are shown in red. On the right of each profile, we show the rotation cycles while on their left we depict the $3\sigma$ error bar.}
    \label{fig:ZDI_LSD}
\end{figure*}

\begin{figure*} \ContinuedFloat  
\centering
    \begin{subfigure}{0.33\textwidth}
         \centering
         \includegraphics[scale=0.15,trim={0cm 0cm 0cm 0cm},clip]{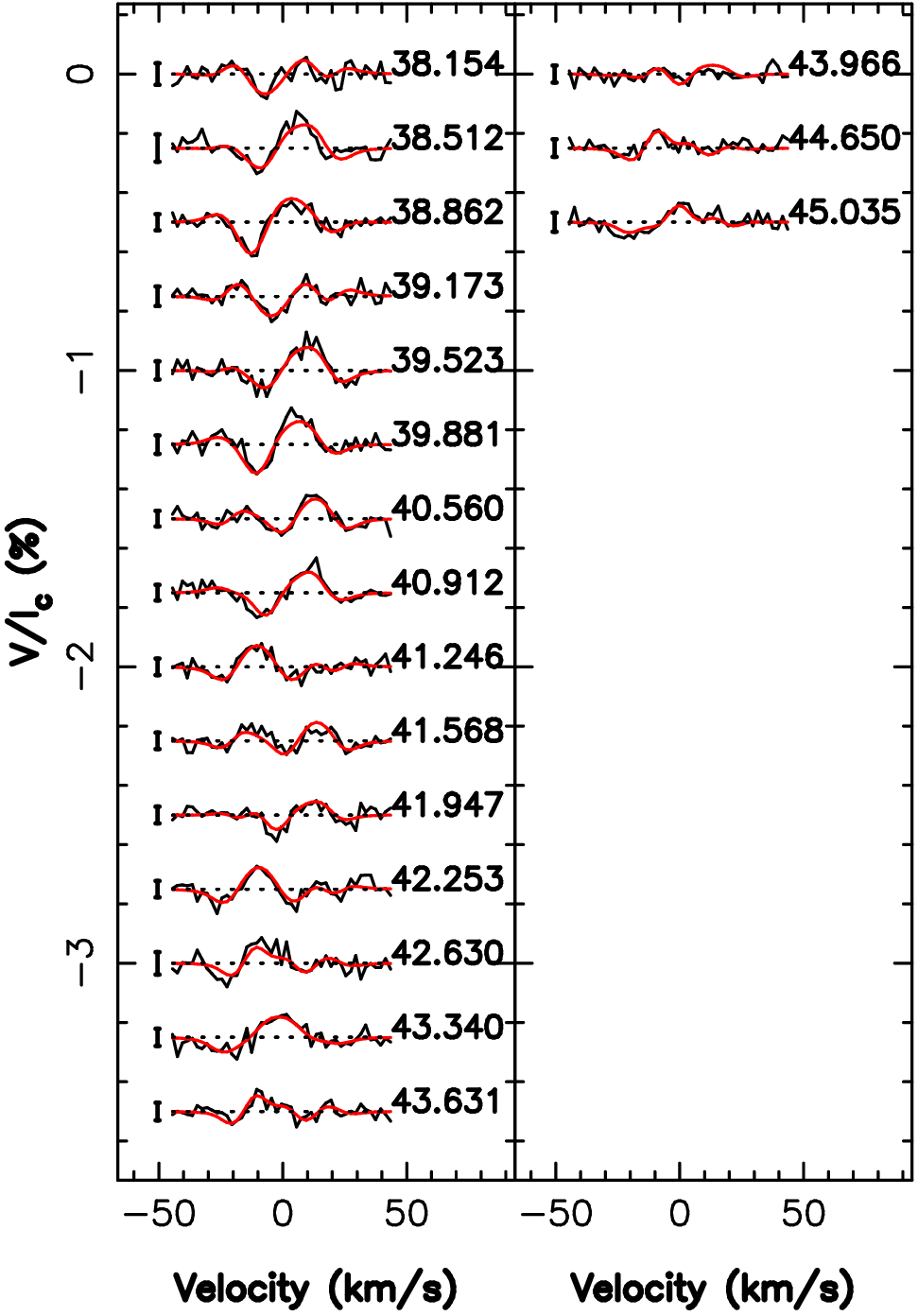}
         \caption{2020 Sep - Oct}
    \end{subfigure}
    \hfill
    \begin{subfigure}{0.33\textwidth}
         \centering
         \includegraphics[scale=0.15,trim={0cm 0cm 0cm 0cm},clip]{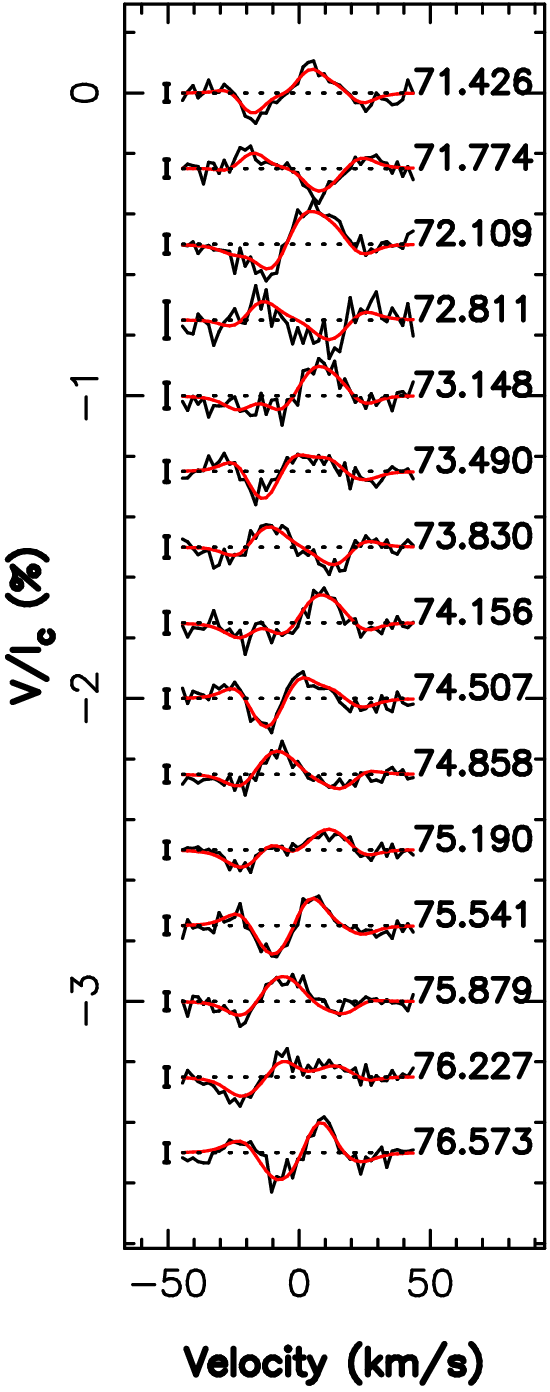}
         \caption{2020 Dec - 2021 Jan}
    \end{subfigure}
    \hfill
    \begin{subfigure}{0.33\textwidth}
         \centering
         \includegraphics[scale=0.15,trim={0cm 0cm 0cm 0cm},clip]{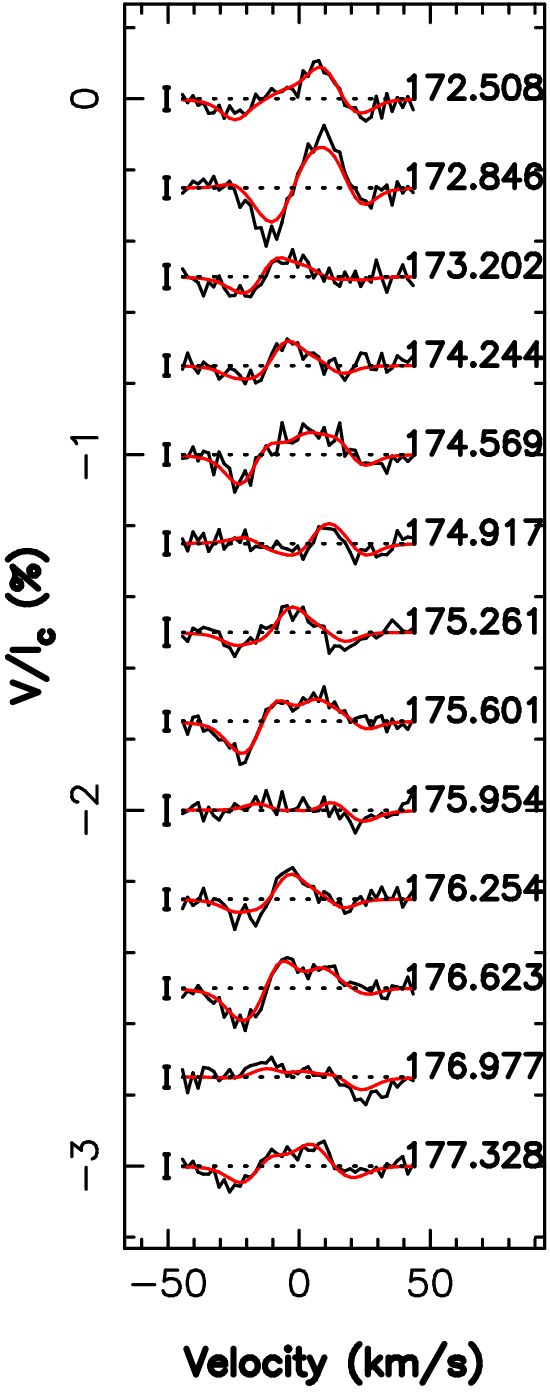}
         \caption{2021 Oct}
    \end{subfigure}
    %\hfill
    %\hspace*{-0.4cm}

    \begin{subfigure}{0.33\textwidth}
         \centering
         \includegraphics[scale=0.15,trim={0cm 0cm 0cm 0cm},clip]{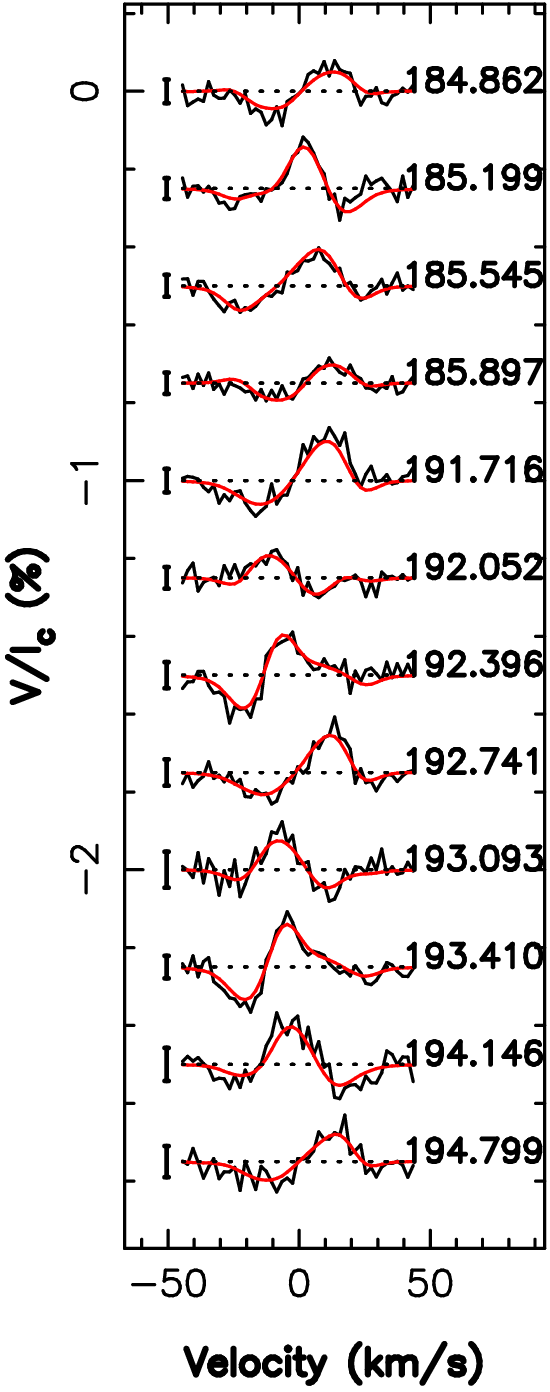}
         \caption{2021 Nov - Dec}
    \end{subfigure}
    \hfill
    \begin{subfigure}{0.33\textwidth}
         \centering
         \includegraphics[scale=0.15,trim={0cm 0cm 0cm 0cm},clip]{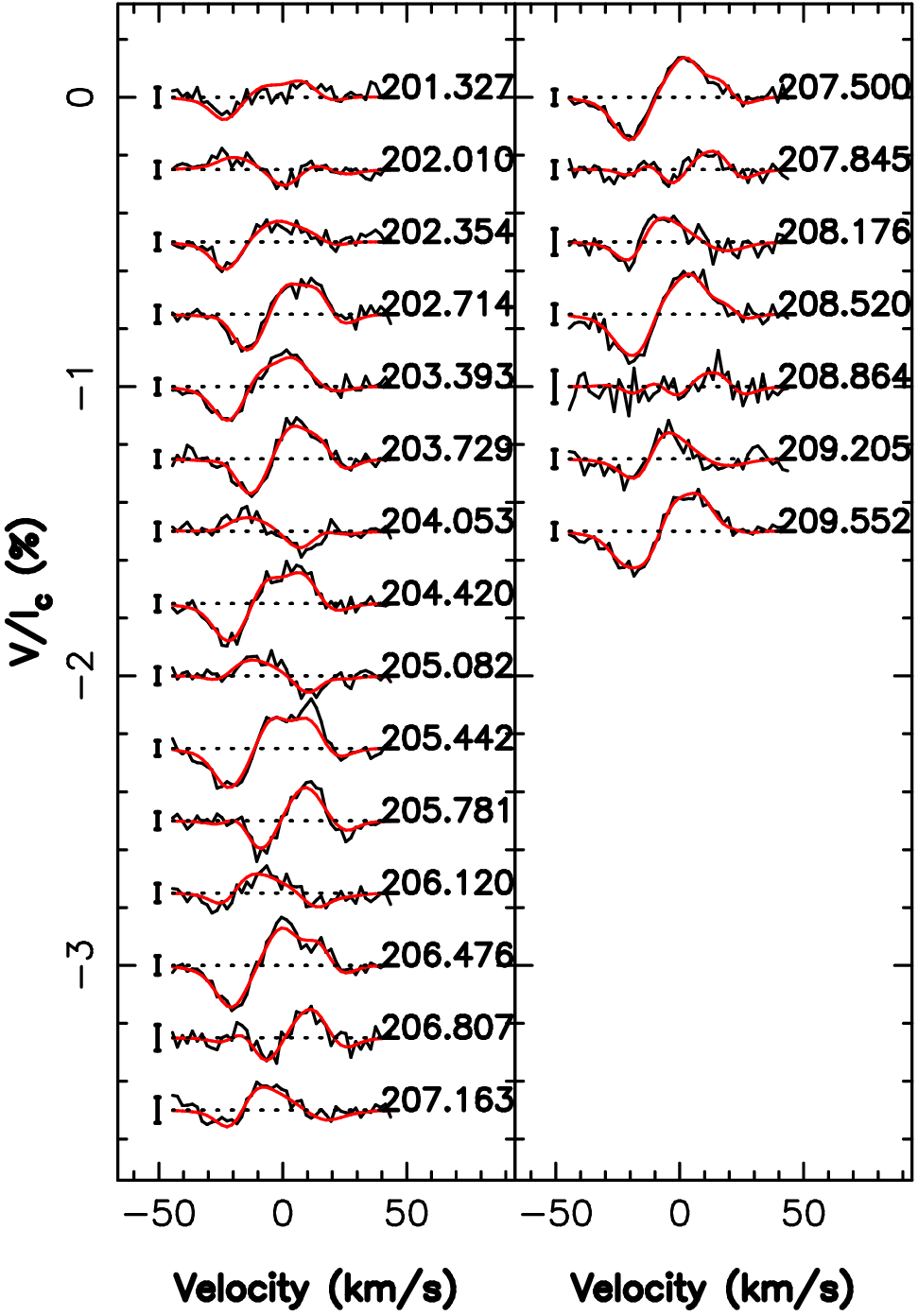}
         \caption{2022 Jan}
    \end{subfigure}
    \hfill
    \begin{subfigure}{0.33\textwidth}
         \centering
         \includegraphics[scale=0.15,trim={0cm 0cm 0cm 0cm},clip]{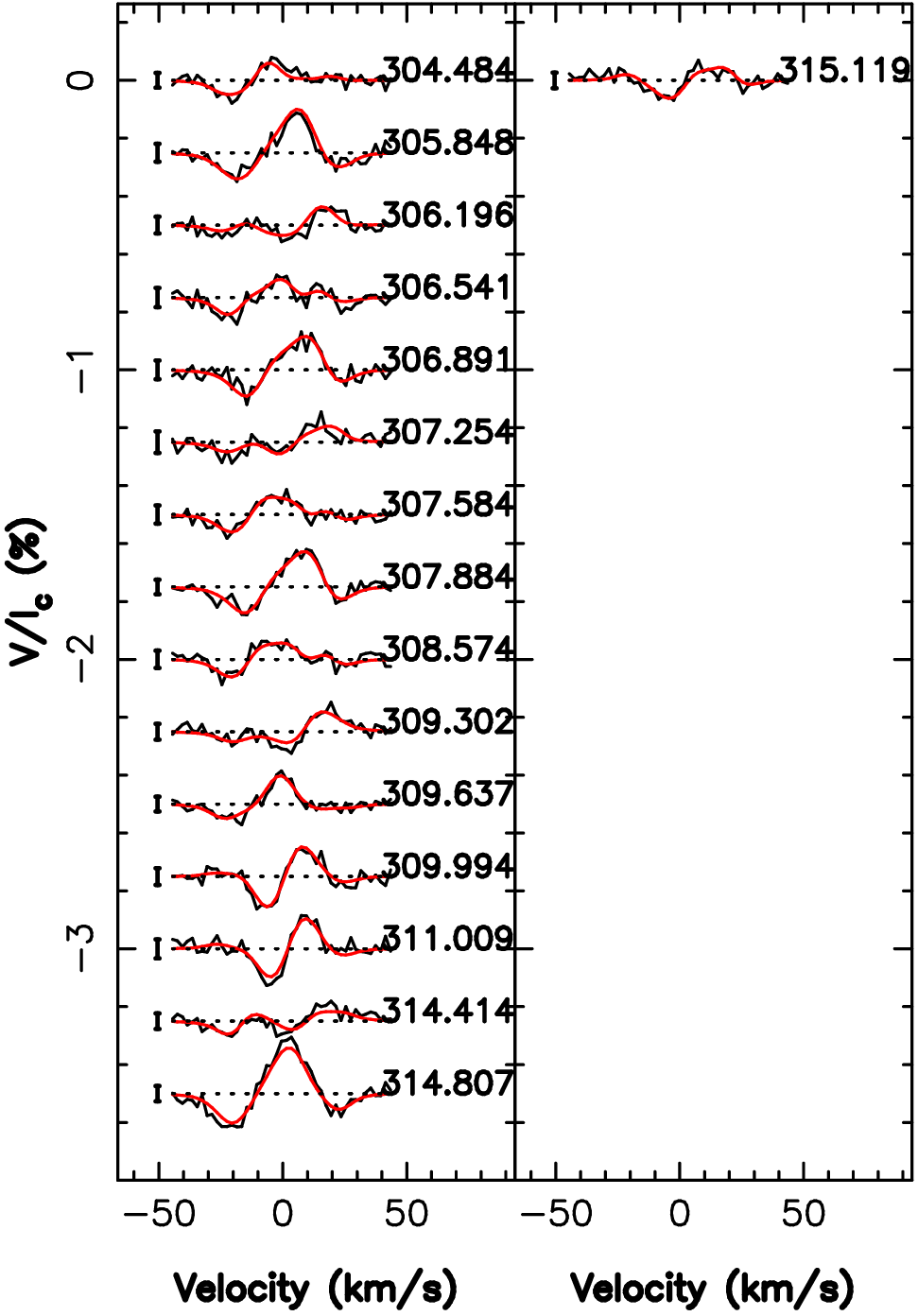}
         \caption{2022 Nov}
    \end{subfigure}
    \hfill
    %\hspace*{-0.4cm}
    \begin{subfigure}{0.33\textwidth}
         \centering
         \includegraphics[scale=0.15,trim={0cm 0cm 0cm 0cm},clip]{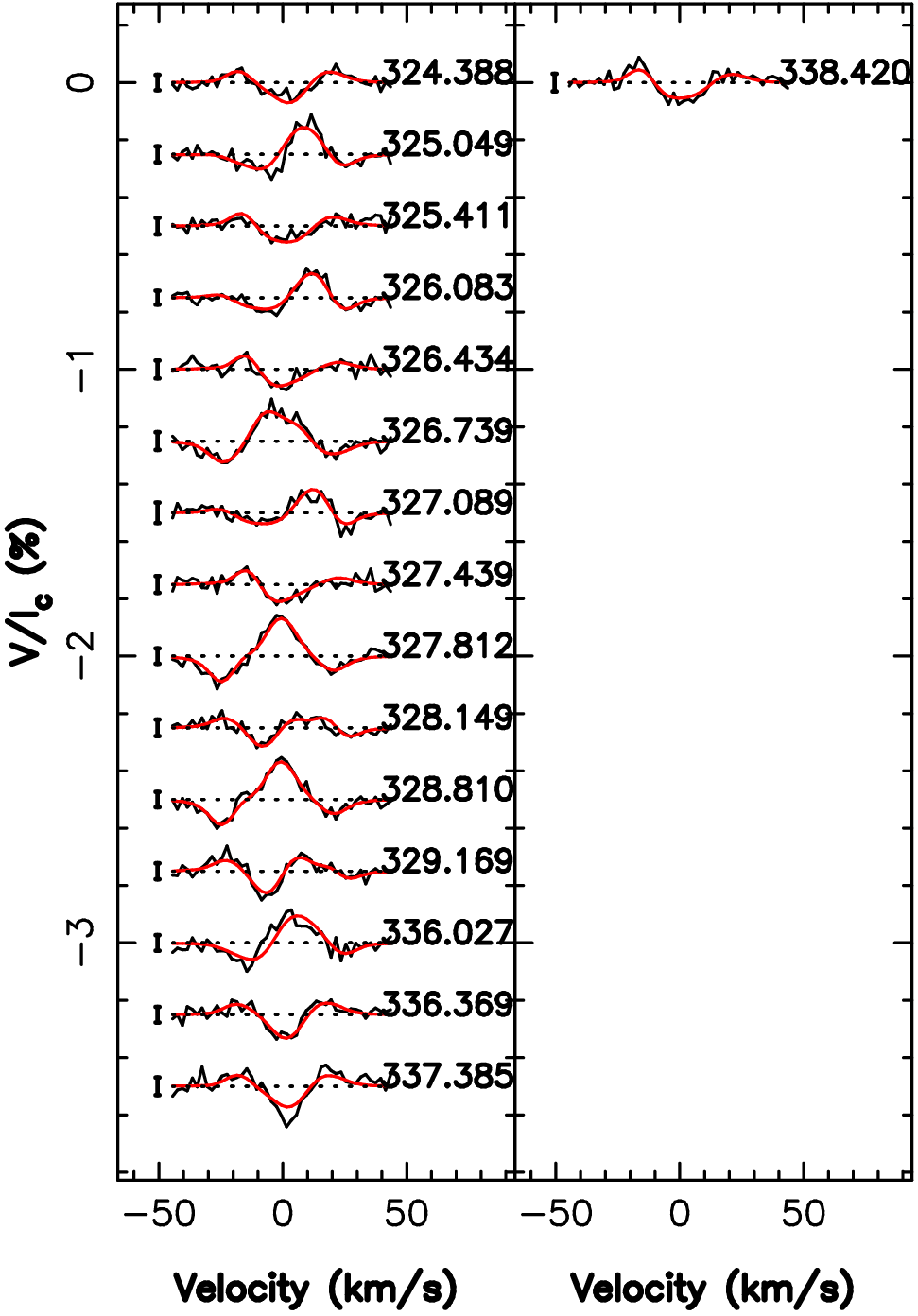}
         \caption{2023 Jan}
    \end{subfigure}

    \caption{Continued.}
    
\end{figure*}

\section{TIMeS reconstruction}
\label{sec:TIMeS_reconstruction_appendix} 
\begin{figure*} %\ContinuedFloat
    \centering
    
    \begin{subfigure}{\textwidth}
         \centering
         \includegraphics[scale=0.15,trim={0cm 0cm 0cm 0cm},clip]{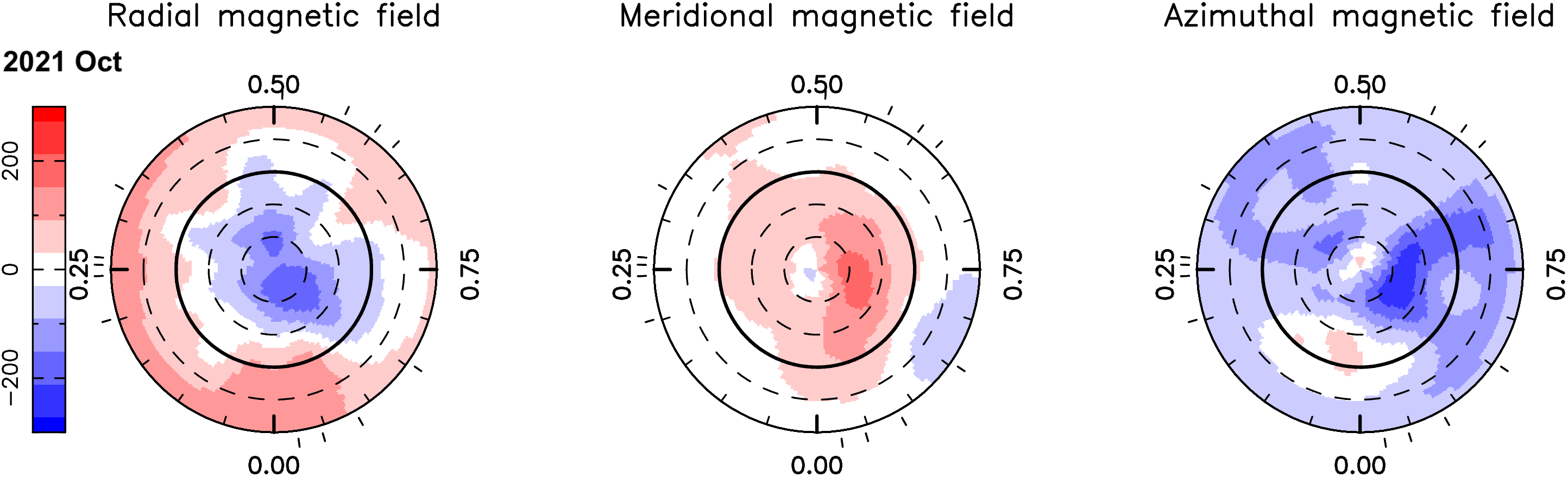}
         
    \end{subfigure}
    \hfill
    %\hspace*{-0.4cm}
    \begin{subfigure}{\textwidth}
         \centering
         \includegraphics[scale=0.15,trim={0cm 0cm 0cm 2.3cm},clip]{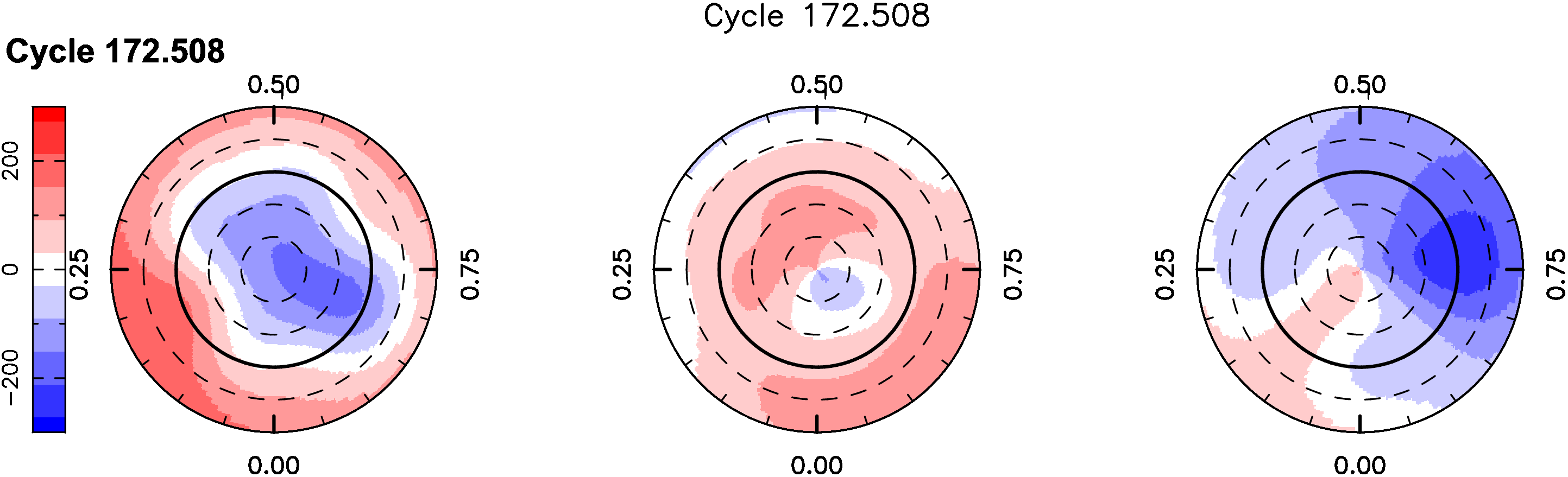}
         
    \end{subfigure}
    \hfill
    %\hspace*{-0.4cm}
    \begin{subfigure}{\textwidth}
         \centering
         \includegraphics[scale=0.15,trim={0cm 0cm 0cm 2.3cm},clip]{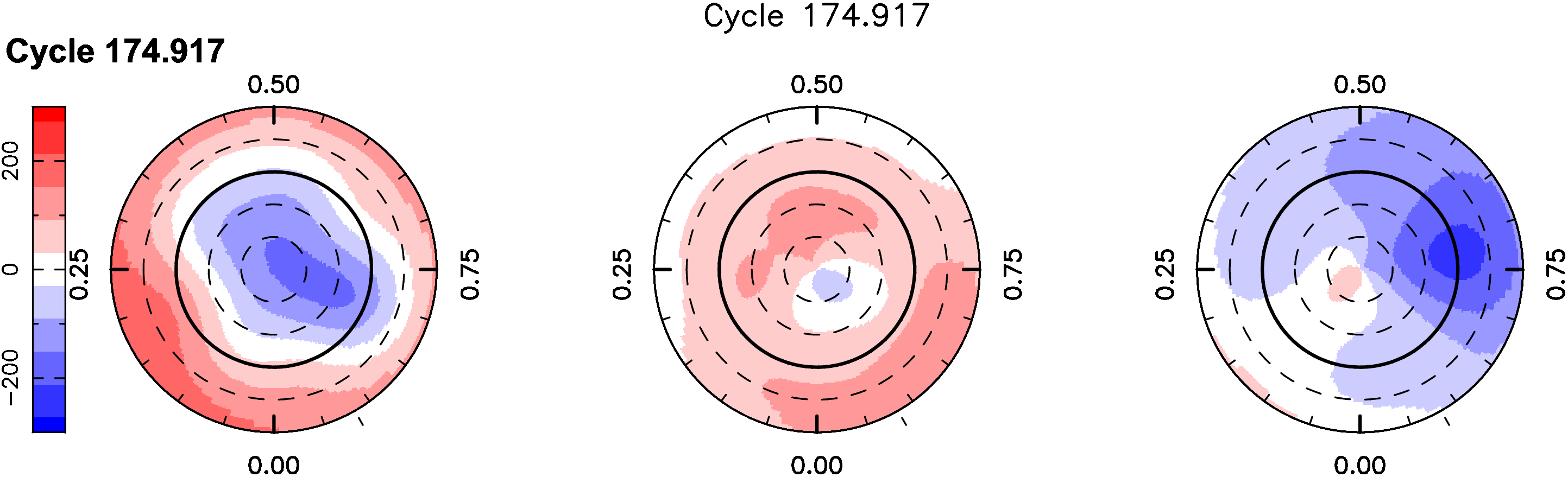}
         
    \end{subfigure}

    \hfill
    %\hspace*{-0.4cm}
    \begin{subfigure}{\textwidth}
         \centering
         \includegraphics[scale=0.15,trim={0cm 0cm 0cm 2.3cm},clip]{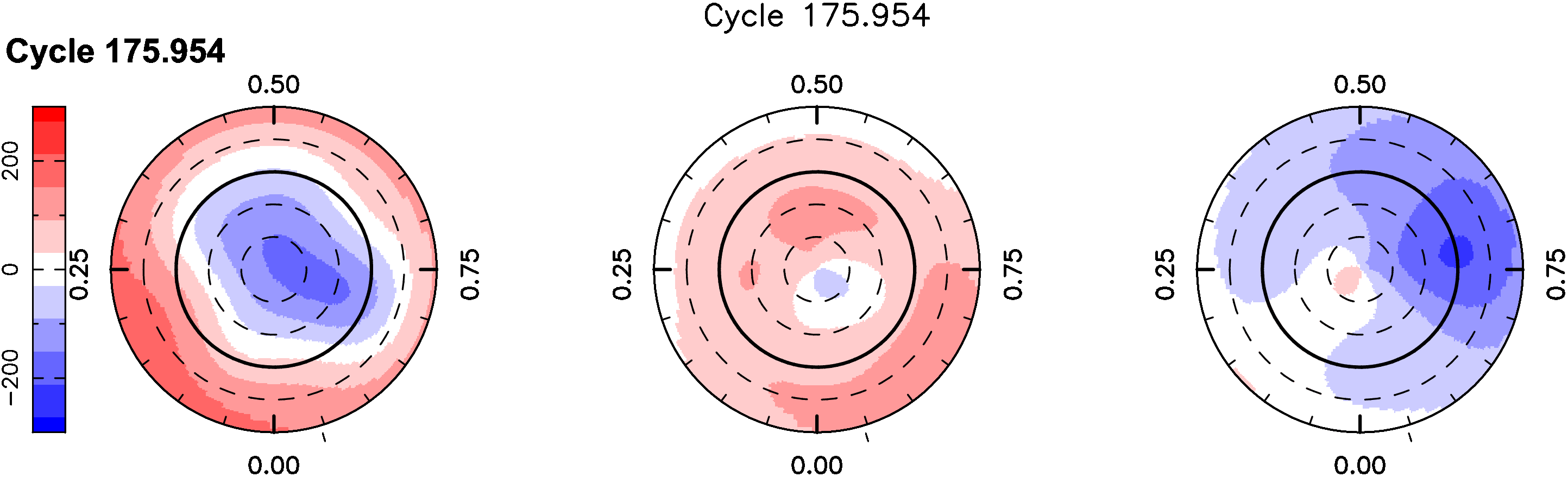}
         
    \end{subfigure}

    \hfill
    %\hspace*{-0.4cm}
    \begin{subfigure}{\textwidth}
         \centering
         \includegraphics[scale=0.15,trim={0cm 0cm 0cm 2.3cm},clip]{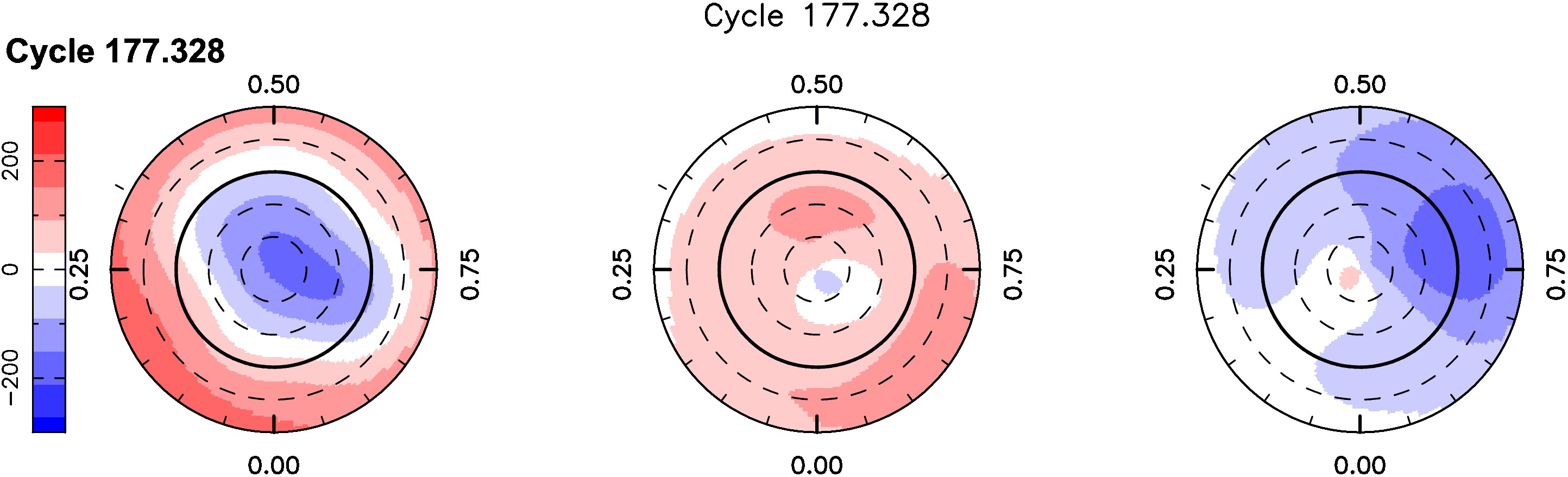}
         
    \end{subfigure}
    
    \caption{Magnetic maps in 2021 Oct reconstructed with ZDI without differential rotation (first row) and TIMeS (second to fifth rows). We show the TIMeS maps at four specific epochs, that are indicated on the left. See Fig.~\ref{fig:zdi_maps} for a detailed description of the Figure.}
    \label{fig:map_TIMeS_oct21}
\end{figure*}

\begin{figure*} %\ContinuedFloat
    \centering
    
    \begin{subfigure}{\textwidth}
         \centering
         \includegraphics[scale=0.15,trim={0cm 0cm 0cm 0cm},clip]{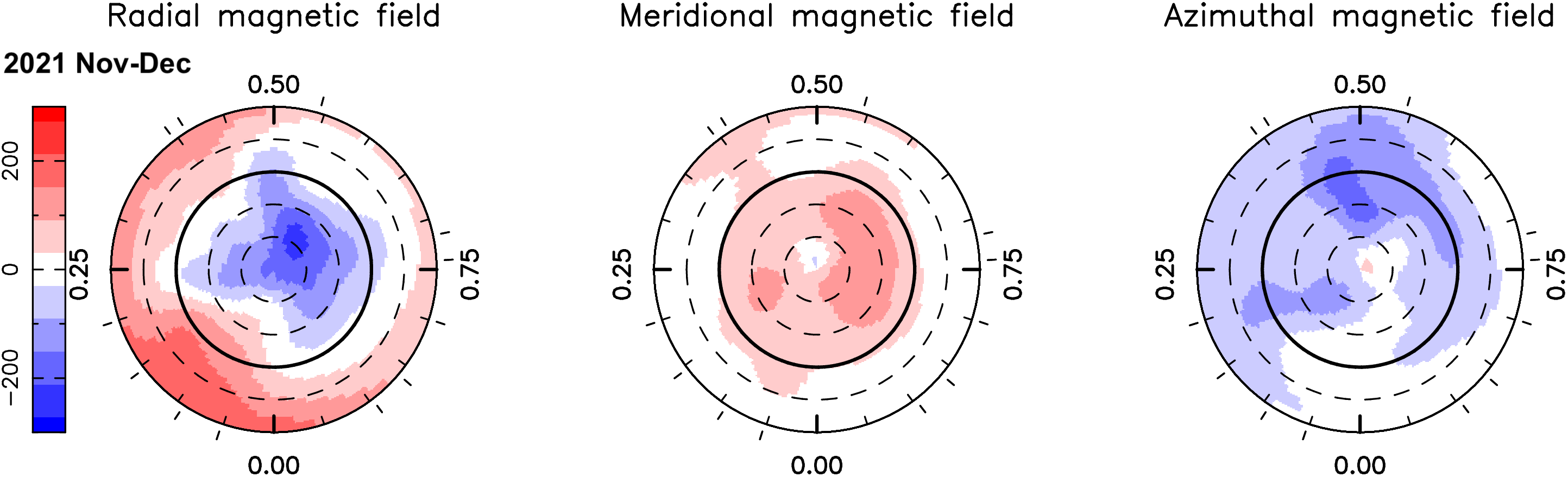}
         
    \end{subfigure}
    \hfill
    %\hspace*{-0.4cm}
    \begin{subfigure}{\textwidth}
         \centering
         \includegraphics[scale=0.15,trim={0cm 0cm 0cm 2.3cm},clip]{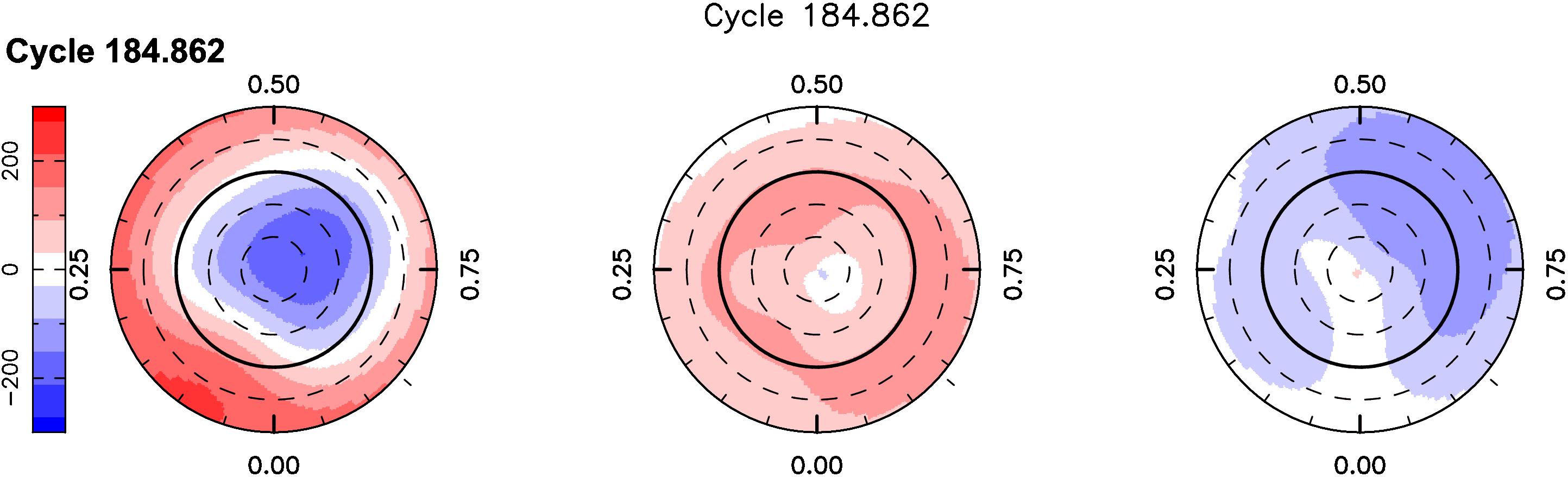}
         
    \end{subfigure}
    \hfill
    %\hspace*{-0.4cm}
    \begin{subfigure}{\textwidth}
         \centering
         \includegraphics[scale=0.15,trim={0cm 0cm 0cm 2.3cm},clip]{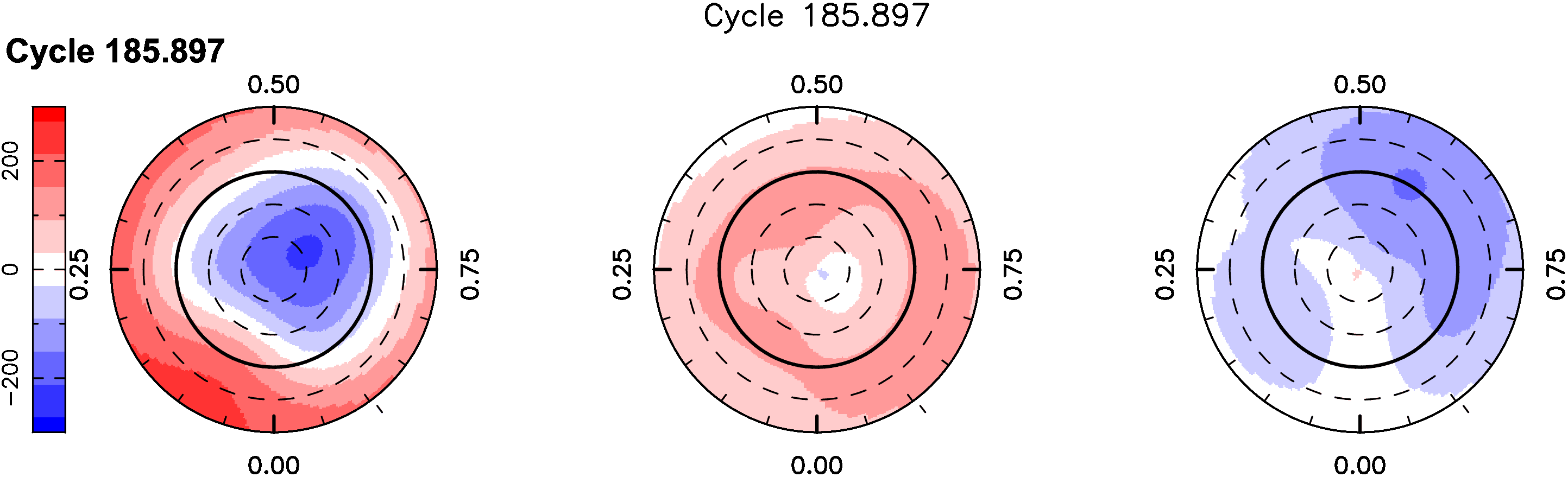}
         
    \end{subfigure}

    \hfill
    %\hspace*{-0.4cm}
    \begin{subfigure}{\textwidth}
         \centering
         \includegraphics[scale=0.15,trim={0cm 0cm 0cm 2.3cm},clip]{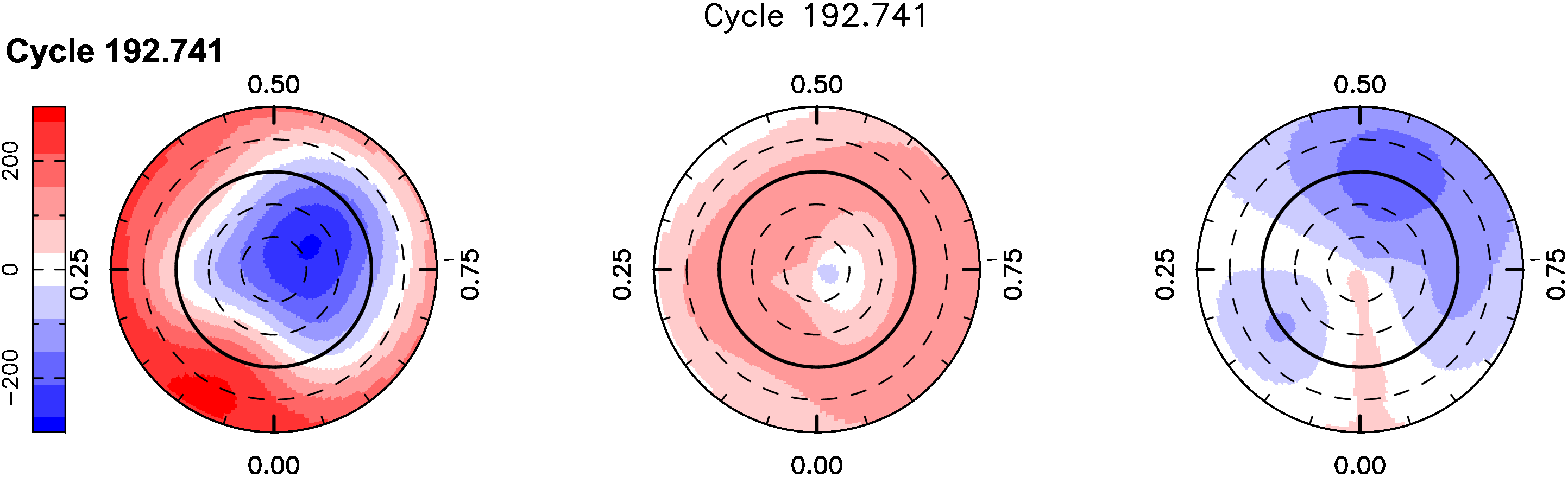}
         
    \end{subfigure}

    \hfill
    %\hspace*{-0.4cm}
    \begin{subfigure}{\textwidth}
         \centering
         \includegraphics[scale=0.15,trim={0cm 0cm 0cm 2.3cm},clip]{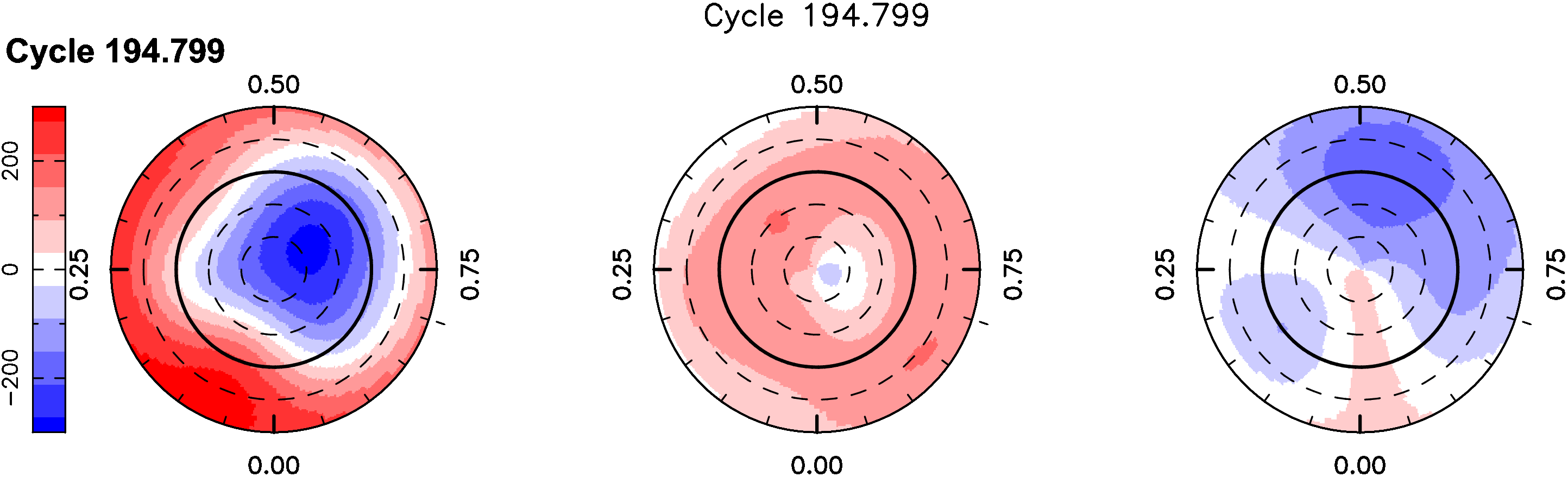}
         
    \end{subfigure}
    
    \caption{Same as Fig.~\ref{fig:map_TIMeS_oct21} for 2021 Nov-Dec.}
    \label{fig:map_TIMeS_novdec21}
\end{figure*}

\begin{figure*} %\ContinuedFloat
    \centering
    
    \begin{subfigure}{\textwidth}
         \centering
         \includegraphics[scale=0.15,trim={0cm 0cm 0cm 0cm},clip]{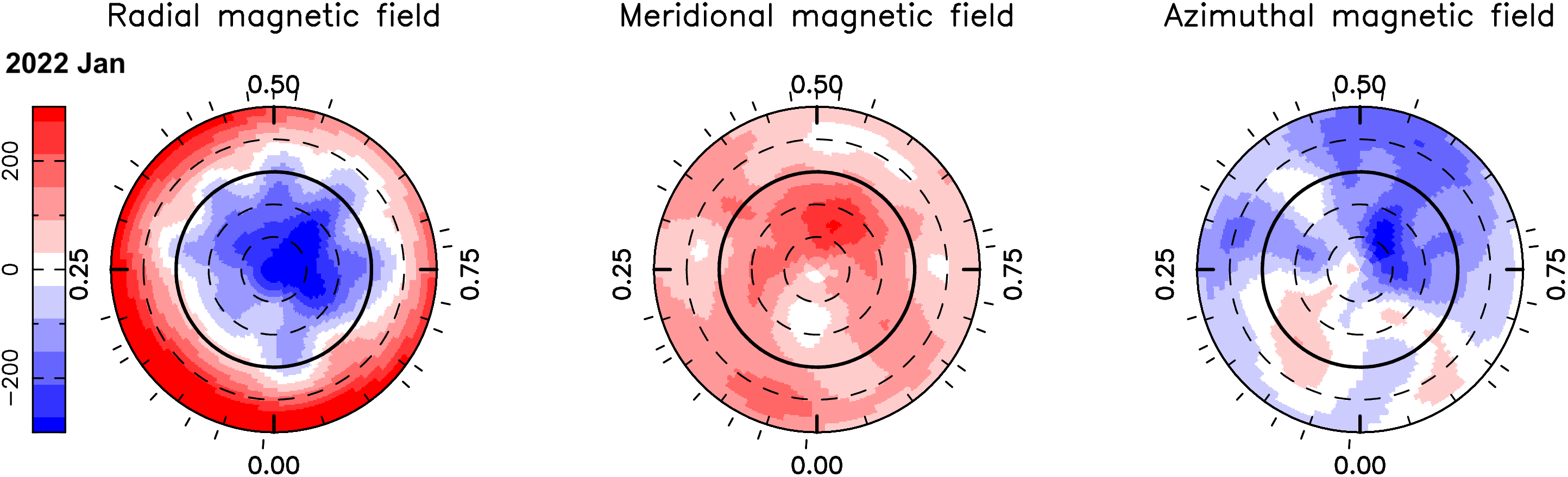}
         
    \end{subfigure}
    \hfill
    %\hspace*{-0.4cm}
    \begin{subfigure}{\textwidth}
         \centering
         \includegraphics[scale=0.15,trim={0cm 0cm 0cm 2.3cm},clip]{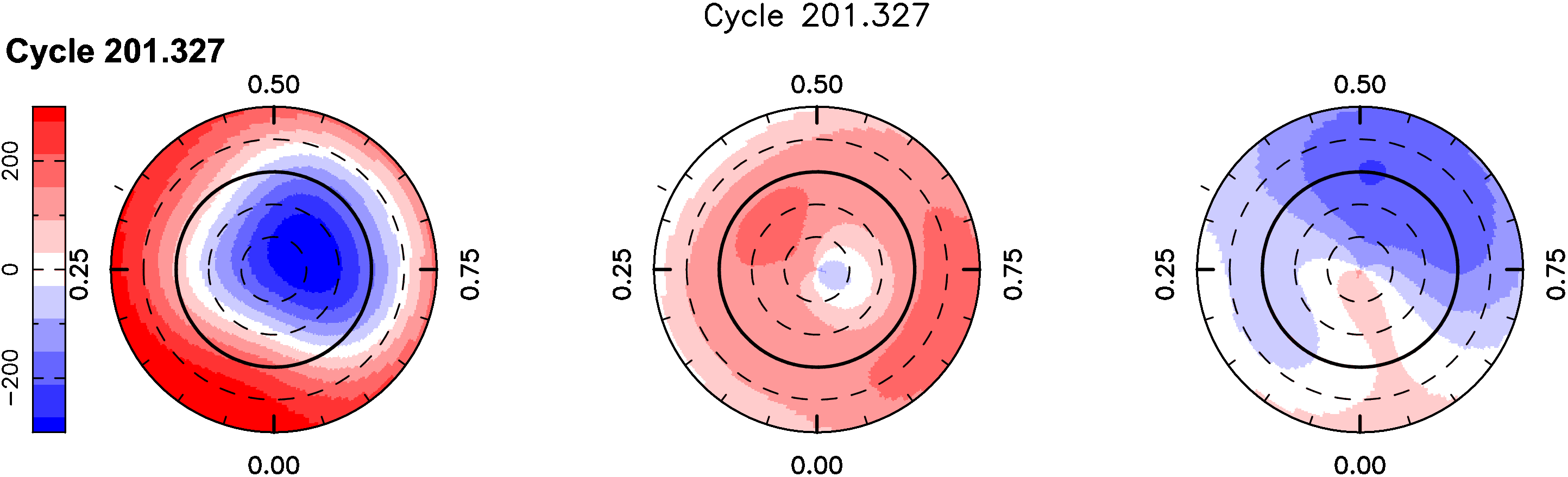}
         
    \end{subfigure}
    \hfill
    %\hspace*{-0.4cm}
    \begin{subfigure}{\textwidth}
         \centering
         \includegraphics[scale=0.15,trim={0cm 0cm 0cm 2.3cm},clip]{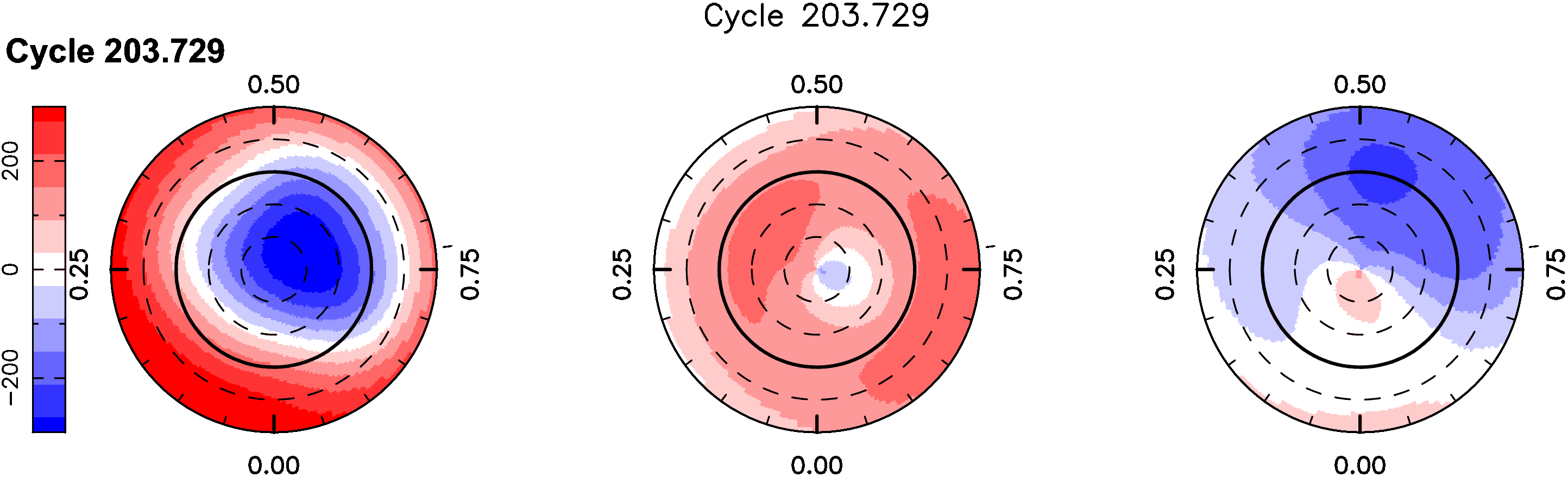}
         
    \end{subfigure}

    \hfill
    %\hspace*{-0.4cm}
    \begin{subfigure}{\textwidth}
         \centering
         \includegraphics[scale=0.15,trim={0cm 0cm 0cm 2.3cm},clip]{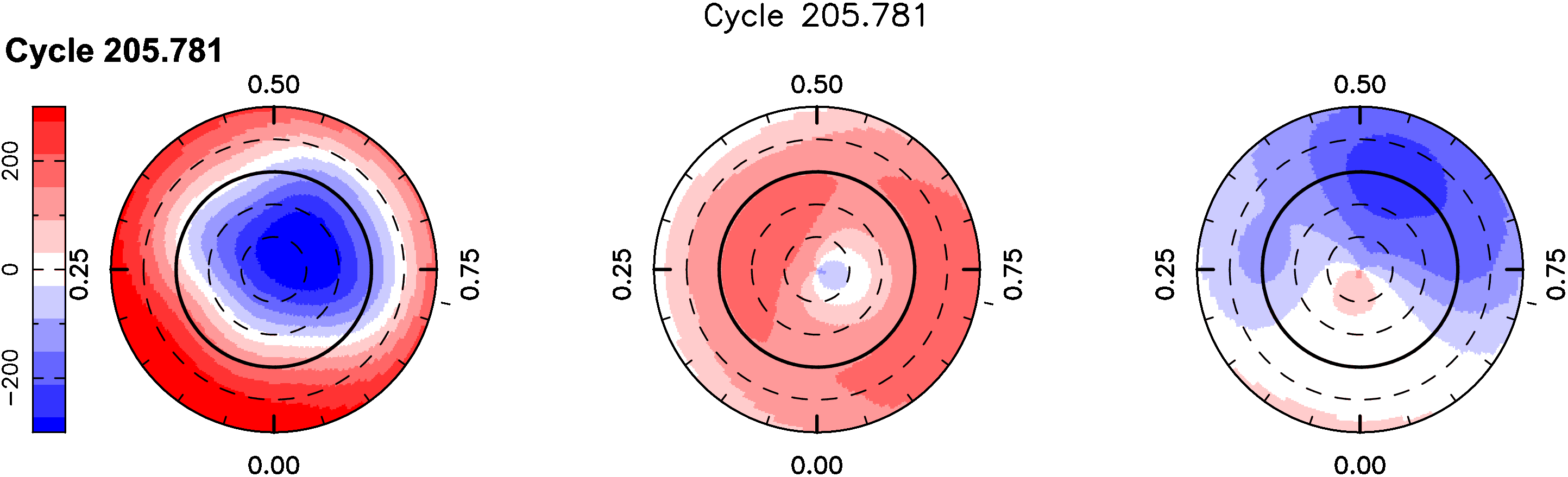}
         
    \end{subfigure}

    \hfill
    %\hspace*{-0.4cm}
    \begin{subfigure}{\textwidth}
         \centering
         \includegraphics[scale=0.15,trim={0cm 0cm 0cm 2.3cm},clip]{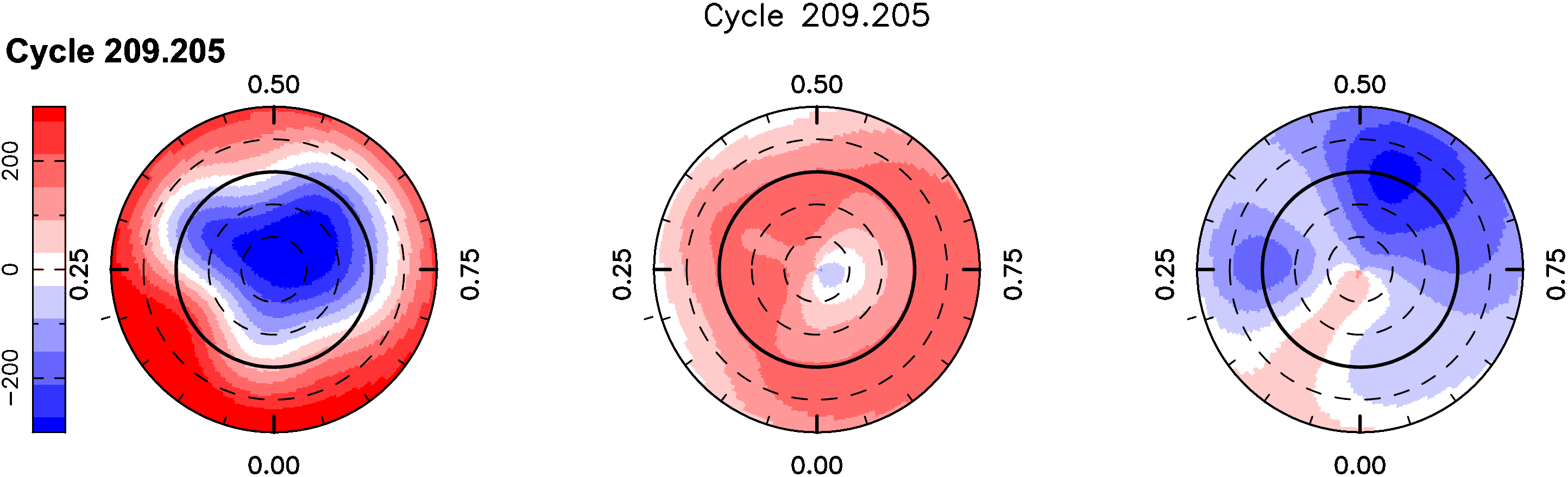}
         
    \end{subfigure}
    
    \caption{Same as Fig.~\ref{fig:map_TIMeS_oct21} for 20212 Jan.}
    \label{fig:map_TIMeS_jan22}
\end{figure*}

\begin{figure*} %\ContinuedFloat
    \centering
    
    \begin{subfigure}{\textwidth}
         \centering
         \includegraphics[scale=0.15,trim={0cm 0cm 0cm 0cm},clip]{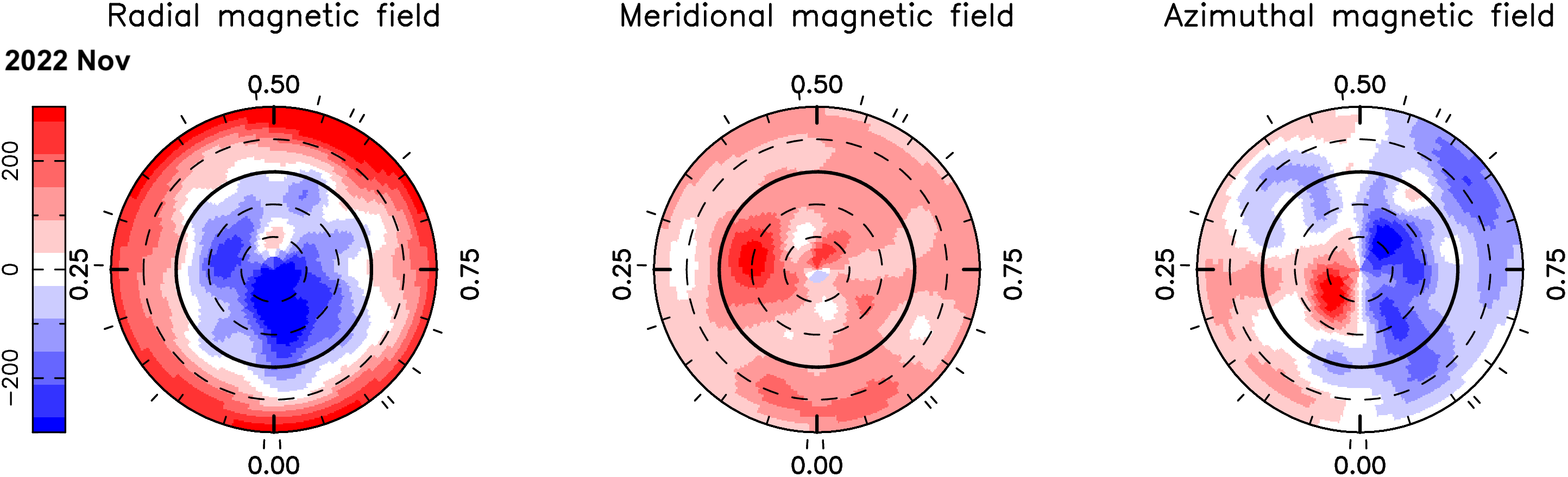}
         
    \end{subfigure}
    \hfill
    %\hspace*{-0.4cm}
    \begin{subfigure}{\textwidth}
         \centering
         \includegraphics[scale=0.15,trim={0cm 0cm 0cm 2.3cm},clip]{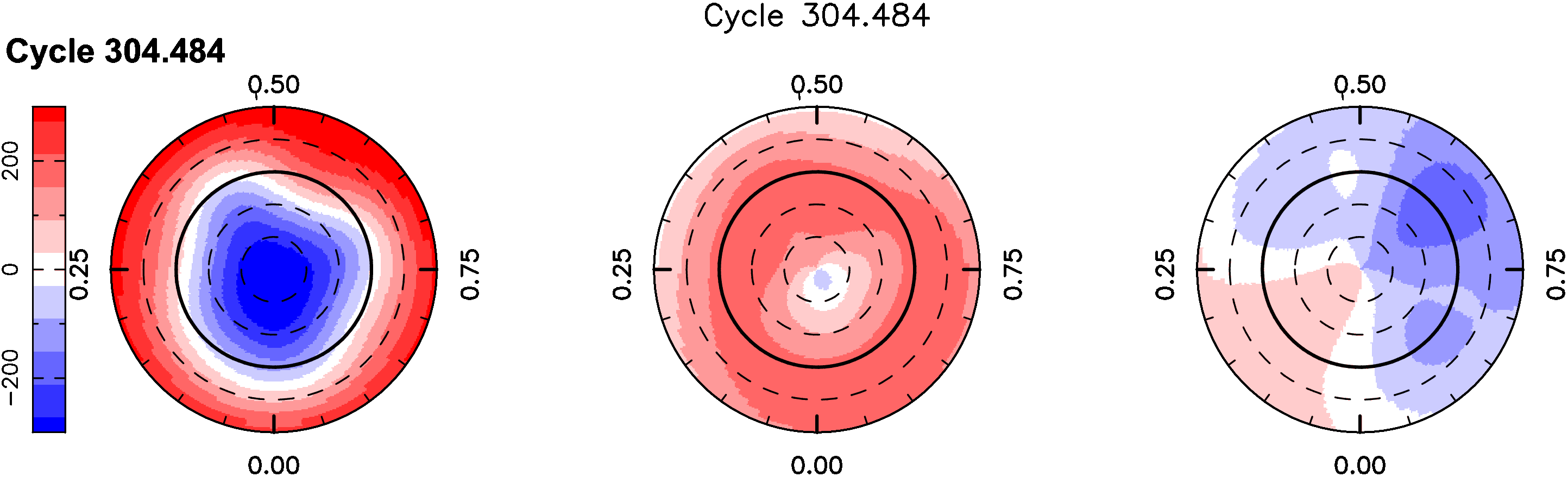}
         
    \end{subfigure}
    \hfill
    %\hspace*{-0.4cm}
    \begin{subfigure}{\textwidth}
         \centering
         \includegraphics[scale=0.15,trim={0cm 0cm 0cm 2.3cm},clip]{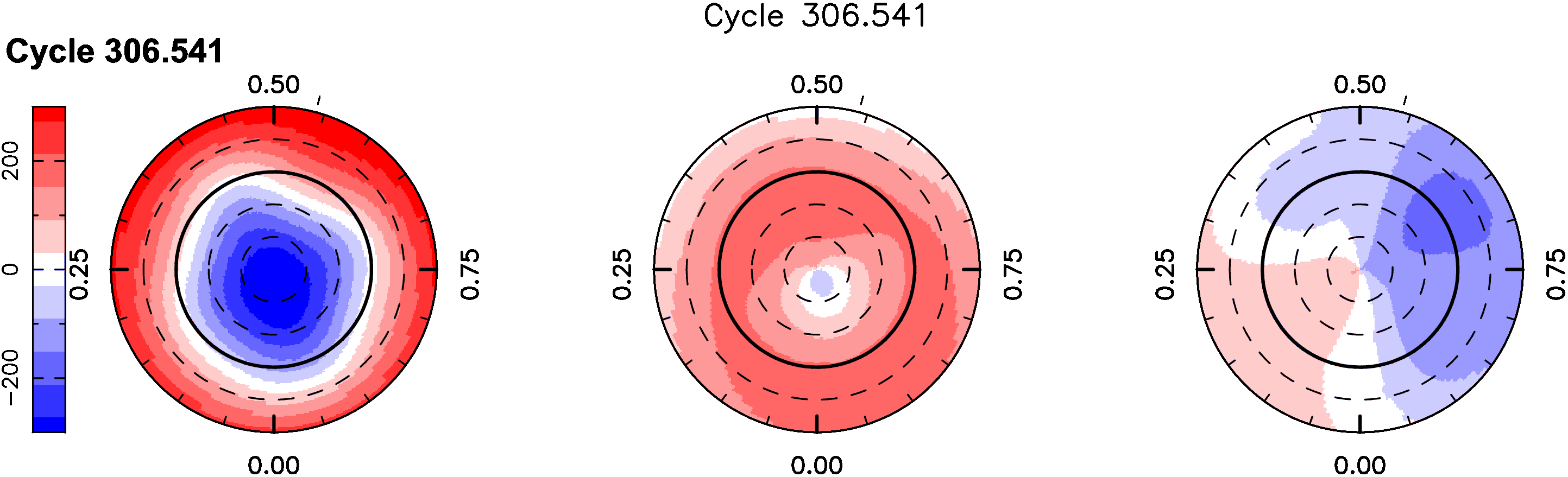}
         
    \end{subfigure}

    \hfill
    %\hspace*{-0.4cm}
    \begin{subfigure}{\textwidth}
         \centering
         \includegraphics[scale=0.15,trim={0cm 0cm 0cm 2.3cm},clip]{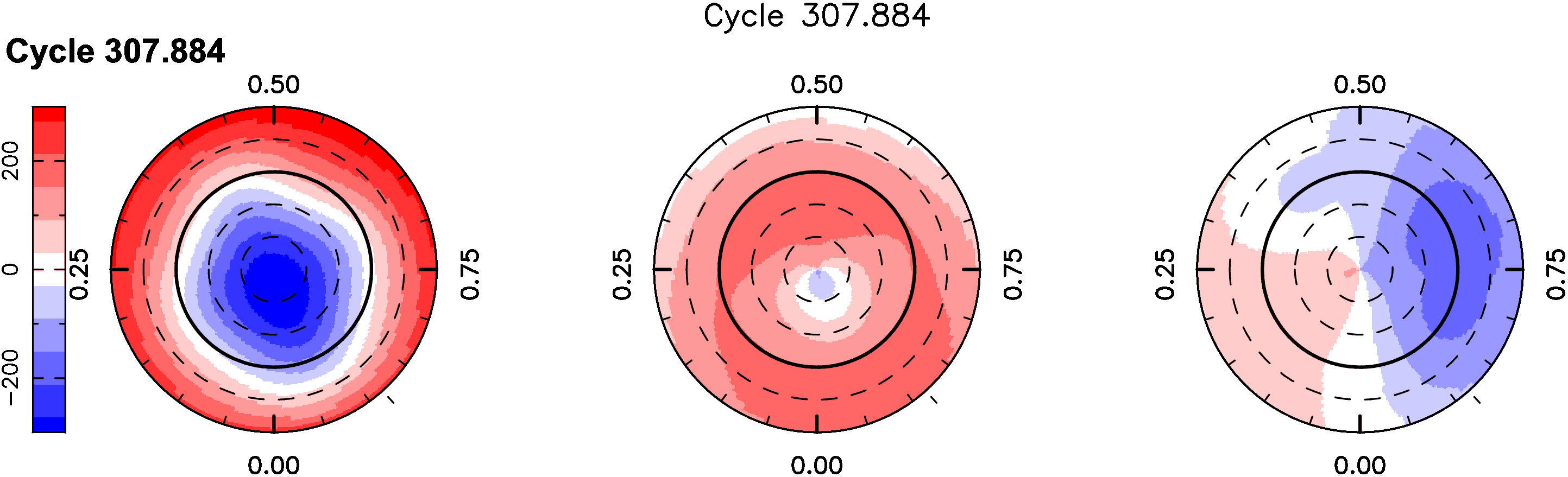}
         
    \end{subfigure}

    \hfill
    %\hspace*{-0.4cm}
    \begin{subfigure}{\textwidth}
         \centering
         \includegraphics[scale=0.15,trim={0cm 0cm 0cm 2.3cm},clip]{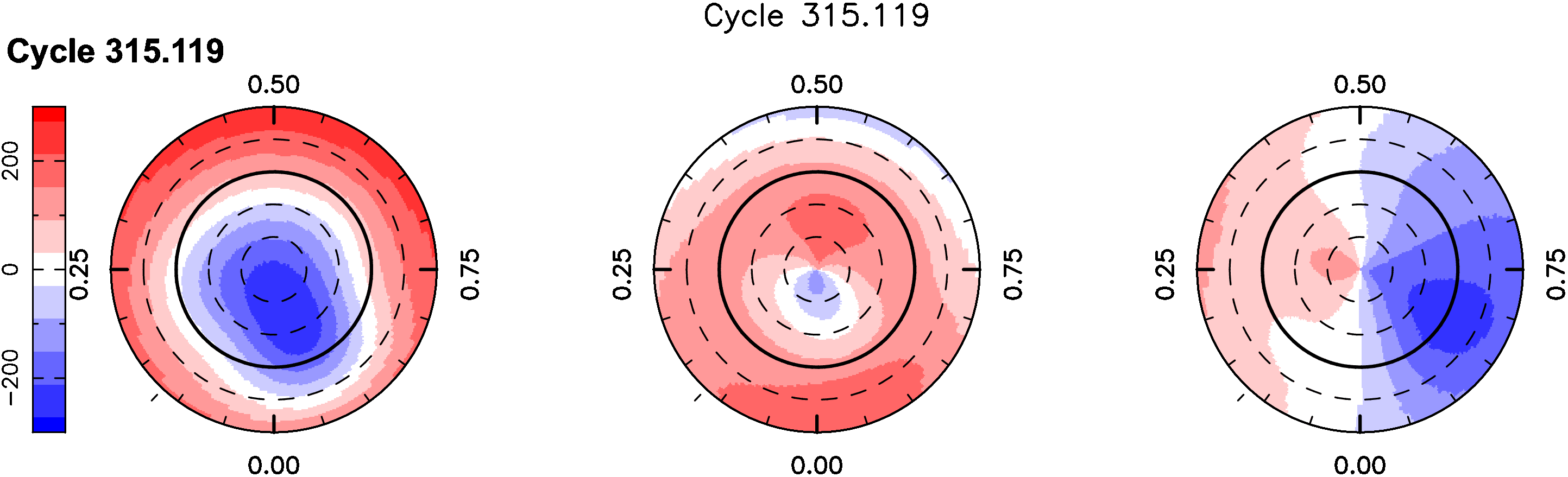}
         
    \end{subfigure}
    
    \caption{Same as Fig.~\ref{fig:map_TIMeS_oct21} for 2022 Nov.}
    \label{fig:map_TIMeS_nov22}
\end{figure*}

\begin{figure*} %\ContinuedFloat
    \centering
    
    \begin{subfigure}{\textwidth}
         \centering
         \includegraphics[scale=0.15,trim={0cm 0cm 0cm 0cm},clip]{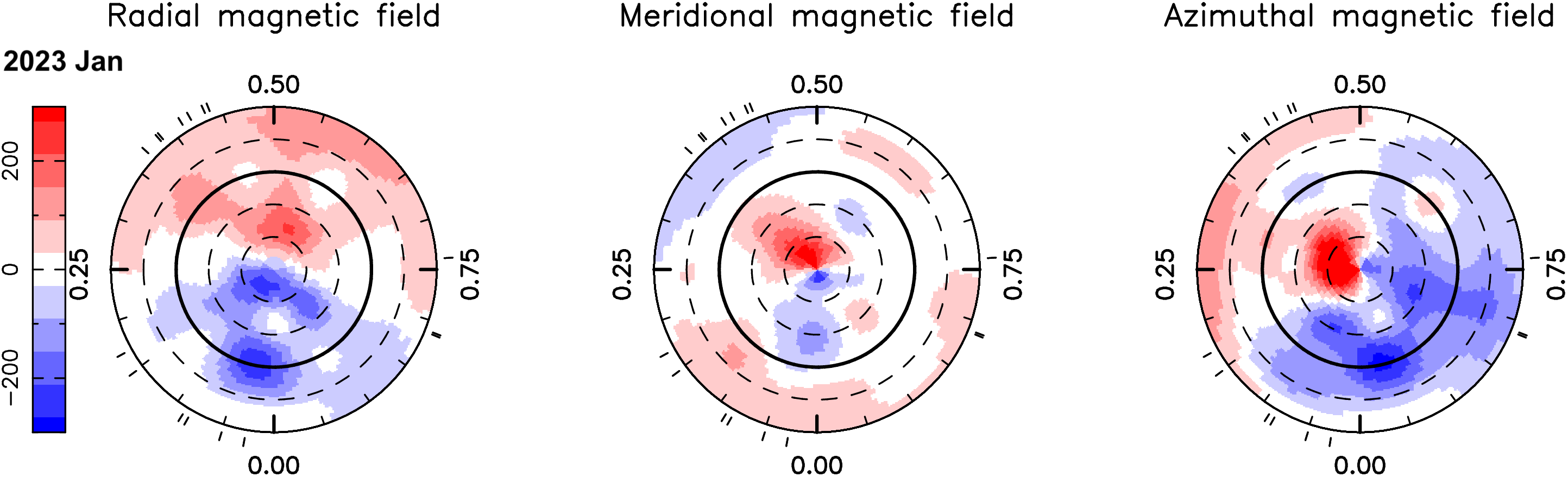}
         
    \end{subfigure}
    \hfill
    %\hspace*{-0.4cm}
    \begin{subfigure}{\textwidth}
         \centering
         \includegraphics[scale=0.15,trim={0cm 0cm 0cm 2.3cm},clip]{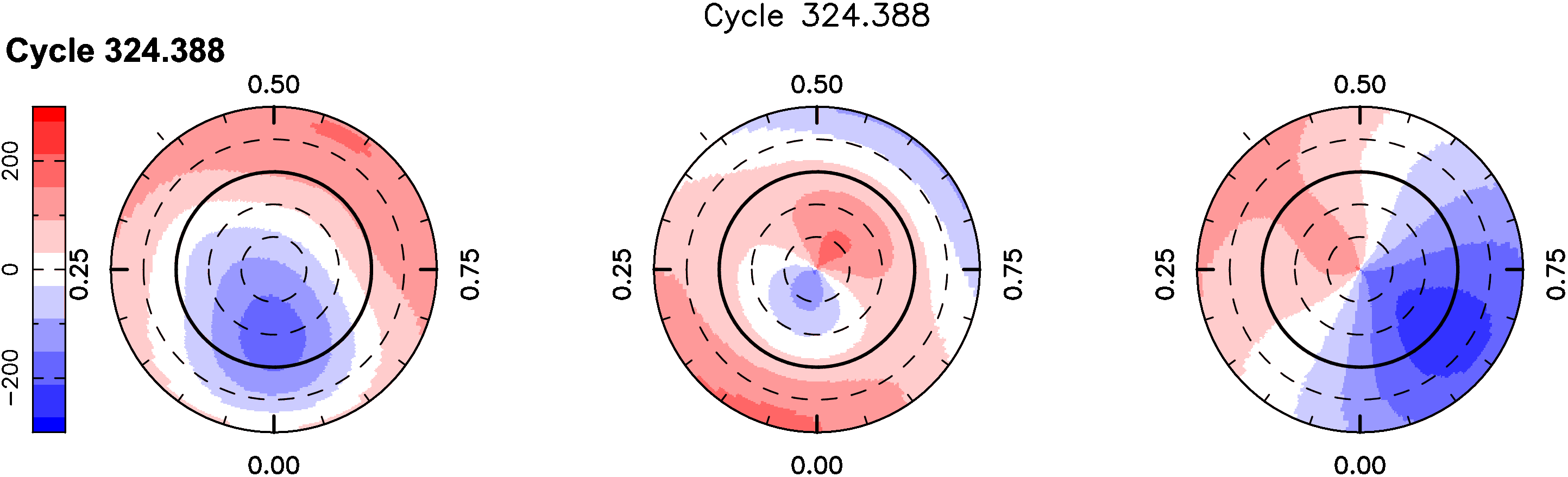}
         
    \end{subfigure}
    \hfill
    %\hspace*{-0.4cm}
    \begin{subfigure}{\textwidth}
         \centering
         \includegraphics[scale=0.15,trim={0cm 0cm 0cm 2.3cm},clip]{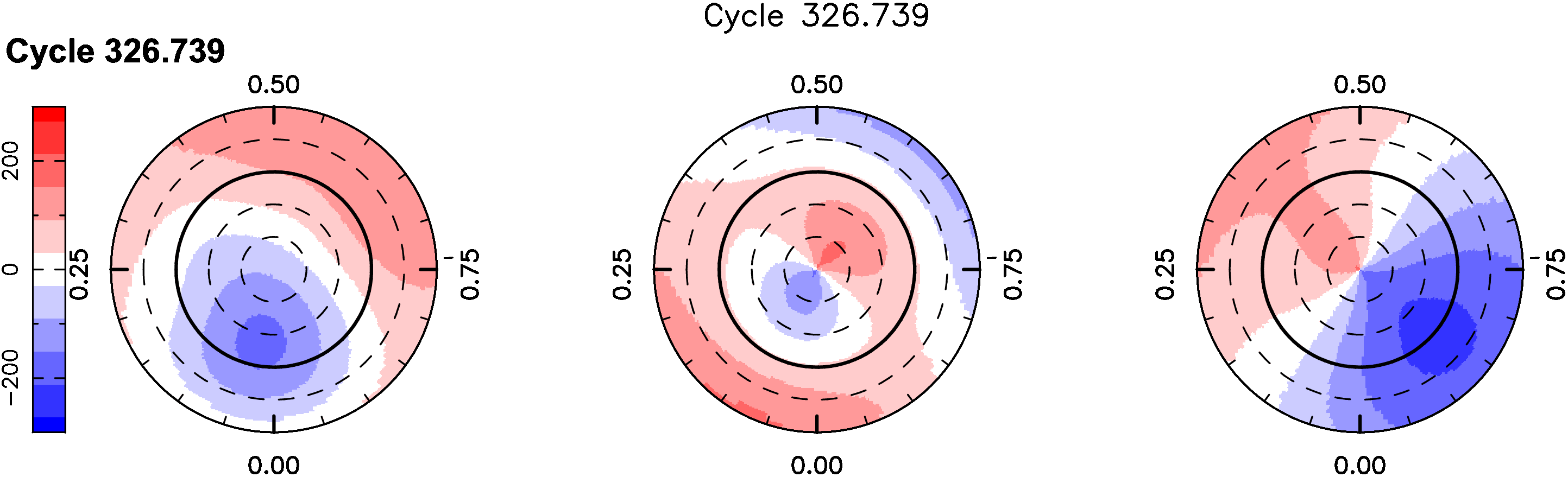}
         
    \end{subfigure}

    \hfill
    %\hspace*{-0.4cm}
    \begin{subfigure}{\textwidth}
         \centering
         \includegraphics[scale=0.15,trim={0cm 0cm 0cm 2.3cm},clip]{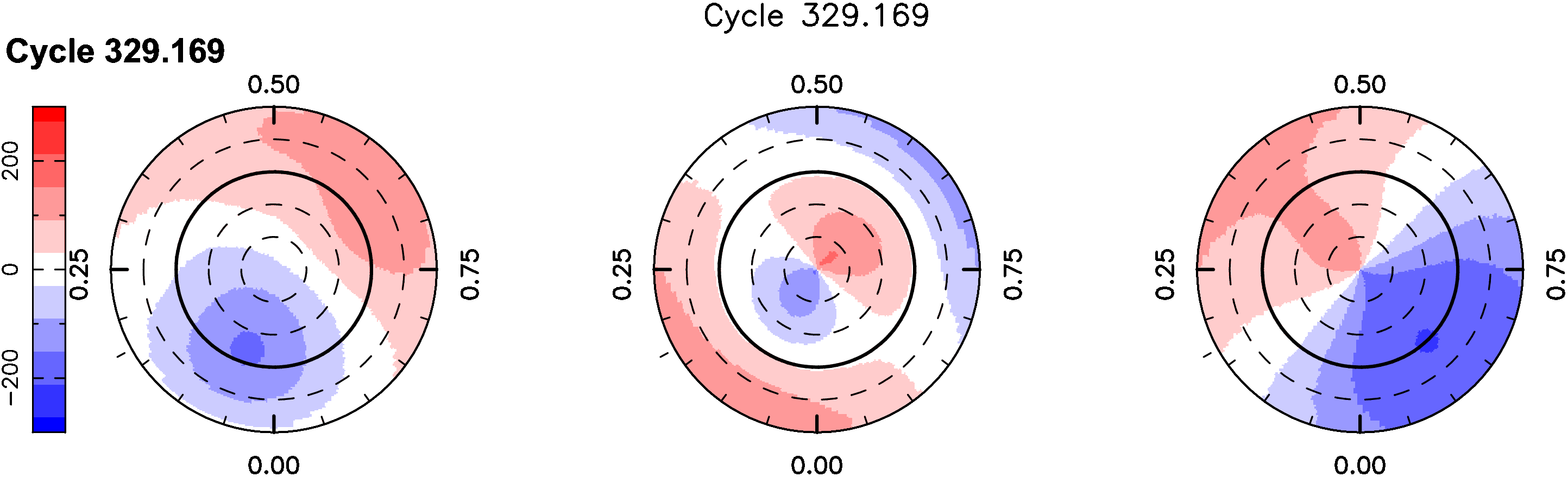}
         
    \end{subfigure}

    \hfill
    %\hspace*{-0.4cm}
    \begin{subfigure}{\textwidth}
         \centering
         \includegraphics[scale=0.15,trim={0cm 0cm 0cm 2.3cm},clip]{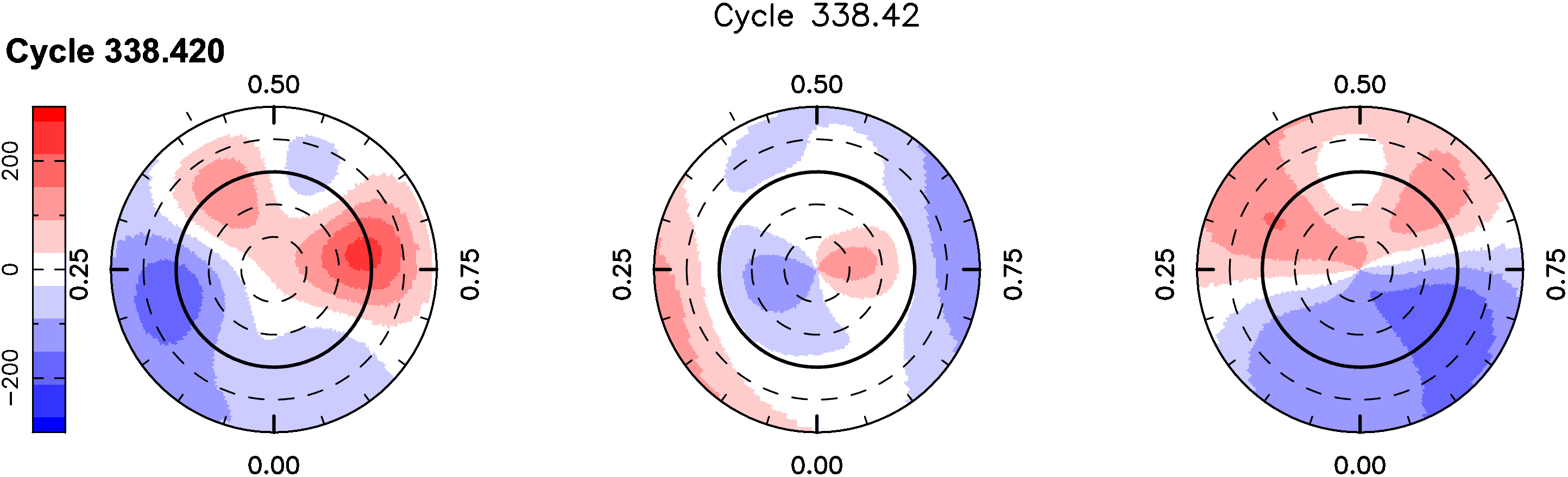}
         
    \end{subfigure}
    
    \caption{Same as Fig.~\ref{fig:map_TIMeS_oct21} for 2023 Jan.}
    \label{fig:map_TIMeS_jan23}
\end{figure*}

\begin{figure*}
    \centering
    \hspace*{-0.5cm}
    \includegraphics[scale=0.2, trim={0cm 0cm 0cm 0cm},clip]{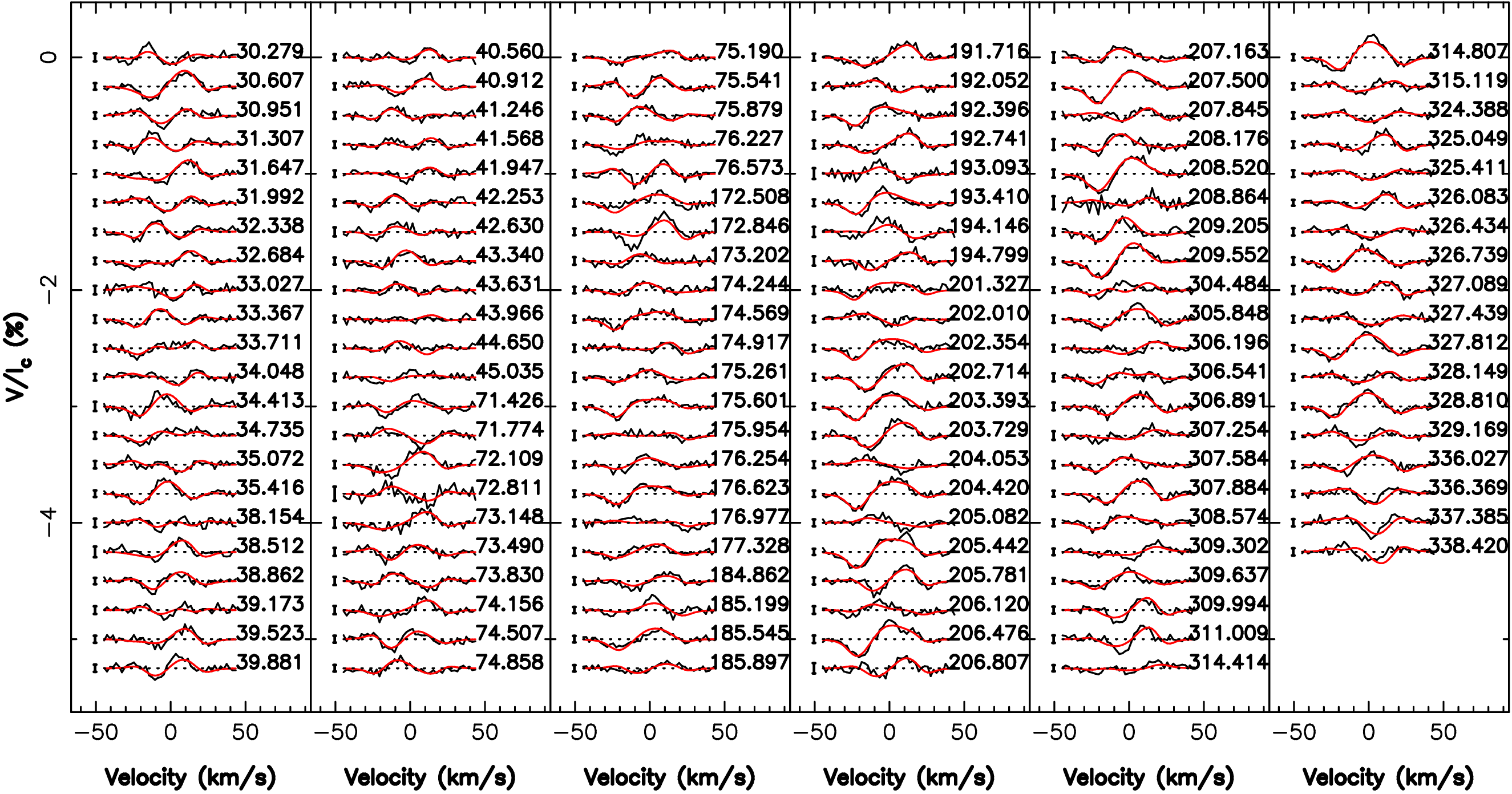}
    \caption{Stokes~$V$ LSD profiles used in the TIMeS reconstruction. The observed data are shown in black while the synthetic model is shown in red. The $\chi^2_r$ between both sets of profiles reaches $1.45$.}
    \label{fig:TIMeS_LSD}
\end{figure*}

\begin{figure*}
    \centering
    \includegraphics[scale=0.3, trim={5cm 8cm 6cm 8cm},clip]{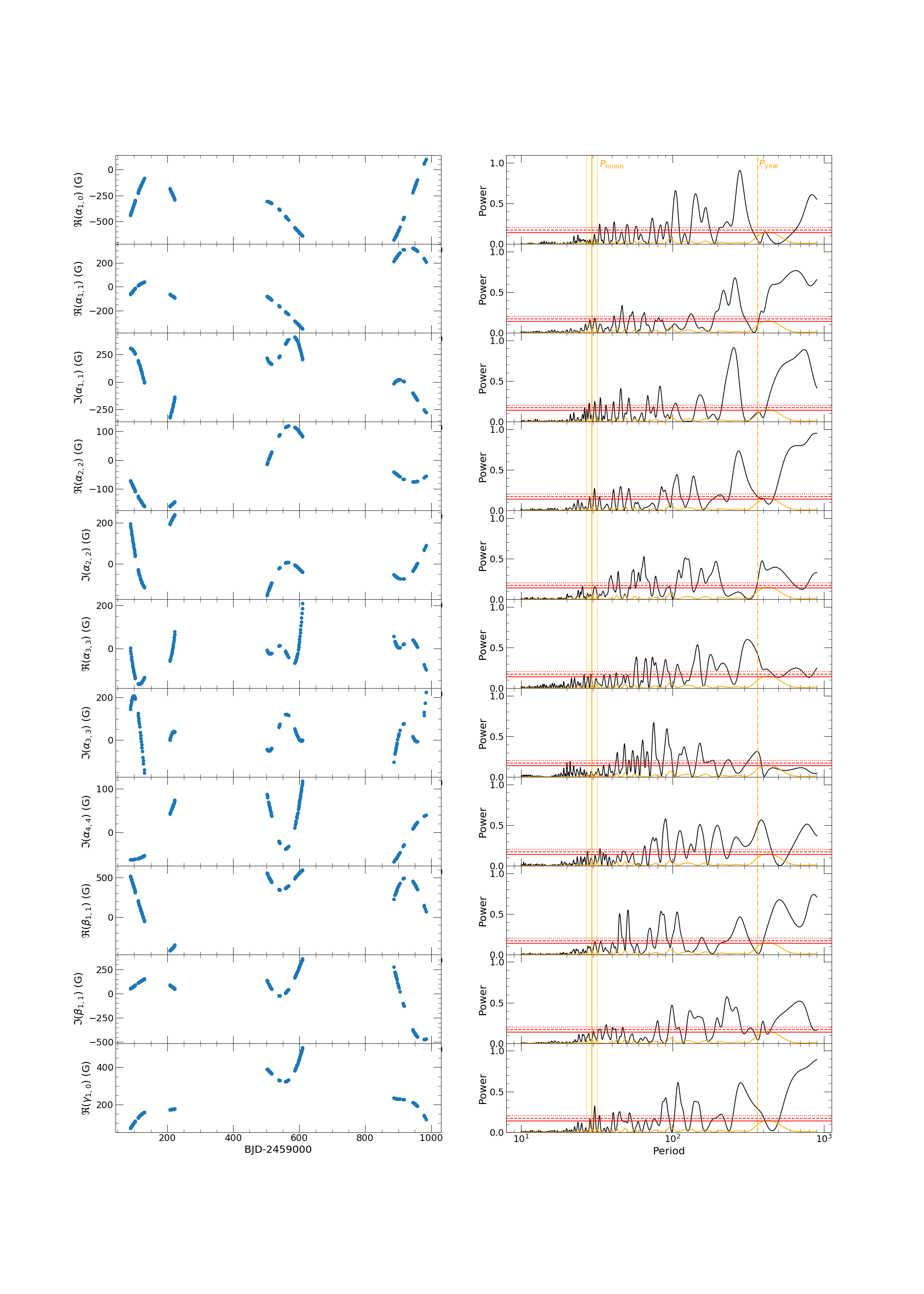}
    \caption{Reconstructed coefficients with TIMeS using Stokes~$V$ LSD profiles between 2020 and 2023. \textit{Left column}: temporal evolution of the 10 coefficients describing the reconstructed magnetic topology, with $\alm$ and $\blm$ characterizing the poloidal component and $\glm$ representing the toroidal component ($\ell$ and $m$ being the degree and order of the associated spherical harmonic mode in the expansion). \textit{Right column}: Periodograms associated with the time series of the reconstructed coefficients, computed using the \textsc{astropy python}. The dotted, dashed and solid red lines show the FAP levels associated with a probability of 1, 0.1 and 0.01\%, respectively. \benjamin{The orange curve depicts the window function, the orange solid and dotted vertical lines correspond to the synodic period of the Moon and its 1-yr aliases, respectively, and the orange dashed vertical line outlines the 1-yr period. }}
    \label{fig:periodogram_TIMeS}
\end{figure*}

We show the magnetic maps reconstructed with TIMeS in Figs.~\ref{fig:map_TIMeS_oct21} to \ref{fig:map_TIMeS_jan23} and the Stokes~$V$ LSD profiles fitted with this method in Fig.~\ref{fig:TIMeS_LSD}.In Fig.~\ref{fig:periodogram_TIMeS}, we show the temporal evolution of each of the coefficients describing the magnetic topology found with the TIMeS method.

\section{MCMC results}
\label{sec:mcmc_results}
\begin{figure*}
    \centering
    \includegraphics[scale=0.3]{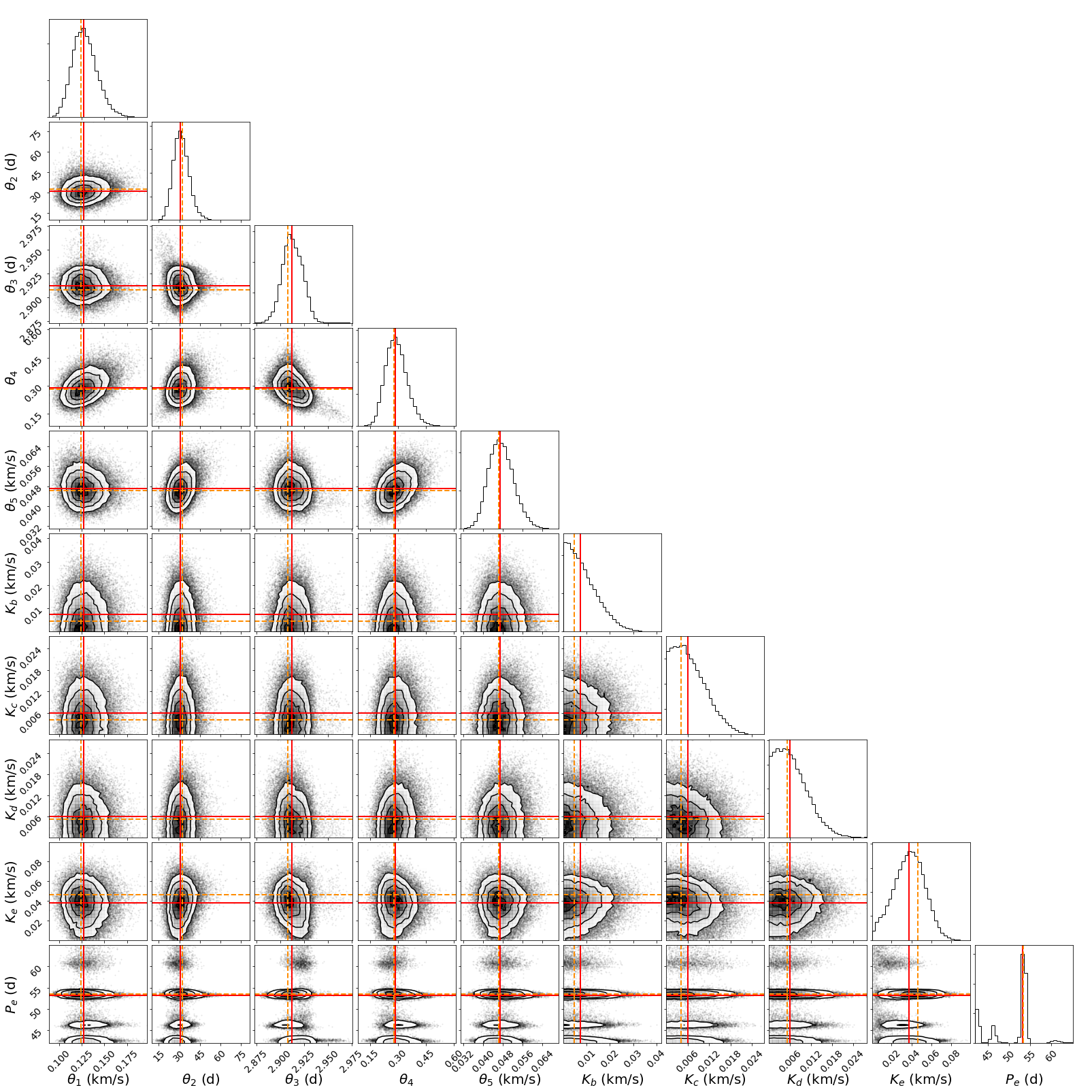}
    \caption{\jf{Corner plot of the posterior distributions of the hyperparameters of quasi-periodic GP and the planet parameters obtained through a MCMC approach assuming a uniform prior between 42 and 65~d for $P_e$. The red lines show the medians of the posterior distributions, while the orange dashed lines correspond to the value that maximizes the log likelihood function, chosen as the optimal parameter. This plot was created using the \textsc{corner python} module \citep{corner}.} }
    \label{fig:corner_plot_rv_uniform}
\end{figure*}

\begin{figure*}
    \centering
    \includegraphics[scale=0.3]{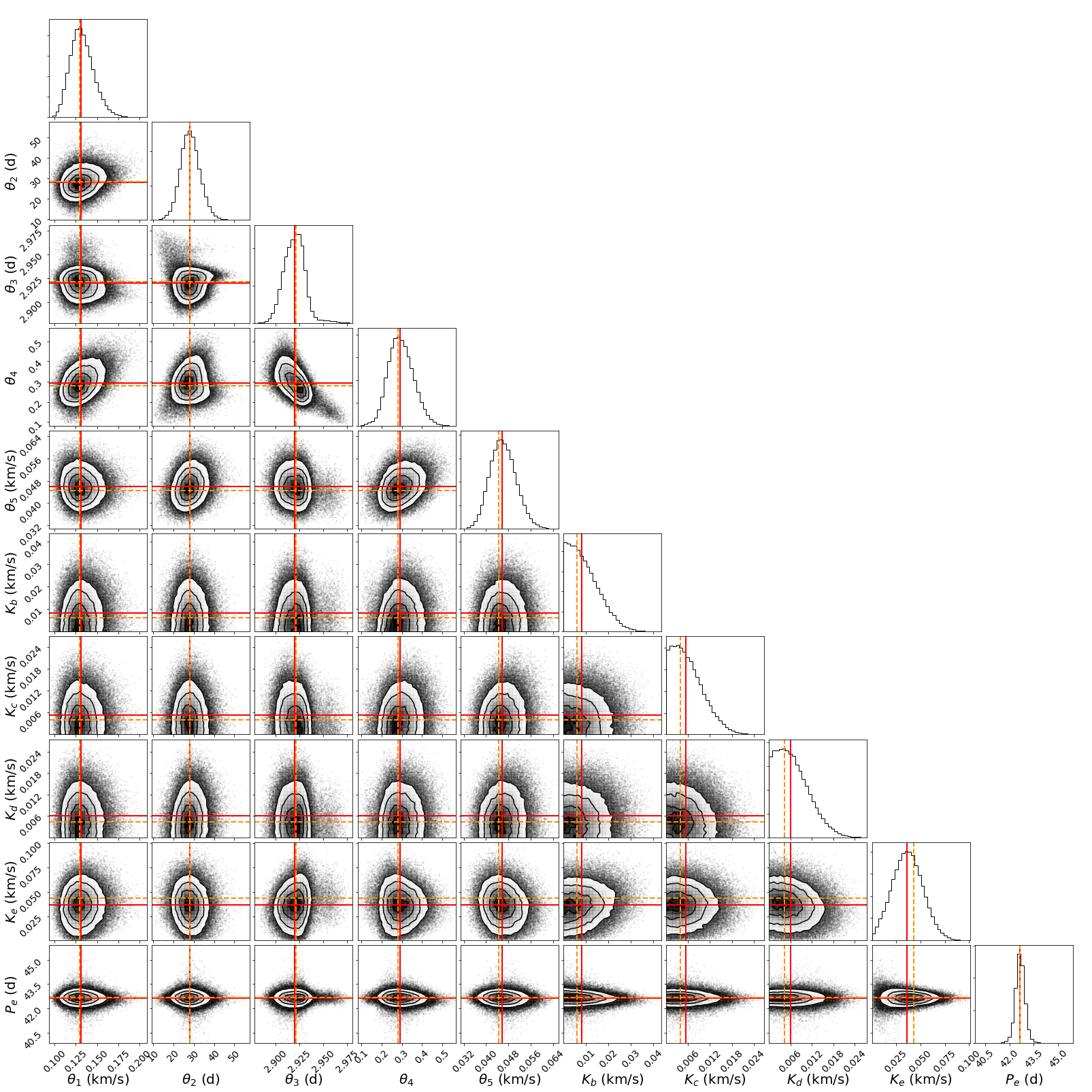}
    \caption{\benjamin{Same as Fig.~\ref{fig:corner_plot_rv_uniform} with a narrow prior centered on 42.7~d for $P_e$.} }
    \label{fig:corner_plot_rv_42.7}
\end{figure*}

\begin{figure*}
    \centering
    \includegraphics[scale=0.3]{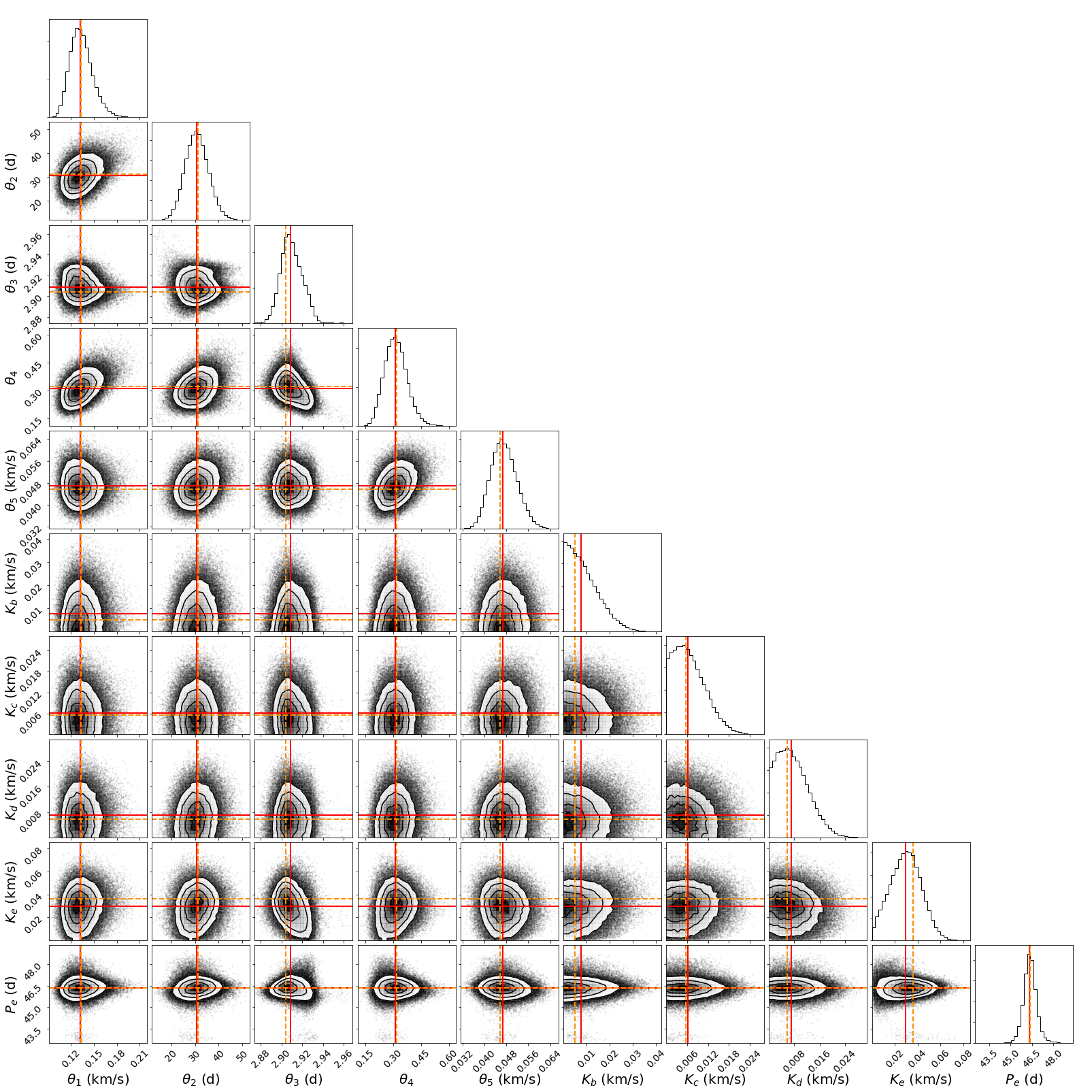}
    \caption{\benjamin{Same as Fig.~\ref{fig:corner_plot_rv_uniform} with a narrow prior centered on 46.4~d for $P_e$.}}
    \label{fig:corner_plot_rv_46.4}
\end{figure*}

\begin{figure*}
    \centering
    \includegraphics[scale=0.3]{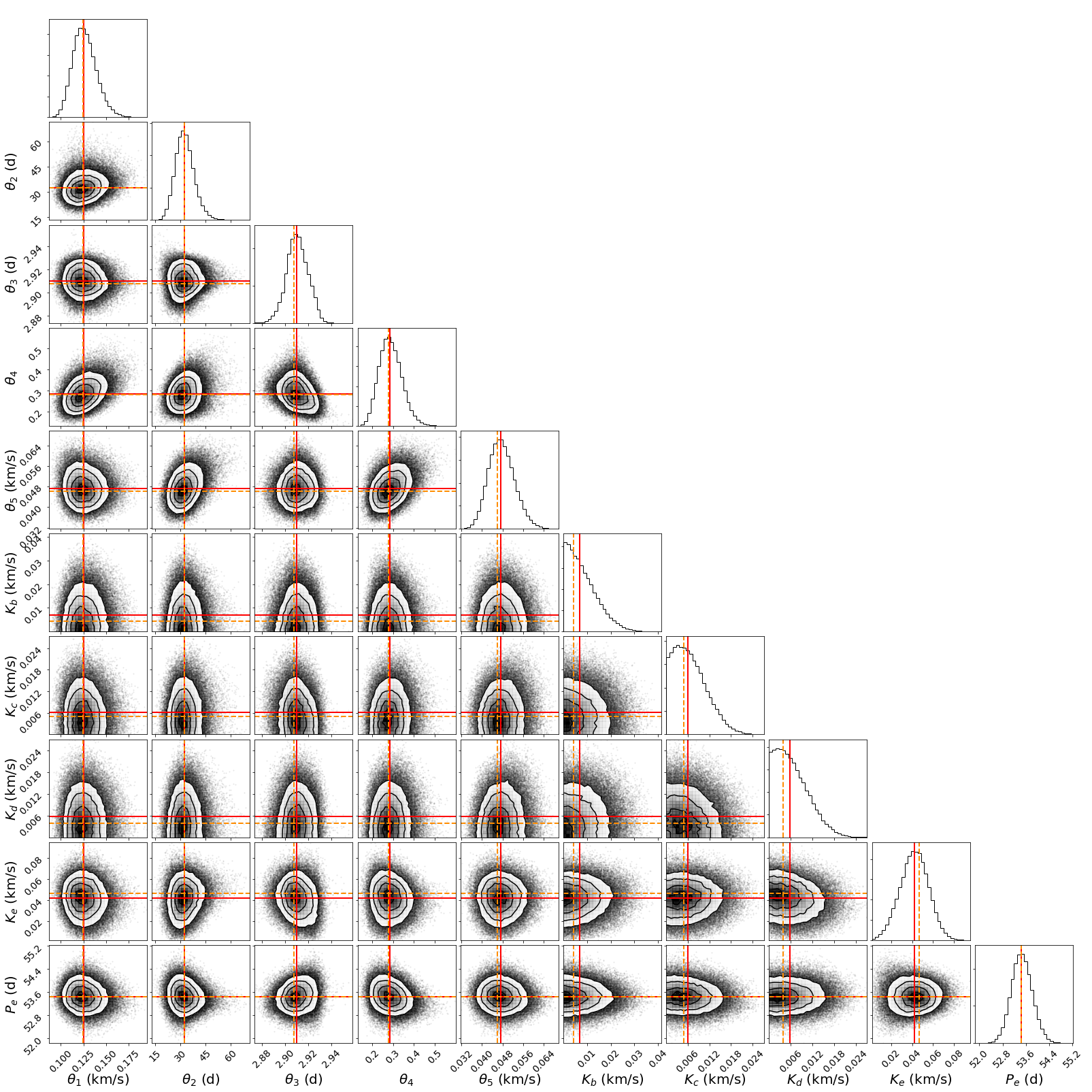}
    \caption{\benjamin{Same as Fig.~\ref{fig:corner_plot_rv_uniform} with a narrow prior centered on 53.5~d for $P_e$.} }
    \label{fig:corner_plot_rv_53.5}
\end{figure*}

\begin{figure*}
    \centering
    \includegraphics[scale=0.3]{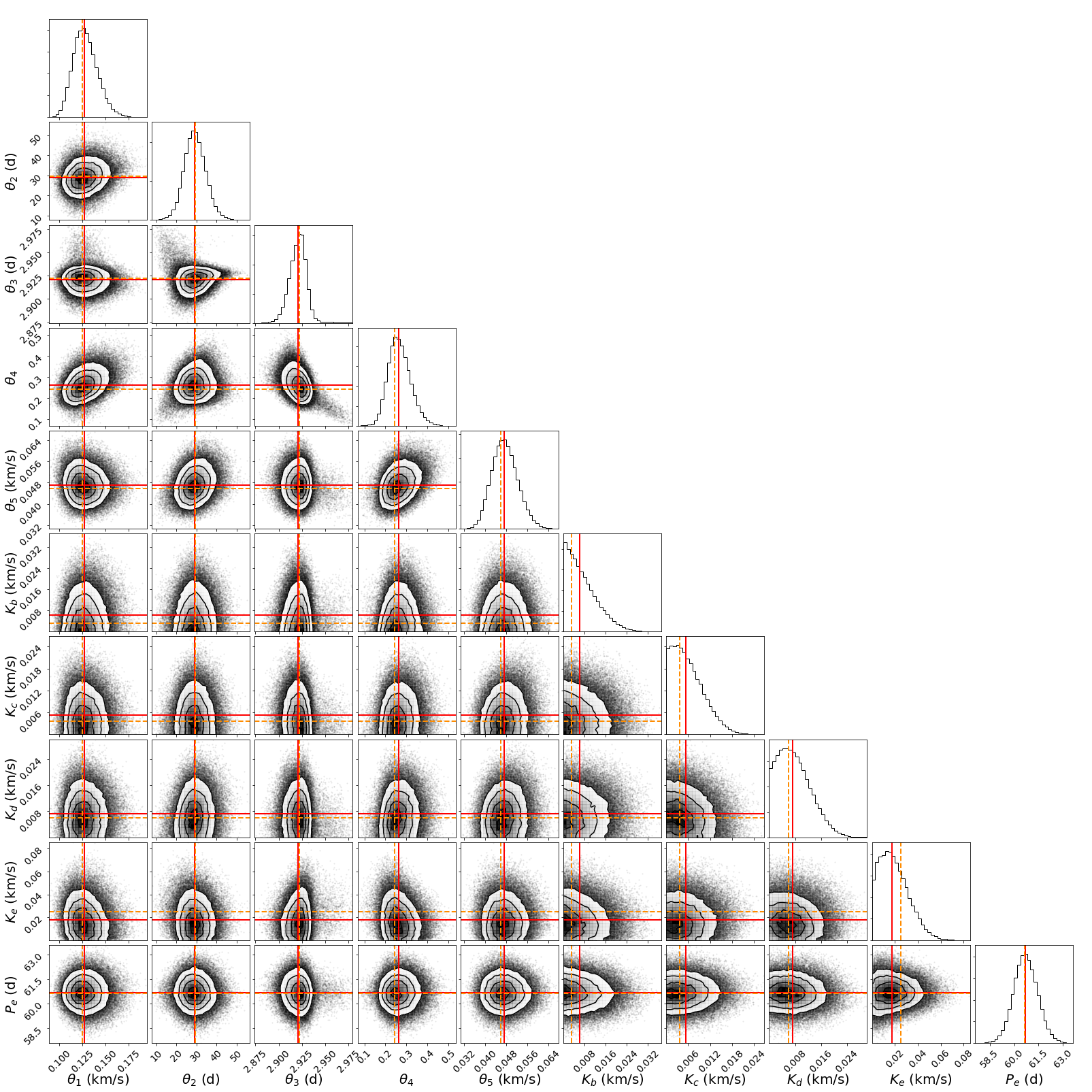}
    \caption{\benjamin{Same as Fig.~\ref{fig:corner_plot_rv_uniform} with a narrow prior centered on 60.6~d for $P_e$.} }
    \label{fig:corner_plot_rv_60.6}
\end{figure*}

\benjamin{In Figs.~\ref{fig:corner_plot_rv_uniform}, \ref{fig:corner_plot_rv_42.7}, \ref{fig:corner_plot_rv_46.4}, \ref{fig:corner_plot_rv_53.5} and \ref{fig:corner_plot_rv_60.6}, we show the corner plots of the posterior distributions when using a uniform prior on the orbital period of planet e, and when using a narrow prior on this parameter, centered at $42.9$~d, $46.1$~d, $53.5$~d or $60.5$~d, respectively, assuming circular orbits for all the four planets of the system.

We also list the less likely results provided by the MCMC approach when using the orbital periods derived by \cite{feinstein22} in Tables~\ref{tab:RV_results_feinstein_less_likely} and \ref{tab:RV_results_feinstein_not_likely}.}

\begin{table*}
	\centering
	\caption{\benjamin{MCMC results for the analysis of the RV data for the 6 planet e orbital periods derived by \citet{feinstein22} identified as plausible from our data. 1$^{\rm st}$ and 2$^{\rm nd}$ columns list the parameters and their associated priors corresponding to a model featuring an activity signal and four planet-related RV signatures. The 6 last columns show the best value for each parameter, in the case of a narrow prior centred on different values of $P_{e_0}$ for the orbital period of planet e. \jf{The knee of the modified Jeffreys priors is set to the typical RV uncertainty, noted $\sigma_{RV}$.} The log BF is computed with respect to the marginal logarithmic likelihood associated with $P_{e_0}=53.0039$~d (see Table~\ref{tab:RV_results_feinstein}).}  }
	\label{tab:RV_results_feinstein_less_likely}
    \resizebox{\textwidth}{!}{
	\begin{tabular}{llccccccc}
    \hline
    Parameter & Prior & $P_{e_0}=61.1583$~d &$P_{e_0}=59.6294$~d  & $P_{e_0}=47.7035$~d &  $P_{e_0}=62.7678$~d &  $P_{e_0}=45.0033$~d &  $P_{e_0}=48.6770$~d   \\
    \hline

    $\theta_1$ (\kms) & mod Jeffreys ($\sigma_{RV}$) & \jf{$0.13\pm0.01$}  & \jf{$0.13\pm0.01$}  & \jf{$0.13\pm0.01$}  & \jf{$0.13\pm0.01$} & \jf{$0.13\pm0.01$} & \jf{$0.13\pm0.01$}   \\
    
    $\theta_2$ (d) & log Gaussian ($\log 30$, $\log 2$) & \jf{$30^{+6}_{-5}$} & \jf{$30^{+6}_{-5}$} & \jf{$29^{+5}_{-4}$} & \jf{$30^{+6}_{-5}$} & \jf{$30^{+6}_{-5}$} & \jf{$30^{+6}_{-5}$}   \\
    
    $\theta_3$ (d) & Gaussian (2.91, 0.1) & \jf{$2.922\pm0.008$} & \jf{$2.924\pm0.008$} & \jf{$2.918\pm0.009$} & \jf{$2.922\pm0.009$} & \jf{$2.923\pm0.009$} & \jf{$2.925\pm0.009$}   \\
    
    $\theta_4$ & Uniform (0, 3) & \jf{$0.25\pm0.05$} & \jf{$0.25\pm0.04$} & \jf{$0.27\pm0.05$} & \jf{$0.26\pm0.05$} & \jf{$0.26\pm0.06$} & \jf{$0.25\pm0.05$} \\
    
    $\theta_5$ (\ms) & mod Jeffreys ($\sigma_{RV}$) & \jf{$46\pm5$} & \jf{$46\pm5$} & \jf{$46\pm5$} & \jf{$46\pm5$} & \jf{$47\pm5$} & \jf{$45\pm5$}  \\ \hline

    $K_b$ (\ms) & mod Jeffreys ($\sigma_{RV}$) & \jf{$2.9 \pm 11.9$} & \jf{$3.8 \pm 11.2$}  & \jf{$4.8 \pm 10.8$} & \jf{$3.5 \pm 11.0$} & \jf{$3.5 \pm 10.6$} & \jf{$4.4 \pm 10.6$}  \\[1mm]
    
    $P_b$ (d) & Fixed from \citet{feinstein22} & 24.1315 & 24.1315 & 24.1315 & 24.1315 & 24.1315 & 24.1315   \\
    
    $T_b$ (2459000+) & Fixed from \citet{feinstein22} & 481.0902 & 481.0902 & 481.0902 & 481.0902 & 481.0902 & 481.0902   \\
    
    $R_b$ (\rjup) & Fixed from \citet{feinstein22} & 0.85 & 0.85 & 0.85 & 0.85 & 0.85 & 0.85   \\
    
    $M_b$ (\mjup) & Derived from $K_b$, $P_b$ and $M_*$ & \jf{$0.05\pm0.20$} & \jf{$0.06\pm0.19$}  & \jf{$0.08\pm0.18$} & \jf{$0.06\pm0.18$} & \jf{$0.06\pm0.18$} & \jf{$0.07\pm0.18$}  \\[1mm]
    $\rho_b$ (\gcm)& Derived from $M_b$ and $R_b$  & \jf{$0.09\pm0.40$}  & \jf{$0.13\pm0.38$} & \jf{$0.16\pm0.36$} & \jf{$0.12\pm0.37$} & \jf{$0.12\pm0.36$} & \jf{$0.15\pm0.36$} \\ \hline

    $K_c$ (\ms) & mod Jeffreys ($\sigma_{RV}$) & \jf{$3.4\pm6.5$} & \jf{$3.2\pm6.3$} & \jf{$4.2\pm6.5$}  & \jf{$3.5\pm6.5$} & \jf{$3.6\pm6.4$} & \jf{$3.3 \pm 6.5$}  \\[1mm]
    
    $P_c$ (d) & Fixed from \citet{feinstein22} & 8.2438 & 8.2438 & 8.2438 & 8.2438 & 8.2438 & 8.2438   \\
    
    $T_c$ (2459000+) & Fixed from \citet{feinstein22} & 481.1664 & 481.1664 & 481.1664 & 481.1664 & 481.1664 & 481.1664   \\
    
    $R_c$ (\rjup)& Fixed from \citet{feinstein22} & 0.45 & 0.45 & 0.45 & 0.45 & 0.45 & 0.45  \\
    
    $M_c$ (\mjup) & Derived from $K_c$, $P_c$ and $M_*$ & \jf{$0.04\pm0.07$} & \jf{$0.04\pm0.07$}  & \jf{$0.05\pm0.07$} & \jf{$0.04\pm0.07$} & \jf{$0.04\pm0.07$} & \jf{$0.04\pm0.07$}  \\[1mm]
    $\rho_c$ (\gcm)& Derived from $M_c$ and $R_c$  & \jf{$0.55\pm1.03$}  & \jf{$0.50\pm1.00$} & \jf{$0.65\pm1.03$} & \jf{$0.56\pm1.03$} & \jf{$0.57\pm1.01$} & \jf{$0.51\pm1.02$} \\ \hline

    $K_d$ (\ms) & mod Jeffreys ($\sigma_{RV}$) & \jf{$6.1\pm6.9$} & \jf{$6.4\pm6.7$} & \jf{$5.8\pm6.8$} & \jf{$5.9\pm6.7$} & \jf{$6.8\pm7.0$} & \jf{$5.6 \pm 6.6$}  \\[1mm]
    
    $P_d$ (d) & Fixed from \citet{feinstein22} &  12.3960 &  12.3960 & 12.3960  & 12.3960 & 12.3960 & 12.3960    \\
    
    $T_d$ (2459000+) & Fixed from \citet{feinstein22} & 478.4149 & 478.4149 & 478.4149 & 478.4149 & 478.4149 & 478.4149   \\
    
    $R_d$ (\rjup)& Fixed from \citet{feinstein22} & 0.55 & 0.55 & 0.55 & 0.55 & 0.55 & 0.55  \\
    
    $M_d$ (\mjup) & Derived from $K_d$, $P_d$ and $M_*$ & \jf{$0.08\pm0.09$} & \jf{$0.08\pm0.09$}  & \jf{$0.07\pm0.09$} & \jf{$0.07\pm0.08$} & \jf{$0.09\pm0.09$} & \jf{$0.07\pm0.08$}  \\[1mm]
    $\rho_d$ (\gcm)& Derived from $M_d$ and $R_d$  & \jf{$0.60\pm0.68$}  & \jf{$0.63\pm0.67$} & \jf{$0.58\pm0.68$} & \jf{$0.58\pm0.67$} & \jf{$0.67\pm0.69$} & \jf{$0.56\pm0.66$} \\ \hline

    $K_e$ (\ms) & mod Jeffreys ($\sigma_{RV}$) & \jf{$24^{+21}_{-11}$} & \jf{$21^{+24}_{-11}$}  & \jf{$17^{+17}_{-8.6}$} & \jf{$15^{+20}_{-8.7}$} & \jf{$14^{+17}_{-7.8}$} & \jf{$16^{+22}_{-9.3}$}  \\[1mm]
    
    $P_e$ (d) & Gaussian ($P_{e_0}$, 0.0001) & $61.1583\pm0.0001$ &$59.6294\pm0.0001$ & $47.7035\pm0.0001$ &  $62.7678\pm0.0001$ &  $45.0033\pm0.0001$ &  $48.6770\pm0.0001$  \\
    
    $T_e$ (2459000+) & Fixed from \citet{feinstein22} & 481.7967 & 481.7967 & 481.7967  & 481.7967 & 481.7967 & 481.7967   \\
    
    $R_e$ (\rjup)& Fixed from \citet{feinstein22} & 0.89 & 0.89 & 0.89 & 0.89 & 0.89 & 0.89  \\
    
    $M_e$ (\mjup) & Derived from $K_e$, $P_e$ and $M_*$ & \jf{$0.55^{+0.47}_{-0.25}$}  & \jf{$0.47^{+0.55}_{-0.25}$} & \jf{$0.36^{+0.36}_{-0.18}$} & \jf{$0.35^{+0.45}_{-0.20}$} & \jf{$0.29^{+0.36}_{-0.16}$} & \jf{$0.34^{+0.45}_{-0.20}$} \\[1mm]
    $\rho_e$ (\gcm)& Derived from $M_e$ and $R_e$ & \jf{$0.96^{+0.85}_{-0.46}$}  & \jf{$0.83^{+0.96}_{-0.45}$} & \jf{$0.63^{+0.64}_{-0.33}$} & \jf{$0.61^{+0.81}_{-0.36}$} & \jf{$0.51^{+0.63}_{-0.29}$} & \jf{$0.60^{+0.80}_{-0.36}$} \\ \hline

    $\chi^2_r$ &  & \jf{5.77} & \jf{5.77} & \jf{5.64} & \jf{5.80} & \jf{5.98} & \jf{5.63}   \\
    
    RMS (\ms) & & \jf{34.4} & \jf{34.5} &  \jf{34.0} & \jf{34.4} & \jf{35.0} & \jf{34.0}   \\ 
    
    $\log \mathcal{L_M}$ & & \jf{$-219.6$} & \jf{$-220.3$} & \jf{$-220.7$}  & \jf{$-220.9$} & \jf{$-221.0$} & \jf{$-221.0$} \\
    
    log BF & & \jf{$-3.4$} & \jf{$-3.4$} & \jf{$-4.5$} & \jf{$-4.7$} & \jf{$-4.8$} & \jf{$-4.8$}  \\ \hline
    %log BF & & -1.3 & -1.7 & 0 \\ \hline
    \end{tabular}
    }
\end{table*}

\begin{table*}
	\centering
	\caption{\benjamin{MCMC results for the analysis of the RV data for the 7 planet e orbital periods derived by \citet{feinstein22} considered as the less likely ones from our data. 1$^{\rm st}$ and 2$^{\rm nd}$ columns list the parameters and their associated priors corresponding to a model featuring an activity signal and four planet-related RV signatures. The 7 last columns show the best value for each parameter, in the case of a narrow prior centred on different values of $P_{e_0}$ for the orbital period of planet e. \jf{The knee of the modified Jeffreys priors is set to the typical RV uncertainty, noted $\sigma_{RV}$.} The log BF is computed with respect to the marginal logarithmic likelihood associated with $P_{e_0}=53.0039$~d (see Table~\ref{tab:RV_results_feinstein}).}  }
	\label{tab:RV_results_feinstein_not_likely}
    \resizebox{\textwidth}{!}{
	\begin{tabular}{llccccccc}
    \hline
    Parameter & Prior & $P_{e_0}=51.8516$~d &  $P_{e_0}=44.1699$~d &$P_{e_0}=49.6911$~d  & $P_{e_0}=55.4692$~d &  $P_{e_0}=50.7484$~d & $P_{e_0}=58.1750$~d & $P_{e_0}=56.7899$~d  \\
    \hline

    $\theta_1$ (\kms) & mod Jeffreys ($\sigma_{RV}$) & \jf{$0.13\pm0.01$} & \jf{$0.13\pm0.01$}  & \jf{$0.13\pm0.01$}  & \jf{$0.13\pm0.01$}  & \jf{$0.13\pm0.01$} & \jf{$0.13\pm0.01$} & \jf{$0.13\pm0.01$} \\
    
    $\theta_2$ (d) & log Gaussian ($\log 30$, $\log 2$) & \jf{$30^{+6}_{-5}$}  & \jf{$29^{+6}_{-5}$} & \jf{$29^{+6}_{-5}$} & \jf{$30^{+6}_{-5}$} & \jf{$29^{+6}_{-5}$} & \jf{$30^{+6}_{-5}$} & \jf{$30^{+6}_{-5}$}  \\
    
    $\theta_3$ (d) & Gaussian (2.91, 0.1) & \jf{$2.922\pm0.009$} & \jf{$2.922\pm0.009$} & \jf{$2.923\pm0.009$} & \jf{$2.924\pm0.008$} & \jf{$2.923\pm0.009$} & \jf{$2.924\pm0.008$} & \jf{$2.924\pm0.008$}  \\
    
    $\theta_4$ & Uniform (0, 3) & \jf{$0.26\pm0.06$} & \jf{$0.26\pm0.05$} & \jf{$0.25\pm0.05$} & \jf{$0.25\pm0.05$} &  \jf{$0.25\pm0.05$} &  \jf{$0.25\pm0.05$} & \jf{$0.25\pm0.05$} \\
    
    $\theta_5$ (\ms) & mod Jeffreys ($\sigma_{RV}$) & \jf{$46\pm5$} & \jf{$46\pm5$} & \jf{$46\pm5$} & \jf{$46\pm5$} & \jf{$46\pm5$} & \jf{$46\pm5$} & \jf{$46\pm5$}  \\ \hline

    $K_b$ (\ms) & mod Jeffreys ($\sigma_{RV}$) & \jf{$3.6 \pm 10.9$} & \jf{$3.5 \pm 11.3$} & \jf{$3.8 \pm 10.2$}  & \jf{$3.7 \pm 10.9$} & \jf{$3.6 \pm 11.4$} & \jf{$3.7 \pm 11.3$} & \jf{$3.7 \pm 11.1$}  \\[1mm]
    
    $P_b$ (d) & Fixed from \citet{feinstein22} & 24.1315 & 24.1315 & 24.1315 & 24.1315 & 24.1315 & 24.1315 & 24.1315 \\
    
    $T_b$ (2459000+) & Fixed from \citet{feinstein22} & 481.0902 & 481.0902 & 481.0902 & 481.0902 & 481.0902 & 481.0902 & 481.0902 \\
    
    $R_b$ (\rjup) & Fixed from \citet{feinstein22} & 0.85 & 0.85 & 0.85 & 0.85 & 0.85 & 0.85 & 0.85 \\
    
    $M_b$ (\mjup) & Derived from $K_b$, $P_b$ and $M_*$ & \jf{$0.06\pm0.18$} & \jf{$0.05\pm0.19$}  & \jf{$0.06\pm0.17$} & \jf{$0.06\pm0.18$} & \jf{$0.06\pm0.19$} & \jf{$0.06\pm0.19$} & \jf{$0.06\pm0.19$}  \\[1mm]
    $\rho_b$ (\gcm)& Derived from $M_b$ and $R_b$  & \jf{$0.12\pm0.37$}  & \jf{$0.12\pm0.38$} & \jf{$0.13\pm0.35$} & \jf{$0.13\pm0.37$} & \jf{$0.12\pm0.39$} & \jf{$0.13\pm0.38$} & \jf{$0.13\pm0.37$} \\ \hline

    $K_c$ (\ms) & mod Jeffreys ($\sigma_{RV}$) & \jf{$4.0 \pm 6.5$} & \jf{$3.4\pm6.4$} & \jf{$3.1\pm6.2$} & \jf{$3.5\pm6.5$}  & \jf{$3.7\pm6.5$} & \jf{$3.4\pm6.1$} & \jf{$3.6\pm6.2$}  \\[1mm]
    
    $P_c$ (d) & Fixed from \citet{feinstein22} & 8.2438 & 8.2438 & 8.2438 & 8.2438 & 8.2438 & 8.2438 & 8.2438 \\
    
    $T_c$ (2459000+) & Fixed from \citet{feinstein22} & 481.1664 & 481.1664 & 481.1664 & 481.1664 & 481.1664 & 481.1664 & 481.1664 \\
    
    $R_c$ (\rjup)& Fixed from \citet{feinstein22} & 0.45 & 0.45 & 0.45 & 0.45 & 0.45 & 0.45 & 0.45 \\
    
    $M_c$ (\mjup) & Derived from $K_c$, $P_c$ and $M_*$ & \jf{$0.04\pm0.07$} & \jf{$0.04\pm0.07$}  & \jf{$0.03\pm0.07$} & \jf{$0.04\pm0.07$} & \jf{$0.04\pm0.07$} & \jf{$0.04\pm0.07$} & \jf{$0.04\pm0.07$}  \\[1mm]
    $\rho_c$ (\gcm)& Derived from $M_c$ and $R_c$  & \jf{$0.63\pm1.03$}  & \jf{$0.54\pm1.02$} & \jf{$0.49\pm0.98$} & \jf{$0.56\pm1.02$} & \jf{$0.59\pm1.03$} & \jf{$0.55\pm0.97$} & \jf{$0.56\pm0.98$} \\ \hline

    $K_d$ (\ms) & mod Jeffreys ($\sigma_{RV}$) & \jf{$6.0 \pm 6.8$} & \jf{$7.0\pm6.4$} & \jf{$6.3\pm6.7$} & \jf{$6.1\pm6.7$} & \jf{$6.3\pm6.7$} & \jf{$6.5\pm6.4$} & \jf{$5.9\pm6.8$}  \\[1mm]
    
    $P_d$ (d) & Fixed from \citet{feinstein22} &  12.3960 &  12.3960 & 12.3960 & 12.3960 & 12.3960  & 12.3960 & 12.3960 \\
    
    $T_d$ (2459000+) & Fixed from \citet{feinstein22} & 478.4149 & 478.4149 & 478.4149 & 478.4149 & 478.4149 & 478.4149 & 478.4149 \\
    
    $R_d$ (\rjup)& Fixed from \citet{feinstein22} & 0.55 & 0.55 & 0.55 & 0.55 & 0.55 & 0.55 & 0.55 \\
    
    $M_d$ (\mjup) & Derived from $K_d$, $P_d$ and $M_*$ & \jf{$0.08\pm0.09$} & \jf{$0.09\pm0.08$}  & \jf{$0.08\pm0.08$} & \jf{$0.08\pm0.08$} & \jf{$0.08\pm0.08$} & \jf{$0.08\pm0.08$} & \jf{$0.08\pm0.09$}  \\[1mm]
    $\rho_d$ (\gcm)& Derived from $M_d$ and $R_d$  & \jf{$0.60\pm0.68$}  & \jf{$0.69\pm0.65$} & \jf{$0.63\pm0.66$} & \jf{$0.60\pm0.67$} & \jf{$0.63\pm0.67$} & \jf{$0.64\pm0.64$} & \jf{$0.59\pm0.68$} \\ \hline

    $K_e$ (\ms) & mod Jeffreys ($\sigma_{RV}$) & \jf{$12^{+17}_{-7.2}$} & \jf{$9.5^{+14.4}_{-5.7}$} & \jf{$8.6^{+13.7}_{-5.3}$}  & \jf{$8.6^{+15.0}_{-5.5}$} & \jf{$6.5^{+11.0}_{-4.1}$} & \jf{$6.8^{+11.8}_{-4.3}$} & \jf{$4.0^{+7.4}_{-2.6}$}  \\[1mm]
    
    $P_e$ (d) & Gaussian ($P_{e_0}$, 0.0001) & $51.8516\pm0.0001$ & $44.1699\pm0.0001$ & $49.6911\pm0.0001$ & $55.4692\pm0.0001$  & $50.7484\pm0.0001$ & $58.1750\pm0.0001$ & $56.7899\pm0.0001$  \\
    
    $T_e$ (2459000+) & Fixed from \citet{feinstein22} & 481.7967 & 481.7967 & 481.7967  & 481.7967 & 481.7967  & 481.7967 & 481.7967  \\
    
    $R_e$ (\rjup)& Fixed from \citet{feinstein22} & 0.89 & 0.89 & 0.89 & 0.89 & 0.89 & 0.89 & 0.89\\
    
    $M_e$ (\mjup) & Derived from $K_e$, $P_e$ and $M_*$ & \jf{$0.26^{+0.38}_{-0.16}$}  & \jf{$0.19^{+0.29}_{-0.12}$} & \jf{$0.18^{+0.28}_{-0.12}$} & \jf{$0.19^{+0.33}_{-0.12}$} & \jf{$0.14^{+0.23}_{-0.08}$} & \jf{$0.15^{+0.26}_{-0.09}$} & \jf{$0.08^{+0.17}_{-0.05}$} \\[1mm]
    $\rho_e$ (\gcm)& Derived from $M_e$ and $R_e$ & \jf{$0.46^{+0.66}_{-0.27}$}  & \jf{$0.34^{+0.51}_{-0.21}$} & \jf{$0.32^{+0.50}_{-0.20}$} & \jf{$0.34^{+0.58}_{-0.22}$} & \jf{$0.24^{+0.41}_{-0.16}$} & \jf{$0.26^{+0.46}_{-0.17}$} & \jf{$0.16^{+0.28}_{-0.11}$} \\ \hline

    $\chi^2_r$ & & \jf{5.81} & \jf{5.73} & \jf{5.65} & \jf{5.73} & \jf{5.70} & \jf{5.73} & \jf{5.71}  \\
    
    RMS (\ms) & & \jf{34.5} & \jf{34.3} & \jf{34.0} &  \jf{34.3} & \jf{34.2} & \jf{34.3}  & \jf{34.2}    \\ 
    
    $\log \mathcal{L_M}$ & & \jf{$-221.3$} & \jf{$-221.6$} & \jf{$-221.6$} & \jf{$-221.8$}  & \jf{$-222.1$} & \jf{$-222.1$} & \jf{$-222.6$} \\
    
    log BF & & \jf{$-5.1$} & \jf{$-5.4$} & \jf{$-5.4$} & \jf{$-5.6$} & \jf{$-5.9$} & \jf{$-5.9$} & \jf{$-6.4$}  \\ \hline
    %log BF & & -1.3 & -1.7 & 0 \\ \hline
    \end{tabular}
    }
\end{table*}

% Don't change these lines
\bsp	% typesetting comment
\label{lastpage}
\end{document}